\documentclass[prd,aps,floatfix]{revtex4}

\usepackage{graphicx}
\usepackage{amsmath}
\usepackage{amssymb}
\usepackage{longtable}

\newcommand{\be}{\begin{equation}}
\newcommand{\ee}{\end{equation}}
\newcommand{\bea}{\begin{eqnarray}}
\newcommand{\eea}{\end{eqnarray}}
\newcommand{\bi}{\begin{itemize}}
\newcommand{\ei}{\end{itemize}}
\newcommand{\ben}{\begin{enumerate}}
\newcommand{\een}{\end{enumerate}}
\newcommand{\bt}{\begin{tabbing}}
\newcommand{\et}{\end{tabbing}}
\newcommand{\nn}{\nonumber}

\newcommand{\calO}{{\mathcal O}}

\begin{document}

%
%

\title{
Light hadron spectroscopy with two flavors of 
$O(a)$-improved dynamical quarks
}

\author{
   JLQCD Collaboration:
   S.~Aoki$^1$, 
   R.~Burkhalter$^{1,2}$\thanks{present address : KPMG Consulting AG, Badenerstrasse 172, 8804 Zurich, Switzerland}, 
   M.~Fukugita$^{3}$, 
   S.~Hashimoto$^{4}$, 
   K-I.~Ishikawa$^{2}$, 
   N.~Ishizuka$^{1,2}$, 
   Y.~Iwasaki$^{1,2}$, 
   K.~Kanaya$^{1,2}$, 
   T.~Kaneko$^{4}$, 
   Y.~Kuramashi$^{4}$, 
   M.~Okawa$^5$, 
   T.~Onogi$^6$, 
   N.~Tsutsui$^{4}$, 
   A.~Ukawa$^{1,2}$, 
   N.~Yamada$^{4}$, 
   and T.~Yoshi\'e$^{1,2}$
}

\affiliation{
$^1$Institute of Physics, University of Tsukuba, Tsukuba, Ibaraki
    305-8571, Japan \\
$^2$Center for Computational Physics, University of Tsukuba, 
    Tsukuba, Ibaraki 305-8577, Japan \\
$^3$Institute for Cosmic Ray Research, University of Tokyo, 
    Kashiwa 277-8582, Japan \\
$^4$High Energy Accelerator Research Organization
(KEK), Tsukuba, Ibaraki 305-0801, Japan \\
$^5$Department of Physics, Hiroshima University, Higashi-Hiroshima, 
    Hiroshima 739-8526, Japan \\
$^6$Yukawa Institute for Theoretical Physics, Kyoto University,
    Kyoto 606-8502, Japan
}

\date{\today}

\begin{abstract}
   We present a high statistics study of the light hadron spectrum 
   and quark masses in QCD with two flavors of dynamical quarks. 
   Numerical simulations are carried out using the plaquette 
   gauge action and the $O(a)$-improved Wilson quark action 
   at $\beta=5.2$, where the lattice spacing is found to be 
   $a=0.0887(11)$~fm from $\rho$ meson mass, 
   on a $20^3\times 48$ lattice. 
   At each of five sea quark masses 
   corresponding to $m_{\rm PS}/m_{\rm V} \simeq 0.8$\,--\,0.6, 
   we generate 12000 trajectories 
   using the symmetrically preconditioned Hybrid Monte Carlo algorithm.
   Finite spatial volume effects are investigated employing 
   $12^3 \times 48$, $16^3 \times 48$ lattices.
   We also perform a set of simulations in quenched QCD 
   with the same lattice actions at a similar lattice spacing 
   to those for the full QCD runs.
   In the meson sector we find clear evidence of sea quark effects.  
   The $J$ parameter increases for lighter sea quark masses, and 
   the full QCD meson masses are systematically closer
   to experiment than in quenched QCD.
   Careful finite-size studies are made to ascertain that these are not 
   due to finite-size effects.  
   Evidence of sea quark effects is less clear in the baryon sector 
   due to larger finite-size effects. 
   We also calculate light quark masses and find 
   $m_{ud}^{\overline{\rm MS}}(\mbox{2~GeV})
   =3.223\left(^{+0.046}_{-0.069}\right)$~MeV and  
   $m_{s}^{\overline{\rm MS}}(\mbox{2~GeV})
   =84.5\left(^{+12.0}_{-1.7}\right)$~MeV
   which are about 20\,\% smaller than in quenched QCD.
\end{abstract}


\maketitle


\section{Introduction}
\label{sec:introduction}


%
%
Lattice QCD calculation of the light hadron mass spectrum 
has witnessed significant progress in recent years
\cite{Spectrum.review.lat00,Spectrum.review.lat01,KS-Spectrum.review.lat01}. 
%
%
In the quenched approximation in which the quark vacuum polarization
effects are ignored,  
%
%
the CP-PACS Collaboration performed 
a precise calculation of hadron masses,
in which the estimated accuracy reached the level of a few percent 
in the continuum limit \cite{Spectrum.Nf0.CP-PACS}.
They found that 
the quenched spectrum shows a significant and systematic deviation 
from experiment; 
the $K^*$--$K$ hyperfine splitting is smaller by about 10\,\% 
than experiment. The decuplet baryon mass splittings are also small, 
and the octet baryon masses are themselves smaller than experiment. 

Since this work, 
the focus of efforts has shifted toward full QCD simulations 
including vacuum polarization effects of dynamical quarks.  
A number of simulations now exist, pursued by 
the SESAM-T$\chi$L~\cite{Spectrum.Nf2.SESAM,Spectrum.Nf2.TchiL,Spectrum.Nf2.SESAM-TchiL}, 
UKQCD~\cite{Spectrum.Nf2.UKQCD.csw176,Spectrum.Nf2.UKQCD}, 
CP-PACS~\cite{mq.Nf2.CP-PACS,Spectrum.Nf2.CP-PACS} and 
QCDSF-UKQCD~\cite{Spectrum.Nf2.QCDSF} Collaborations for two flavors 
using the Wilson-type quark action, and by the MILC 
Collaboration~\cite{Spectrum.Nf2.MILC,Spectrum.Nf3.MILC}
for two and three flavors using the Kogut-Susskind(KS) quark action. 
In particular the CP-PACS 
calculation~\cite{mq.Nf2.CP-PACS,Spectrum.Nf2.CP-PACS}
made a first attempt toward execution of the chiral and 
continuum extrapolations within the single set of simulations, 
as pioneered by the GF11 Collaboration~\cite{Spectrum.Nf0.GF11} 
in their quenched spectrum study. 
The chief finding of this work was that the $K^*$--$K$ hyperfine 
splitting agrees much better with experiment in two-flavor full QCD 
than in quenched QCD, 
and that light quark masses decrease by about 25\% 
by the inclusion of dynamical $u$ and $d$ quarks. 

A subtle point with the CP-PACS results is that 
the dynamical sea quark effects become manifest 
only after the continuum extrapolation. 
Further studies are required to consolidate 
effects of dynamical sea quarks. 
The CP-PACS simulation used a renormalization group (RG) improved 
gauge action \cite{RGaction}, but the $O(a)$-improved Wilson quark 
action \cite{SWaction} with only 
tadpole-improved \cite{tadpole_improvement} clover coefficient.
This leaves scaling violation of $O(g^2a)$.  
The use of non-perturbatively determined clover coefficient 
removing all of $O(a)$ errors should be much better to control the 
continuum extrapolation \cite{NPimprovement}. 
Studies along this direction were previously carried out
by the UKQCD and QCDSF Collaborations using the plaquette gauge action.
However, 
sea quark effects are not clear in their results of hadron masses,
albeit encouraging evidence is seen in the static quark potential
\cite{Spectrum.Nf2.UKQCD,Spectrum.Nf2.QCDSF}.


In the present work,
we explore sea quark effects in hadron and quark masses 
in two-flavor QCD 
using the plaquette gauge action and 
the non-perturbatively $O(a)$-improved Wilson quark action.
Our simulations are performed at a single lattice spacing
$a^{-1}\! \simeq \!2$~GeV at $\beta=5.2$ 
using a $20^3\times 48$ lattice.  
We also carry out calculations in quenched QCD
with the same action and similar simulation parameters 
to those in full QCD 
in order to make a direct comparison between full and quenched QCD.
Preliminary results of these calculations have been reported 
in Refs.~\cite{Spectrum.Nf2.JLQCD.lat00,Spectrum.Nf2.JLQCD.lat01,Spectrum.Nf2.JLQCD.lat02.Kaneko,Spectrum.Nf2.JLQCD.lat02.Hashimoto}.

We pay particular attention to two points 
which are important for an unambiguous identification 
of sea quark effects. 
One is the finite spatial volume effect
whose magnitude is believed to be
more pronounced in full QCD simulations 
than in quenched QCD \cite{FSE.Z3}.
An increase of hadron masses due to this effect 
could mimic sea quark effects.  
There are only a few studies of finite-size effects 
in full QCD for the Wilson-type quark action 
\cite{Spectrum.Nf2.TchiL,Spectrum.Nf2.UKQCD.csw176}.
This leads us to perform a systematic investigation
of finite-size effects employing $12^3 \times 48$, 
$16^3 \times 48$ and $20^3 \times 48$ lattices.


Another point is the chiral extrapolation. 
With currently available computer power and simulation algorithms,
the sea quark mass which can be explored 
with the Wilson-type quark action is limited to 
values corresponding to $m_{\rm PS,sea}/m_{\rm V,sea}\!\gtrsim\!0.6$.
The long extrapolation to the physical $u$ and $d$ quark masses
may involve sizable systematic errors, potentially blurring, 
or artificially enhancing, sea quark effects.
This can be avoided if one examines sea quark effects 
at the quark masses actually simulated.  
In this study, therefore, 
sea quark effects are examined in detail
not only at the physical quark mass but also at our simulation points.


We have also made efforts 
to accumulate high statistics of 12000 trajectories 
each at five values of sea quark masses.  
Our implementation of the symmetric 
preconditioning of the lattice clover-Dirac operator
\cite{Even-Odd2,PHMC.JLQCD}
speeded up the configuration generation by a factor two 
by allowing a doubly larger step size 
over the even-odd preconditioning.


This paper is organized as follows.
We describe details of configuration generation 
in full and quenched QCD in Sec.~\ref{sec:sim}.
Method of measurement of hadron masses and the static
quark potential is explained in Sec.~\ref{sec:meas}.
Finite-size effects on hadron masses are discussed 
in Sec.~\ref{sec:FSE}. 
Section~\ref{sec:chiralfit} is devoted to detailed description of 
the chiral extrapolation of our hadron mass data.
We examine sea quark effects in light hadron masses
in Sec.~\ref{sec:SQE}.
Results of the decay constants and quark masses 
are presented in Secs.~\ref{sec:f} and \ref{sec:mq}.
Our conclusion is given in Sec.~\ref{sec:conclusion}.

\section{Simulation Method}
\label{sec:sim}

\subsection{Simulation parameters and algorithm}
\label{sec:sim:param}

We carry out numerical simulations of lattice QCD 
with two flavors of degenerate dynamical quarks
which are identified with the up and down quarks.
%
%
We use the standard plaquette action for gauge fields defined by 
\bea
   S_g & = & \frac{\beta}{6} 
             \sum_{x,\mu \nu} \mbox{Tr } U_{x,\mu\nu},
   \label{eqn:sim:param:Sg}
\eea
where $U_{x,\mu\nu}$ is the product of gauge link variables 
$U_{x,\mu}$ around the plaquette given by 
\bea
   U_{x,\mu\nu} & = & U_{x,\mu} 
                      U_{x+\hat{\mu},\nu}
                      U_{x+\hat{\nu},\mu}^{\dagger}
                      U_{x,\nu}^{\dagger}.
   \label{eqn:sim:param:plaq}
\eea
%
%
The $O(a)$-improved Wilson action \cite{SWaction} defined by 
\bea
   S_q & = & \sum_{x,y} \bar{q}_x D_{xy} q_y 
   \\
   D_{xy} 
   & = & 
   \delta_{xy} -K \sum_{\mu} \left\{ \left( 1 - \gamma_{\mu} \right)
                                     U_{x,\mu} \delta_{x+\hat{\mu},y}
                                    +\left( 1 + \gamma_{\mu} \right)
                                     U_{x,\mu}^{\dagger} 
                                     \delta_{x,y+\hat{\mu}}
                             \right\}
  -\frac{1}{2} K c_{\rm SW} \sigma_{\mu\nu} F_{x,\mu\nu} \delta_{xy}
   \label{eqn:sim:param:Sq}
\eea
is used for the quark part.
The field strength tensor on the lattice is defined by 
\bea
   F_{x,\mu\nu} 
   & = & 
   \frac{1}{8i} \left\{ \left( U_{x,\mu\nu}   + U_{x,-\mu\nu}
                              +U_{x,-\mu-\nu} + U_{x,\mu-\nu}
                        \right)
                       -\left(\mbox{h.c.}\right)
                \right\},
   \label{eqn:sim:param:Fmunu}
\eea
where $\left(\mbox{h.c.}\right)$ denotes the hermitian conjugate 
of the preceding bracket,
and $\sigma_{\mu\nu}=(i/2)\left[\gamma_{\mu},\gamma_{\nu}\right]$.
%
%
The clover coefficient $c_{\rm SW}$ is set to 
the non-perturbative value determined by the ALPHA
Collaboration \cite{NPcsw.Nf2.ALPHA}.


%
%
Our simulations are performed at a single value of 
$\beta\!=\!5.2$.
The lattice spacing fixed from $m_{\rho}$
at the physical sea quark mass is found to be 0.0887(11)~fm.
%
%
Our value of $\beta$ is slightly off the range 
$\beta\!=\!12.0$\,--\,5.4
where the ALPHA Collaboration carried out 
a non-perturbative determination of $c_{\rm SW}$.
We set $c_{\rm SW}=2.02$ by extrapolating 
their parametrization formula of the non-perturbative $c_{\rm SW}$
as a function of the bare coupling.
We performed an independent non-perturbative 
determination of $c_{\rm SW}$ at $\beta\!=\!5.2$ and 
confirmed that our preliminary result
$c_{\rm SW}\!=\!1.98(7)$ is consistent with 2.02
within the error~\cite{NPcsw.Nf2.JLQCD}.
%
%


We employ three lattice sizes that differ in spatial volumes,
$N_s^3 \times N_t \! = \! 12^3\times48$, $16^3\times48$ 
and $20^3\times48$.
The hadron spectrum and quark masses at the physical point
are calculated using the data on the largest lattice.
The data on the two smaller lattices are used to 
investigate finite-size effects on hadron masses.


On each lattice size, 
we adopt five values of the sea quark mass 
corresponding to the hopping parameter
$K_{\rm sea}\!=\!0.1340$, 0.1343, 0.1346, 0.1350 and 0.1355.
This choice covers 
$m_{\rm PS,sea}/m_{\rm V,sea}=0.6$\,--\,0.8,
and enables us to extrapolate our data to the physical sea quark mass.
These simulation parameters are summarized 
in Table~\ref{tab:sim:param:param_all}.


We note that the UKQCD Collaboration also performed 
a set of simulations using the same lattice action 
at $a^{-1}\!\simeq\!2$~GeV~\cite{Spectrum.Nf2.UKQCD}.
There are, however, some differences in the choice of 
$\beta$ and $K_{\rm sea}$:
The UKQCD simulations shift $\beta$ with the sea quark mass
keeping the Sommer scale $r_0/a$~\cite{r0} fixed, 
while our simulations are performed at fixed $\beta$.
Another difference is the range of the sea quark mass
covered in the two simulations.
We explore light sea quark masses down to 
$m_{\rm PS,sea}/m_{\rm V,sea}\!\simeq\!0.6$,
whilst UKQCD's lightest point is around 
$m_{\rm PS,sea}/m_{\rm V,sea}\!\simeq\!0.7$.
Although the UKQCD Collaboration made another simulation 
at a smaller sea quark mass 
$m_{\rm PS,sea}/m_{\rm V,sea}\!\simeq\!0.6$
at a spatial extent of $N_s\!=\!16$ ($N_s a\!\simeq\!1.6$~fm),  
finite-size effects seem to be significant there 
(see discussion in Sec.~\ref{sec:FSE}).

%
%
Gauge configurations are generated using 
the Hybrid Monte Carlo (HMC) algorithm~\cite{HMC.Duane,HMC.Gottlieb}.
We use simulation programs 
with three variants of HMC
for the $O(a)$-improved Wilson action:
\bi
   \item HMC with the even/odd preconditioning~\cite{Even-Odd.D}
         only for the inversion of the quark matrix $D_{xy}$.
         This algorithm is used in the simulations 
         on the $16^3 \times 48$ lattice.
   \item HMC with the asymmetric preconditioning for the lattice 
         action (A-HMC)~\cite{Even-Odd1,Even-Odd2,PHMC.JLQCD}.
         Whole simulations on the $12^3 \times 48$ lattice 
         are performed with this algorithm.
   \item HMC with the symmetric preconditioning for the action 
         (S-HMC)~\cite{Even-Odd2,PHMC.JLQCD},
         which shows the best performance among the three algorithms.
\ei
Our main simulation on the $20^3\times 48$ lattices is
initially started with the A-HMC algorithm, 
but is later switched to the S-HMC to speed up the calculations.
The trajectory length in each HMC step 
is fixed to the unit length.
We use the conventional leap-frog integration scheme 
for the molecular dynamics equation. 
The step size $\Delta \tau$ is chosen to achieve 
an acceptance of 60\,--\,80\%.

%
%
The even/odd preconditioned BiCGStab algorithm \cite{BiCGStab}
is used for the quark matrix inversion to solve the equation 
$D_{xy}G_{y}\!=\!B_{x}$.
We take the stopping condition of the form $|| DG - B || < \Delta$ 
in the HMC program.
A modified form $|| DG - B ||/||B|| < \Delta$ 
is used in the A-HMC and S-HMC programs.
The value of $\Delta$ in the evaluation of the fermionic force
is determined so that the reversibility over 
unit length is satisfied to a relative level better than 
$10^{-13}$ for the Hamiltonian.
We use a stricter stopping condition
in the calculation of the Hamiltonian 
in the Metropolis accept/reject test.
Table~\ref{tab:sim:param:param_all} shows 
our choice of $\Delta$ 
together with the average number of the BiCGStab iteration
in the quark matrix inversion for 
the force calculation, $N_{\rm inv}$.

We accumulate 12000 HMC trajectories at each sea quark mass 
on the $20^3 \times 48$ lattice.
The statistics on smaller lattices are 3000 trajectories.
Measurements of light hadron masses and the static quark 
potential are carried out at every 10 HMC trajectories.
Details of the measurement method 
will be described in the next section.


%
%
All simulations are performed on the parallel computer
HITACHI SR8000 model F1 installed at KEK.
This machine consists of 100 nodes and 
has a peak speed of 1.2 TFLOPS and 448 GB of main memory in total.
The CPU time needed per unit HMC trajectory on the full machine
is listed in Table~\ref{tab:sim:param:param_all}.
The total time for configuration generation 
on each lattice size is 8.6 days on $12^3 \times 48$,
58 days on $16^3 \times 48$, and 130 days on $20^3 \times 48$ lattices.
Additional 100 days are spent for the measurement 
of the hadron masses and the static potential.

\subsection{Simulation in quenched QCD}
\label{sec:sim:quenched}

While many calculations of the hadron spectrum
have been performed in quenched QCD, 
comparisons between our full QCD results and quenched results
from other simulations may be subject to systematic uncertainties
due to the difference in the simulation details.
We therefore carry out a set of quenched calculations of the hadron 
spectrum using the same lattice actions and simulation parameters 
as those for full QCD runs.

Our simulations are performed at $\beta=6.0$, 
where the lattice spacing fixed from $m_{\rho}$ equals 
0.1074(14)~fm.
We take $c_{\rm SW}\!=\!1.769$
which is the value determined non-perturbatively 
by the ALPHA Collaboration~\cite{NPcsw.Nf0.ALPHA}.
Three lattice sizes $12^3\times48$, $16^3\times48$ 
and $20^3\times48$ are employed 
in order to investigate finite-size effects.

Gauge configurations are generated with a combination of
the heat-bath and the over-relaxation algorithms.
We call four heat-bath sweeps with a succeeding
over-relaxation step an iteration.
We accumulate statistics of 60000 iterations on each lattice size. 
Hadron masses and the static potential are calculated 
at every 200 iterations.


\section{Measurement}
\label{sec:meas}

\subsection{Hadron Masses}
\label{sec:meas:had}

%
%


In measurements in full QCD,
we use six values of the valence quark mass
corresponding to the hopping parameter
$K_{{\rm val},i}(i\!=\!1,\ldots6)=0.1340$, 0.1343, 0.1346, 0.1350, 0.1355 and 0.1358,
which cover the range of 
$m_{\rm PS,val}/m_{\rm V,val}\!\simeq\!0.5$\,--\,0.8.
At each sea quark mass, therefore,
there is one value of $K_{{\rm val},i}$, which equals $K_{\rm sea}$
and is identified as the light quark mass.
Other five values of $K_{{\rm val},i}$ correspond to 
the mass of strange quarks treated in the quenched approximation.
%
%
In the following, we use the abbreviation ``diagonal data'' 
to represent hadron correlators or masses
with a quark mass combination in which 
all valence quark masses are equal to the sea quark mass.


We employ meson operators defined by 
\bea
   M (x) = \bar{q}^{(f)}_{x} \Gamma q^{(g)}_{x}, \hspace{5mm}
   \Gamma = I, \gamma_5, \gamma_{\mu}, 
            \gamma_5 \gamma_{\mu},
   \label{meas:had:meson:op}
\eea 
where $f$ and $g$ are flavor indices and 
$x$ is the coordinates on the lattice.
Meson correlators $\langle M(x) M(0)^{\dagger} \rangle$
are calculated for the following eleven combinations of valence 
quark masses
\bea
   \begin{array}{l}
   (K_{{\rm val},i},K_{{\rm val},i}) \ (i=1,\ldots 6), 
   \\[2mm]
   (K_{\rm sea},K_{{\rm val},i}) \ 
   (i=1,\ldots 6, \, K_{\rm sea} \neq K_{{\rm val},i}).
   \end{array}
   \label{meas:had:meson:Kval}
\eea
The former is identified with a degenerate light or strange meson
and the latter with a non-degenerate light-strange meson.
This choice of the valence quark masses enables us to
calculate the full spectrum of strange and non-strange mesons.


For baryons, we use the same operators 
as those employed in Ref.~\cite{Spectrum.Nf2.CP-PACS}.
Namely, the octet baryon operator is defined as
\bea
   O^{fgh}(x)
 = \epsilon^{abc}
   \left( q_x^{(f)a \, T} C \gamma_5 q_x^{(g)b} \right) 
   q_{x}^{(h)c},
   \label{meas:has:octet:op}
\eea
where $a, b, c$ are color indices and 
$C\!=\!\gamma_4 \gamma_2$ is the charge conjugation matrix.
We measure baryon correlators with two types of flavor structure
($\Sigma$- and $\Lambda$-like baryons),
\bea
   \Sigma  & : & - \frac{1}{\sqrt{2}}
                   \left( O^{[fh]g} + O^{[gh]f} \right),
   \\
   \Lambda & : &   \frac{1}{\sqrt{6}}
                   \left( O^{[fh]g} - O^{[gh]f} -2O^{[fg]h} 
                   \right),
   \label{meas:had:octet:flv}
\eea
where $O^{[fg]h} \! = \! O^{fgh} - O^{gfh}$.
Decuplet baryon correlators are calculated 
using an operator defined by 
\bea
   D^{fgh}(x)
 = \epsilon^{abc}
   \left( q_x^{(f)a \, T} C \gamma_{\mu} q_x^{(g)b} \right) 
   q_{x}^{(h)c}
   \label{meas:had:decuplet:op}
\eea
with symmetrized flavor structure
\bea
   \begin{array}{l}
      O^{fff}, 
      \hspace{5mm}
      \frac{1}{\sqrt{3}}\left( D^{ffg} + D^{fgf} + D^{gff} \right), 
      \\[2mm]
      \frac{1}{\sqrt{6}}\left( D^{fgh} + D^{hfg} + D^{ghf} 
                           +D^{fhg} + D^{gfh} + D^{hgf} 
                        \right).
   \end{array}
   \hspace{5mm}
   \label{meas:had:decpulet:flv}
\eea
We take quark mass combinations of 
$(K_{{\rm val},i}, K_{{\rm val},i}, K_{{\rm val},i}) \ (i\!=\!1,\ldots
6)$,
$(K_{\rm sea}, K_{{\rm val},i}, K_{{\rm val},i})$ and 
$(K_{\rm sea}, K_{\rm sea}, K_{{\rm val},i}) \ 
 (i\!=\!1,\ldots 6, K_{\rm sea} \neq K_{{\rm val},i})$
for the baryon correlators.


In order to construct the smeared hadron operators,
we measure the wave function of the pseudoscalar (PS) meson
\bea
   \phi({\bf r}) 
   & = & 
   \frac{ \sum_{\bf x} 
          \left\langle 
             \bar{q}({\bf x},t) \gamma_5 q({\bf x}+{\bf r},t)
             P({\bf 0},0)^{\dagger}
          \right\rangle 
        } 
        { \sum_{\bf x} 
          \left\langle 
             P({\bf x},t) P({\bf 0},0)^{\dagger} 
          \right\rangle
        },
   \label{meas:had:smr:wavefunc}
\eea
where $P$ is the PS meson operator, Eq.~(\ref{meas:had:meson:op}),
with $\Gamma\!=\!\gamma_5$ and $t$ fixed to 12.
The measurement is performed at each sea quark mass and lattice size
using a subset of gauge configurations 
(30 configurations every 100 trajectories).
We parameterize $\phi({\bf r})$ using a polynomial approximation 
$\phi({\bf r}) = 1 + \sum_{n=1,8} c_n |{\bf r}|^n$
and use it as the smearing function.
We employ three types of the meson operator : 
i) local operator, 
ii) smeared operator 
    $M(x) = \sum_{\bf r} 
            \phi({\bf r}) \bar{q}({\bf x},t) \Gamma 
                          q({\bf x}+{\bf r},t)
    $,
iii) doubly smeared operator 
    $M(x) = \sum_{{\bf r},{\bf r}^{\prime}}
            \phi({\bf r}) \phi({\bf r}^{\prime})
            \bar{q}({\bf x}+{\bf r},t) \Gamma 
            q({\bf x}+{\bf r}^{\prime},t).
    $
Additionally, we use ``triply smeared operator''
\bea
    O^{fgh}(x)
    & = &
    \sum_{{\bf r}_1,{\bf r}_2,{\bf r}_3}
    \phi({\bf r}_1) \phi({\bf r}_2) \phi({\bf r}_3) 
    \epsilon^{abc} 
    \left( q^{a \, T}({\bf x}+{\bf r}_1,t) 
    C \Gamma 
    q^{b}({\bf x}+{\bf r}_2,t) 
    \right)
    q^{c}({\bf x}+{\bf r}_3,t)
   \label{meas:had:smr:triply}
\eea
for baryons.
Hadron correlators are measured with
a) point source and sink operators, 
b) smeared source and point sink, 
and 
c) smeared source and sink operators.
We fix configurations to the Coulomb gauge,
since b) and c) are not gauge invariant.


We observe that,
when valence quarks are lighter than sea quarks,
the hadron correlator takes an exceptionally large value 
on a small number of configurations.
This might be caused by a fluctuation of 
the lowest eigenvalue of the Dirac operator of 
the $O(a)$-improved Wilson action.
If the PS meson correlator on the $i$-th gauge configuration 
takes a value larger than 20 times the statistical average,
which is evaluated without that configuration,
at a certain time slice 
\bea
   \langle P(x) P(0)^{\dagger} \rangle_{i}
   >
   \frac{20}{N_{\rm conf}-1} \sum_{k=1,k \neq i}^{N_{\rm conf}}
   \langle P(x) P(0)^{\dagger} \rangle_{k}
   \label{meas:had:excep:criterion}
\eea
where $N_{\rm conf}$ is the total number of configurations,
we consider it as an exceptional configuration and 
remove it from the following analysis.
The number of the removed configurations is given 
in Table~\ref{tab:sim:param:param_all}.


In order to reduce the statistical fluctuation of hadron correlators
on the $20^3 \times 48$ lattice,
we repeat the measurement for two choices of 
the location of the hadron source,
$t_{\rm src}\!=\!1$ and $N_t/2\!+\!1(=25)$, 
and take the average over the two sources:
\bea
   \frac{1}{2} 
       \left( \left\langle M(t_{\rm src}+t)
                           M(t_{\rm src})^{\dagger} 
              \right\rangle_{t_{\rm src}\!=\!1}
             +\left\langle M(t_{\rm src}+t)
                           M(t_{\rm src})^{\dagger} 
              \right\rangle_{t_{\rm src}\!=\!25}
       \right).
   \label{meas:had:averageexcep:prop}
\eea
We find that this procedure reduces
the statistical error of hadron correlators by typically 20\,\%,
which suggests that the statistics is increased effectively 
by a factor of 1.5.
For further reduction of the statistical fluctuation,
we take the average over three polarization states
for vector mesons, two spin states for octet baryons
and four spin states for decuplet baryons.

\begin{figure}[htbp]
   \includegraphics[width=70mm]{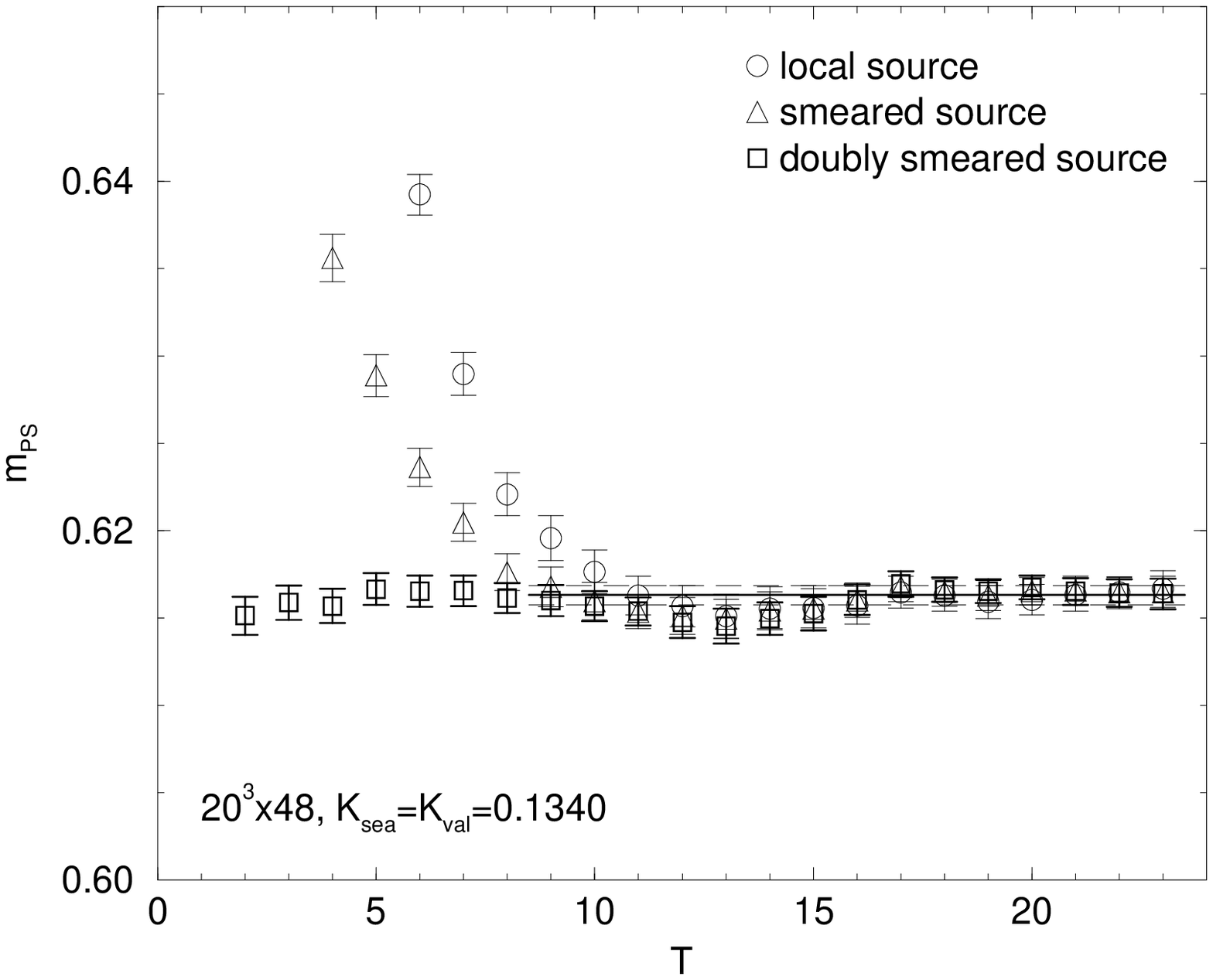}
   \includegraphics[width=70mm]{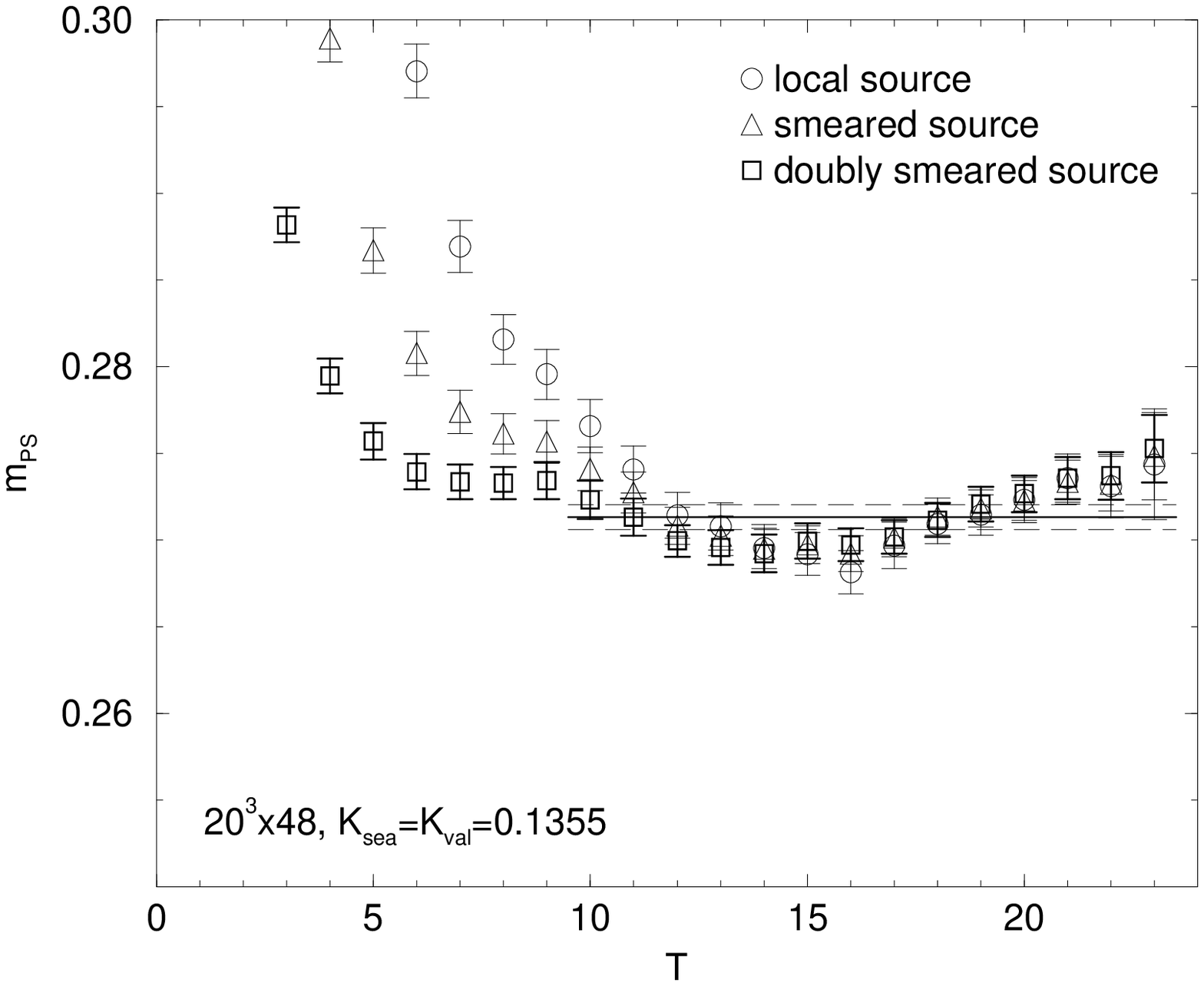}
   \caption{
      Effective mass of PS meson 
      at $K_{\rm sea}\!=\!K_{\rm val}\!=\!0.1340$ (left figure) 
      and 0.1355 (right figure) 
      on $20^3 \times 48$ lattice in full QCD.
      We use the local sink operator for all data.
   }
   \label{fig:meas:had:em_20x48_PS}
\end{figure}
\begin{figure}[htbp]
   \includegraphics[width=70mm]{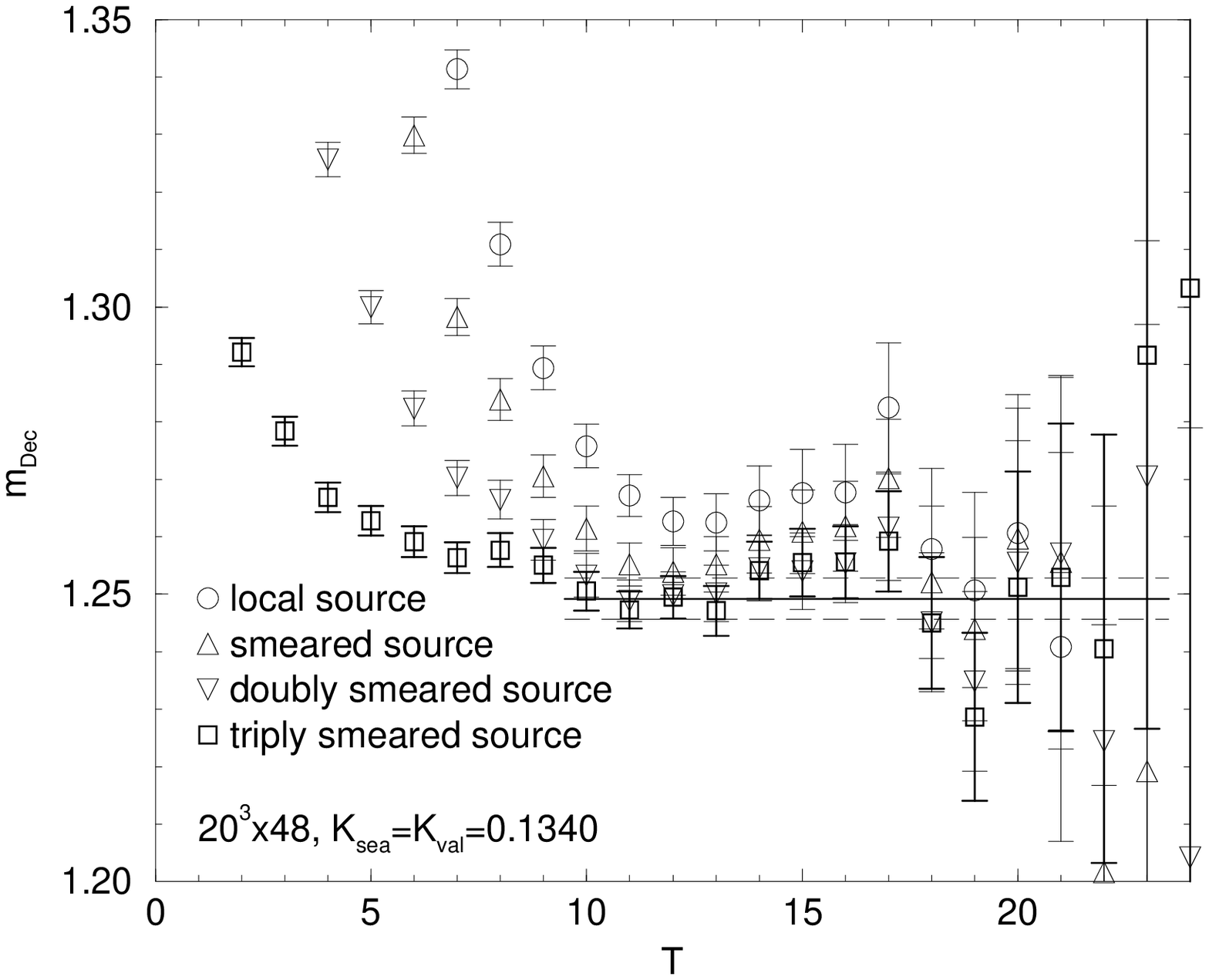}
   \includegraphics[width=70mm]{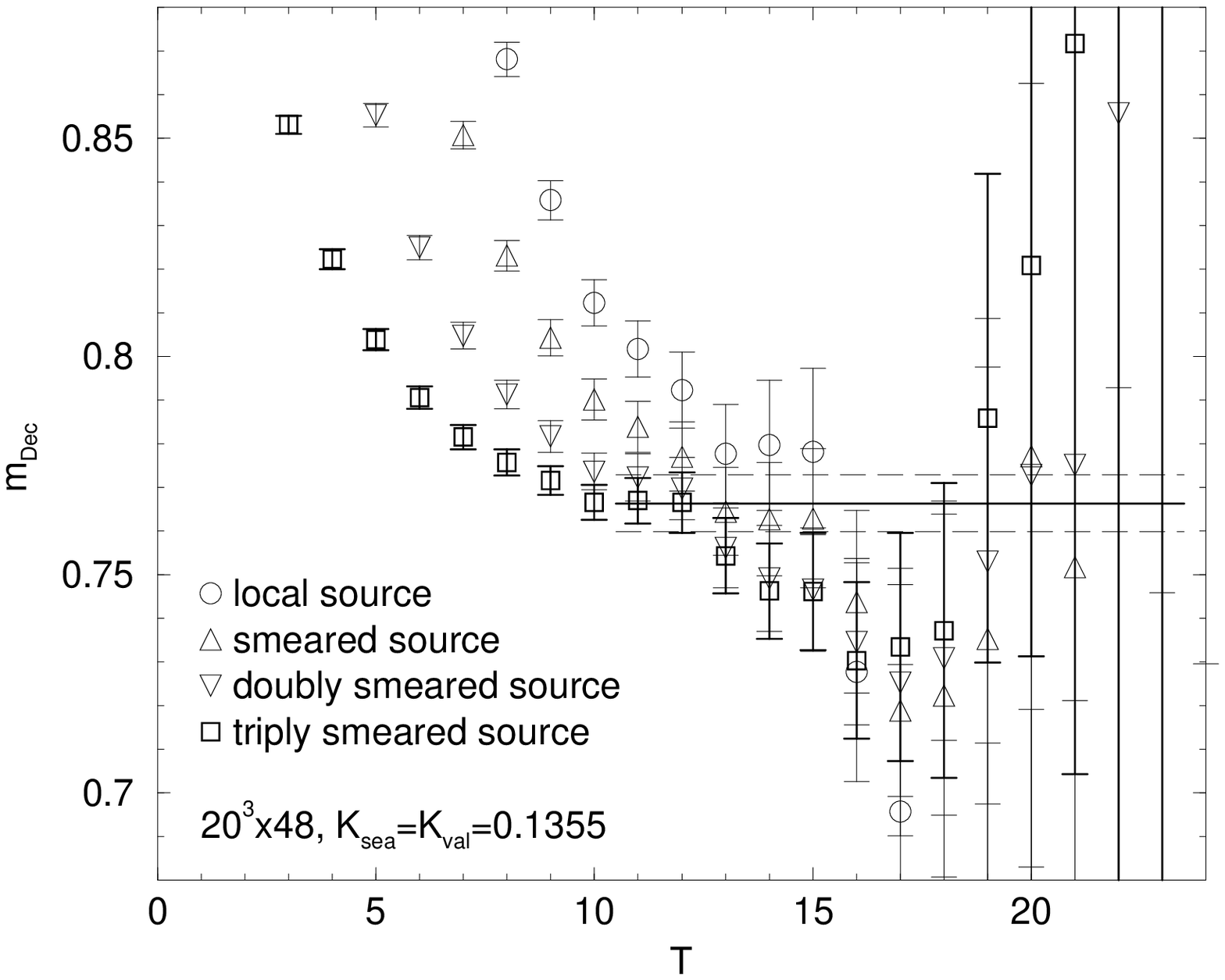}
   \caption
   {
      Effective mass of decuplet baryon 
      at $K_{\rm sea}\!=\!K_{\rm val}\!=\!0.1340$ (left figure) 
      and 0.1355 (right figure) 
      on $20^3 \times 48$ lattice in full QCD.
      We use the local sink operator for all data.
   }
   \label{fig:meas:had:em_20x48_Dec}
\end{figure}

Figures~\ref{fig:meas:had:em_20x48_PS}\,--\,\ref{fig:meas:had:em_20x48_Dec}
show examples of effective mass plots.
We find that the best plateau of the effective mass is obtained 
from hadron correlators with the point sink 
and the doubly smeared source for mesons 
and the triply smeared one for baryons.
Therefore, hadron masses are extracted from these types of correlators.


We carry out $\chi^2$ fits to hadron correlators
by taking account of correlations among different time slices.
A single hyperbolic cosine form is assumed for mesons,
and a single exponential form for baryons.
The lower cut of the fit range $t_{\rm min}$ 
is determined by inspecting stability of the fitted mass.
The upper cut ($t_{\rm max}$) dependence of the fit results 
is small and, therefore, we fix $t_{\rm max}$ to $N_t/2$
for all hadrons.
Our choice of fit ranges and resulting hadron masses are summarized
in Tables~\ref{tab:meas:had:meson_12x48}\,--\,\ref{tab:meas:had:baryon_20x48} in Appendix~A.
%
%
Statistical errors of hadron masses are estimated 
with the jack-knife procedure.
We adopt the bin size of 100 trajectories by inspecting 
the bin size dependence of the jack-knife error as discussed 
in Sec.\ref{sec:meas:autocorr}.

\begin{figure}[htbp]
   \includegraphics[width=70mm]{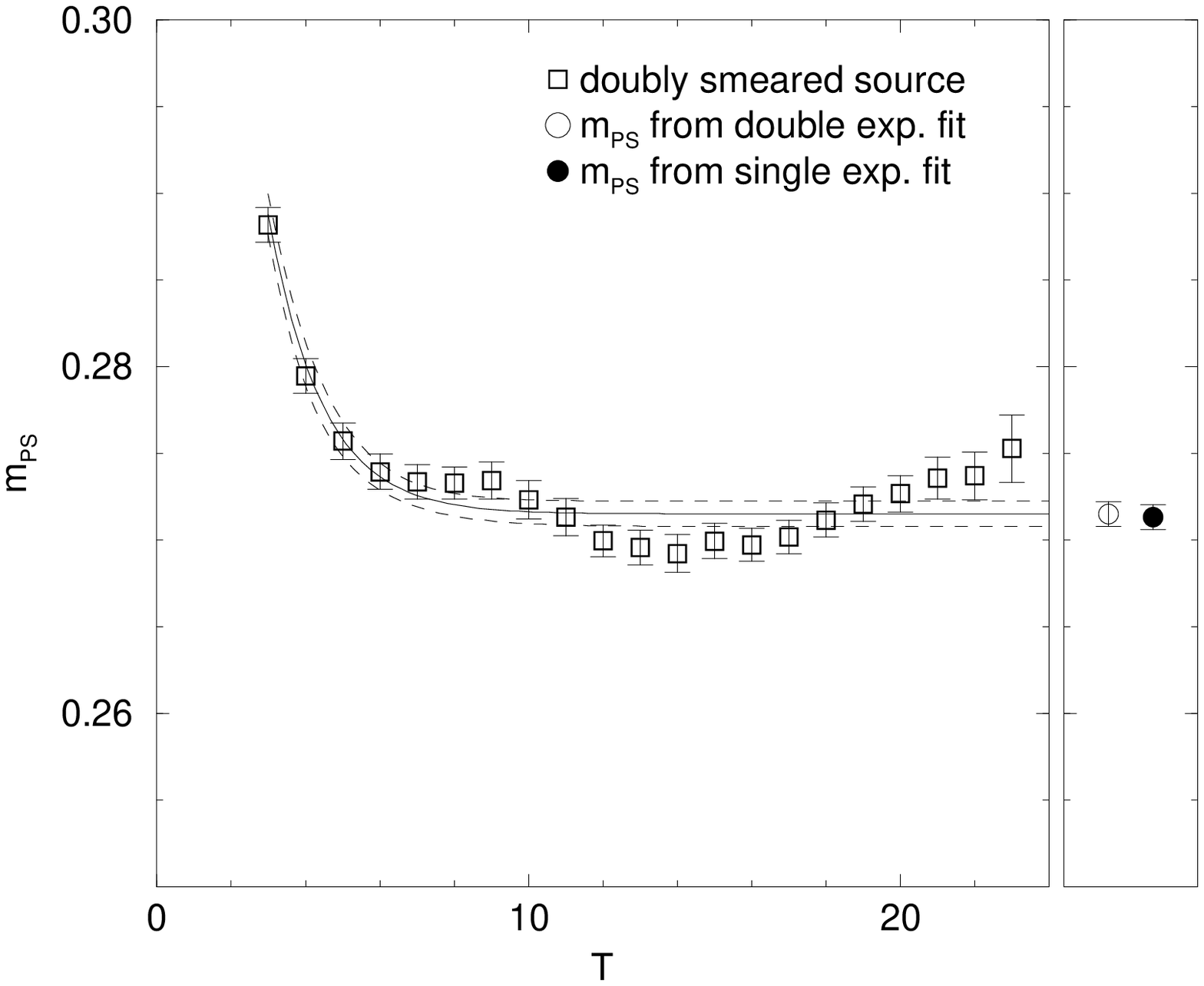}
   \includegraphics[width=70mm]{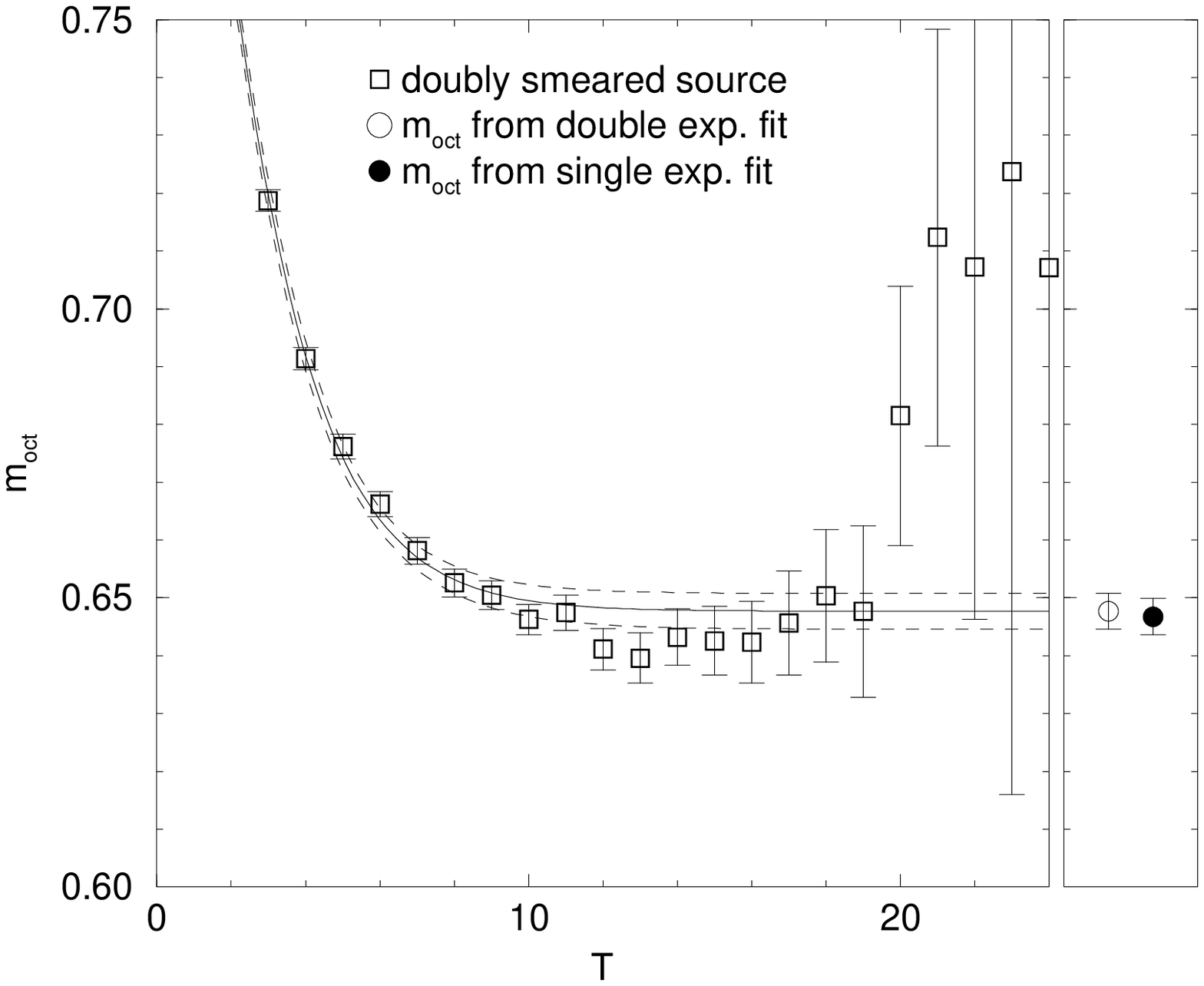}
   \caption{
      Double exponential fits to PS meson (left figure) and 
      octet baryon masses (right figure) at $K_{\rm sea}\!=\!0.1355$.
      Right panel in each figure shows fitted masses determined from 
      double exponential (open symbol) and single exponential fit
      (filled symbol).
      The local sink operator is used for all data.
   }
   \label{fig:meas:had:em_20x48_dexp}
\end{figure}

In Fig.~\ref{fig:meas:had:em_20x48_dexp}, 
we test double exponential fits to extract hadron masses 
at $K_{\rm sea}\!=\!K_{\rm val}\!=\!0.1355$.
While these fits are unstable and lead to 
a large error for the mass of the first excited state,
the result for the ground state mass is consistent with 
that from the single exponential fit.
The situation is similar at other sea and valence quark masses.
This suggests that 
the hadron masses in Tables~\ref{tab:meas:had:meson_12x48}\,--\,\ref{tab:meas:had:baryon_20x48}
and the light hadron spectrum calculated from 
these results have small  contamination from excited states.


Hadron correlators in quenched QCD are calculated 
in an analogous manner.
We use six values of $K_{\rm val}$,
0.13260, 0.13290, 0.13331, 0.13384, 0.13432 and 0.13465,
corresponding to 
$m_{\rm PS,val}/m_{\rm V,val}\!\simeq\!0.50$\,--\,0.80
and the hadron operators and smearing procedure same as those 
in the full QCD study.
%
%
A difference is that we can take more combinations
of valence quark masses than in full QCD, 
since any value of the six valence quark masses 
can be identified with either light or strange quark mass.
We take all combinations 
$(K_{{\rm val},i},K_{{\rm val},j}) \ (i,j=1,..,6)$
for mesons, 
and somewhat restricted choices 
$(K_{{\rm val},i},K_{{\rm val},i},K_{{\rm val},j}) \ (i,j=1,..,6)$
for baryons.
%
%
Statistical errors are estimated with the jack-knife procedure
with bin size of 200 iterations.
The exceptional configurations are discarded with the same criterion 
as defined in Eq.~(\ref{meas:had:excep:criterion}).
Results of hadron masses
are collected in Tables~\ref{tab:meas:had:meson_q12x48}\,--\,\ref{tab:meas:had:baryon_q20x48}
in Appendix~A.

\subsection{Static quark potential}
\label{sec:meas:pot}

 
We calculate the static quark potential in order to determine
the Sommer scale~\cite{r0}
which we use in our analysis of hadron masses.
For this purpose, the temporal Wilson loops $W(r,t)$ 
up to $t\!=\!16$ and $r\!=\!(\sqrt{3}N_s/2)$ 
are measured both in full and quenched QCD simulations.
We apply the smearing procedure of Ref.~\cite{Potential.Nf0.Bali} 
up to twelve steps and the measurements are carried out
every four steps.

\begin{figure}[htbp]
   \includegraphics[width=70mm]{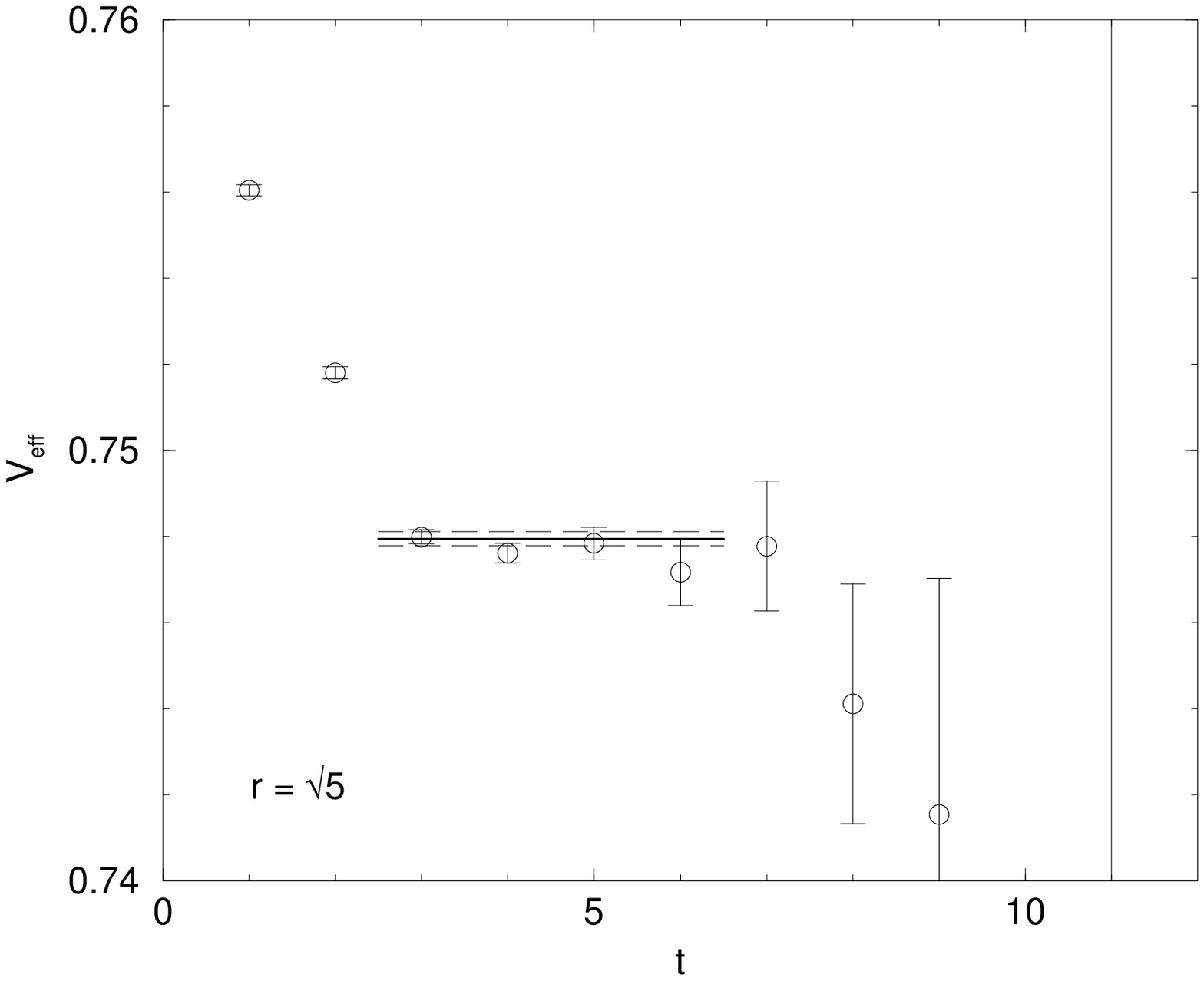}
   \includegraphics[width=70mm]{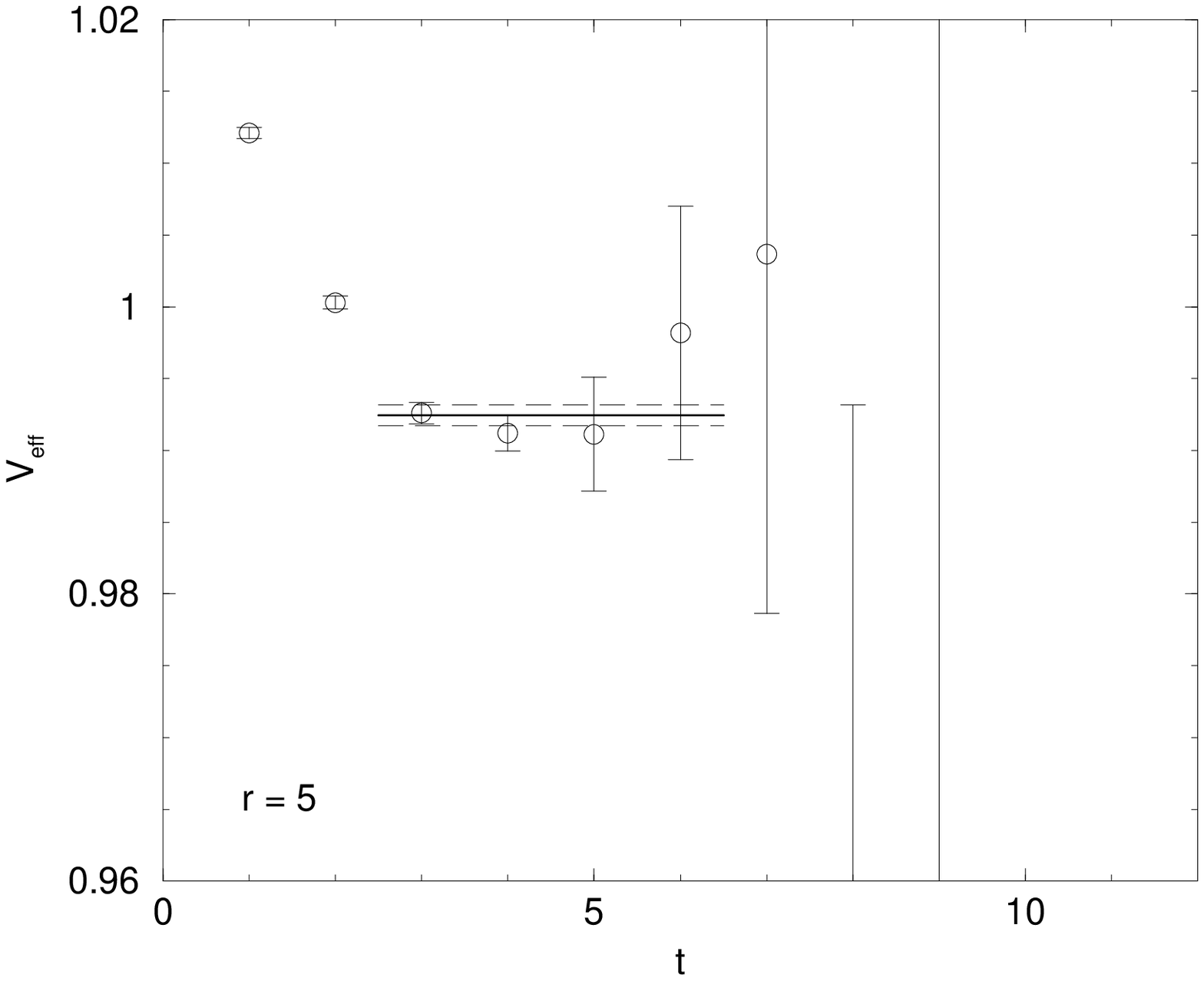}
   \caption
   {
      Effective potential energies 
      $V_{\rm eff}(r,t)$ 
      as a function of temporal separation $t$ 
      at $K_{\rm sea}\!=\!0.1350$ on $20^3 \times 48$ lattice.
   }
   \label{fig:meas:pot:em}
\end{figure}

%
%
The static quark potential $V(r)$ is determined from 
the correlated fit of the form
\bea
   W(r,t) & = & C(r) \exp \left[ - V(r)t \right].
   \label{eqn:meas:pot:fit1}
\eea
We take the fit range $[t_{\rm min},t_{\rm max}]\!=\![3,7]$ 
in all simulations in full and quenched QCD 
by inspecting the $t$ dependence of the effective potential
\bea
   V_{\rm eff}(r,t) \!=\! \ln\left[ W(r,t)/W(r,t+1) \right].
   \label{meas:pot:effV}
\eea
Examples of $V_{\rm eff}$ are plotted in Fig.~\ref{fig:meas:pot:em}.
For each $r$, the number of smearing steps is fixed to 
its optimum value at which 
the overlap to the ground state $C(r)$ takes the largest value.

\begin{figure}[htbp]
   \includegraphics[width=70mm]{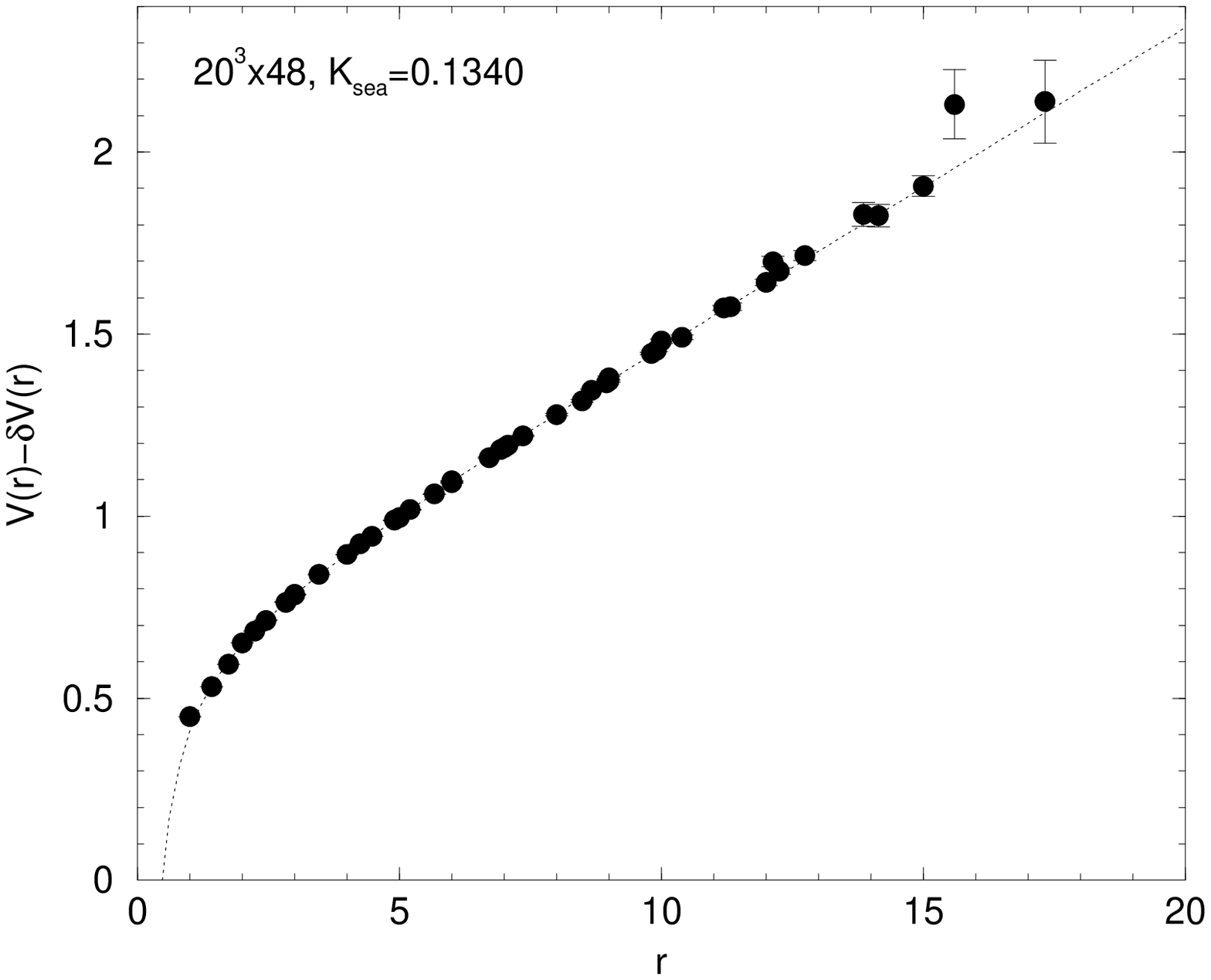}
   \includegraphics[width=70mm]{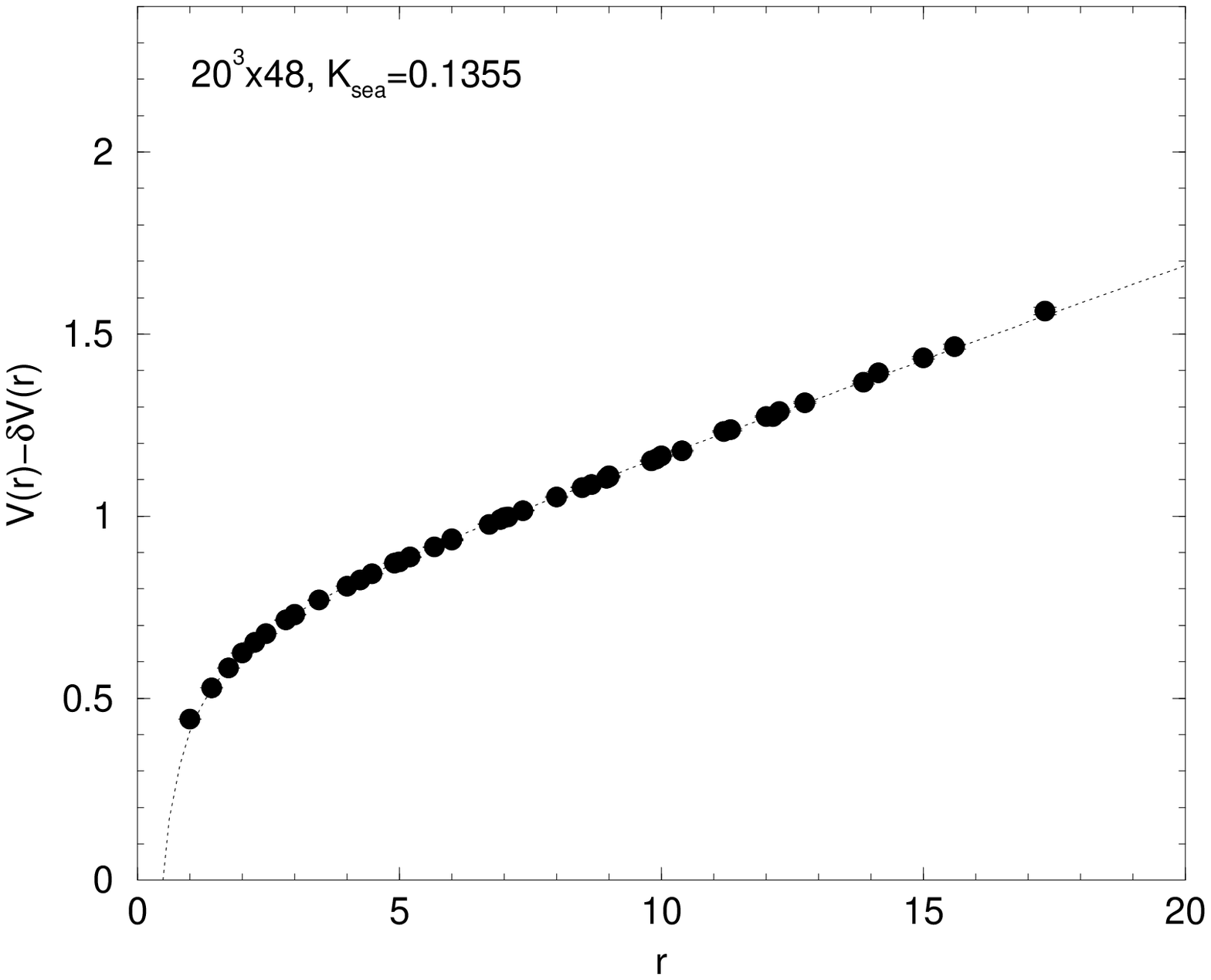}
   \caption
   {
      Static quark potential on $20^3 \times 48$ lattice.
      Left and right figures show data 
      at $K_{\rm sea}\!=\!0.1340$ and 0.1355, respectively.
   }
   \label{fig:meas:pot:VvsR}
\end{figure}
\begin{figure}[htbp]
   \includegraphics[width=70mm]{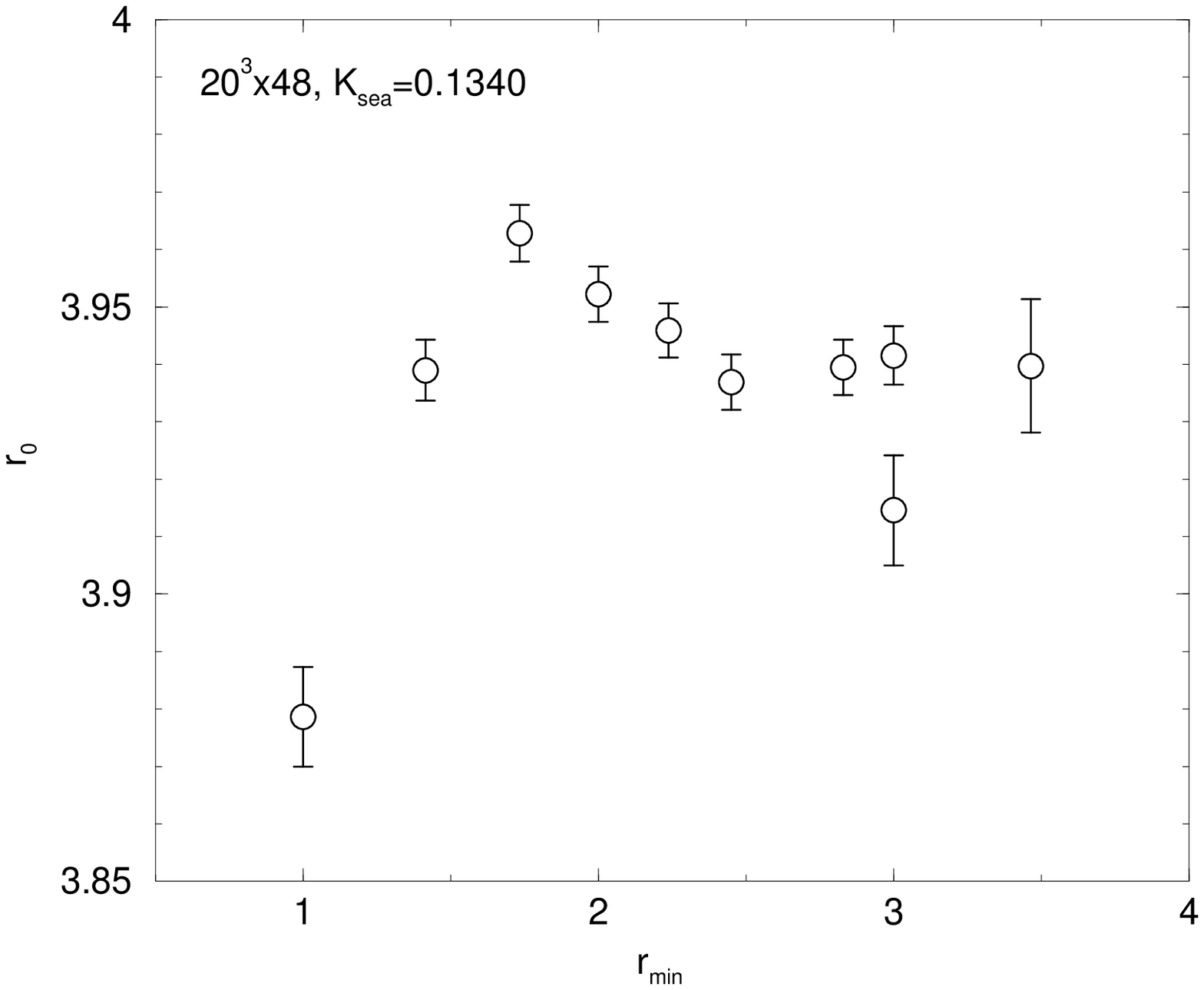}
   \includegraphics[width=70mm]{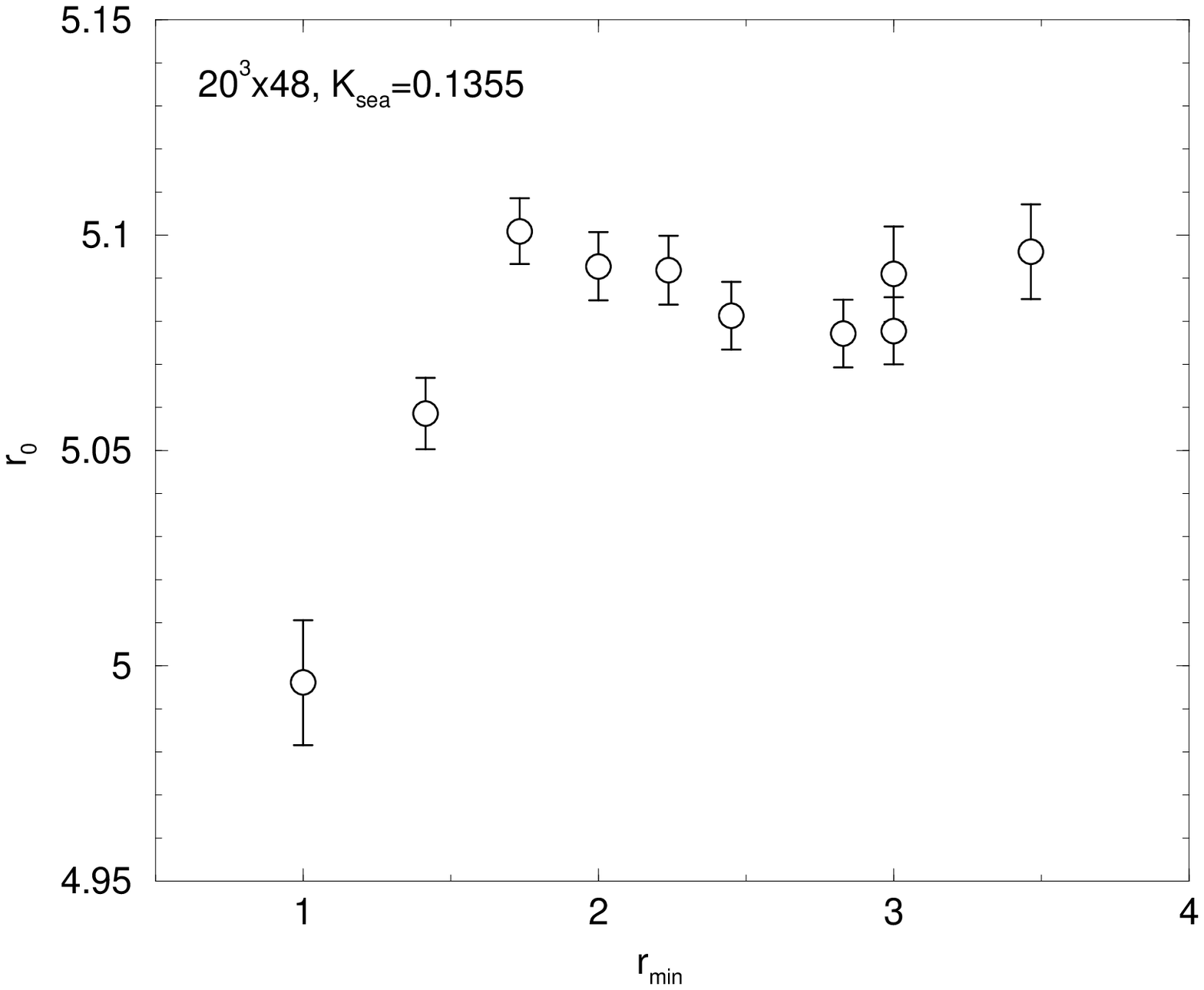}
   \caption
   {
      Sommer scale on $20^3 \times 48$ lattice   
      as a function of $r_{\rm min}$.
      Left and right figures show data 
      at $K_{\rm sea}\!=\!0.1340$ and 0.1355 
   }
   \label{fig:meas:pot:r0_vs_Rmin}
\end{figure}

%
As shown in Fig.\ref{fig:meas:pot:VvsR},
we do not observe any clear indication of the string breaking.
Therefore $V(r)$ is fitted to the form
\bea
   V(r)        
   & = & 
   V_0 - \frac{\alpha}{r} 
       +\sigma r
       +\delta V(r),
   \label{eqn:meas:pot:fit2}
\eea
where $\delta V(r)$ is 
the lattice correction to the Coulomb term calculated perturbatively 
from one lattice-gluon exchange diagram
\cite{deltaV}
\bea
   \delta V(r) 
   & = & 
 - g_c \left(G({\bf r})-\frac{1}{r} \right),
   \\
   G({\bf r})  
   & = & 4 \pi \sum_{\bf k}
           \frac{\cos\left[ {\bf k} {\bf r} \right]}
                {4 \sum_{i=1}^{3} \sin^2\left[k_i/2 \right]}.
   \label{eqn:meas:pot:deltaV}
\eea
The Sommer scale $r_0$ defined through~\cite{r0} 
\bea
   r_0^2 \left. \frac{d V(r)}{d r} \right|_{r=r_0}
   =
   1.65
   \label{eqn:meas:pot:r0_def}
\eea
is then determined from 
the parametrization of the corrected potential $V(r)\!-\!\delta V(r)$:
\bea
   r_{0} & = & \sqrt{\frac{1.65-\alpha}{\sigma}}.
   \label{eqn:meas:pot:r0_fit}
\eea
%
%
The lower cut of the fit range in Eq.~(\ref{eqn:meas:pot:fit2}), 
$r_{\rm min}$, is determined 
by inspecting the $r_{\rm min}$ dependence of $r_0$.
We observe that $r_0$ is relatively stable
for $r_{\rm min} \in [\sqrt{2},2\sqrt{2}]$ 
as shown in Fig.\ref{fig:meas:pot:r0_vs_Rmin}.
With $r_{\rm min}\!<\!\sqrt{2}$, $\chi^2/{\rm dof}$ takes
an unacceptably large value 
due to the violation of rotational symmetry,
while $\alpha$ becomes ill-determined with
$r_{\rm min}\!>\!2\sqrt{2}$.
We therefore take $r_{\rm min}\!=\!\sqrt{5}$.
While the $r_{\rm max}$ dependence of $r_0$ is rather mild,
the covariance matrix becomes ill-determined with 
$r_{\rm max}$ greater than $9\sqrt{2}$ on 
$20^3\times48$, $7\sqrt{2}$ on $16^3\times48$ 
and $6\sqrt{2}$ on $12^3\times48$.
We therefore fix $r_{\rm max}$ to these values.


We repeat the fits, Eqs.~(\ref{eqn:meas:pot:fit1}) 
and (\ref{eqn:meas:pot:fit2}),
with other choices of the range:
$t_{\rm min}=4$  or $r_{\rm min} \in [\sqrt{2},2\sqrt{2}]$.
The largest deviations in the fit parameters and $r_0$ 
are included into their systematic errors.
Other systematic errors due to the choice of $t_{\rm max}$,
the optimum number of the smearing steps and $r_{\rm max}$ are
small and ignored.
Fit parameters in Eq.~(\ref{eqn:meas:pot:fit2}) and $r_0$
are summarized in Table~\ref{tab:meas:pot:r_0:full} for full QCD,
and in Table~\ref{tab:meas:pot:r_0:qQCD} for quenched QCD.

\subsection{Autocorrelation} 
\label{sec:meas:autocorr}


The autocorrelation in our full QCD data is studied 
by calculating the cumulative autocorrelation time 
\bea
   \tau^{\rm cum}_{\calO}(\Delta t_{\rm max}) 
  =\frac{1}{2}  
   +\sum_{\Delta t\!=\!1}^{\Delta t_{\rm max}}
    \rho_{\calO}(\Delta t),
   \label{eqn:meas:autocoor:tau_cum}
\eea
where $\rho_{\calO}(t)$ is the autocorrelation function
\bea
   &&
   \rho_{\calO}(\Delta t) 
   = 
   \frac{\Gamma_{\calO}(\Delta t)}
         {\Gamma_{\calO}(0)}, \\
   &&
   \Gamma_{\calO}(\Delta t) 
   = 
   \langle \left( \calO(t)-\langle \calO \rangle
           \right)
           \left( \calO(t+\Delta t)
                 -\langle \calO \rangle
           \right)
   \rangle
   \label{eqn:meas:autocoor:rho}
\eea
and we take $\Delta t_{\rm max}\!=\!200$.

In Table~\ref{tab:meas:autocorr:tau_cum},
we summarize $\tau^{\rm cum}_{\calO}$ in the A-HMC and S-HMC
simulations on the $20^3 \times 48$ lattices for three quantities:
i) the plaquette which is measured at every trajectory,
ii) the PS meson propagator at $t\!=\!12$,
iii) the temporal Wilson loop with $(r,t)\!=\!(5,4)$.
%
%
The results do not show any systematic differences 
in $\tau^{\rm cum}_{\calO}$ between the A-HMC and S-HMC runs.
%
%
The plaquette shows the largest autocorrelation 
with $\tau^{\rm cum}_{\rm plaq}\!=\!10$\,--\,30, 
which is similar to those found in the UKQCD 
simulation~\cite{Spectrum.Nf2.UKQCD} 
using the same lattice action and similar simulation parameters.
%
%
We obtain smaller values of $\tau^{\rm cum}_{\calO}$ 
for the other two quantities.
This is contrary to a naive expectation that 
these long-distance observables
have a longer autocorrelation than the local quantity
like the plaquette. 
This suggests that 
the size of noise arising from short correlation modes 
is larger than that of the longest mode in these observables
and our statistics are not sufficient to extract 
$\tau^{\rm cum}_{\calO}$ of the longest but weak mode.


The statistical error including the effect of autocorrelation 
is given by $\sqrt{2\tau^{\rm cum}_{\calO}}$ 
times the naive error.
Therefore, the above observation tells us that 
the bin size in the jack-knife procedure of 60 HMC trajectories 
or larger is a safe choice 
to take account of the autocorrelation in our data.

\begin{figure}[htbp]
   \includegraphics[width=70mm]{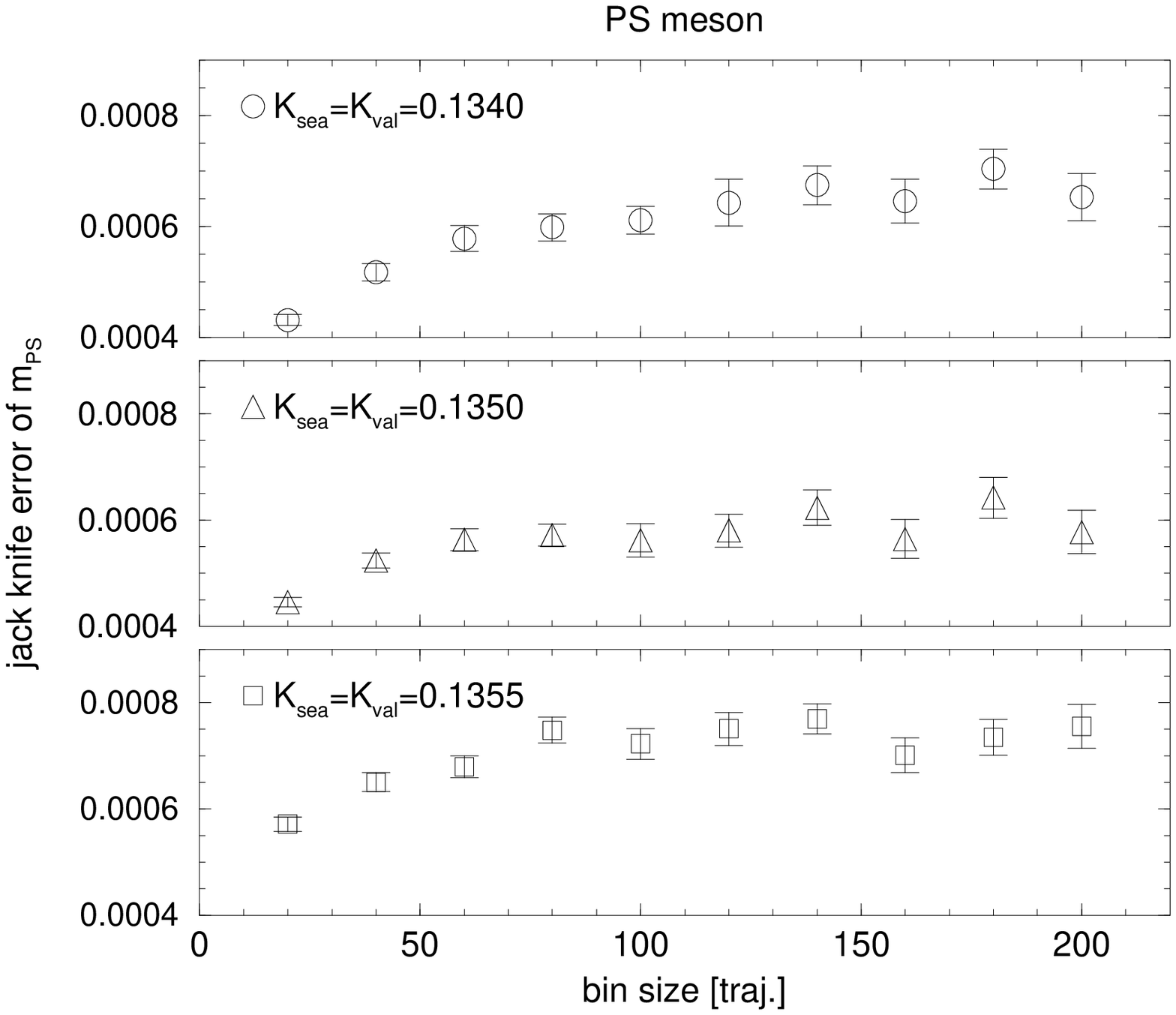}
   \includegraphics[width=70mm]{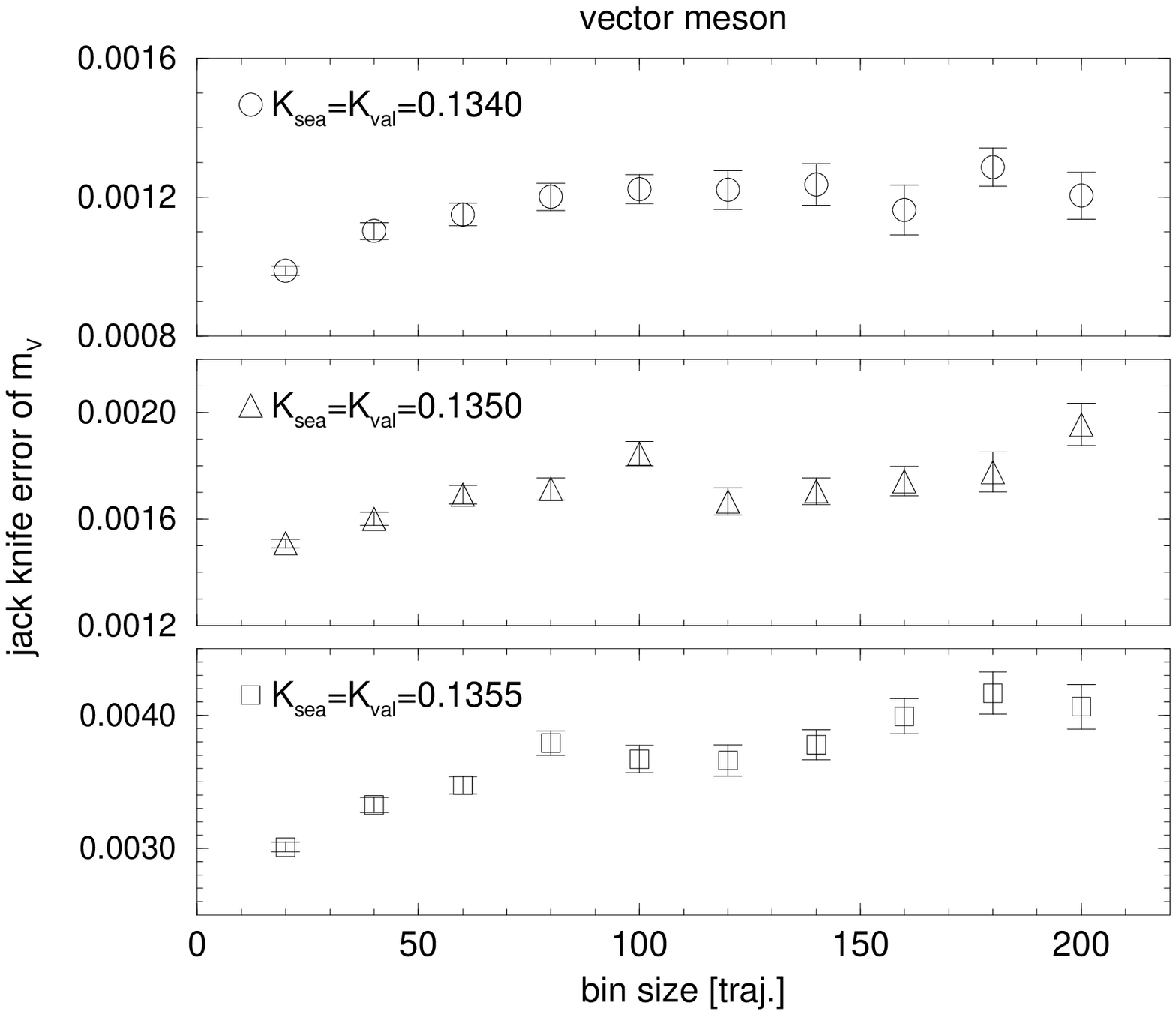}
   \includegraphics[width=70mm]{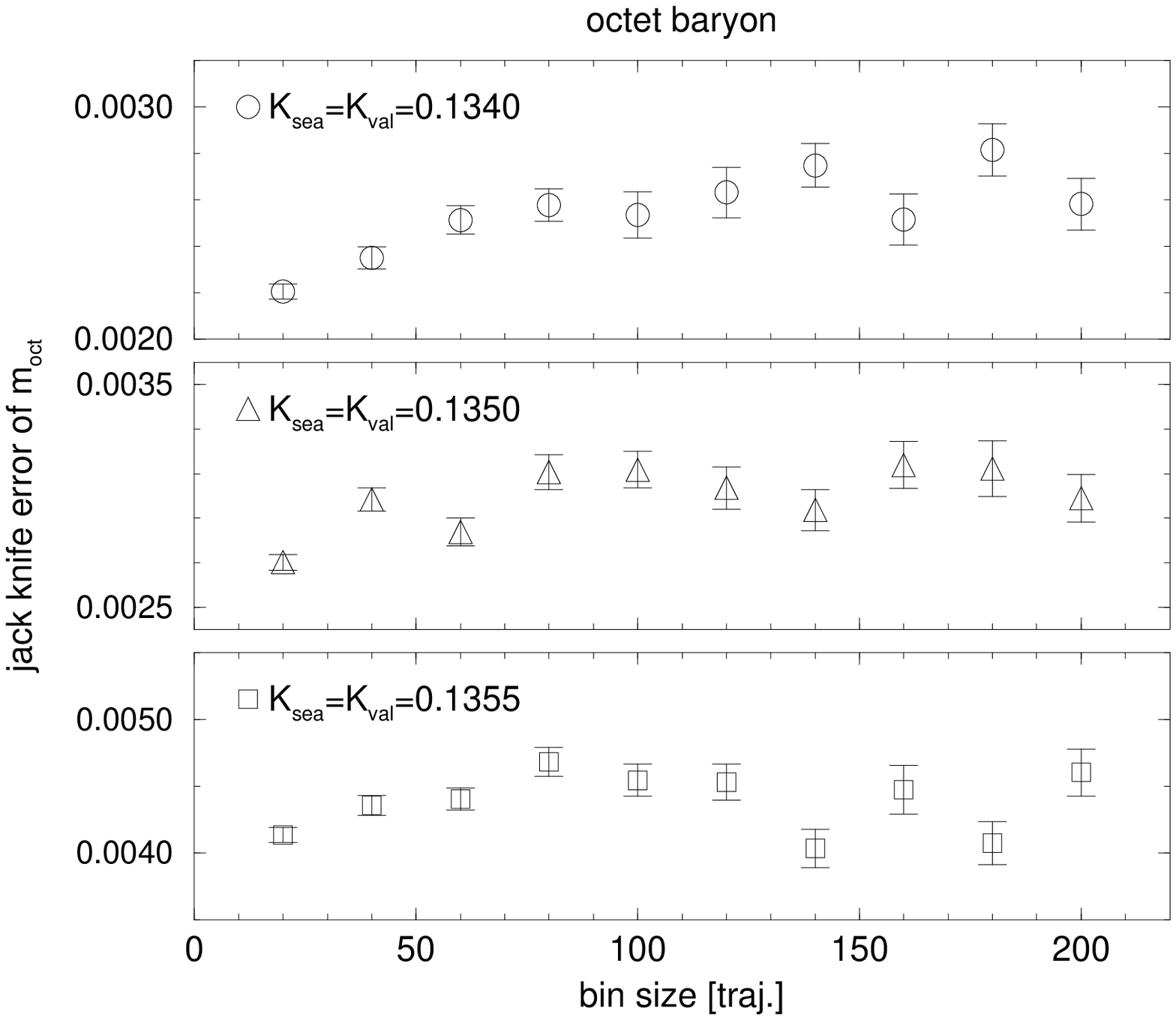}
   \includegraphics[width=70mm]{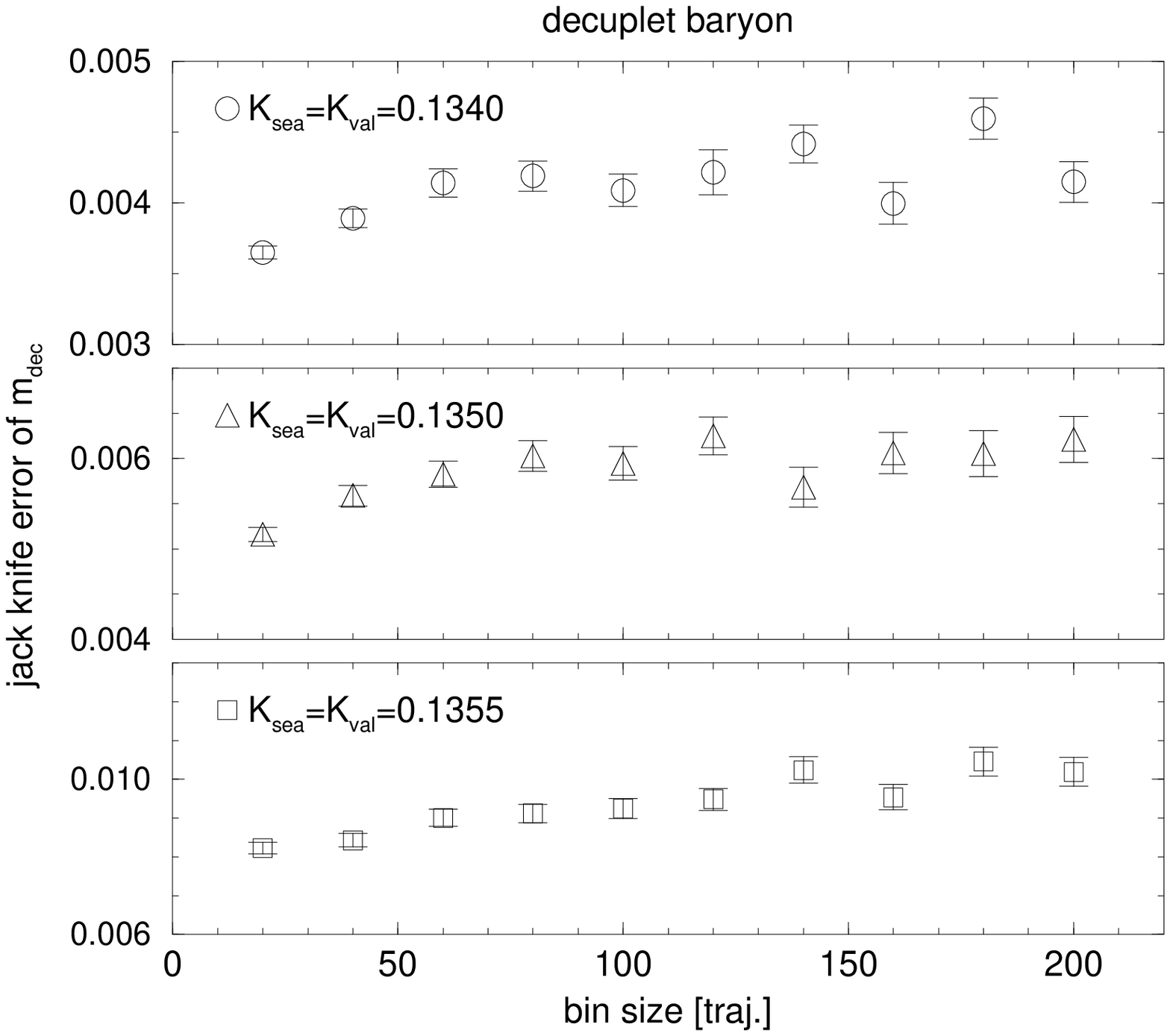}
   \caption
   {
      Bin size dependence of jack-knife error of hadron masses
      on $20^3 \times 48$ lattice in full QCD.
   }
   \label{fig:meas:autocorr:jke:had:full}
\end{figure}

\begin{figure}[htbp]
   \includegraphics[width=70mm]{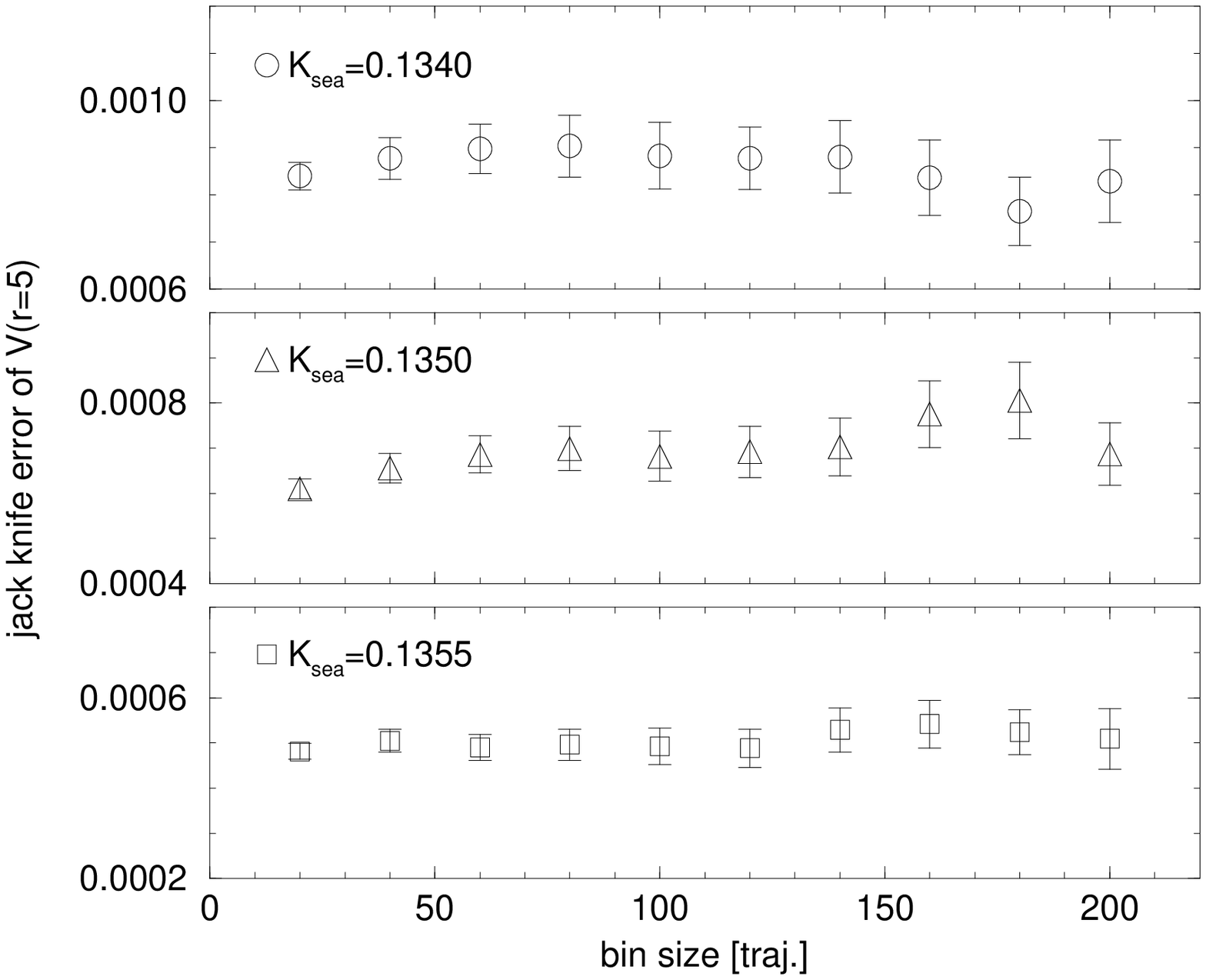}
   \caption
   {
      Bin size dependence of jack-knife error of static potential
      at $r=5$ on $20^3 \times 48$ lattice in full QCD.
   }
   \label{fig:meas:autocorr:jke:pot:full}
\end{figure}

The bin size dependence of the jack-knife error of 
hadron masses and the static potential is plotted 
in Figs.~\ref{fig:meas:autocorr:jke:had:full} and
\ref{fig:meas:autocorr:jke:pot:full}.
We use errors obtained from uncorrelated fits 
because, with large bin sizes, 
the number of bins would not be sufficiently large to determine the
covariance matrix reliably.
For both hadron masses and the static potential,
the jack-knife error reaches its plateau at 
bin size of 50--100 trajectories, 
which is roughly consistent with the above estimate
from $\tau^{\rm cum}_{\calO}$.
The situation is similar on smaller volumes
$16^3 \times 48$ and $12^3 \times 48$.
We therefore take the bin size of 100 trajectories 
in the error analysis in full QCD.

\begin{figure}[htbp]
   \includegraphics[width=70mm]{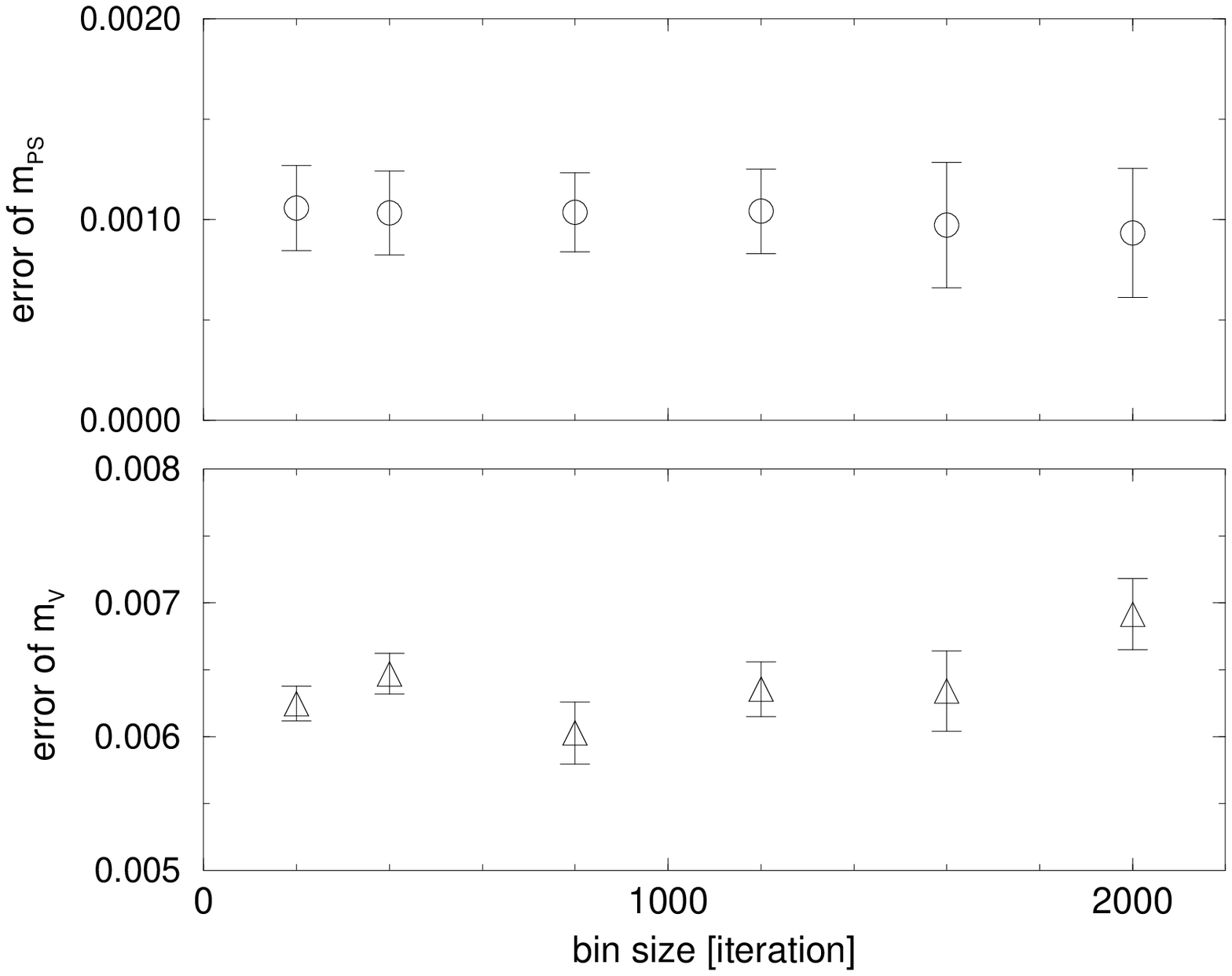}
   \includegraphics[width=70mm]{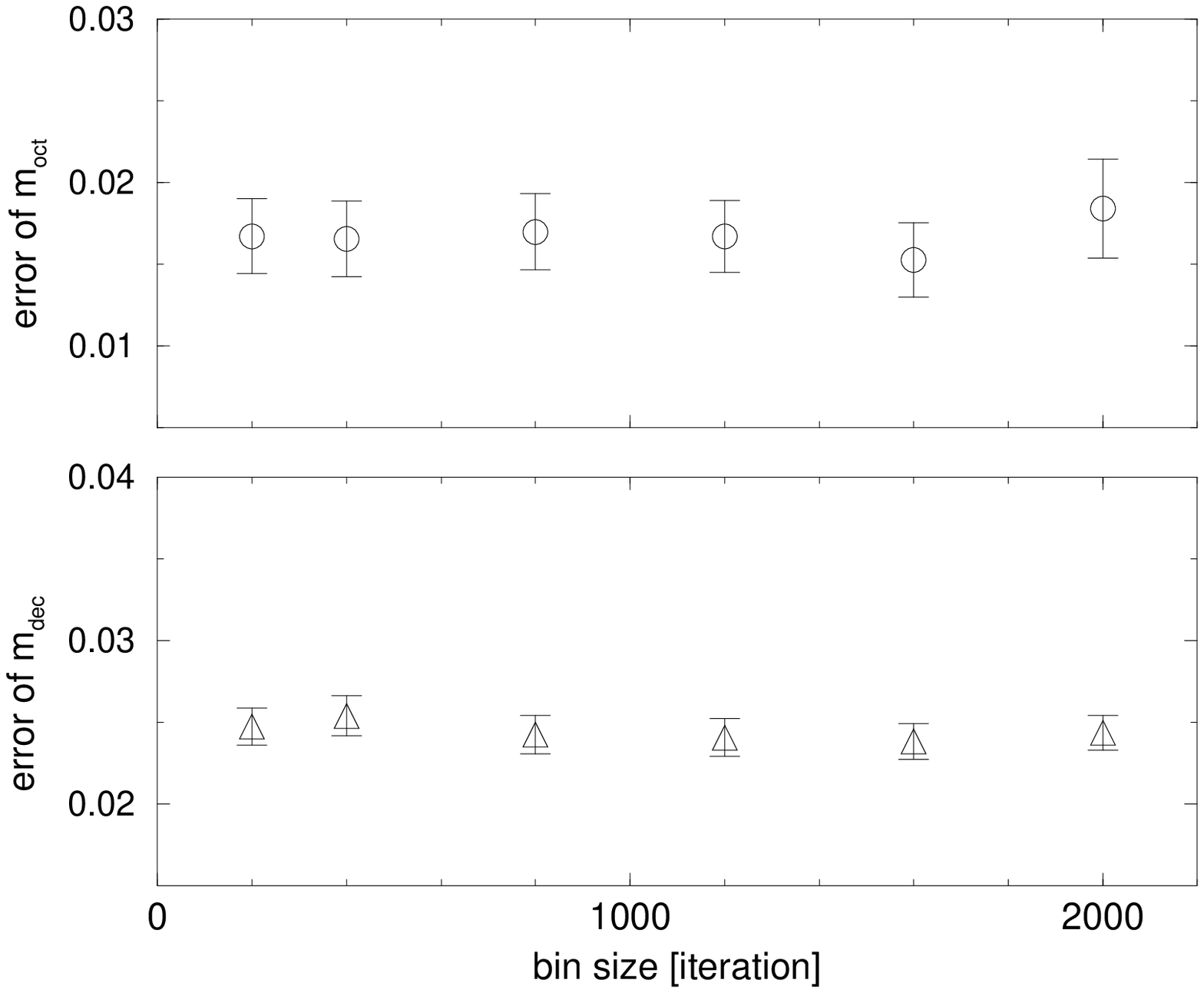}
   \caption
   {
      Bin size dependence of jack-knife error of meson (left figures)
      and baryon masses (right figures) with $K_{\rm val}\!=\!0.13432$,
      which corresponds to $m_{\rm PS,val}/m_{\rm V,val}\!\simeq\!0.6$,
      on $20^3 \times 48$ lattice in quenched QCD.
   }
   \label{fig:meas:autocorr:jke:had:qQCD}
\end{figure}

\begin{figure}[htbp]
   \includegraphics[width=70mm]{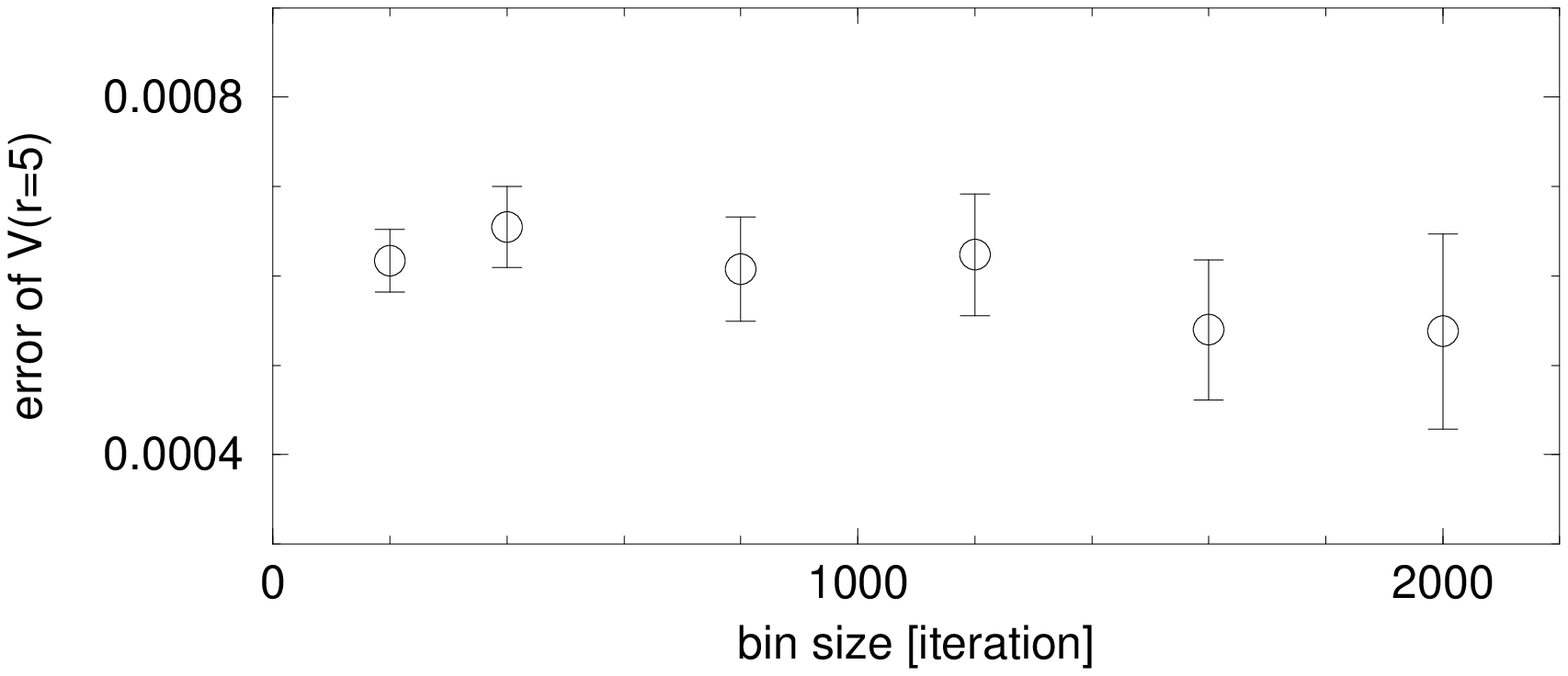}
   \caption
   {
      Bin size dependence of jack-knife error of static potential
      at $r\!=\!5$ on $20^3 \times 48$ lattice in quenched QCD.
   }
   \label{fig:meas:autocorr:jke:pot:qQCD}
\end{figure}

We also investigate the bin size dependence of the jack-knife error 
in quenched QCD. As shown in Figs.~\ref{fig:meas:autocorr:jke:had:qQCD}
and \ref{fig:meas:autocorr:jke:pot:qQCD},
the bin size of 200 iterations is reasonable.


%
%
Another point of interest is 
the sea quark mass dependence of the autocorrelation.
A natural expectation is that 
smaller sea quark mass leads to a larger correlation length 
and hence a longer autocorrelation.
This expectation is supported 
by the CP-PACS observation in Ref.~\cite{Spectrum.Nf2.CP-PACS},
where they used the RG-improved gauge and clover quark actions.
However, our result of $\tau_{\rm plaq}^{\rm cum}$ 
in the S-HMC simulations,
which is determined more precisely than that for the A-HMC
due to the higher statistics,
shows the contrary sea quark mass dependence:
$\tau_{\rm plaq}^{\rm cum}$ decreases as the sea quark mass decreases.
This is consistent 
with the UKQCD's observation in Ref.~\cite{Spectrum.Nf2.UKQCD}.
%
%
We also note that 
$\tau_{\rm plaq}^{\rm cum}$ in our simulations is much larger than 
in the CP-PACS's runs 
particularly at the heaviest sea quark masses
$m_{\rm PS,sea}/m_{\rm V,sea}\!\simeq\!0.8$ ($K_{\rm sea}\!=\!0.1340$).
%
%

In our determination of non-perturbative $c_{\rm SW}$ 
at $\beta\!=\!5.2$~\cite{NPcsw.Nf2.JLQCD}, 
we find that the expectation value of the plaquette 
varies rapidly around $K_{\rm sea}\simeq 0.132$, 
where the plaquette shows the strongest autocorrelation
in the investigated region $K\!\in\![0.100,0.136]$.
Since such a behavior, somewhat similar to a phase transition, 
is not observed at higher $\beta$,
%
%
we consider the unexpected behavior of the plaquette
to be an artifact due to finite lattice spacing.
This artifact is probably absent or well suppressed 
with the CP-PACS's choice of the improved actions.
At sufficiently small lattice spacings, we then expect that 
$\tau^{\rm cum}_{\rm plaq}$ shows 
the natural sea quark mass dependence,
namely larger $\tau^{\rm cum}_{\rm plaq}$ for lighter sea quark masses.
We also expect that, even at $\beta\!=\!5.2$, 
$\tau^{\rm cum}_{\rm plaq}$ will increase 
if the sea quark mass becomes sufficiently small
compared to that corresponding to $K_{\rm sea}\!=\!0.132$.


\section{Finite-size Effects}
\label{sec:FSE}


Finite-size effects (FSE) are one of the major sources of 
systematic errors in lattice calculations. 
Since our largest volume size $\simeq\!(\mbox{1.8~fm})^3$
is still not so large, it is important to check FSE in our data.
We discuss how much FSE is present 
in our data on the largest lattice 
using data on three spatial volumes $12^3 \times 48$, 
$16^3 \times 48$ and $20^3 \times 48$.

\begin{figure}[htbp]
   \includegraphics[width=70mm]{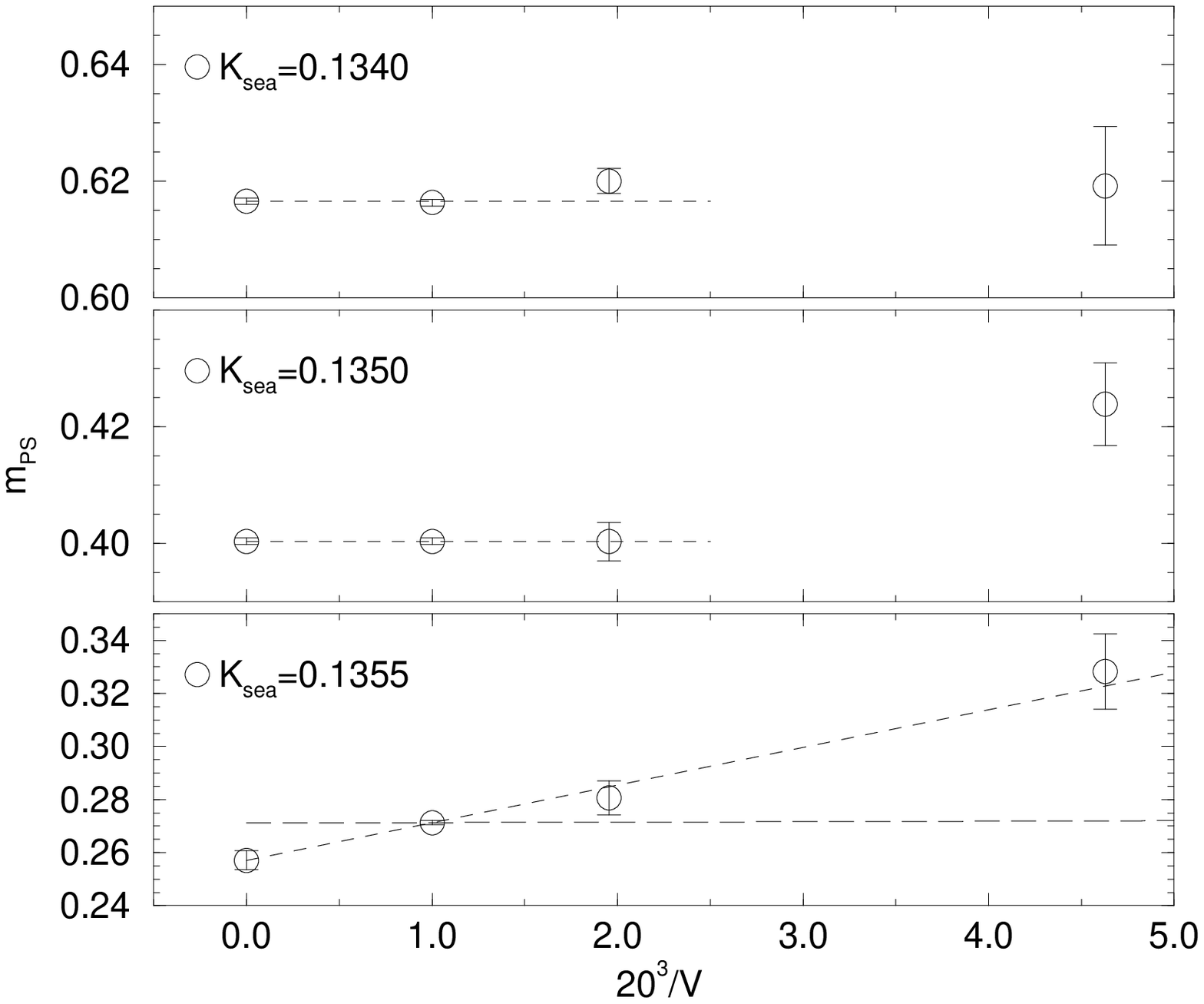}
   \includegraphics[width=70mm]{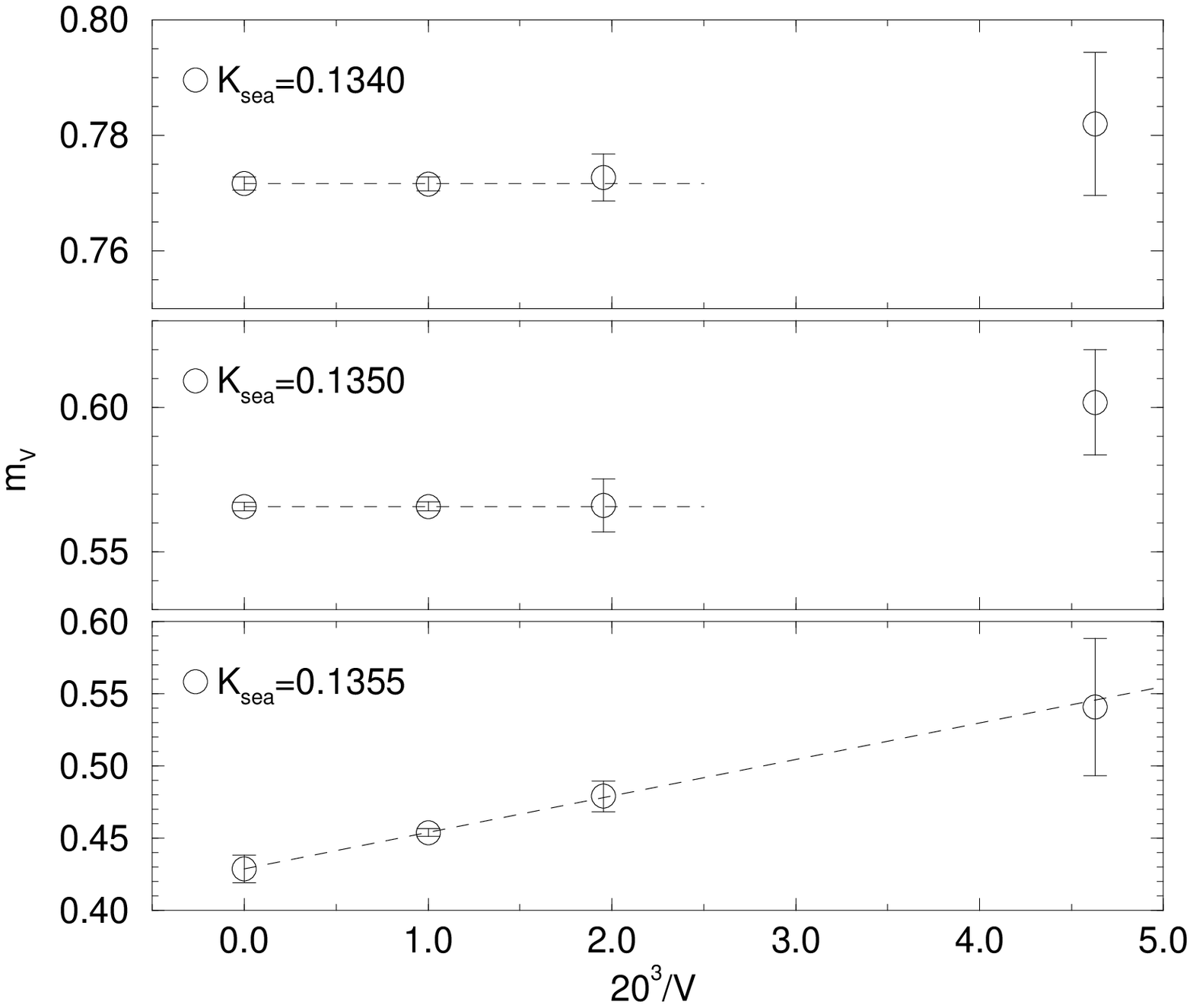}
   \caption
   {
      Diagonal data of 
      PS (left figure) and vector meson masses 
      (right figure) as a function of $20^3/V$.
      Meson masses in the infinite volume limit 
      at $K_{\rm sea}\!=\!0.1340$ and 0.1350 are
      determined by the constant fit to data on two larger volumes, 
      while we assume the linear dependence Eq.~(\ref{eqn:FSE:Vinv})
      at $K_{\rm sea}\!=\!0.1355$.
      We also plot the prediction from the analytic 
      formula\cite{FSE.expL}
      for the PS meson mass at the lightest sea quark mass,
      by long dashed line.
   }
   \label{fig:FSE:Vinv:meson}
\end{figure}

\begin{figure}[htbp]
   \includegraphics[width=70mm]{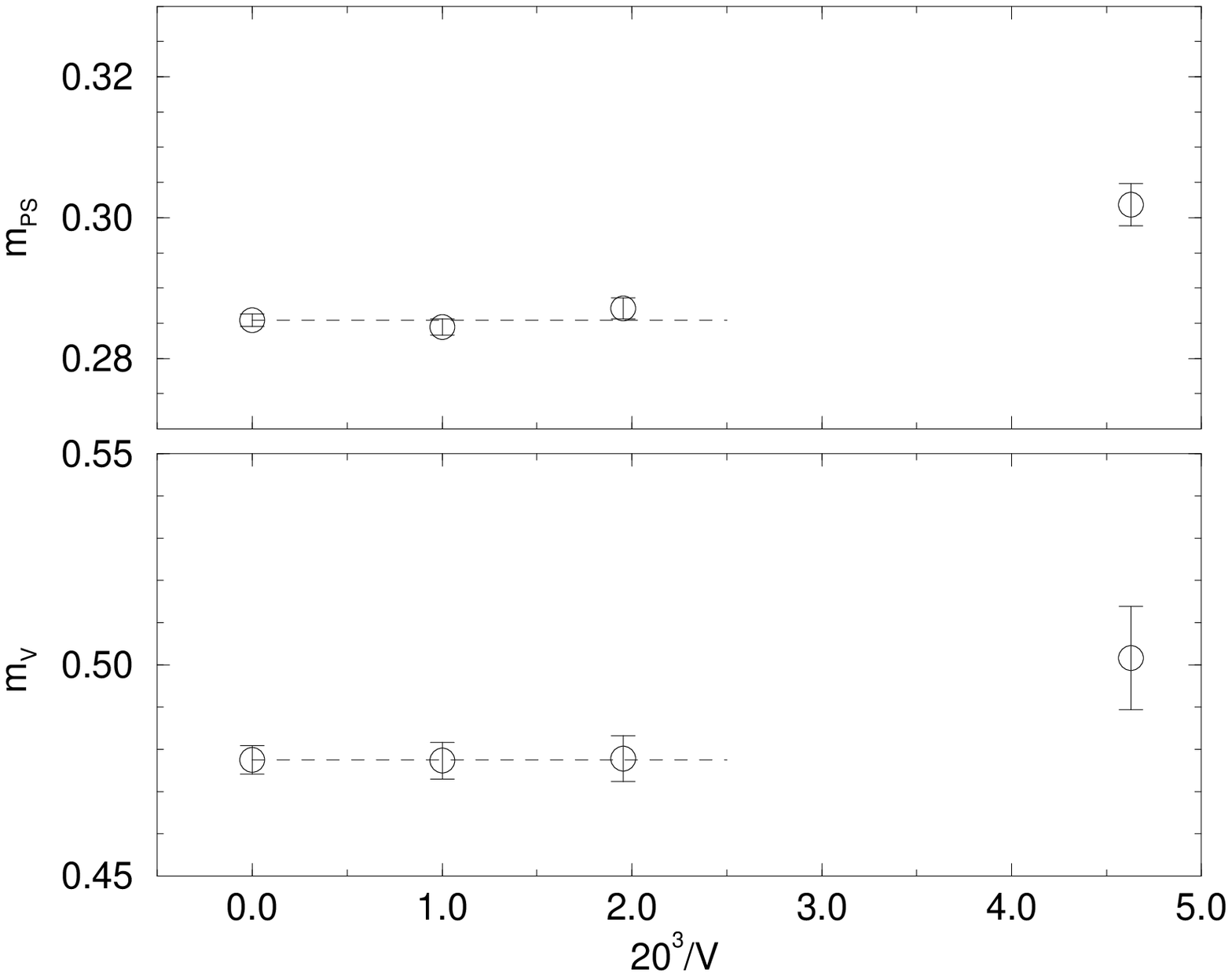}
   \includegraphics[width=70mm]{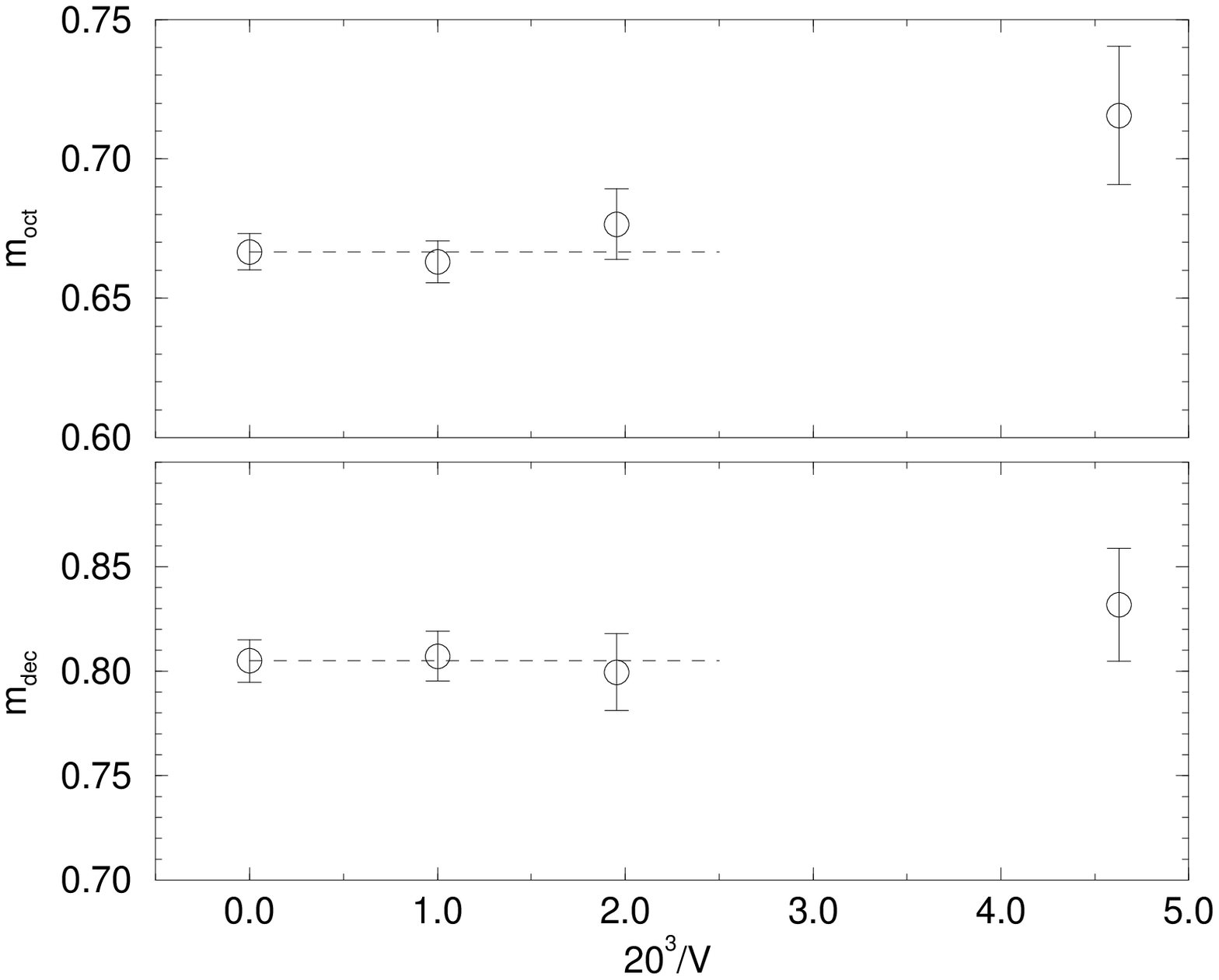}
   \caption
   {
      Volume size dependence of meson (left figure) and baryon masses
      (right figure) at $K_{\rm val}\!=\!0.13432$ 
      in quenched simulations.
   }
   \label{fig:FSE:Vinv:quenched}
\end{figure}

In Figs.~\ref{fig:FSE:Vinv:meson}\,--\,\ref{fig:FSE:Vinv:quenched}, 
we plot diagonal data of hadron masses as a function of 
the spatial volume inverse.
%
For $m_{\rm PS,sea}/m_{\rm V,sea}\!\gtrsim\!0.7$,
including the quenched case,
hadron masses obtained on the $16^3 \times 48$ and $20^3 \times 48$
lattices are consistent with each other within two standard deviations.
%
%
%
On the other hand,
hadron masses 
decrease monotonously up to $V\!=\!20^3$ 
at the lightest sea quark mass corresponding to 
$m_{\rm PS,sea}/m_{\rm V,sea}\!\simeq\!0.6$.

\begin{figure}[htbp]
   \includegraphics[width=70mm]{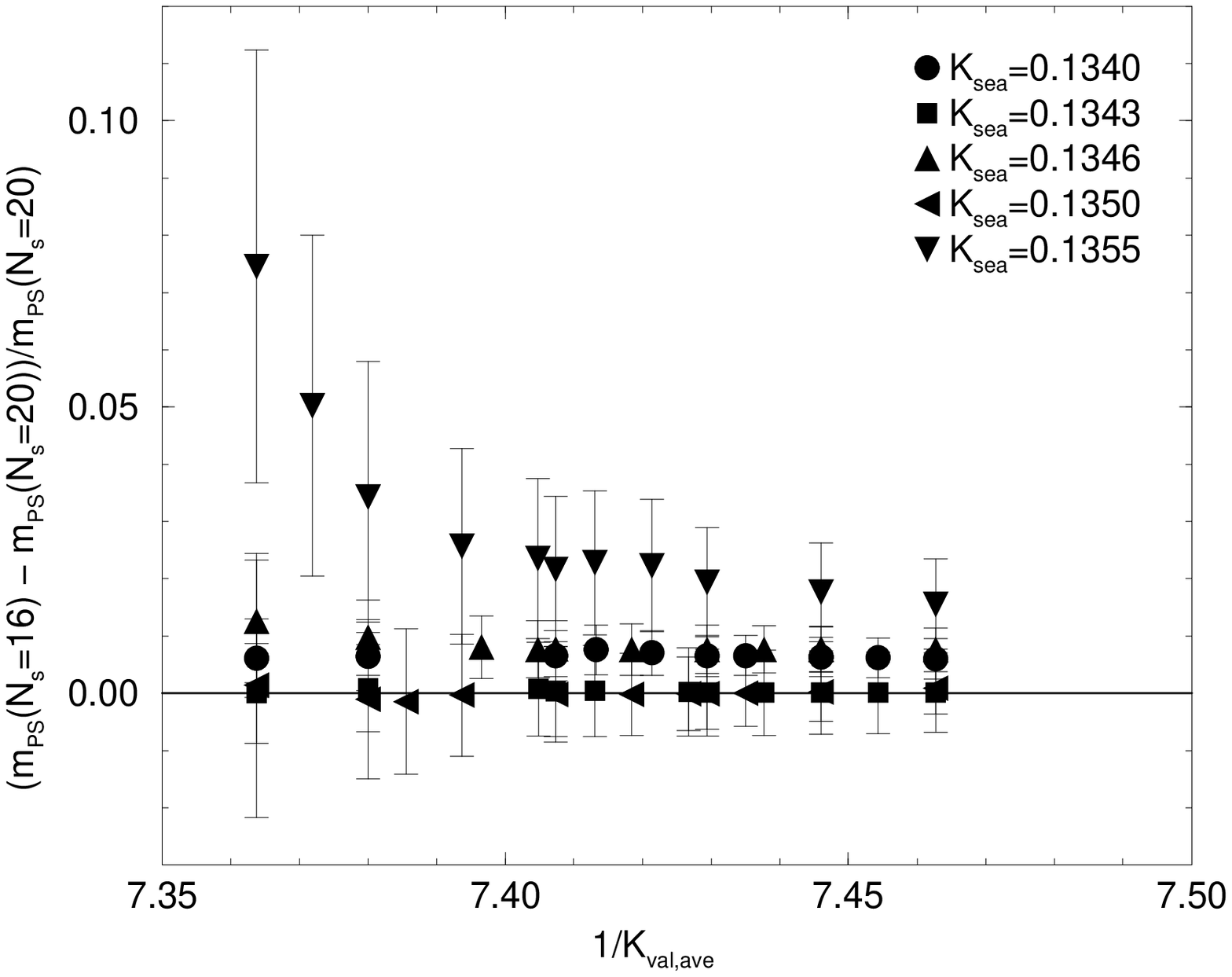}
   \includegraphics[width=70mm]{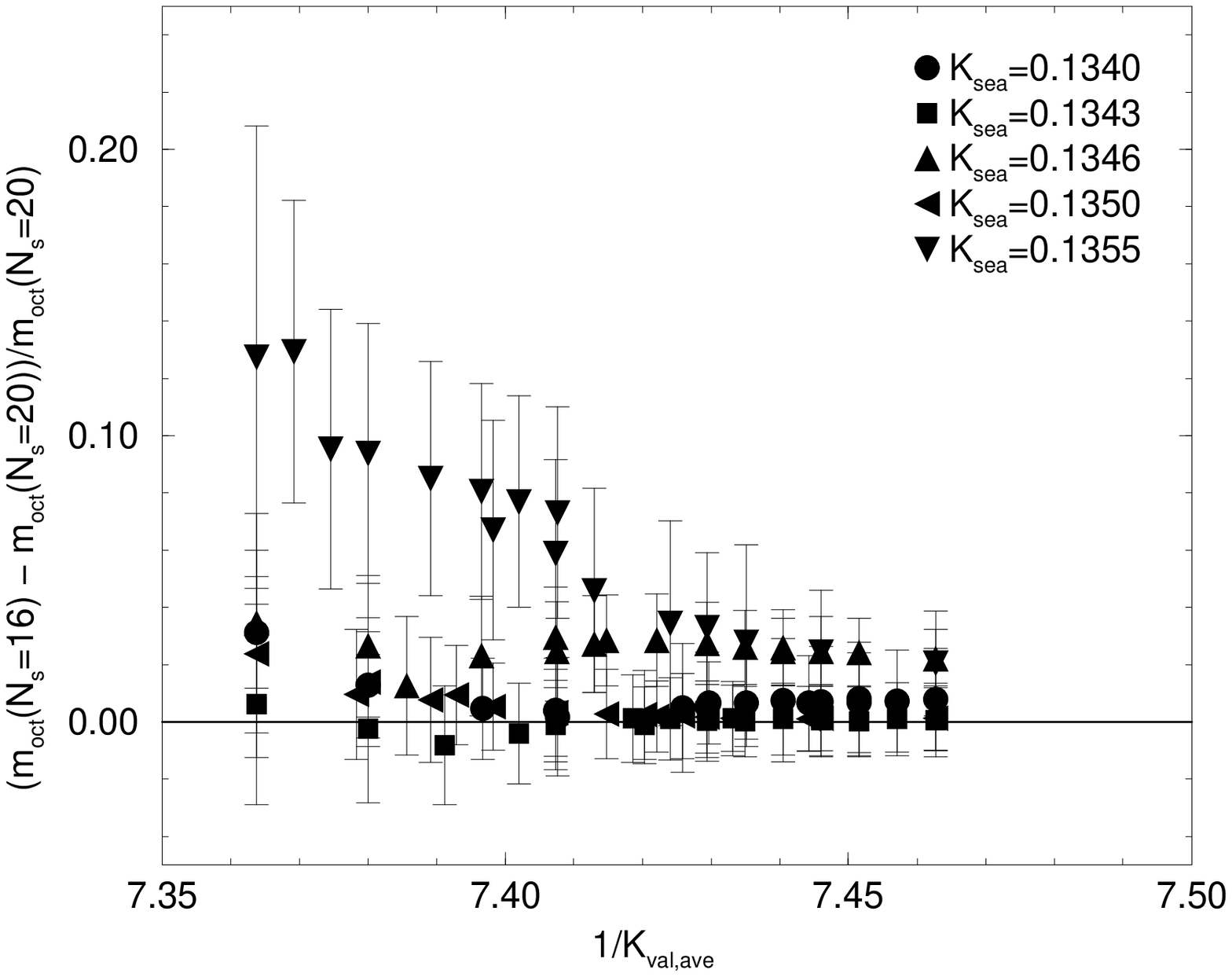}
   \caption
   {
      Valence quark mass dependence of  
      relative difference between hadron masses 
      measured on $16^3$ and $20^3$ lattices. 
      Left and right figure show
      data for the PS meson and the octet baryon, respectively.
      We define $K_{\rm val,ave}$ by
      $1/K_{\rm val,ave}\!=\!(1/K_{\rm val,1}+1/K_{\rm val,2})/2$
      for meson masses $m(K_{\rm sea};K_{\rm val,1},K_{\rm val,2})$ 
      and 
      $1/K_{\rm val,ave}\!=\!(1/K_{\rm val,1}+1/K_{\rm val,2}
                                             +1/K_{\rm val,3})/3$
      for baryon masses 
      $m(K_{\rm sea};K_{\rm val,1},K_{\rm val,2},K_{\rm val,3})$.
   }
   \label{fig:FSE:mval_dep}
\end{figure}

The magnitude of FSE also depends on the valence quark mass.
Figure~\ref{fig:FSE:mval_dep} shows the valence quark mass
dependence of the relative mass shift between the two larger lattices
for PS mesons and octet baryons.
We observe that, except at the lightest sea quark mass,
the mass shift is at most a few percent level
in the whole range of the simulated valence quark mass.
The situation is similar for vector mesons and decuplet baryons.
Therefore, we conclude that the size of FSE on our largest lattice 
is small over our range of valence quark masses 
down to the second lightest sea quark mass.

The mass shift is non-negligible at the lightest sea quark mass.
While the magnitude is of the order of a few percent 
for the heaviest valence quarks,
it clearly increases as the valence quark mass decreases.

We consider that 
the observed FSE is caused by valence quarks
wrapping around the lattice in spatial directions
(namely squeezing of hadrons into the small box)
rather than wrapping of virtual pions.
As shown in Fig.~\ref{fig:FSE:Vinv:meson},
the magnitude of FSE caused by the effects of 
virtual pions (long dashed line)~\cite{FSE.expL} given by 
$m_{\rm PS}(L) \! - \! m_{{\rm PS}}(L\!=\!\infty) \sim 
 \exp \left[ - m_{\rm PS}(L\!=\!\infty)L \right]$
with $L\!=\!N_s a$ is too small compared to observed effects.
%
%
%

A qualitative understanding of the observed FSE is as follows. 
The wrapping of valence quarks is suppressed 
by the center $Z(3)$ symmetry in quenched QCD~\cite{FSE.Z3}.
In full QCD, $Z(3)$ symmetry is broken 
by the wrapping of sea quarks in the spatial directions,
whose magnitude increases toward lighter sea quark.
A possible reason why FSE is significant only at our 
lightest sea quark mass would be that 
the $Z(3)$ breaking turns on rather quickly 
around the lightest sea quarks.
%

\begin{figure}[htbp]
   \includegraphics[width=70mm]{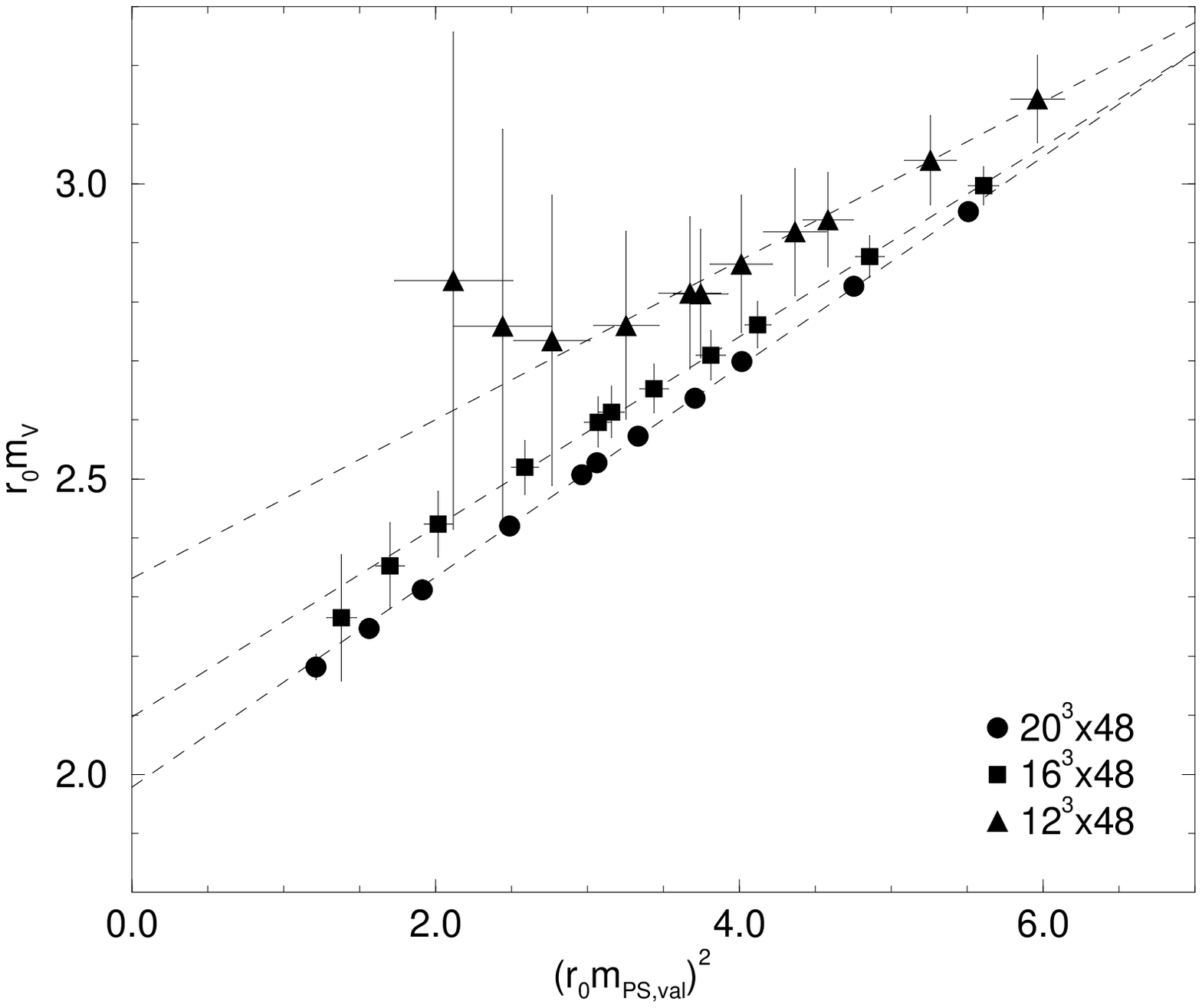}
   \includegraphics[width=70mm]{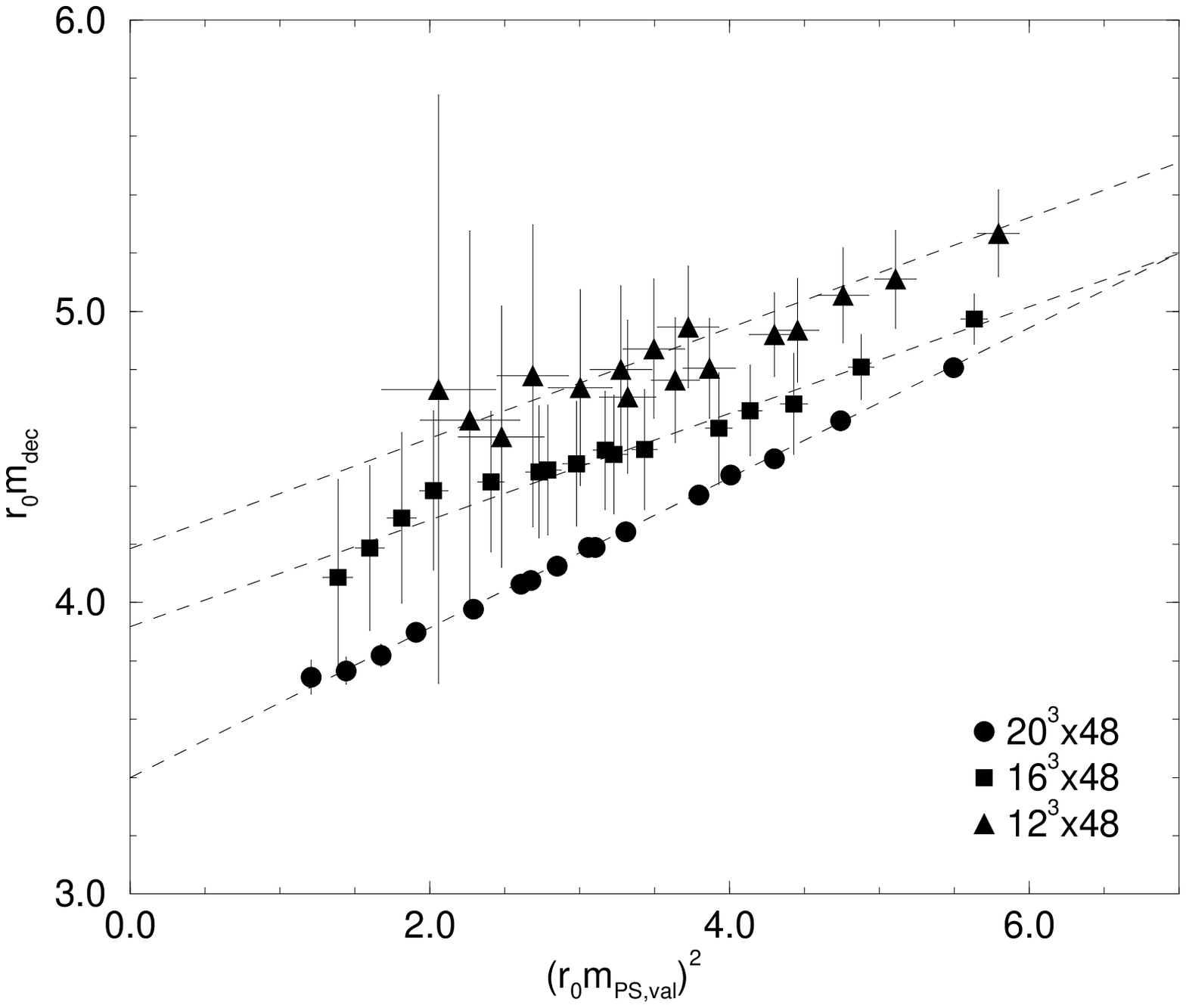}
   \caption
   {
      Vector meson (left figure) and decuplet baryon masses
      (right figure) at $K_{\rm sea}\!=\!0.1355$ 
      on three spatial volumes as a function of PS meson mass squared.
      Linear fit curve to each data set is shown as a guide for eyes.
   }
   \label{fig:FSE:m_vs_mPS2}
\end{figure}

The enhancement of FSE toward the lighter valence quarks
leads to a decrease of the slope 
$dm_{\rm had}/dm_{\rm PS}^2$ in Fig.~\ref{fig:FSE:m_vs_mPS2}
and, hence, underestimation of hadron mass splittings,
such as the $K^*$-$K$ hyperfine splitting.
The mass splittings are expected to be increased
by sea quark effects, 
since these are underestimated in quenched QCD
as well established in Ref.~\cite{Spectrum.Nf0.CP-PACS}.
Therefore, FSE makes sea quark effects less clear.
It is crucial to check how large FSE is in our hadron mass
data at the lightest sea quark mass on the largest lattice.

\begin{figure}[htbp]
   \includegraphics[width=70mm]{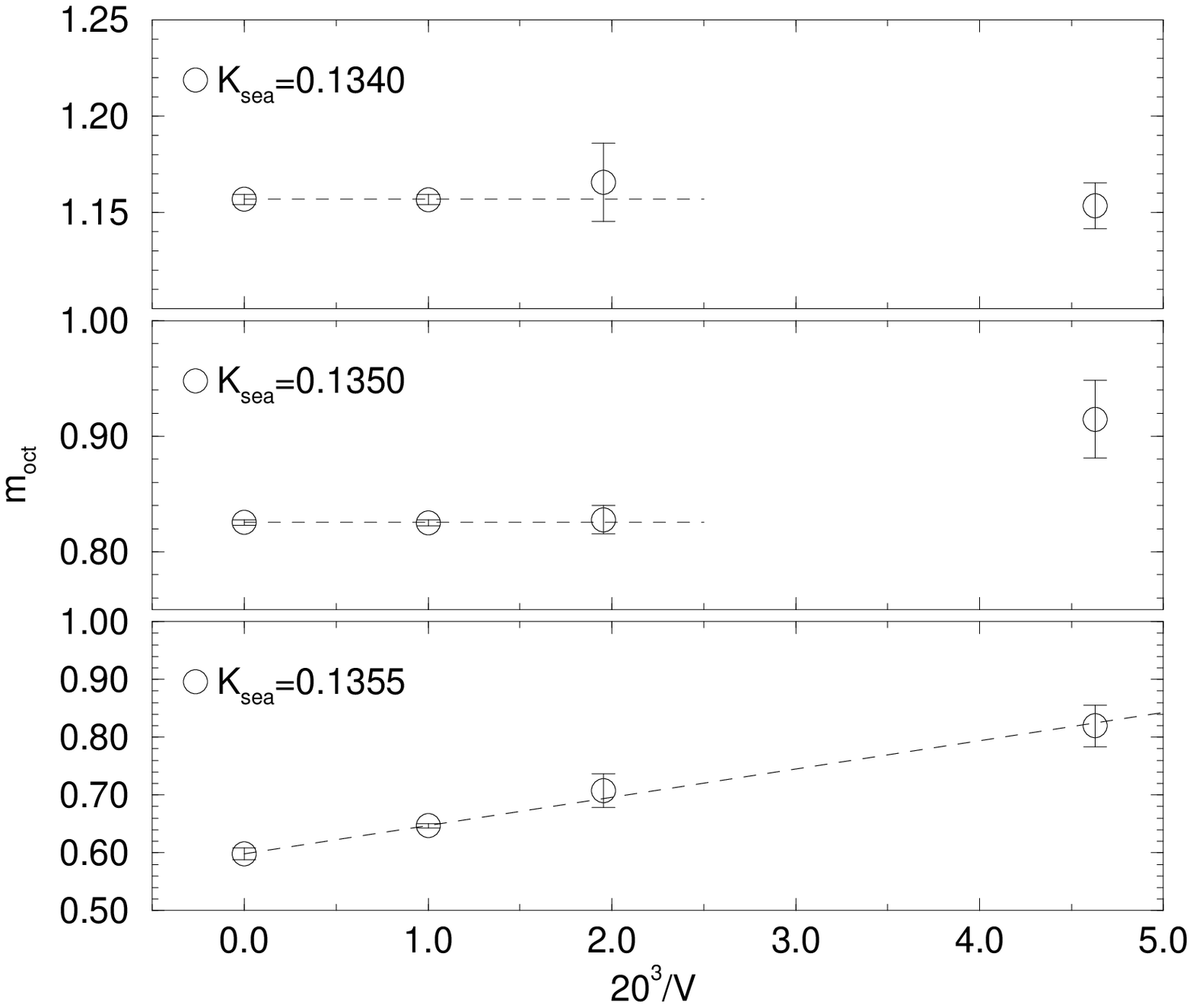}
   \includegraphics[width=70mm]{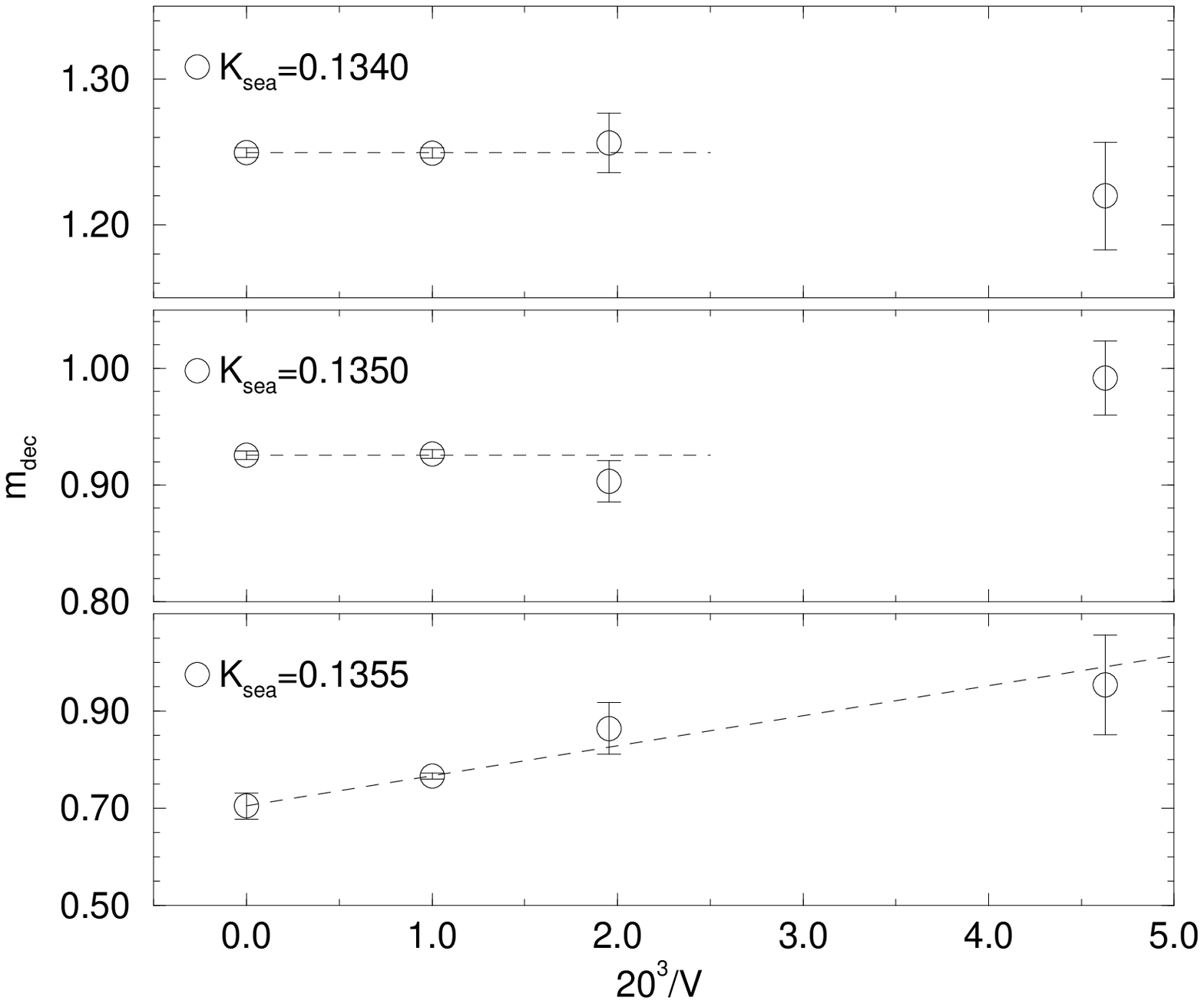}
   \caption
   {
      Diagonal data of 
      octet (left figure) and decuplet baryon masses 
      (right figure) as a function of $20^3/V$.
   }
   \label{fig:FSE:Vinv:baryon}
\end{figure}

\begin{figure}[htbp]
   \includegraphics[width=70mm]{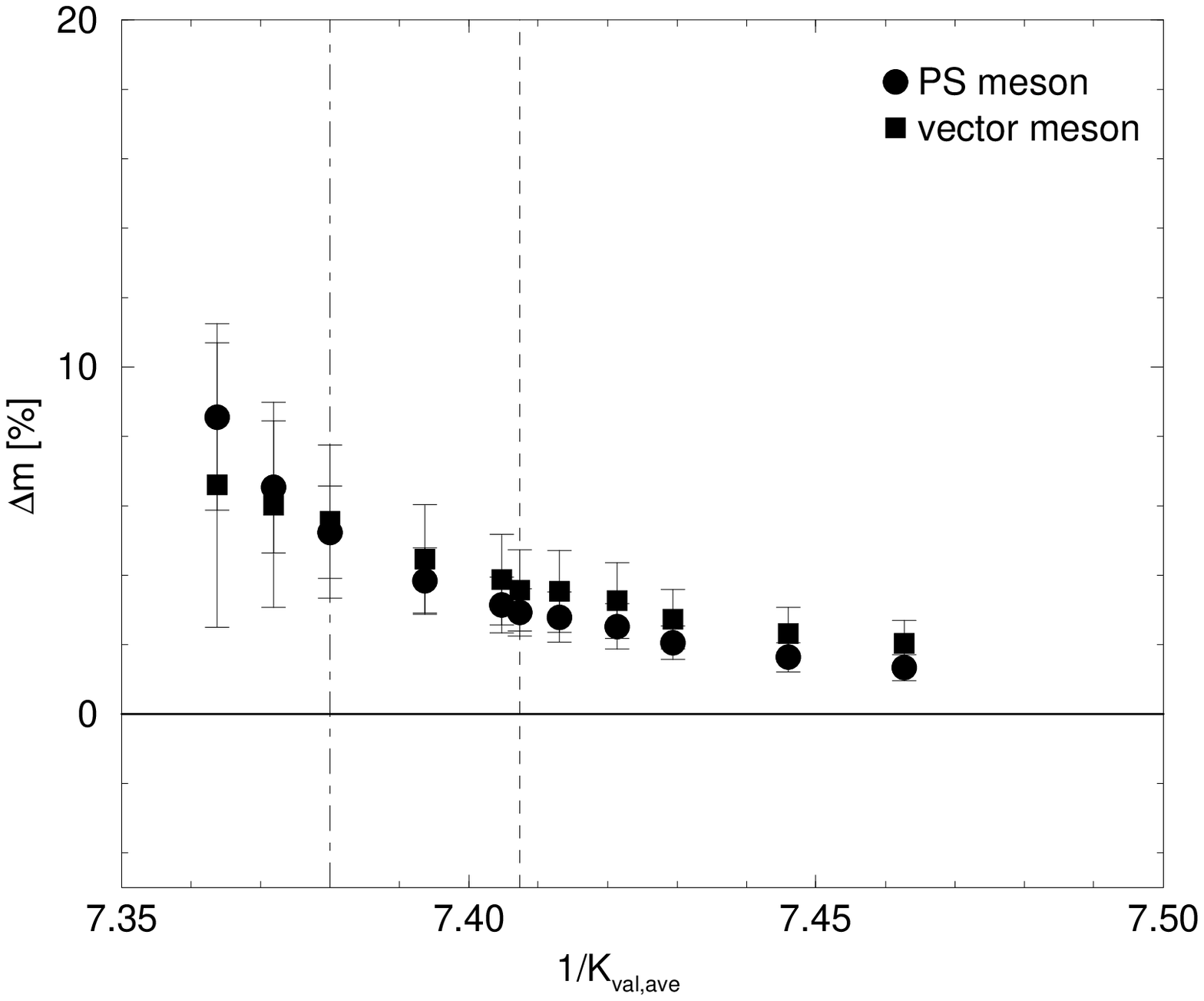}
   \includegraphics[width=70mm]{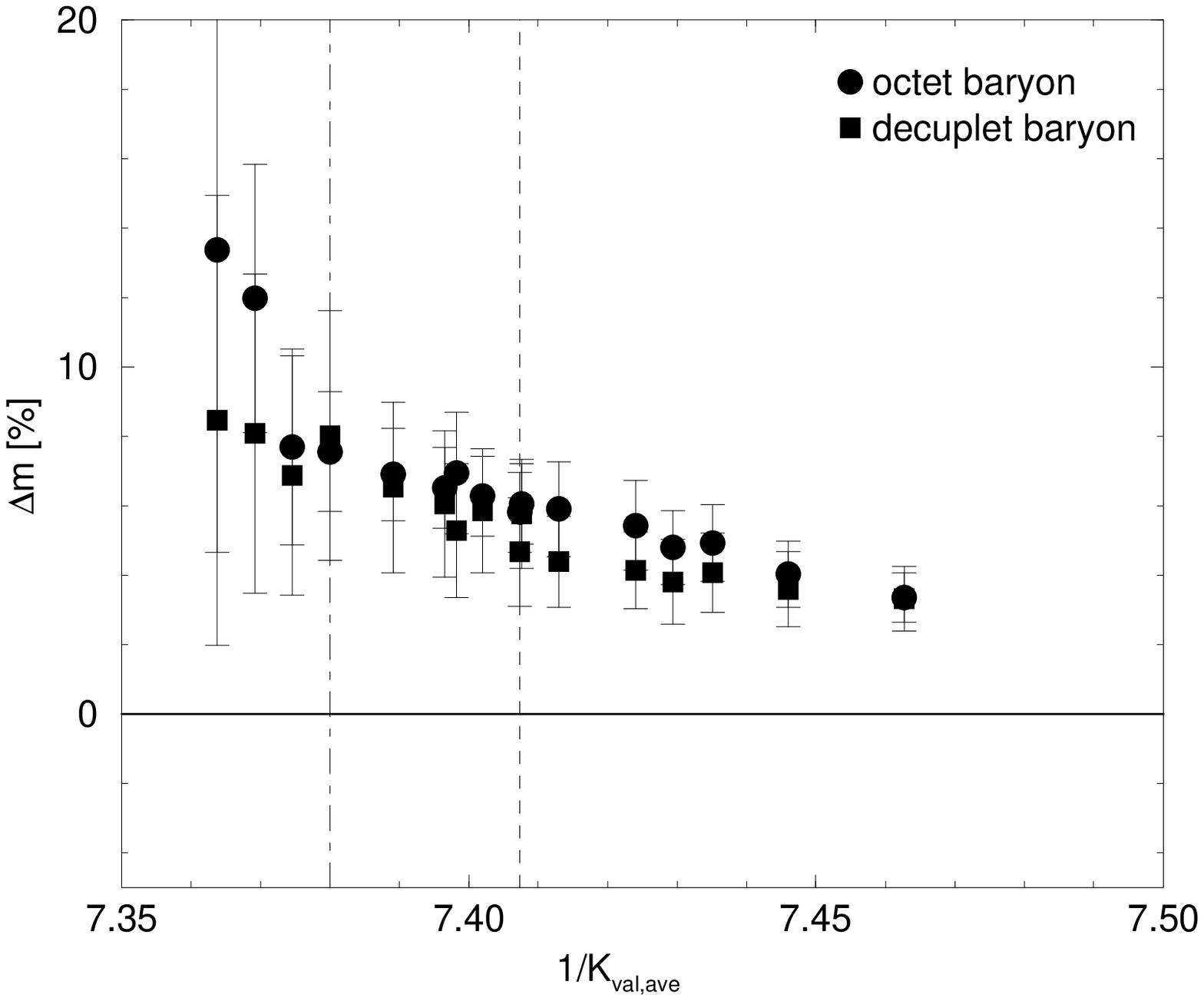}
   \caption
   {
      Relative size of FSE for meson (left figure) and baryon masses 
      (right figure) at $K_{\rm sea}\!=\!0.1355$ on $20^3 \times 48$.
      Dot-dashed and dashed lines show the location where 
      $K_{\rm val,ave}\!=\!K_{\rm sea}$ and 0.1350, respectively.
      The latter roughly corresponds to the strange quark mass.
   }
   \label{fig:FSE:dm}
\end{figure}

Figures~\ref{fig:FSE:Vinv:meson} and \ref{fig:FSE:Vinv:baryon}
show that 
the volume dependence of our data is well described by a power law
\bea
   m_{\rm had}(L) = m_{\rm had}(L\!=\!\infty) + c/L^3
   \label{eqn:FSE:Vinv}
\eea
as found in Ref.~\cite{FSE.Vinv} using the KS fermion.
%
%
The relative size of FSE on the largest lattice
\bea
   \Delta m = \frac{m(L\!=\!20a)\!-\!m_{\rm had}(L\!=\!\infty)} 
                   {m(L\!=\!20a)}
\eea
is estimated from this ansatz
and is plotted in Fig.~\ref{fig:FSE:dm}.
We find that, for PS and vector mesons, 
$\Delta m$ is about 5~\% for diagonal data
and is reduced to a few percent 
at $K_{\rm val}\!=\!0.1350$, which roughly corresponds to 
the strange quark mass $m_s$.
It is expected, however, that 
the volume dependence~(\ref{eqn:FSE:Vinv}) turns into
a milder form $\exp [ - m_{\rm PS} L]$
for sufficiently large volumes.
The actual size of FSE should be smaller than the above 
estimation, say, a few percent.
Since this is smaller than the typical size of quenching errors, 
which is 5\,--\,10\%,
we consider that the examination of sea quark effects 
is feasible in the meson sector,
particularly in strange meson masses.


Finite-size effects are more pronounced for baryon masses
as observed in Fig.~\ref{fig:FSE:dm}.
For diagonal data, 
$\Delta m$ is roughly comparable with 
typical quenching errors in the baryon spectrum
of the order of 5\,--\,10~\%.
%
Sea quark effects in the light baryon masses,
such as $m_N$ and $m_\Delta$,
may become unclear by the contamination of FSE.
We note, however, that $\Delta m$ decreases 
for heavier valence quark masses.
The examination of sea quark effects 
becomes more feasible for strange baryon masses
like $m_{\Xi}$ and $m_{\Omega}$.

\begin{figure}[htbp]
   \includegraphics[width=70mm]{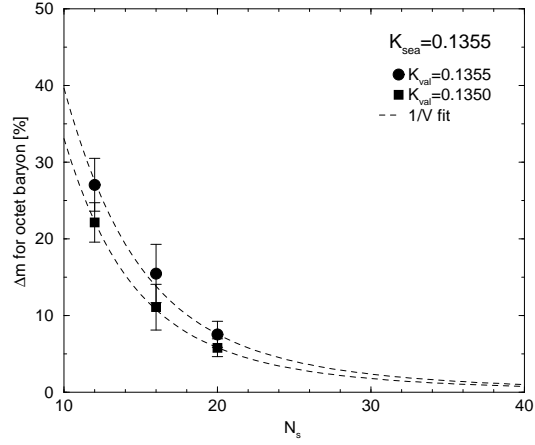}
   \caption
   {
      Magnitude of FSE on octet baryon mass
      at $K_{\rm sea}\!=\!0.1355$
      as a function of spatial linear extent in lattice units.
      Circles are results for diagonal data, while squares
      represent those at $K_{\rm val}\!=\!0.1350$. 
   }
   \label{fig:FSE:dmO_vs_L}
\end{figure}

Figure~\ref{fig:FSE:dmO_vs_L} shows 
$\Delta m$ for the diagonal data of the octet baryon mass
as a function of $N_s$.
We find that the size of $N_s \!\approx\! 30$, 
which corresponds to $L\!\approx\!2.7$~fm,
is required to suppress $\Delta m$ to a few percent level.
The required size becomes slightly smaller, $L\!\approx\!2.4$~fm,
for the valence quark mass around $m_s$.
These sizes are larger than our largest spatial lattice size 
$L\!\simeq\!1.8$~fm.
Further simulations on such large lattices will be needed 
to obtain a definite conclusion on sea quark effects 
in the baryon spectrum.

Tables~\ref{tab:meas:pot:r_0:full} and \ref{tab:meas:pot:r_0:qQCD}
show that FSE in $r_0$ are much smaller than in hadron masses.
While the central value on $20^3$ is 
systematically higher than that on $16^3$
in both full and quenched QCD,
the difference is about 1~\% and not significant with the accuracy of our data.
The size of FSE is small, namely a few percent level,
even on $12^3$.
Therefore 
FSE in $r_0$ on the largest lattice can be safely neglected 
both in full and quenched QCD.


\section{Chiral Extrapolation}
\label{sec:chiralfit}

The hadron spectrum in full QCD is calculated using hadron masses
measured on the $20^3 \times 48$ lattice.
This requires a parametrization of the mass data
as a function of sea and valence quark masses
in order to extrapolate (up-down) or interpolate (strange) quark masses
to their physical values.
We make this parametrization by combined fits to 
masses of a given hadron at all sea quark masses.
%
%
We test the following two methods for the combined fit:
\renewcommand{\labelenumi}{\Alph{enumi}}
\begin{enumerate}
   \item
   The effective lattice spacing, determined from $r_0$ for instance, 
   may vary as a function of the sea quark mass. In order to separate
   this effect from the physical quark mass dependence, 
   we carry out the chiral extrapolation using dimensionless quantities
   such as $(r_0(K_{\rm sea}) 
   m_{\rm had}(K_{\rm sea}; K_{\rm val,1}, K_{\rm val,2}))$,
   where $m_{\rm had}(K_{\rm sea}; K_{\rm val,1}, K_{\rm val,2})$
   represents the measured hadron mass composed of 
   valence quark masses corresponding to $K_{\rm val,1}$ and 
   $K_{\rm val,2}$ 
   on the gauge configurations generated at $K_{\rm sea}$.
   We refer to this way as ``method-A''.

   \item 
   It is also possible to fit hadron masses in lattice units,
   as was done by the SESAM~\cite{Spectrum.Nf2.SESAM} and 
   the CP-PACS~\cite{Spectrum.Nf2.CP-PACS} Collaborations.
   We call this ``method-B''.

\end{enumerate}
\renewcommand{\labelenumi}{\arabic{enumi}}
A detailed description of the two methods
will be given in Secs.~\ref{sec:chiralfit:w_r0} and 
~\ref{sec:chiralfit:wo_r0}.
They should yield a consistent hadron spectrum, 
since fit forms in method-B can be reproduced from those in method-A 
by expanding $r_0$ as a function of sea quark mass.
This consistency is examined in Sec.~\ref{sec:chiralfit:wo_r0}.

\subsection{chiral extrapolation using $r_0$ (method-A)}
\label{sec:chiralfit:w_r0}

\begin{figure}[htbp]
   \includegraphics[width=70mm]{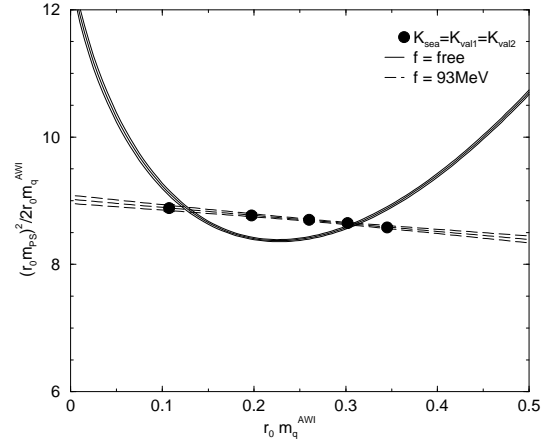}
   \caption
   {
      Test of logarithmic singularity in quark mass dependence 
      of PS meson mass.
      We use the quark mass defined through 
      the axial vector Ward identity in this plot.
      Solid and dashed lines are fit curves of 
      Eq.~(\ref{eqn:chiralfit:w_r0:p4_ChPT_mPS2}) 
      assuming $f$ to be a free parameter or fixed to 
      the experimental value.
   }
   \label{fig:chiralfit:w_r0:chirallog}
\end{figure}

Chiral perturbation theory (ChPT)~\cite{ChPT} provides 
a guide to obtain a controlled chiral limit
of hadron mass data.
For the quark mass dependence of diagonal data of the PS meson mass,
ChPT predicts the presence of logarithmic singularities.
At the one-loop level, the ChPT prediction reads
\bea
   \frac{m_{\rm PS}^2}{2 B_0 m_q} 
   & = &
   1 + \frac{1}{N_f} y \ln \left[ y \right]  
     + A y
   \label{eqn:chiralfit:w_r0:p4_ChPT_mPS2}
\eea
where $y \!=\! 2 B_0 m_q /( 4 \pi f)^2$ and 
$A$ is a linear combination of the low energy constants $\alpha_i$
of ChPT Lagrangian : 
$A = \left( 2\alpha_8 - \alpha_5 \right)
   + N_f \left( 2\alpha_6 - \alpha_4 \right)$.
The mass ratio on the left hand-side, $m_{\rm PS}^2/m_{\rm q}$,
is plotted as a function of $m_q \propto y$
in Fig.~\ref{fig:chiralfit:w_r0:chirallog}.
As we already reported
\cite{Spectrum.Nf2.JLQCD.lat01,Spectrum.Nf2.JLQCD.lat02.Hashimoto},
our data show no hint of the curvature predicted 
by the chiral logarithm.
The fit of Eq.~(\ref{eqn:chiralfit:w_r0:p4_ChPT_mPS2}),
assuming $f$ to be a free parameter,
gives $f \sim 6$~GeV,
which is much larger than its experimental value 93~MeV.
%
%
On the other hand,
the fit gives an unacceptably large $\chi^2/{\rm dof} = O(100)$,
if we fix $f$ to the experimental value.

A similar test using formulae from partially quenched ChPT 
(PQChPT)~\cite{PQChPT} also shows that 
the coefficient of the chiral logarithm term
obtained from our data is much smaller than the prediction
from PQChPT~\cite{Spectrum.Nf2.JLQCD.lat01,Spectrum.Nf2.JLQCD.lat02.Hashimoto,ChPT_test.Nf2.JLQCD}.
A possible reason for the absence of the chiral logarithm is 
that the sea quark mass in our simulations is still too large
and higher order corrections of ChPT should be included
to describe the data.
%


In this study, therefore,
we use simple polynomial fitting forms 
in terms of the quark mass for the chiral extrapolation.
The systematic error due to the chiral extrapolation
is estimated by testing several different polynomial forms.
However, 
the inconsistency between our data and ChPT suggests that 
the extrapolation may have larger uncertainty than this estimation.
This point will be examined in detail in a separate paper
\cite{ChPT_test.Nf2.JLQCD}.

\begin{figure}[htbp]
   \includegraphics[width=70mm]{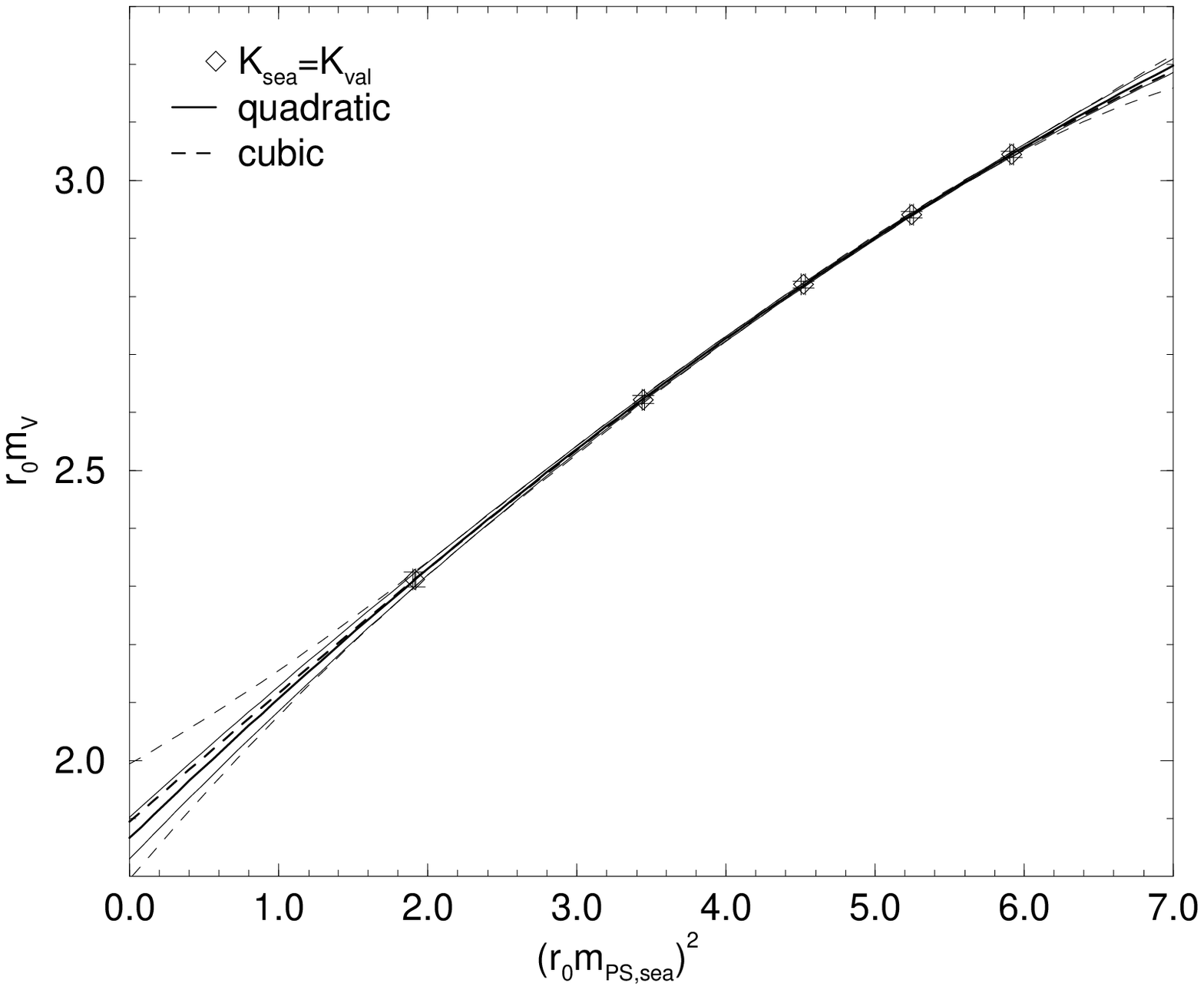}
   \caption
   {
      Comparison of quadratic and cubic diagonal fits in method-A.
   }
   \label{fig:chiralfit:w_r0:diagonal}
\end{figure}

%
%
Since the sea quark mass in our simulations is not so small
as discussed above, it is important to 
check the convergence property of the polynomial expansion 
of hadron masses in our range of the sea quark mass.
We carry out both quadratic and cubic chiral fits to 
diagonal data of PS and vector meson masses:
\bea
   ({r_0}(K_{\rm sea}) \, 
     m_{\rm PS}(K_{\rm sea};K_{\rm sea},K_{\rm sea}))^2
   & = &
   B^{\rm PS}_{\rm diag} \, \mu_{\rm q,sea}
  +C^{\rm PS}_{\rm diag} \, \mu_{\rm q,sea}^2
  +D^{\rm PS}_{\rm diag} \, \mu_{\rm q,sea}^3,
   \label{eqn:chiralfit:w_r0:diag:PSK}
   \\[3mm]
   {r_0}(K_{\rm sea}) \, m_{\rm V}(K_{\rm sea};K_{\rm sea},K_{\rm sea})
   & = &
   A^{\rm V}_{\rm diag} 
  +B^{\rm V}_{\rm diag} \, \mu_{\rm PS,sea}
  +C^{\rm V}_{\rm diag} \, \mu_{\rm PS,sea}^2
  +D^{\rm V}_{\rm diag} \, \mu_{\rm PS,sea}^3,
  \label{eqn:chiralfit:w_r0:diag:VPS}
\eea
where $\mu_{\rm q,diag}$ and $\mu_{\rm PS,sea}$ are 
the quark mass defined through the vector Ward identity (VWI)
and the PS meson mass normalized by $r_0$:
\bea
   \mu_{\rm q,sea}
   & = & 
   r_0(K_{\rm sea}) \, m_{\rm q,sea},
   \label{eqn:chiralfit:diag:mqsea}
   \\
   m_{\rm q,sea}
   & = & 
   \frac{1}{2} 
   \left(\frac{1}{K_{\rm sea}}-\frac{1}{K_{c}}\right),
   \label{eqn:chiralfit:w_r0:notation:mq_diag}
   \\
   \mu_{\rm PS,sea} & = & \left( r_0(K_{\rm sea}) \, m_{\rm PS,sea}\right )^2,
   \label{eqn:chiralfit:w_r0:notation:mu_PSsea}
   \\
   m_{\rm PS,sea}  
   & = & 
   m_{\rm PS}(K_{\rm sea};K_{\rm sea},K_{\rm sea}).
\eea
Fit parameters and $\chi^2/{\rm dof}$
are collected in Tables~\ref{tab:chiralfit:w_r0:PSK_diag}
and \ref{tab:chiralfit:w_r0:VPS_diag}.
The coefficient of the cubic term is small and consistent with zero
for both Eqs.~(\ref{eqn:chiralfit:w_r0:diag:PSK}) and 
(\ref{eqn:chiralfit:w_r0:diag:VPS}).
Consequently, 
the quadratic and cubic 
fits show a good consistency with each other
in the whole range of the quark mass and toward the chiral limit,
as seen in Fig.~\ref{fig:chiralfit:w_r0:diagonal}.
These observations suggest that the polynomial expansion
up to the quadratic order is sufficient to describe 
the quark mass dependence of our data of meson masses 
in the method-A.


We carry out a combined fit to PS meson masses 
as a function of sea and valence quark masses
using the quadratic form 
\bea
   &   &
   \left( {r_0}(K_{\rm sea}) \,
          m_{\rm PS}(K_{\rm sea};K_{\rm val,1},K_{\rm val,2}) \right)^2
   \nn \\
   & = &
   B^{\rm PS}_{\rm s} \, \mu_{\rm q,sea}
  +B^{\rm PS}_{\rm v} \, \mu_{\rm q,val}
  +C^{\rm PS}_{\rm s} \, \mu_{\rm q,sea}^2 
  +C^{\rm PS}_{\rm v} \, \mu_{\rm q,val}^2
  +C^{\rm PS}_{\rm sv} \, \mu_{\rm q,sea} 
                       \, \mu_{\rm q,val},
   \label{eqn:chiralfit:w_r0:PSK}
\eea
where 
$\mu_{\rm q,sea}$ is defined in Eq.~(\ref{eqn:chiralfit:diag:mqsea})
and 
\bea
   \mu_{\rm q,val}   & = & r_0(K_{\rm sea}) \, m_{\rm q,val},
   \\
   m_{\rm q,val}     & = & \frac{1}{2}
                           \left( m_{{\rm q,val},1} +m_{{\rm q,val},2}
                           \right),
   \\
   m_{{\rm q,val},i} & = & \frac{1}{2}
                           \left( 
                              \frac{1}{K_{{\rm val},i}}-\frac{1}{K_c}
                           \right) 
                           \hspace{3mm} (i=1,2).
\eea
The presence of the monomial term in $m_{\rm q,sea}$ means 
that the PS meson mass does not vanish 
in the chiral limit $m_{\rm q,val}\!=\!0$ 
for non-zero values of $m_{\rm q,sea}$.
This is because the value of $K_{\rm val}$ 
where the PS meson mass vanishes 
depends on the sea quark mass 
due to explicit violation of chiral symmetry
with the Wilson-type quark action.
%

\begin{figure}[htbp]
   \includegraphics[width=70mm]{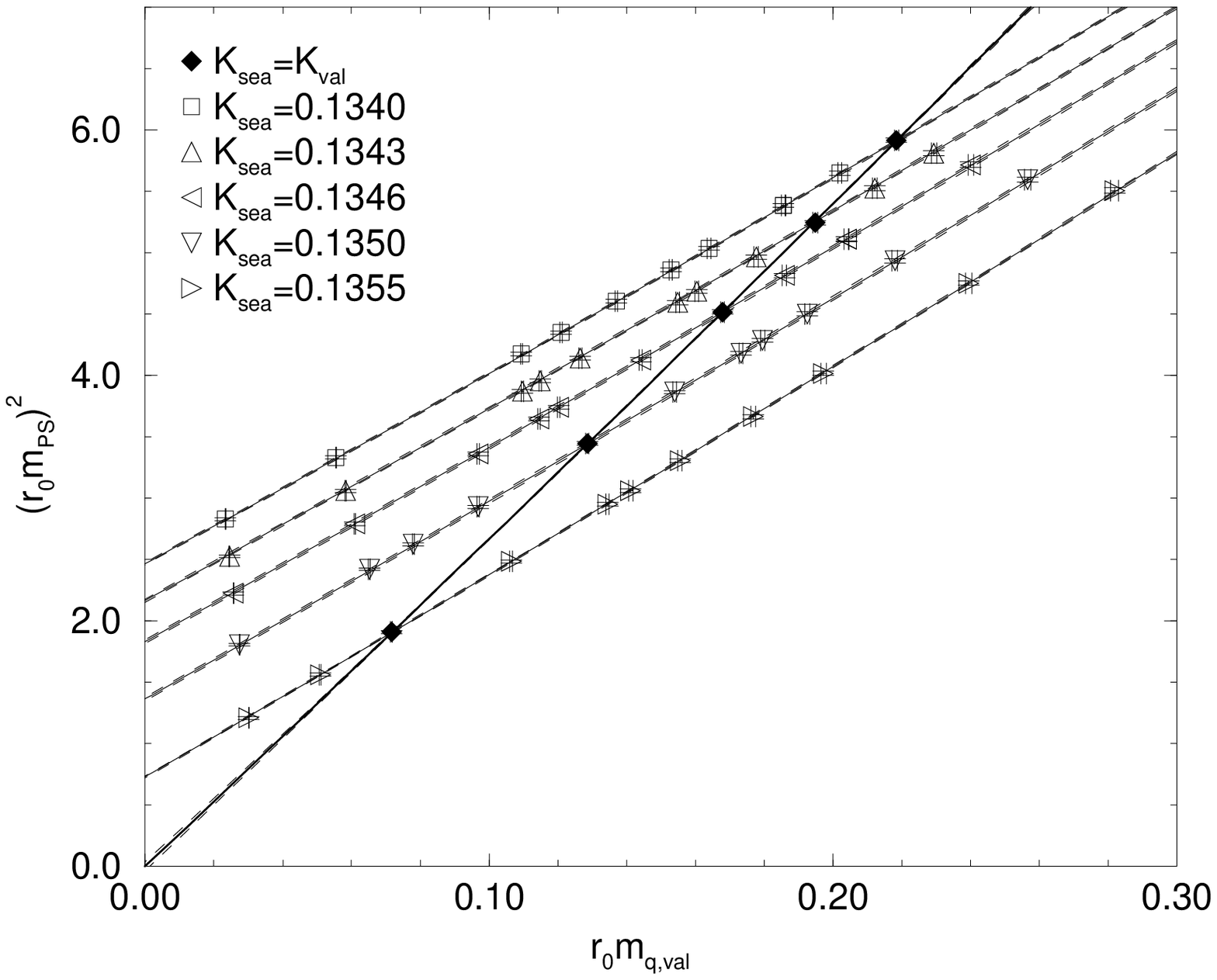}
   \caption
   {
      Combined chiral extrapolation of PS meson masses 
      in terms of VWI quark mass.
   }
   \label{fig:chiralfit:w_r0:PSK}
\end{figure}

We employ uncorrelated fits in the combined chiral extrapolations 
although the data with the same sea quark mass 
are expected to be correlated. 
Therefore, the obtained $\chi^2/{\rm dof}$ can be considered 
only as a guide to judge the quality of the fit.
Figure~\ref{fig:chiralfit:w_r0:PSK} shows 
that this fit form describes our data well.
Parameters of the fit are summarized 
in Table~\ref{tab:chiralfit:w_r0:PSK}.
We note that $K_c$ determined from the diagonal fit 
Eq.~(\ref{eqn:chiralfit:w_r0:diag:PSK})
and the combined fit 
Eq.~(\ref{eqn:chiralfit:w_r0:PSK}) are consistent with each other,
as they should be.

The most general quadratic fit ansatz for the PS meson masses 
should include an additional cross term
\bea
   \left( {r_0}(K_{\rm sea}) \,
          m_{\rm PS}(K_{\rm sea};K_{\rm val,1},K_{\rm val,2}) \right)^2
   & = & 
   \mbox{``r.h.s of Eq.~(\ref{eqn:chiralfit:w_r0:PSK})''}
   + C^{\rm PS}_{\rm vv} \, \mu_{\rm q,val,1} \, \mu_{\rm q,val,2},
   \label{eqn:chiralfit:w_r0:alter:PSK}
\eea
where 
$\mu_{{\rm q,val},i} \!=\! r_0(K_{\rm sea}) \, m_{{\rm q,val},i}$.
However, the coefficient is small 
as shown in fit parameters in Table~\ref{tab:chiralfit:w_r0:PSK}, 
and hence does not change the hadron spectrum. 
We use this fit to estimate the systematic error
due to the choice of the fitting function.

\begin{figure}[htbp]
   \begin{center}
   \includegraphics[width=70mm]{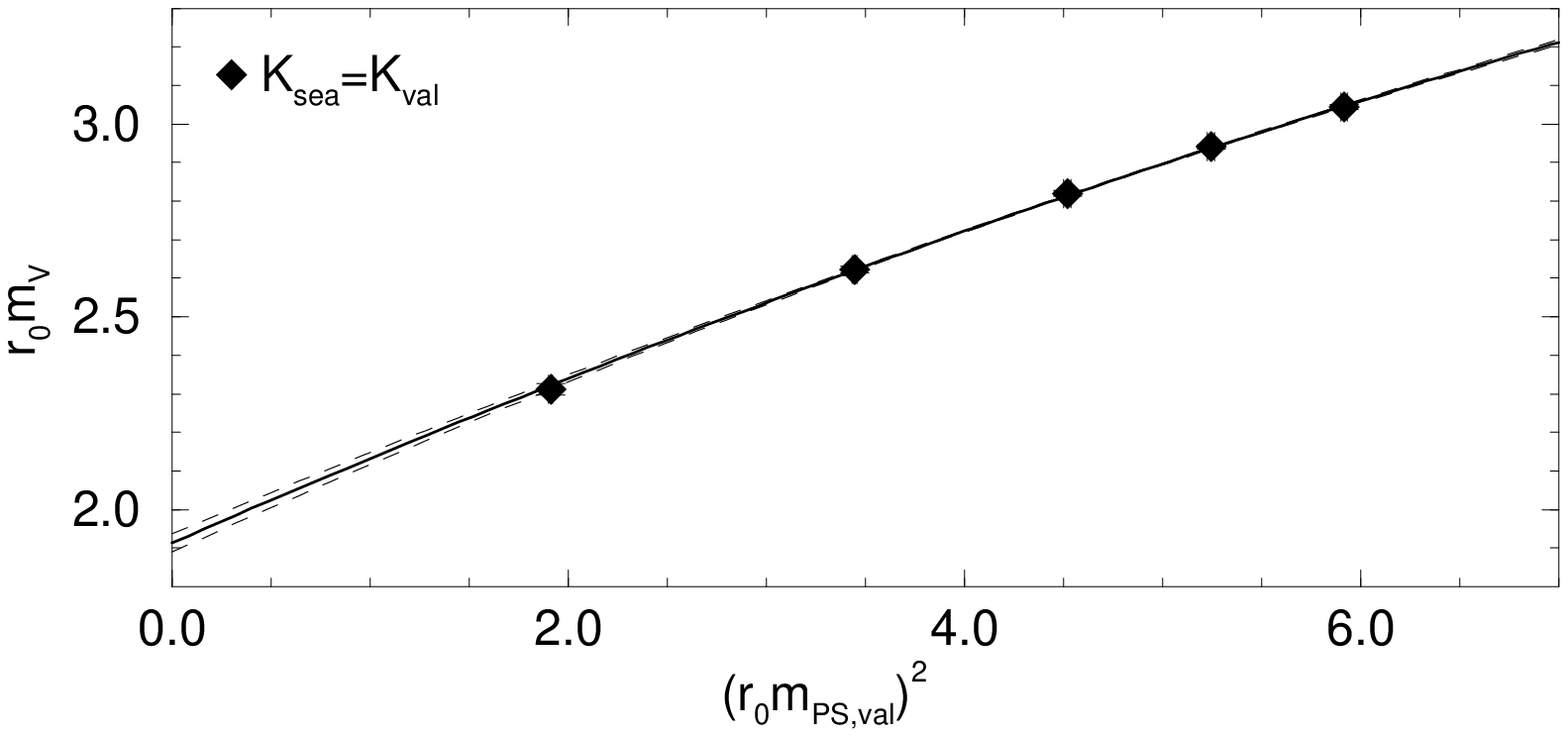}
   \includegraphics[width=70mm]{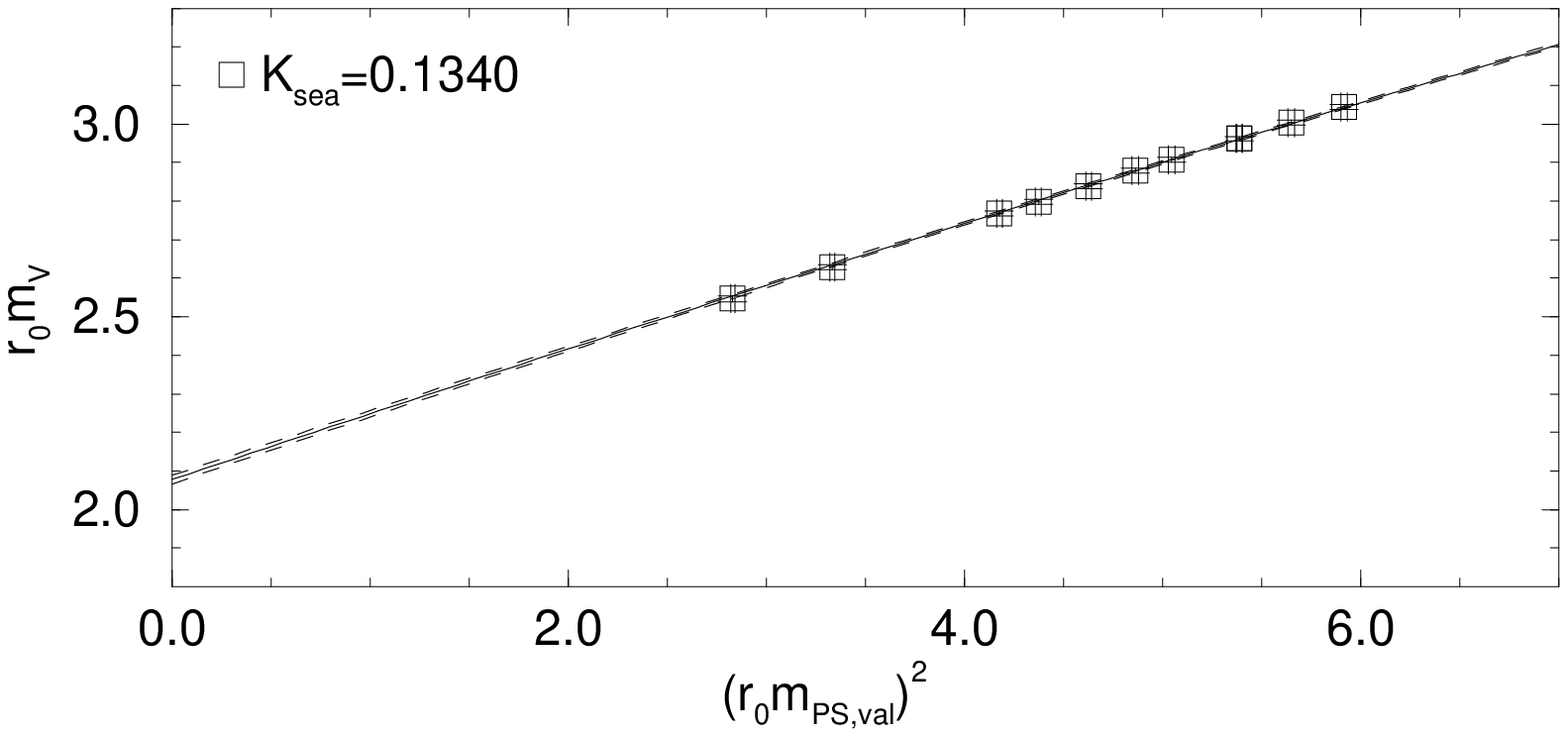}
   \end{center}
   \begin{center}
   \includegraphics[width=70mm]{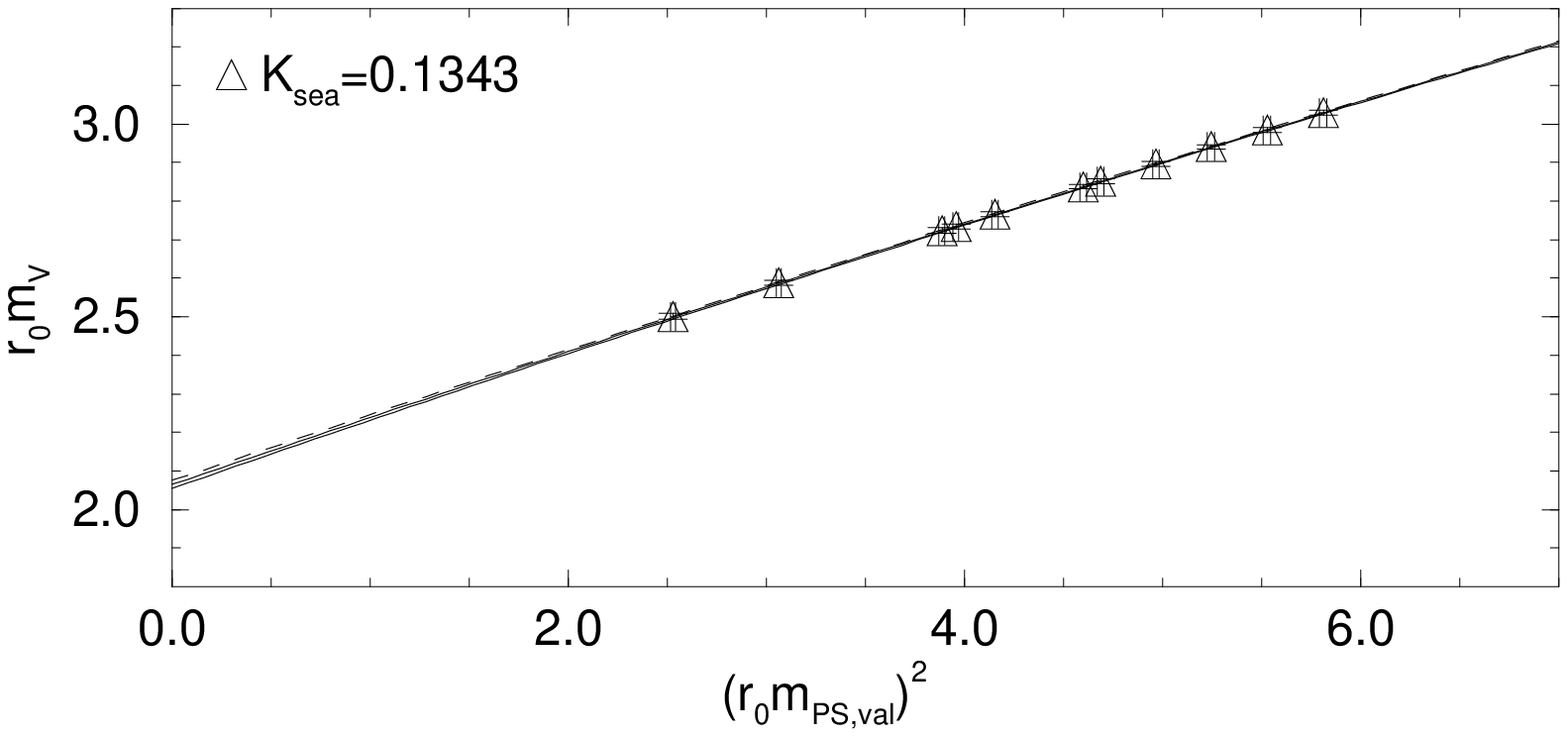}
   \includegraphics[width=70mm]{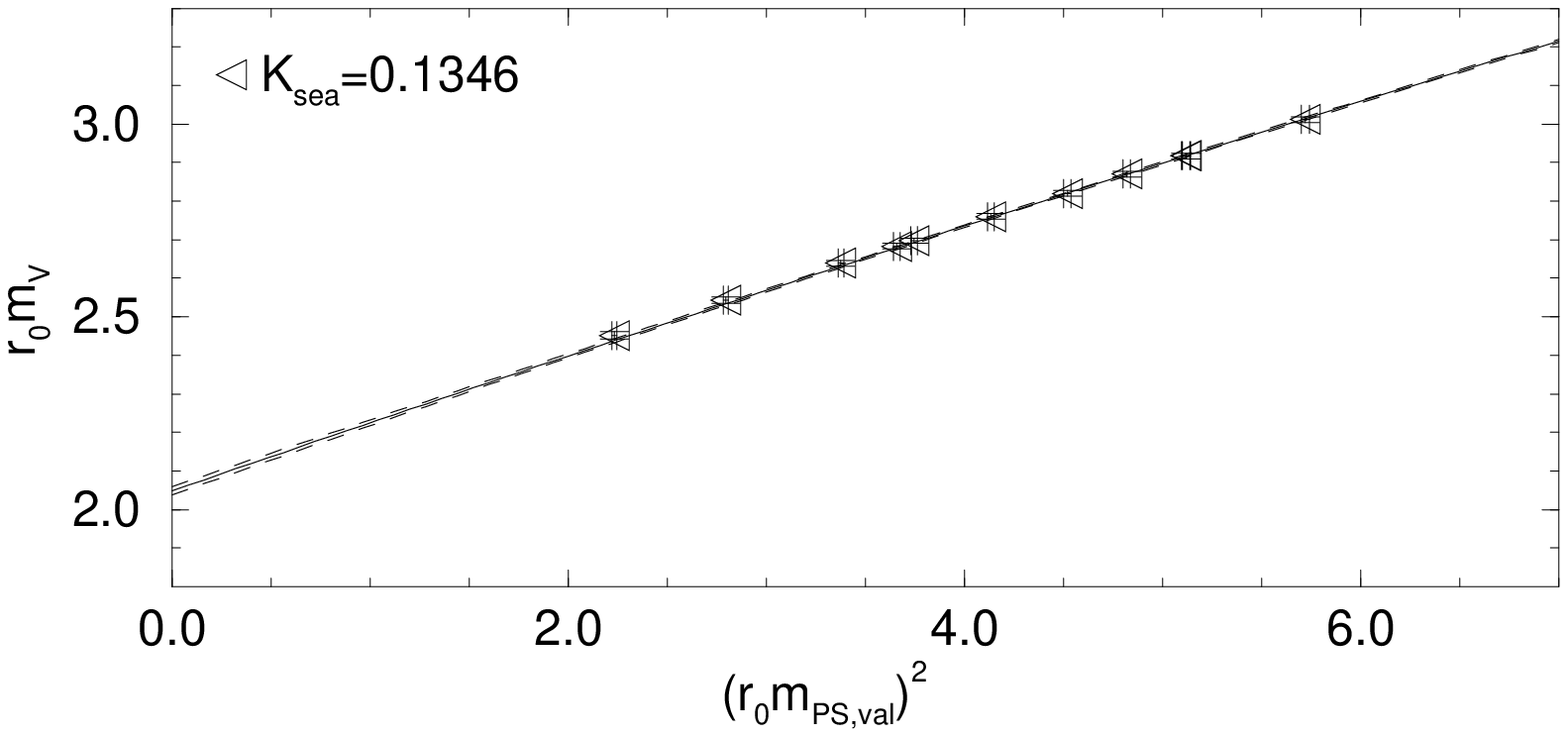}
   \vspace*{0mm}
   \end{center}
   \begin{center}
   \includegraphics[width=70mm]{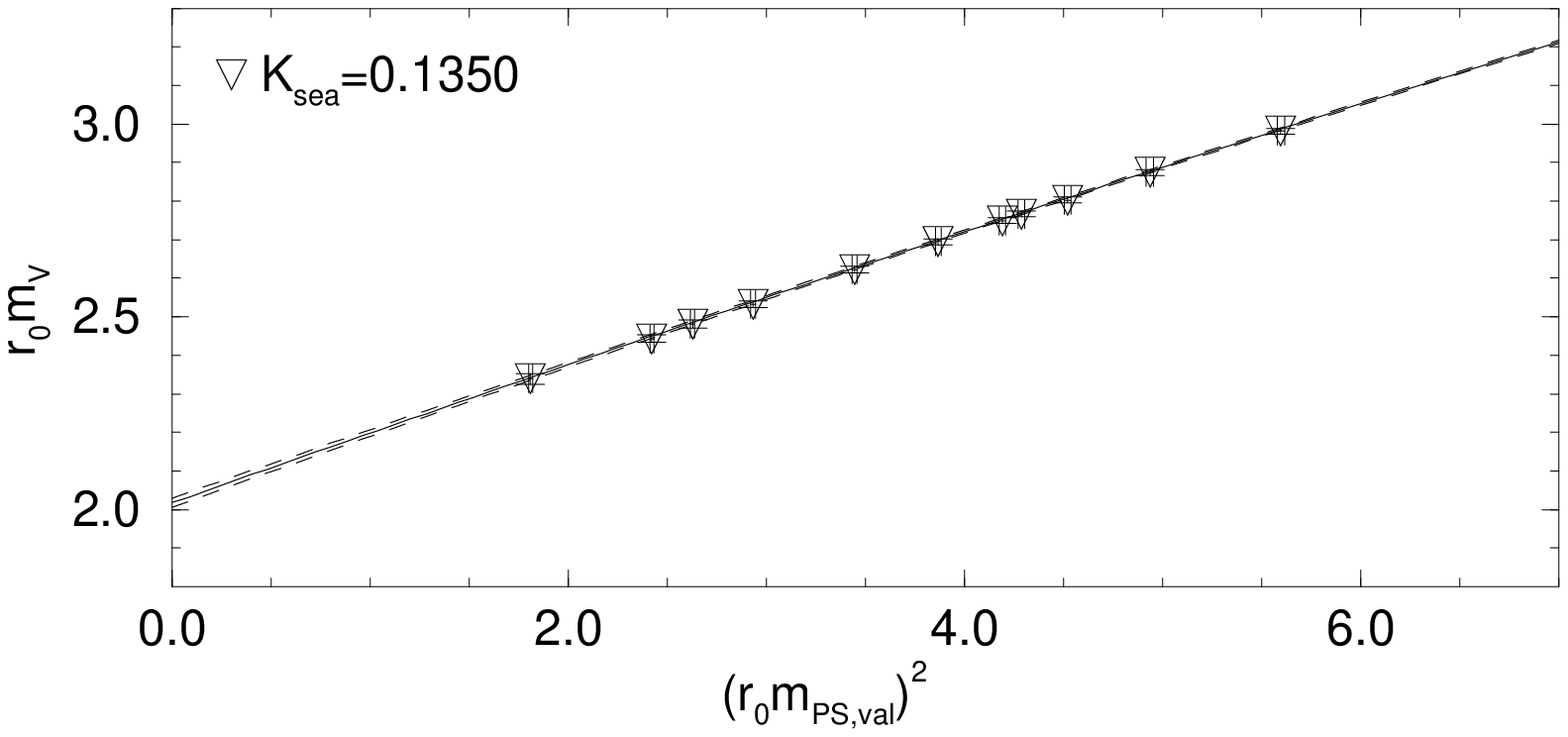}
   \includegraphics[width=70mm]{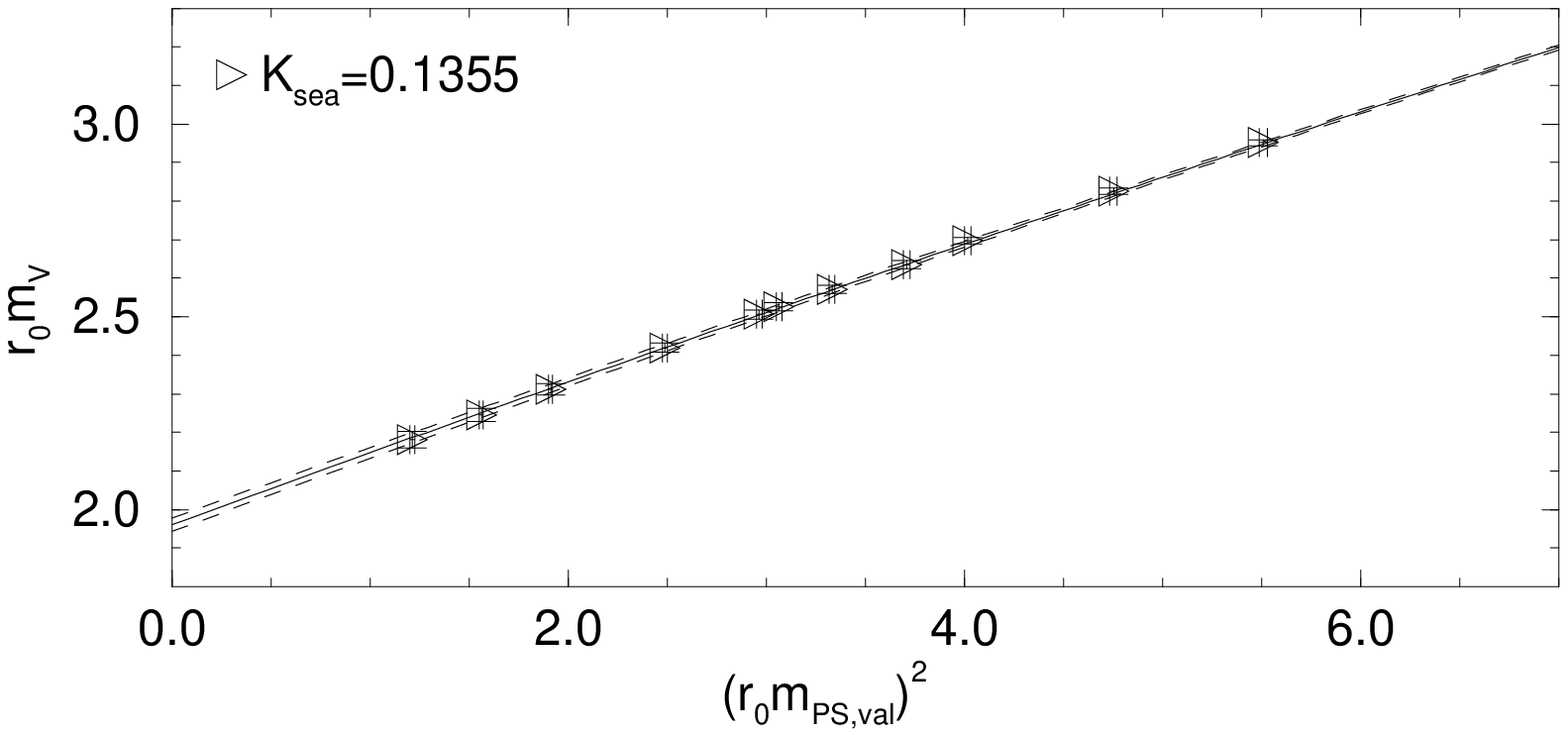}
   \end{center}
   \caption
   {
      Combined chiral extrapolation of vector meson masses.
   }
   \label{fig:chiralfit:w_r0:VPS}
\end{figure}

For the vector meson, 
we find that the following form describes our data well
\bea
   r_0(K_{\rm sea}) \, 
   m_{\rm V}(K_{\rm sea};K_{\rm val,1},K_{\rm val,2})
   & = & 
   A^{\rm V}
  +B^{\rm V}_{\rm s}  \, \mu_{\rm PS,sea}
  +B^{\rm V}_{\rm v}  \, \mu_{\rm PS,val}
  +C^{\rm V}_{\rm sv} \, \mu_{\rm PS,sea}
                      \, \mu_{\rm PS,val},
   \label{eqn:chiralfit:w_r0:VPS}
\eea
where 
$\mu_{\rm PS,sea}$ is defined in 
Eq.~(\ref{eqn:chiralfit:w_r0:notation:mu_PSsea})
and 
\bea
   \mu_{\rm PS,val}
   & = & 
   (r_0(K_{\rm sea}) \, m_{\rm PS,val})^2,
   \label{eqn:chiralfit:w_r0:notation:mu_PSval}
   \\
   m_{\rm PS,val}^2
   & = & 
   \frac{1}{2}\left( m_{\rm PS,val,1}^2 + m_{\rm PS,val,2}^2 \right),
   \label{eqn:chiralfit:w_r0:notation:m_PSval}
   \\
   m_{{\rm PS,val},i}
   & = & 
   m_{\rm PS}(K_{\rm sea};K_{{\rm val},i},K_{{\rm val},i}).
   \label{eqn:chiralfit:w_r0:notation:m_PSval12}
\eea
For a more general fit of form 
\bea
   r_0(K_{\rm sea}) \, 
   m_{\rm V}(K_{\rm sea};K_{\rm val,1},K_{\rm val,2})
   & = &
   \mbox{``r.h.s of Eq.~(\ref{eqn:chiralfit:w_r0:VPS})''}
   + C^{\rm V}_{\rm s} \mu_{\rm PS,sea}^2
   + C^{\rm V}_{\rm v} \mu_{\rm PS,val}^2,
   \label{eqn:chiralfit:w_r0:alter:VPS}
\eea
the additional parameters $C^{\rm V}_{\rm s}$ and $C^{\rm V}_{\rm v}$
are not well-determined
as seen in Table~\ref{tab:chiralfit:w_r0:VPS}.
We use the former fit, 
which is shown in Fig.~\ref{fig:chiralfit:w_r0:VPS},
to calculate the hadron spectrum,
and the latter to estimate systematic error of the chiral extrapolation.


We also carry out a partially quenched fit to vector meson masses 
at each sea quark mass.
We use a linear form, which is 
obtained from Eq.~(\ref{eqn:chiralfit:w_r0:VPS})
by dropping all terms describing the sea quark mass dependence 
\bea
   m_{\rm V}(K_{\rm sea}; K_{\rm val,1}, K_{\rm val,2})
   = 
   A^{\rm V}_{\rm PQ} + B^{\rm V}_{\rm PQ} \, m_{\rm PS,val}^2.
   \label{eqn:chiralfit:w_r0:VPS_PQ}
\eea
Parameters given in Table~\ref{tab:chiralfit:w_r0:VPS_PQ}
are used to calculate the $J$ parameter at each sea quark mass.


The chiral extrapolation of octet baryon masses is carried out 
using a quadratic form based on 
the leading order prediction of ChPT~\cite{ChPT.octet},
which was also used in Ref.~\cite{Spectrum.Nf2.CP-PACS}.
We carry out the simultaneous fit to the $\Sigma$- and $\Lambda$-like 
octet baryon masses using the functions
\bea
   &   &
   r_0(K_{\rm sea}) \
   m_{{\rm oct}, \Sigma}(K_{\rm sea}; K_{\rm val,1}, K_{\rm val,2}, 
                                                     K_{\rm val,2})
   \nn \\
   & = &
   A^{\rm O} 
  +B^{\rm O}_{\rm s} \mu_{\rm PS,sea}
  +(F^{\rm O}_{\rm v}-D^{\rm O}_{\rm v}) \, \mu_{\rm PS,val,1}
  +2 F^{\rm O}_{\rm v} \mu_{\rm PS,val,2}
  +C^{\rm O}_{\rm s} \mu_{\rm PS,sea}^2
  +( C^{\rm O}_{\rm v} 
    +C^{{\rm O},\Sigma}_{\rm v}) \, \mu_{\rm PS,val,1}^2
   \nn \\
   &   &
  +( C^{\rm O}_{\rm v}
    -C^{{\rm O},\Sigma}_{\rm v}) \, \mu_{\rm PS,val,2}^2
  +( C^{\rm O}_{\rm sv}
    +C^{{\rm O},\Sigma}_{\rm sv}) \, \mu_{\rm PS,sea} 
                                  \, \mu_{\rm PS,val,1}
  +( C^{\rm O}_{\rm sv}
    -C^{{\rm O},\Sigma}_{\rm sv}) \, \mu_{\rm PS,sea} 
                                  \, \mu_{\rm PS,val,2},
   \label{eqn:chiralfit:w_r0:OPS:Sigma}
   \\[5mm]
   &   &
   r_0(K_{\rm sea}) \, 
   m_{{\rm oct}, \Lambda}(K_{\rm sea}; K_{\rm val,1}, K_{\rm val, 2}, 
                                                      K_{\rm val, 2})
   \nn \\
   & = &
   A^{\rm O} 
  +B^{\rm O}_{\rm s} \mu_{\rm PS,sea}
  +\left(F^{\rm O}_{\rm v}+\frac{D^{\rm O}_{\rm v}}{3}\right) 
                                           \mu_{\rm PS,val,1}
  +2\left(F^{\rm O}_{\rm v}-\frac{2}{3}D^{\rm O}_{\rm v}\right)
                                           \mu_{\rm PS,val,2}
   \nn \\
   &   &
  +C^{\rm O}_{\rm s} \mu_{\rm PS,sea}^2
  +( C^{\rm O}_{\rm v}
    +C^{{\rm O},\Lambda}_{\rm v}) \, \mu_{\rm PS,val,1}^2
  +( C^{\rm O}_{\rm v}
    -C^{{\rm O},\Lambda}_{\rm v}) \, \mu_{\rm PS,val,2}^2
  +( C^{\rm O}_{\rm sv}
    +C^{{\rm O},\Lambda}_{\rm sv}) \, \mu_{\rm PS,sea}  
                                   \, \mu_{\rm PS,val,1}
   \nn \\
   &   &
  +( C^{\rm O}_{\rm sv}
    -C^{{\rm O},\Lambda}_{\rm sv}) \, \mu_{\rm PS,sea} 
                                   \, \mu_{\rm PS,val,2},
   \label{eqn:chiralfit:w_r0:OPS:Lambda}
\eea
where $\mu_{{\rm PS,val},i} =  (r_0(K_{\rm sea})\, m_{{\rm PS,val},i})^2$.


The decuplet baryon masses are well described by the following form 
\bea
   r_0(K_{\rm sea}) \, 
   m_{\rm dec}(K_{\rm sea}; K_{\rm val,1}, K_{\rm val,2}, 
                                           K_{\rm val,2})
   =
   A^{\rm D} 
  +B^{\rm D}_{\rm s}  \mu_{\rm PS,sea}
  +B^{\rm D}_{\rm v}  \mu_{\rm PS,val}
  +C^{\rm D}_{\rm sv} \, \mu_{\rm PS,sea} \, \mu_{\rm PS,val},
   \label{eqn:chiralfit:w_r0:DPS}
\eea
where $\mu_{\rm PS,val}$ stands for the average of 
three valence quark masses
\bea
   \mu_{\rm PS,val} 
   & = &
   \left( r_0(K_{\rm sea}) \, m_{\rm PS,val} \right)^2,
   \\
   m_{\rm PS,val}^2
   & = &
   \frac{1}{3} \left( m_{\rm PS,val,1}^2 + 2m_{\rm PS,val,2}^2 \right).
   \label{eqn:chiralfit:w_r0:notation:DPS}
\eea

\begin{figure}[htbp]
   \begin{center}
   \includegraphics[width=70mm]{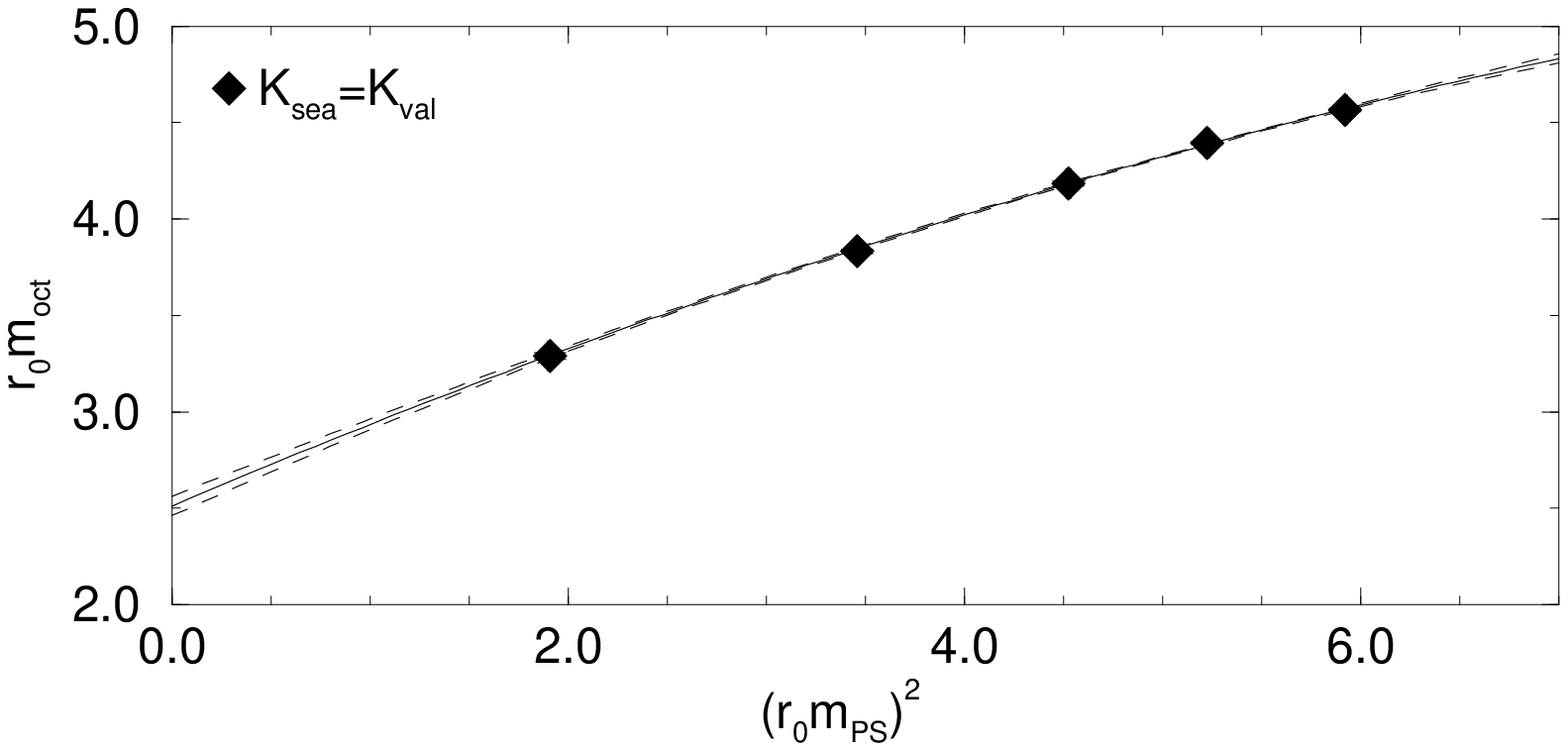}
   \end{center}
   \begin{center}
   \includegraphics[width=70mm]{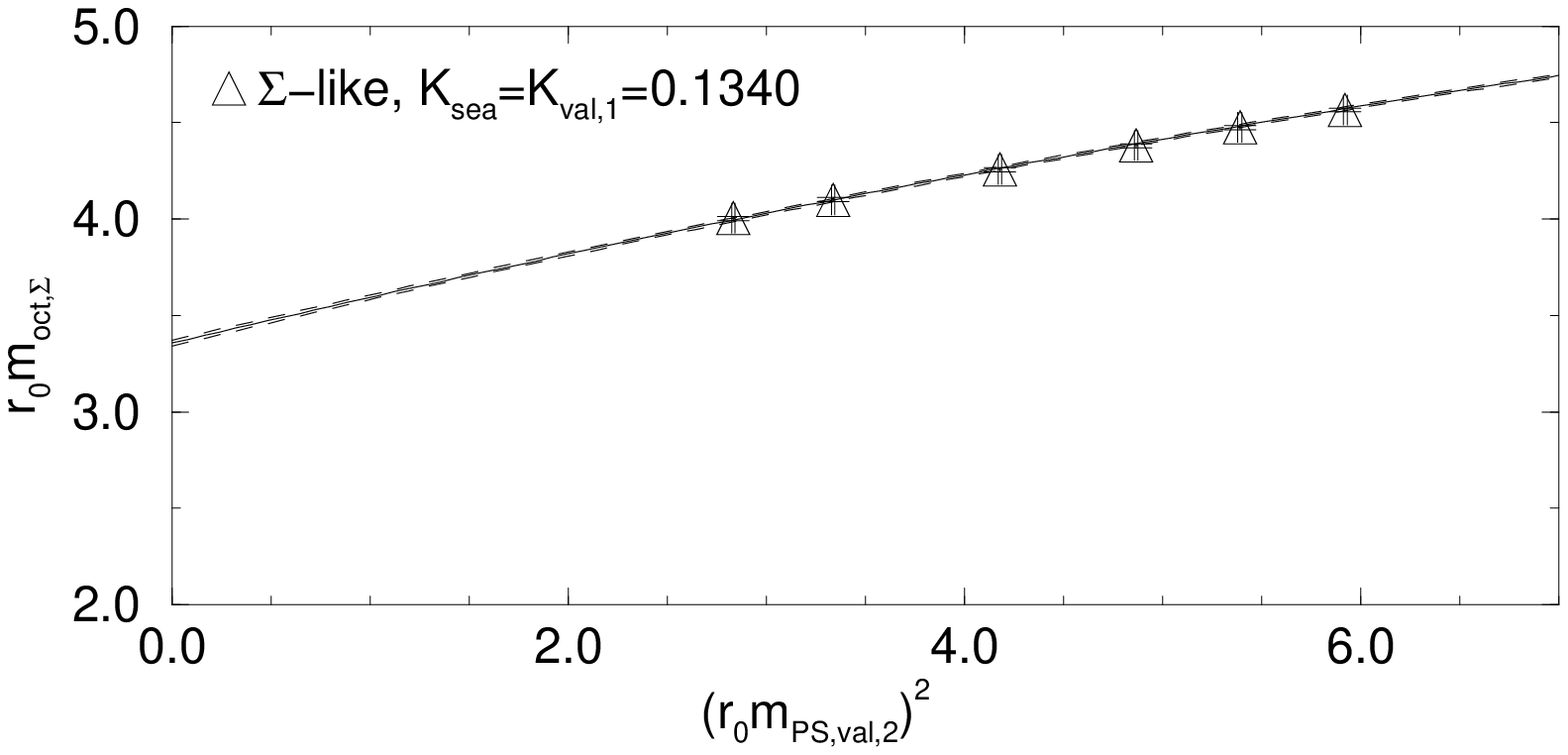}
   \includegraphics[width=70mm]{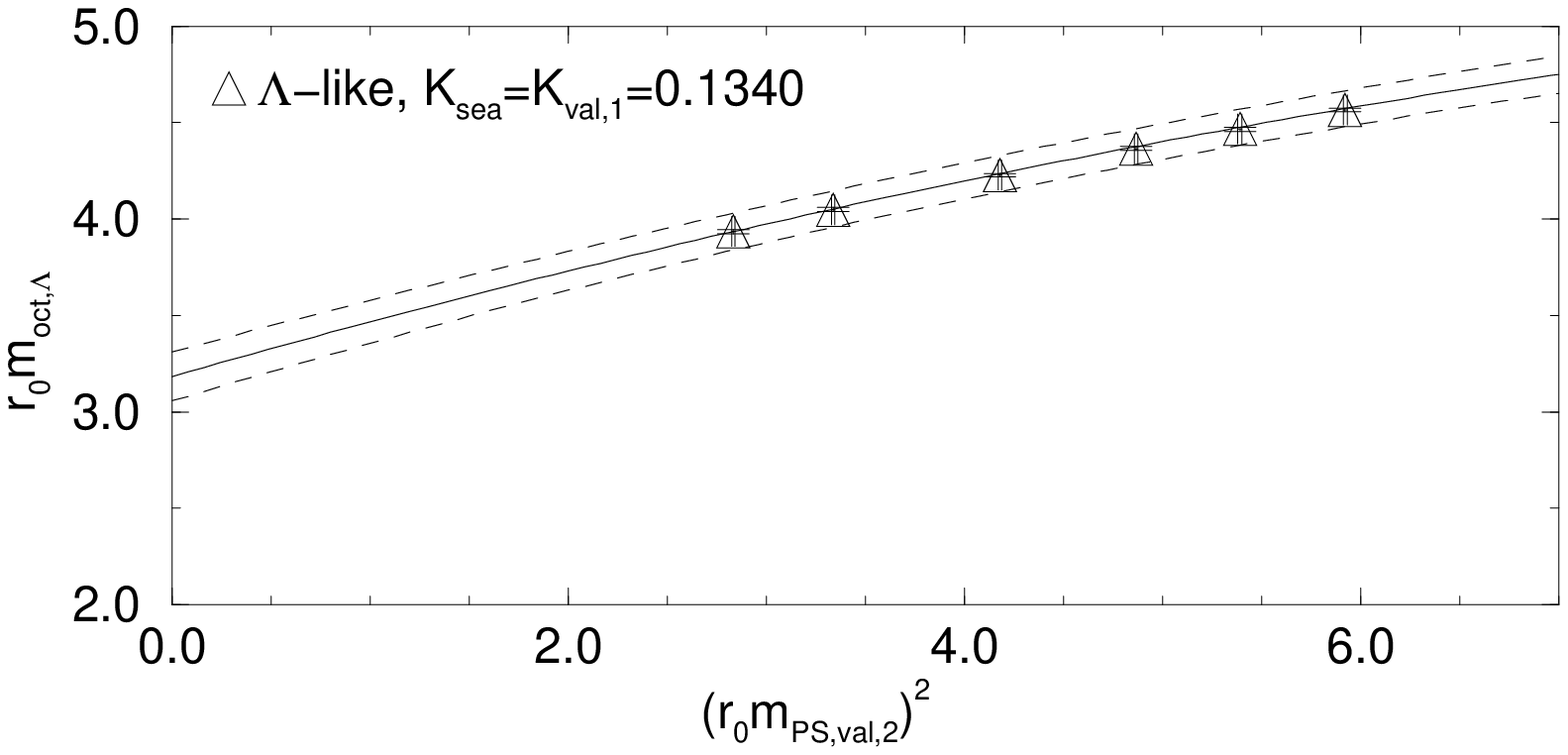}
   \end{center}
   \begin{center}
   \includegraphics[width=70mm]{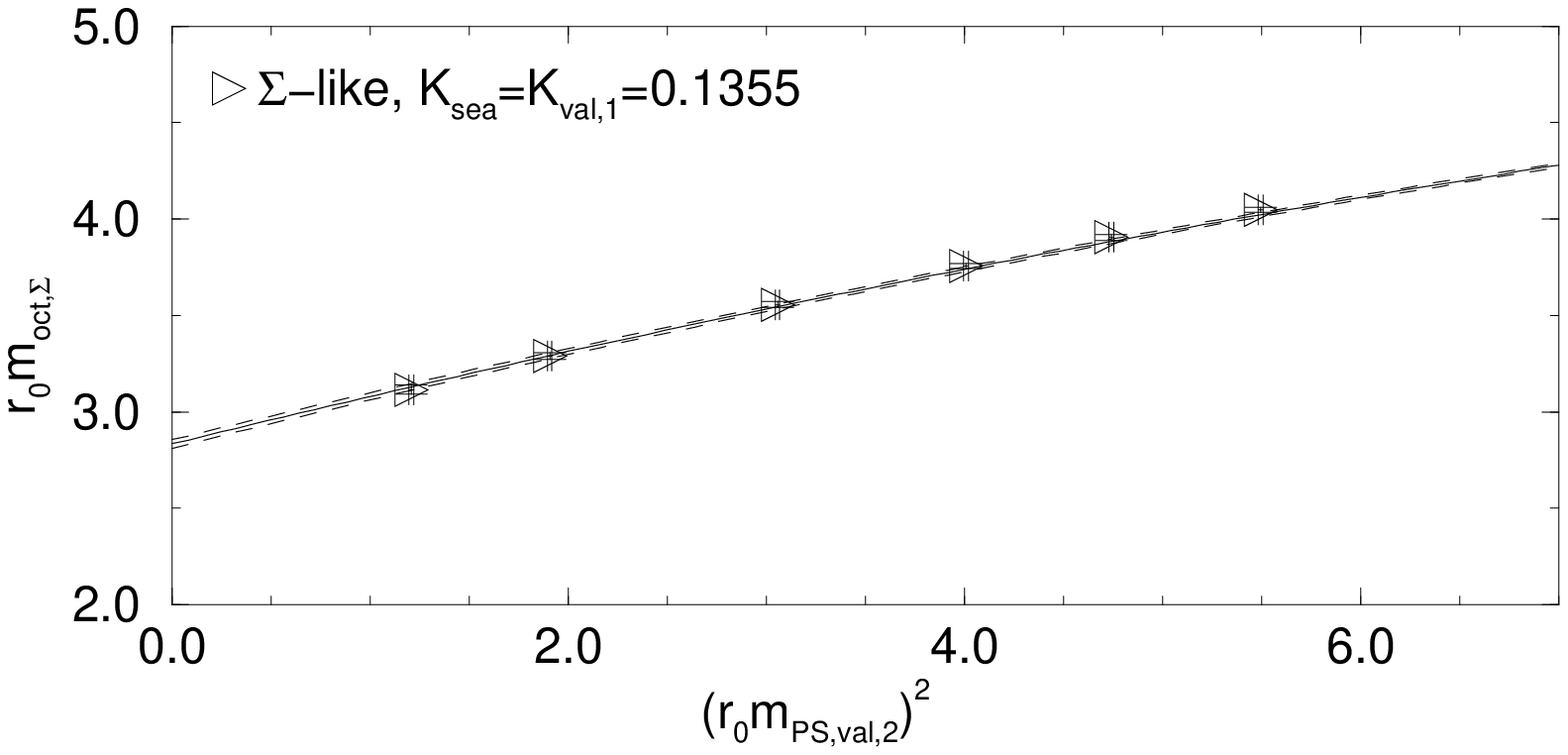}
   \includegraphics[width=70mm]{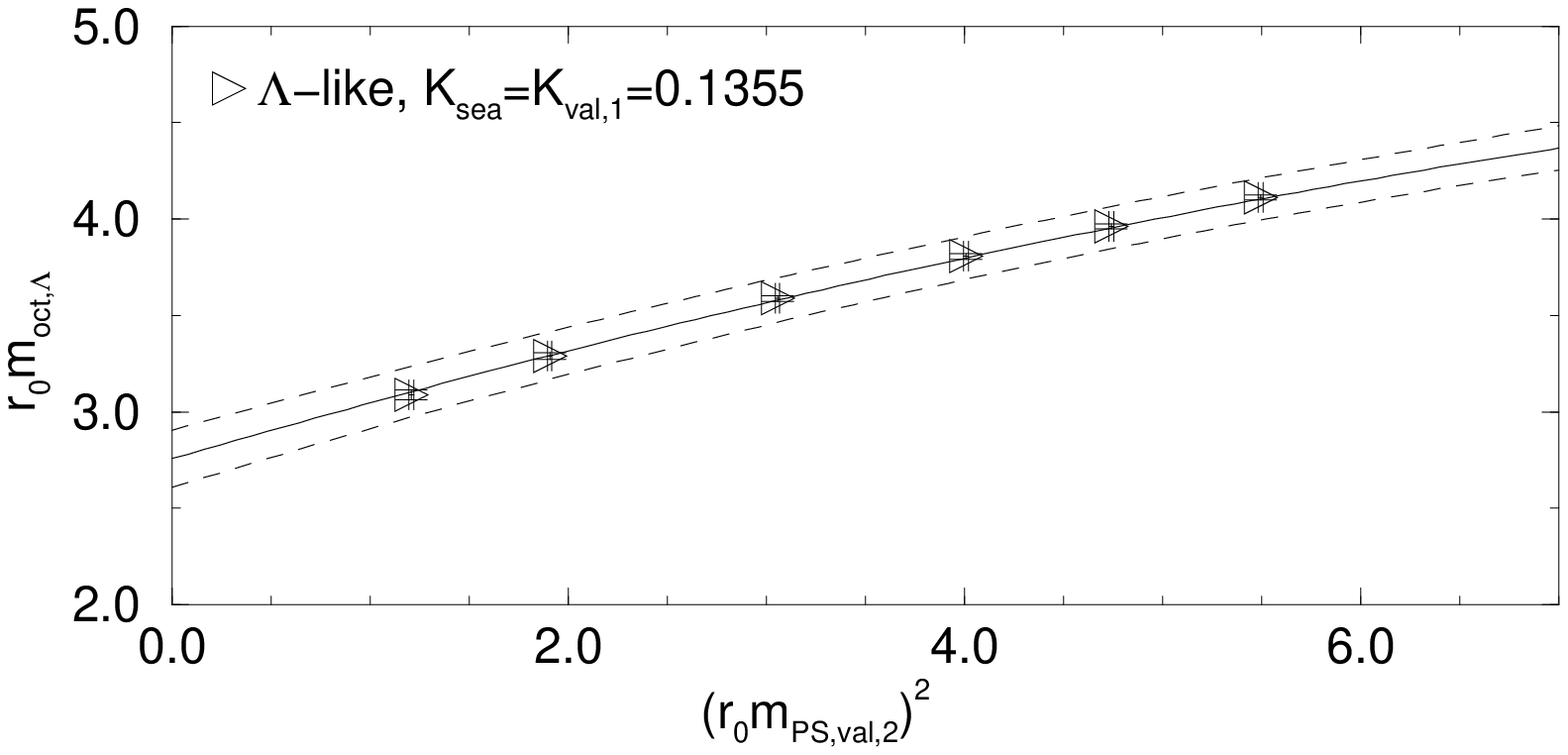}
   \end{center}
   \caption
   {
      Combined chiral extrapolation of octet baryon masses.
      Top figure shows extrapolation of diagonal data.
      Middle and bottom figures are data 
      at $K_{\rm sea}=K_{\rm val,1}=$~0.1340 and 0.1355
      for $\Sigma$-like (left panels) and $\Lambda$-like baryons
      (right panels).
   }
   \label{fig:chiralfit:w_r0:OPS}
\end{figure}

\begin{figure}[htbp]
   \begin{center}
   \includegraphics[width=70mm]{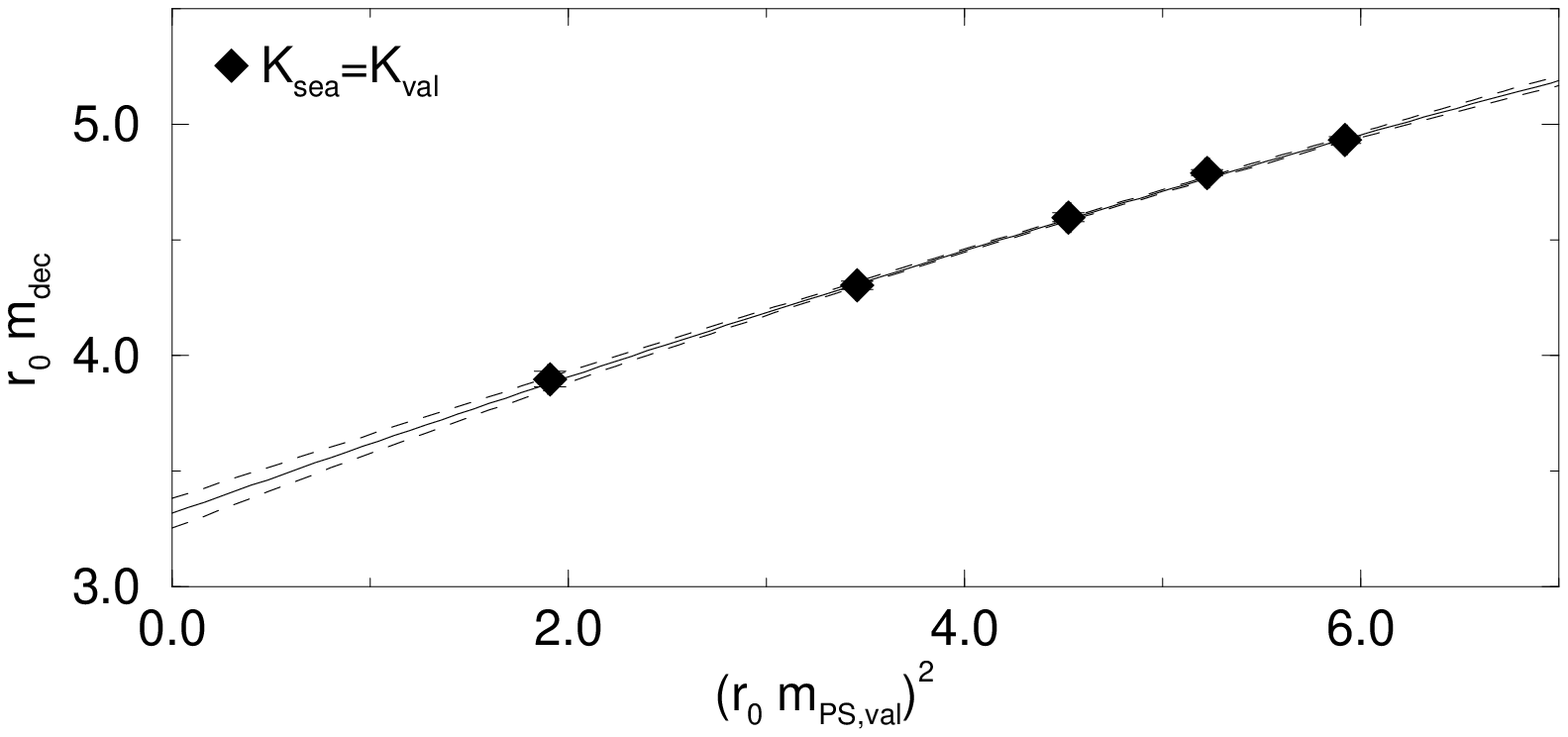}
   \includegraphics[width=70mm]{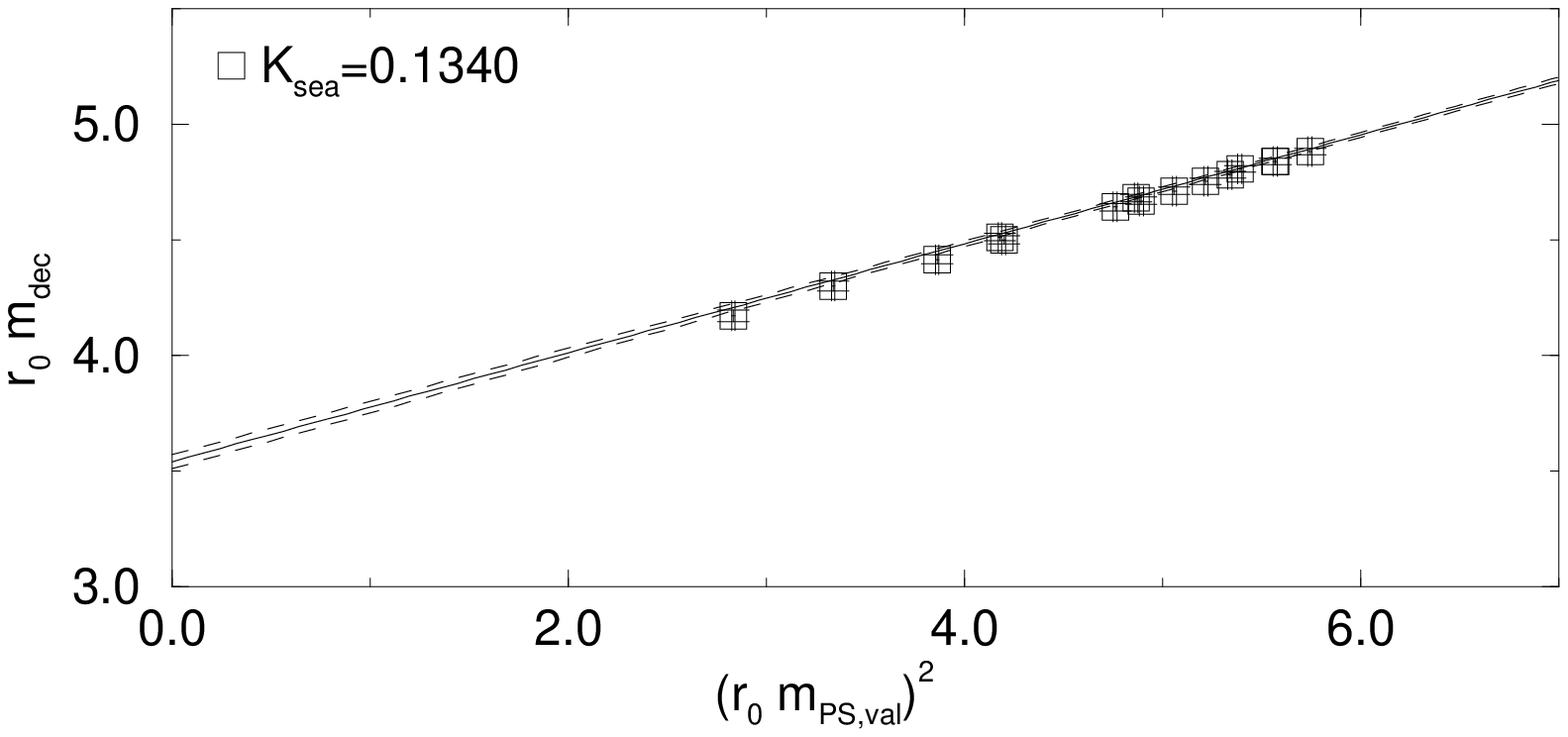}
   \end{center}
   \begin{center}
   \includegraphics[width=70mm]{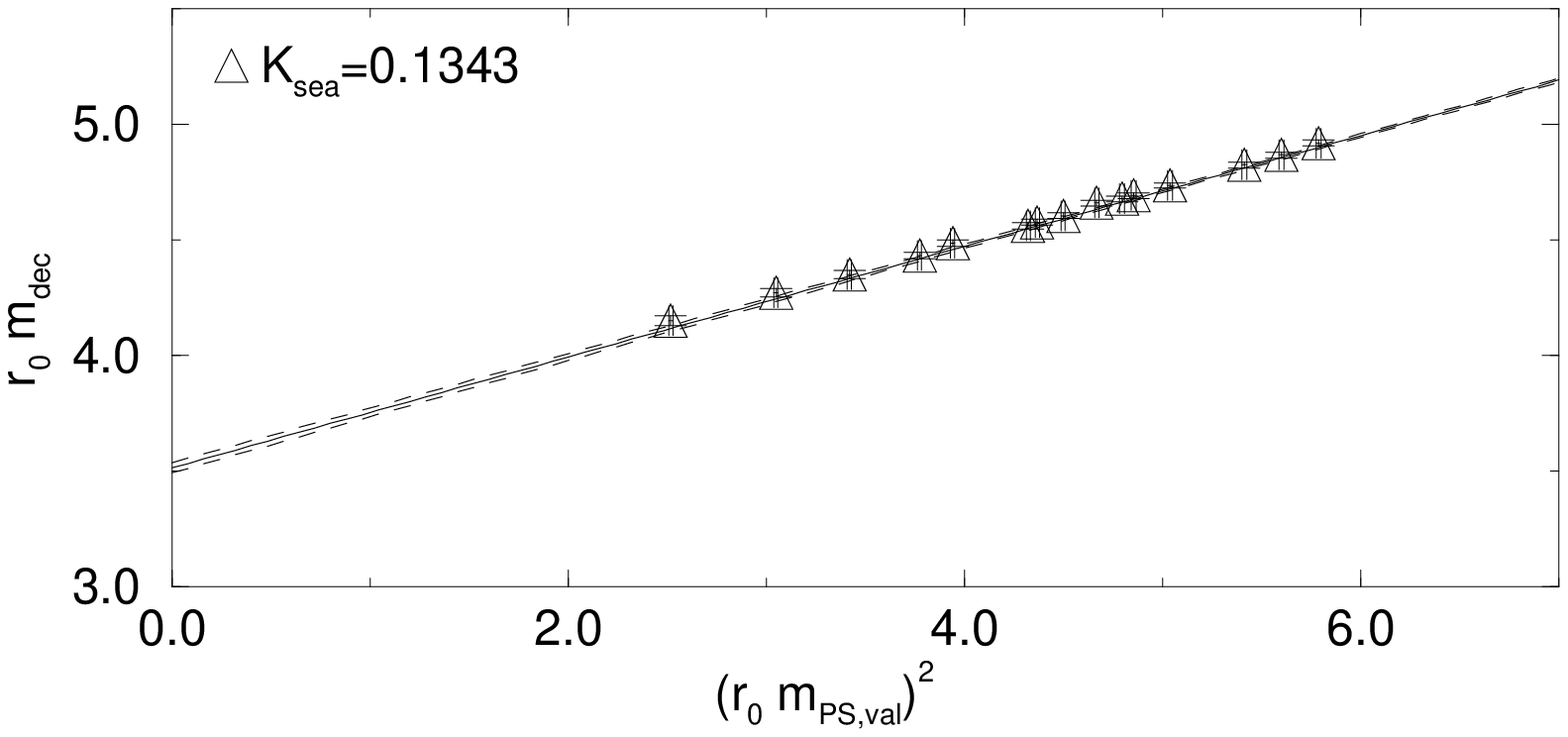}
   \includegraphics[width=70mm]{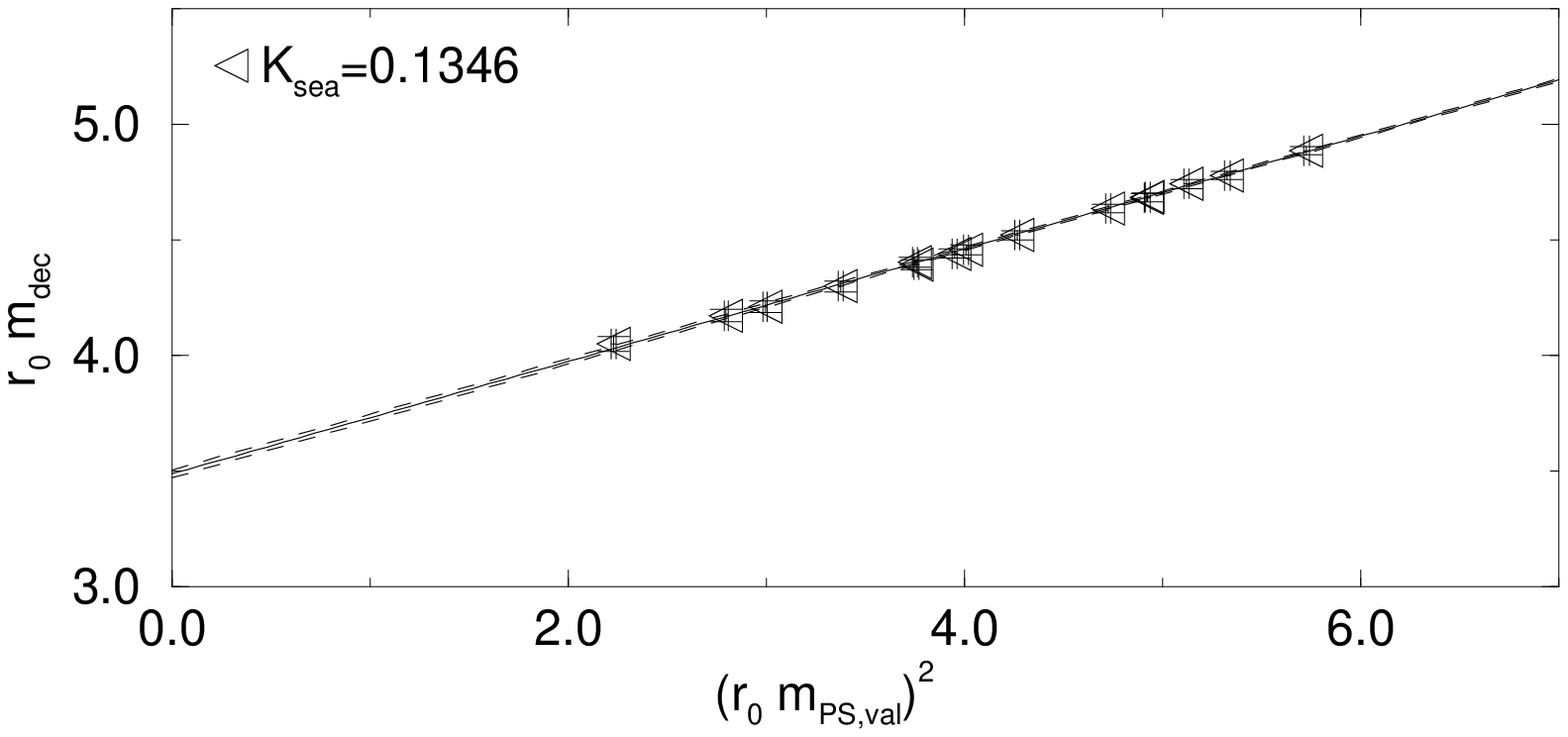}
   \end{center}
   \begin{center}
   \includegraphics[width=70mm]{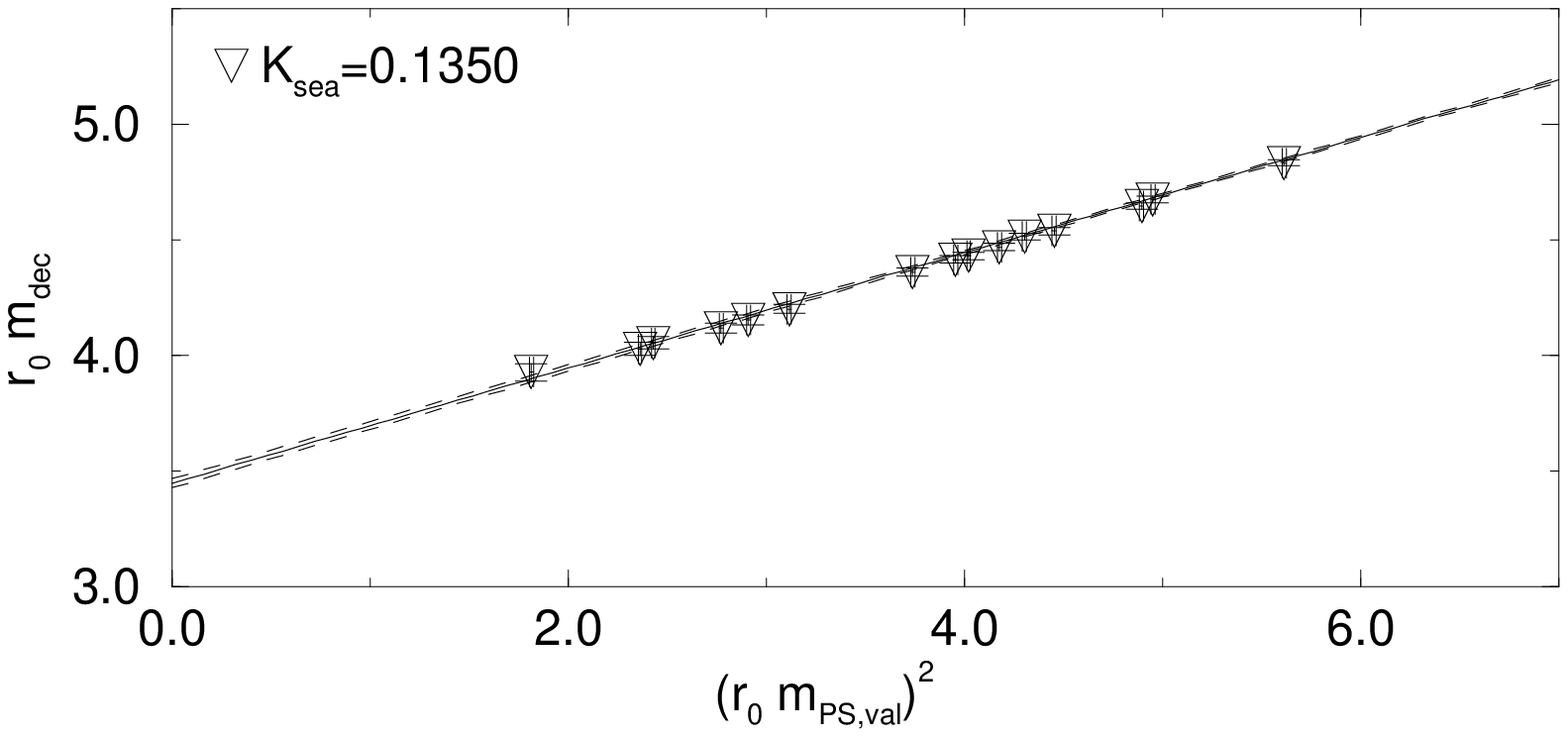}
   \includegraphics[width=70mm]{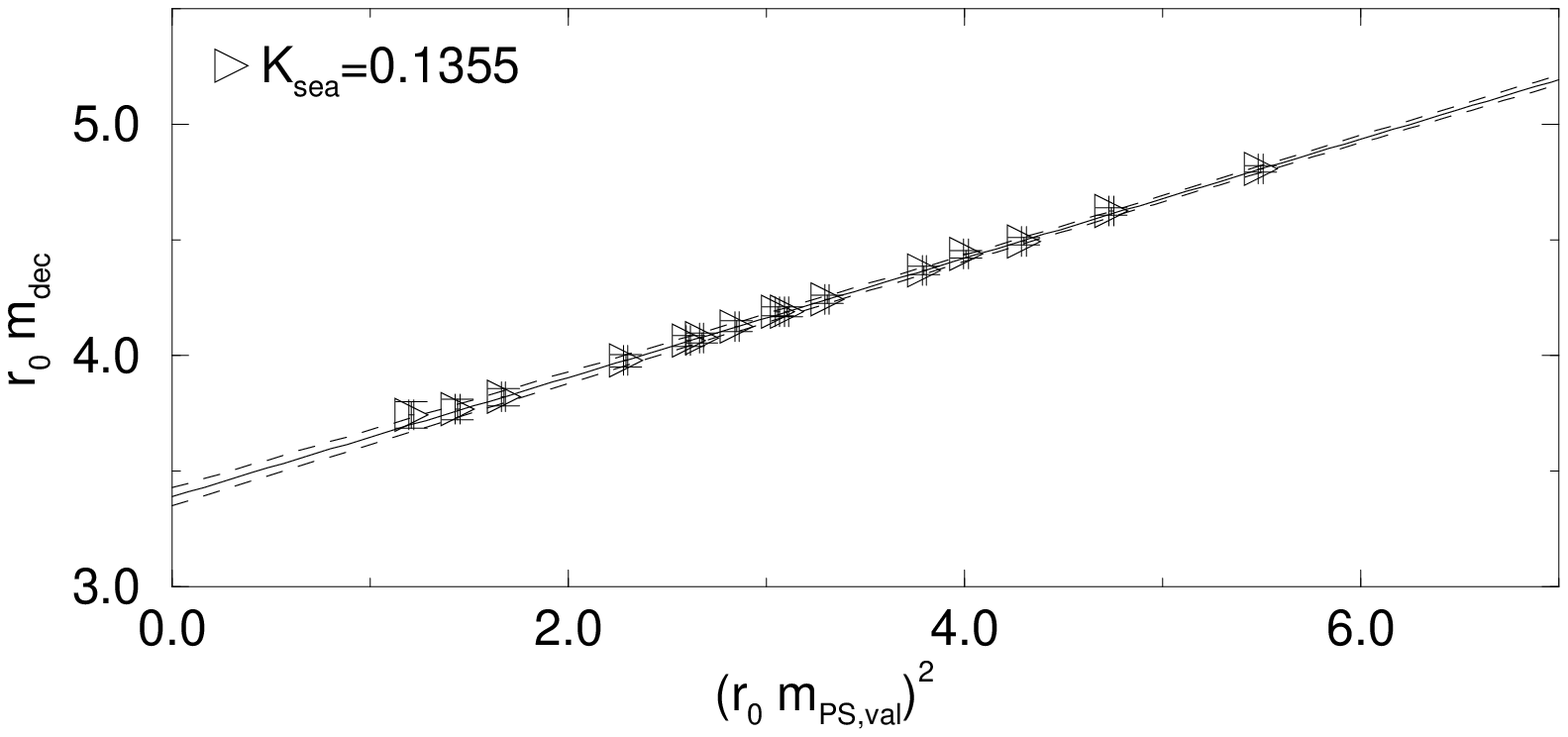}
   \end{center}
   \caption
   {
      Combined chiral extrapolation of decuplet baryon masses.
   }
   \label{fig:chiralfit:w_r0:DPS}
\end{figure}

Figures~\ref{fig:chiralfit:w_r0:OPS} and \ref{fig:chiralfit:w_r0:DPS}
show the fit for octet and decuplet baryon masses.
Parameters are summarized in Tables~\ref{tab:chiralfit:w_r0:OPS}
and ~\ref{tab:chiralfit:w_r0:DPS}.
We also test the following forms to estimate 
systematic error of the baryon spectrum
due to the choice of the fitting form
\bea
   &   &
   r_0(K_{\rm sea}) \
   m_{{\rm oct}}(K_{\rm sea}; K_{\rm val,1}, K_{\rm val,2}, 
                                                     K_{\rm val,2})
   \nn \\
   & = &
   \mbox{``r.h.s of Eqs.~(\ref{eqn:chiralfit:w_r0:OPS:Sigma})
         and (\ref{eqn:chiralfit:w_r0:OPS:Lambda})''}
   + C^{\rm O}_{\rm vv} \, \mu_{\rm PS,val,1} \, \mu_{\rm PS,val,2},
   \label{eqn:chiralfit:w_r0:alter:OPS}
   \\
   &  & 
   r_0(K_{\rm sea}) \, 
   m_{\rm dec}(K_{\rm sea}; K_{\rm val,1}, K_{\rm val,2}, 
                                           K_{\rm val,2})
   \nn \\
   & = &
   \mbox{``r.h.s of Eq.~(\ref{eqn:chiralfit:w_r0:DPS})''}
   + C^{\rm D}_{\rm s} \mu_{\rm PS,sea}^2
   + C^{\rm D}_{\rm v} \mu_{\rm PS,val}^2.
   \label{eqn:chiralfit:w_r0:alter:DPS}
\eea

\begin{figure}[htbp]
   \includegraphics[width=70mm]{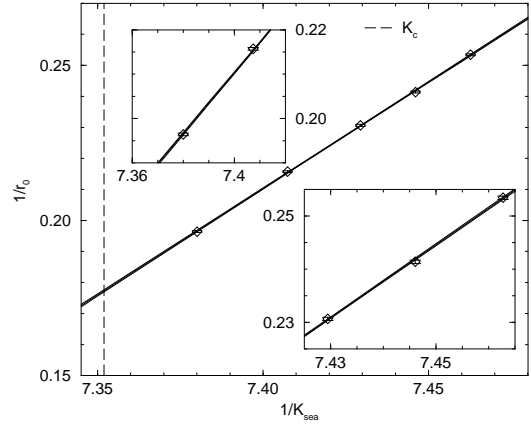}
   \caption
   {
      Chiral extrapolation of $r_0$. 
      The vertical line shows where $K_{\rm sea}\!=\!K_c$.
   }
   \label{fig:chiralfit:w_r0:r0}
\end{figure}

We carry out the chiral extrapolation of $r_0(K_{\rm sea})$
in order to determine $r_0$ at the physical sea quark mass,
which is required to calculate the hadron spectrum in the method-A.
We use a linear form 
\bea
   \frac{1}{r_0} = A_{r_0} + \frac{B_{r_0}}{K_{\rm sea}}.
   \label{eqn:chiralfit:w_r0:r0inv_vs_Kinv}
\eea
As seen in Fig.~\ref{fig:chiralfit:w_r0:r0} and 
Table~\ref{tab:chiralfit:w_r0:r0inv_vs_Kinv},
this fit describes our data well 
and gives a reasonable value of $\chi^2/{\rm dof} \! \sim \! 1.5$.

\subsection{chiral extrapolation in lattice units (method-B)}
\label{sec:chiralfit:wo_r0}


In order to study the convergence properties of polynomial fit forms
in the method-B,
we carry out quadratic and cubic diagonal fits
to PS and vector meson masses
\bea
   m_{\rm PS}(K_{\rm sea};K_{\rm sea},K_{\rm sea})^2
   & = &
   B^{\prime \rm PS}_{\rm diag} \, m_{\rm q,sea}
  +C^{\prime \rm PS}_{\rm diag} \, m_{\rm q,sea}^2
  +D^{\prime \rm PS}_{\rm diag} \, m_{\rm q,sea}^3,
  \label{eqn:chiralfit:wo_r0:diag:PSK}
   \\[3mm]
   m_{\rm V}(K_{\rm sea};K_{\rm sea},K_{\rm sea})
   & = &
   A^{\prime \rm V}_{\rm diag} 
  +B^{\prime \rm V}_{\rm diag} \, m_{\rm PS,sea}^2
  +C^{\prime \rm V}_{\rm diag} \, m_{\rm PS,sea}^4
  +D^{\prime \rm V}_{\rm diag} \, m_{\rm PS,sea}^6.
  \label{eqn:chiralfit:wo_r0:diag:VPS}
\eea
Fit parameters are summarized in 
Tables~\ref{tab:chiralfit:wo_r0:PSK_diag} and 
\ref{tab:chiralfit:wo_r0:VPS_diag}.

\begin{figure}[htbp]
   \includegraphics[width=70mm]{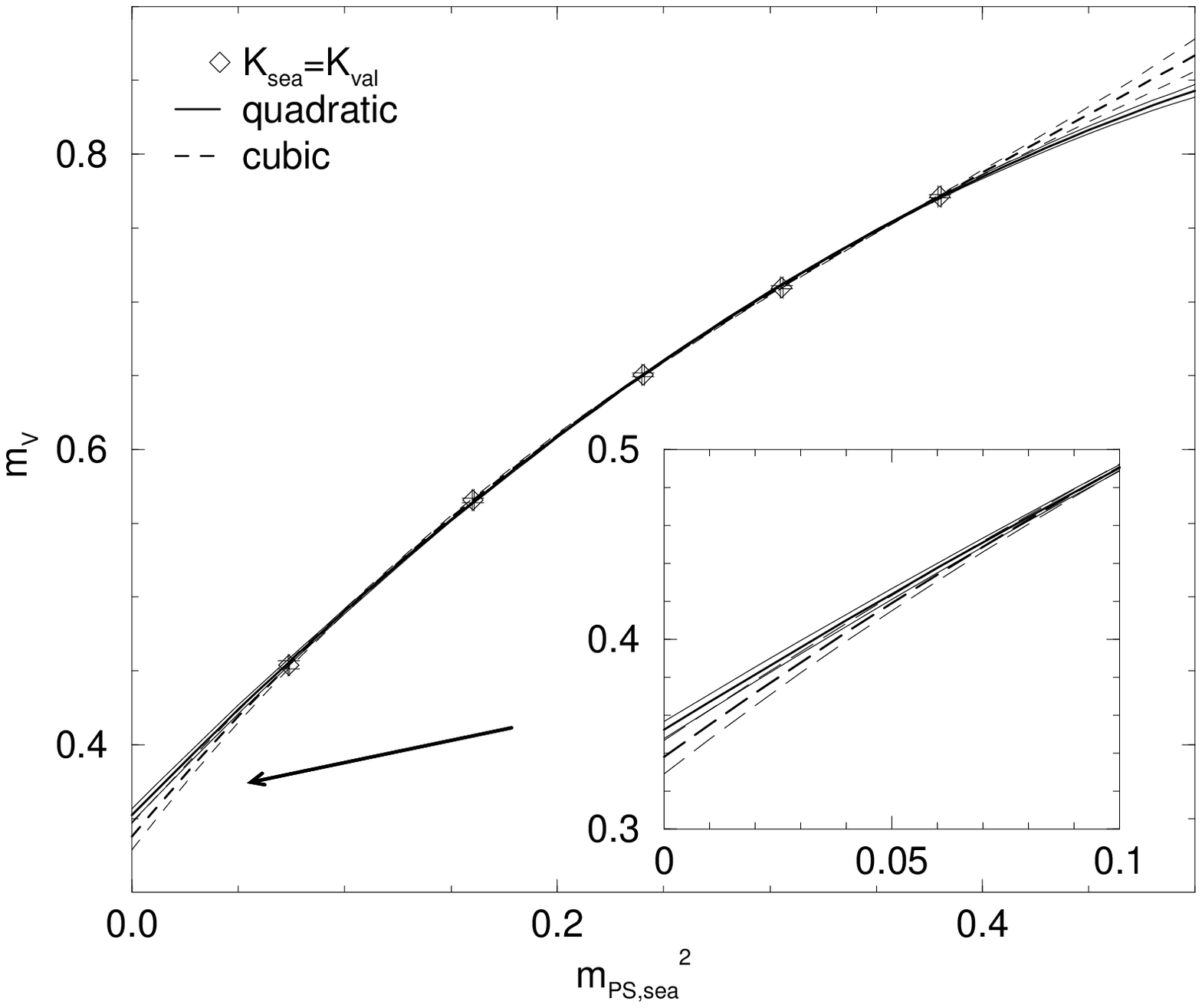}
   \caption
   {
      Comparison of quadratic and cubic diagonal fits in method-B.
   }
   \label{fig:chiralfit:wo_r0:diagonal}
\end{figure}

Fit curves of the quadratic and cubic fits to vector meson masses 
are shown in Fig.~\ref{fig:chiralfit:wo_r0:diagonal}.
While quadratic and cubic fits describe our data reasonably well
at quark masses used in the simulation,
they develop a deviation toward the chiral limit
and for heavy quarks.

\begin{figure}[htbp]
   \includegraphics[width=70mm]{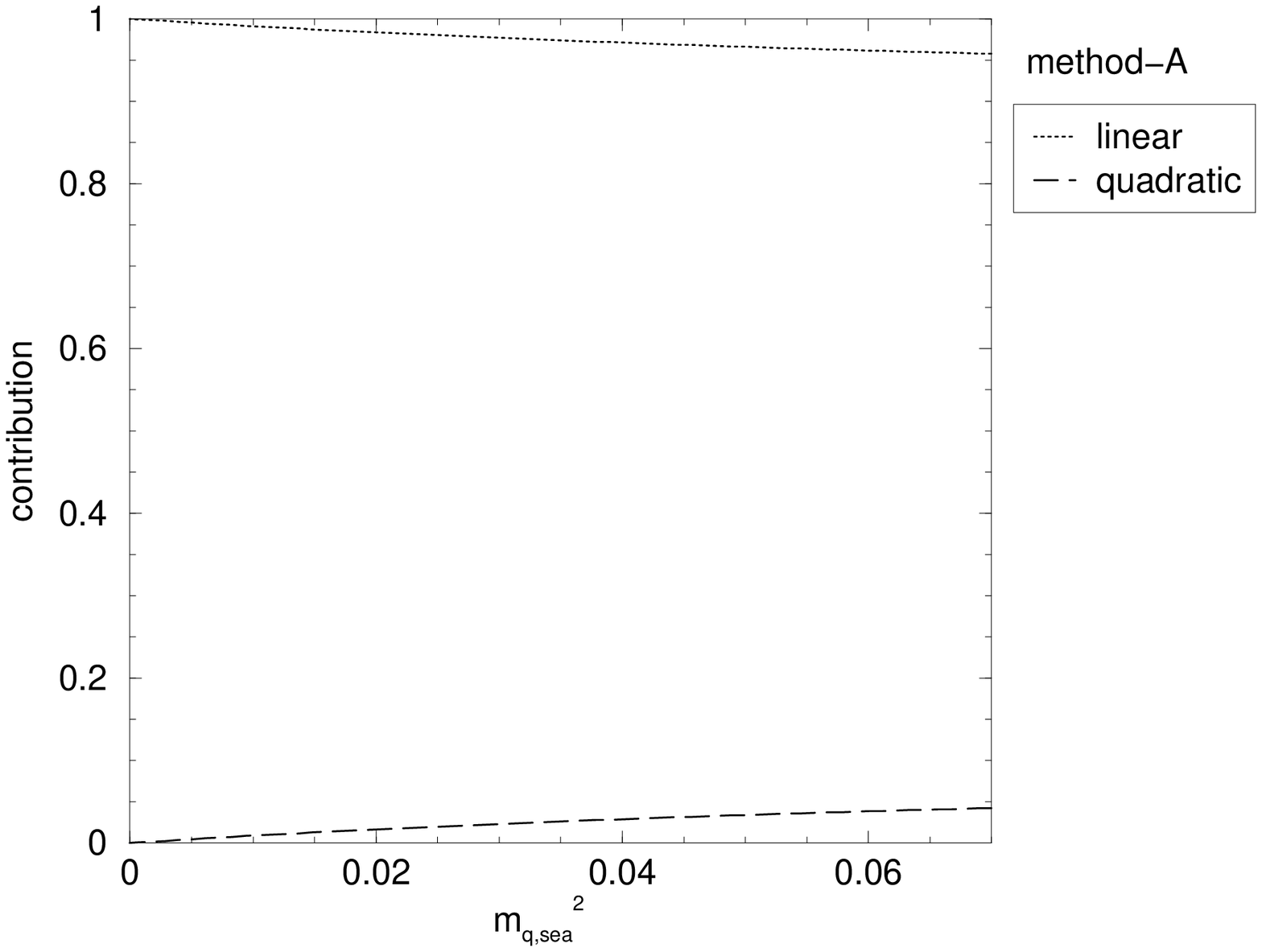}
   \includegraphics[width=70mm]{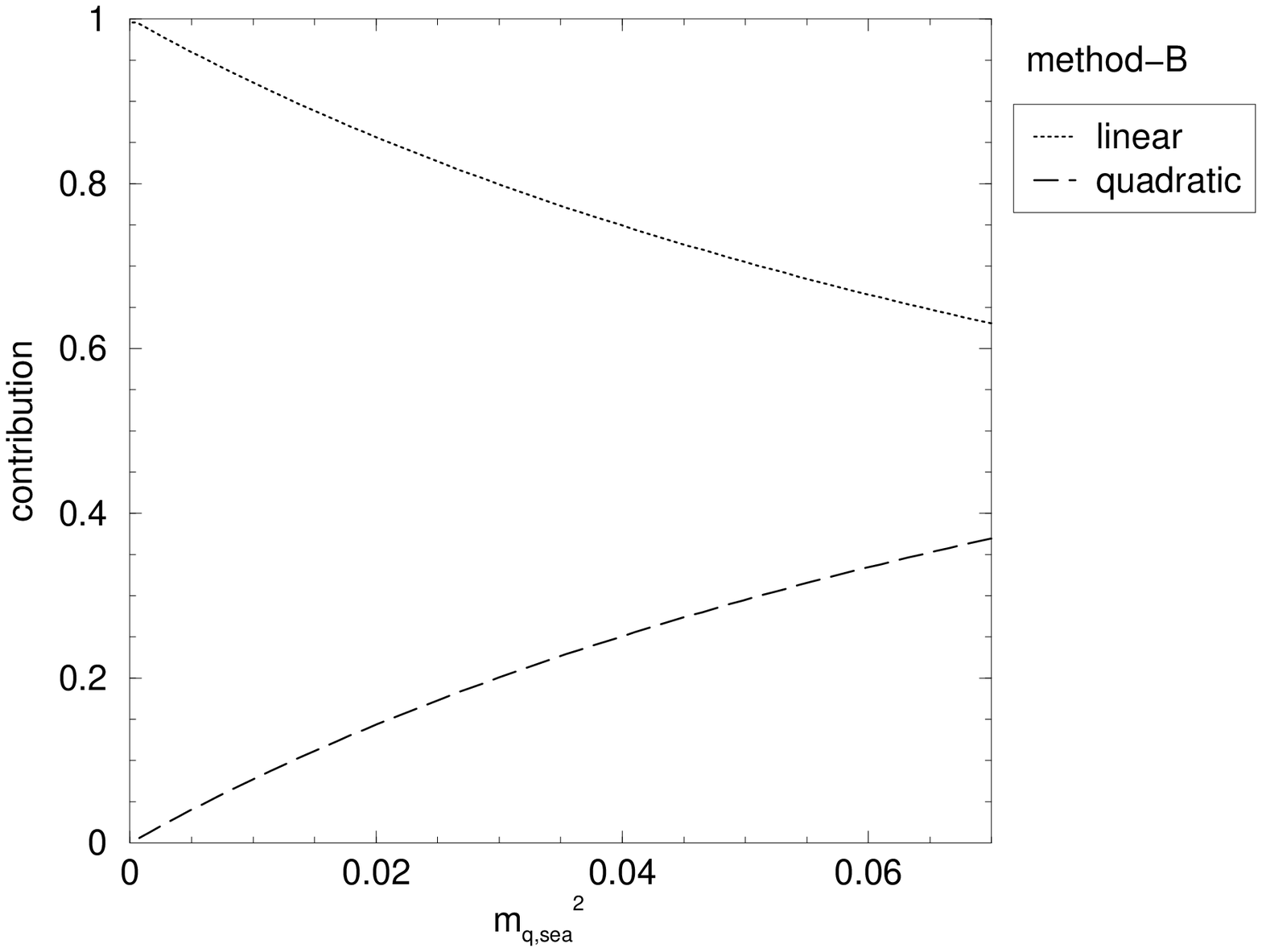}
   \caption
   {
      Relative magnitude of contribution of linear and quadratic 
      terms in quadratic diagonal fit of PS meson masses.
   }
   \label{fig:chiralfit:wo_r0:contrib-PSK}
\end{figure}
\begin{figure}[htbp]
   \includegraphics[width=70mm]{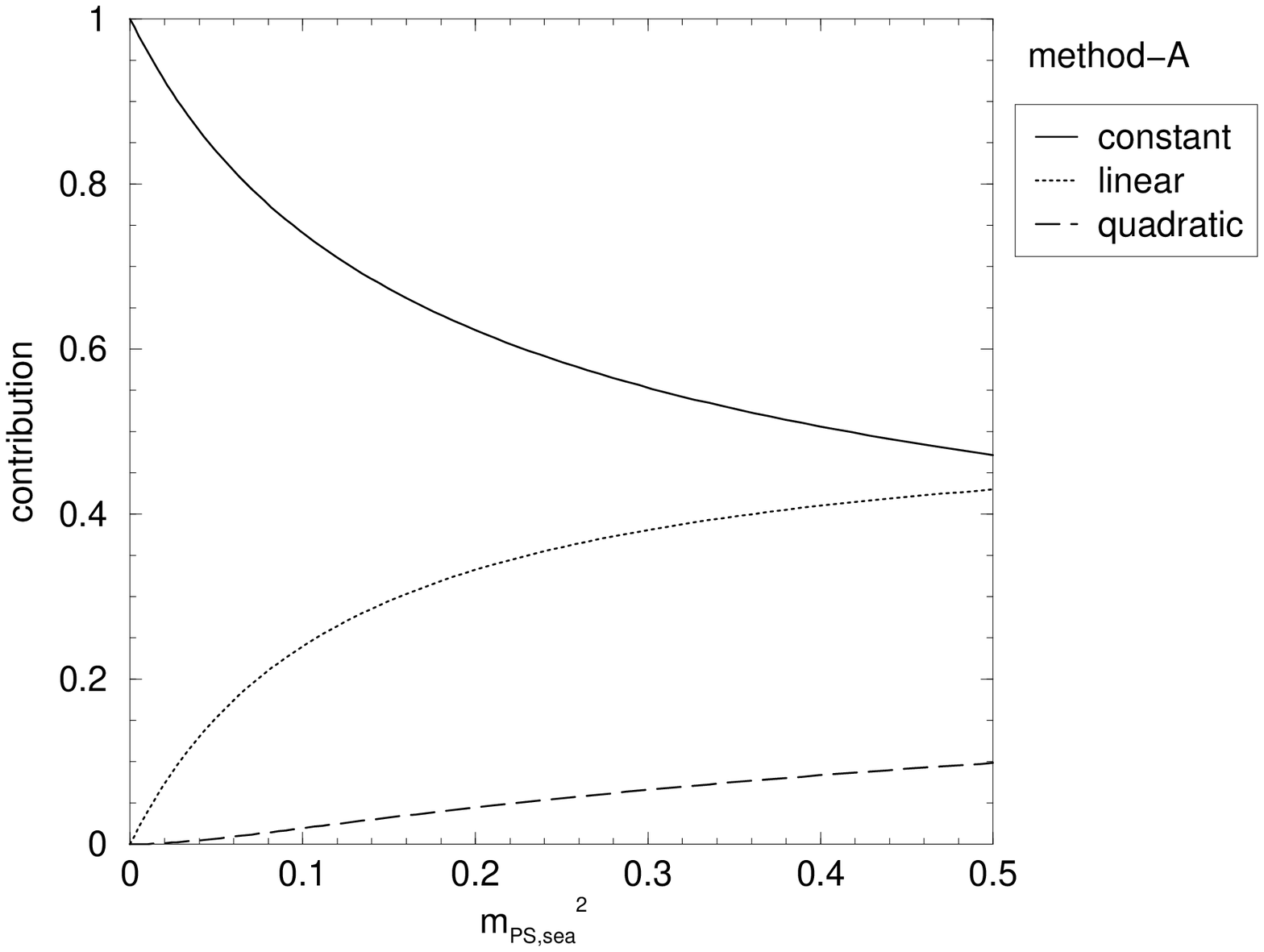}
   \includegraphics[width=70mm]{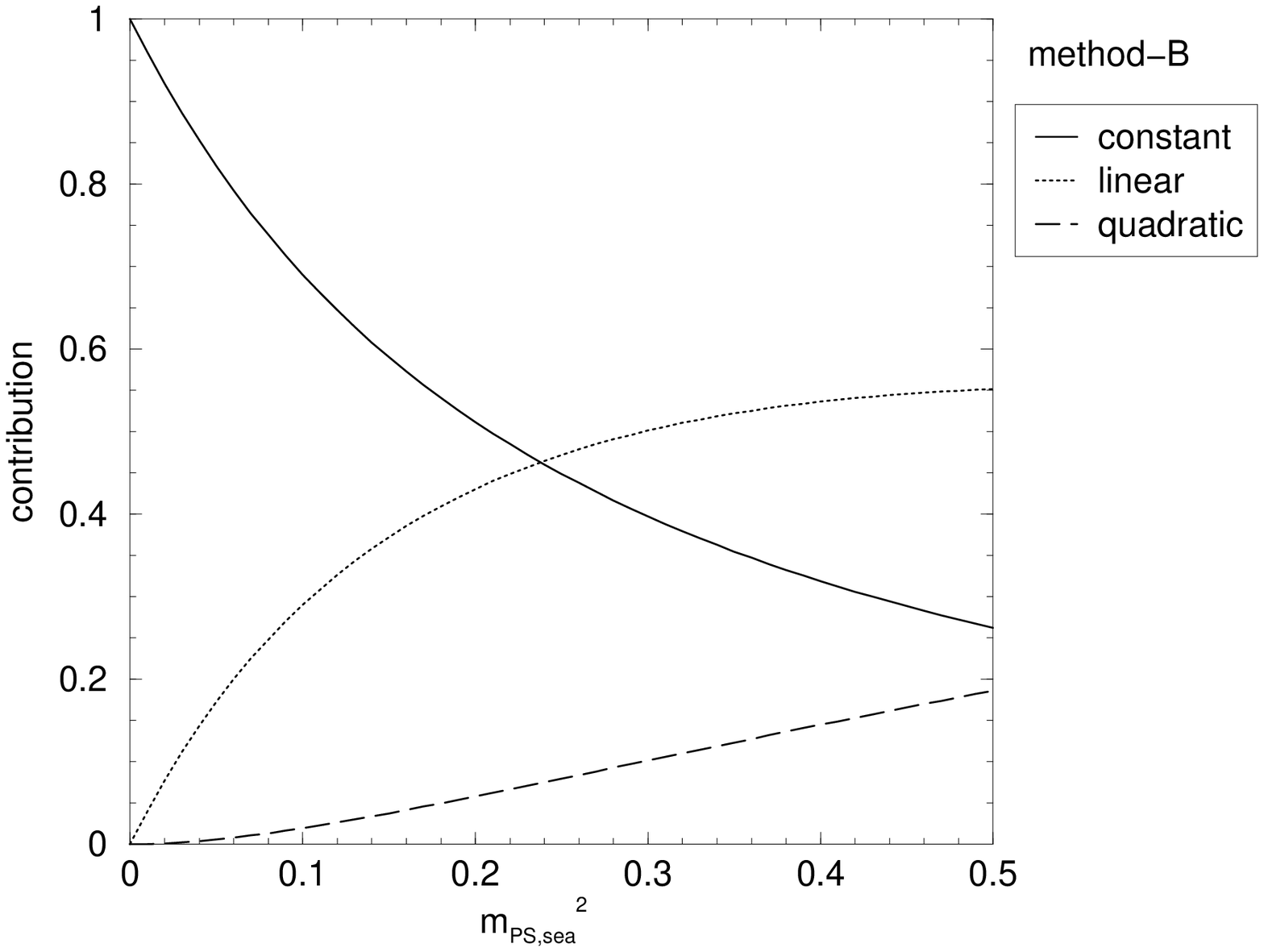}
   \caption
   {
      Relative magnitude of contribution of constant, linear and 
      quadratic terms in quadratic diagonal fit of vector meson masses.
   }
   \label{fig:chiralfit:wo_r0:contrib-VPS}
\end{figure}

Figure~\ref{fig:chiralfit:wo_r0:contrib-PSK} compares
the relative magnitude of the linear and quadratic terms 
in the quadratic diagonal fit to PS meson mass 
in the method-A and B.
As the quark mass increases,
the magnitude of the quadratic contribution
in the method-B increases more rapidly than in the method-A;
it is no longer a small correction
at the simulated quark masses 
($m_{\rm q,diag} \! \simeq \! 0.015$\,--\,0.055).
A similar situation is observed in the chiral fit to 
vector meson masses 
as shown in Fig.~\ref{fig:chiralfit:wo_r0:contrib-VPS}.

\begin{figure}[htbp]
   \includegraphics[width=70mm]{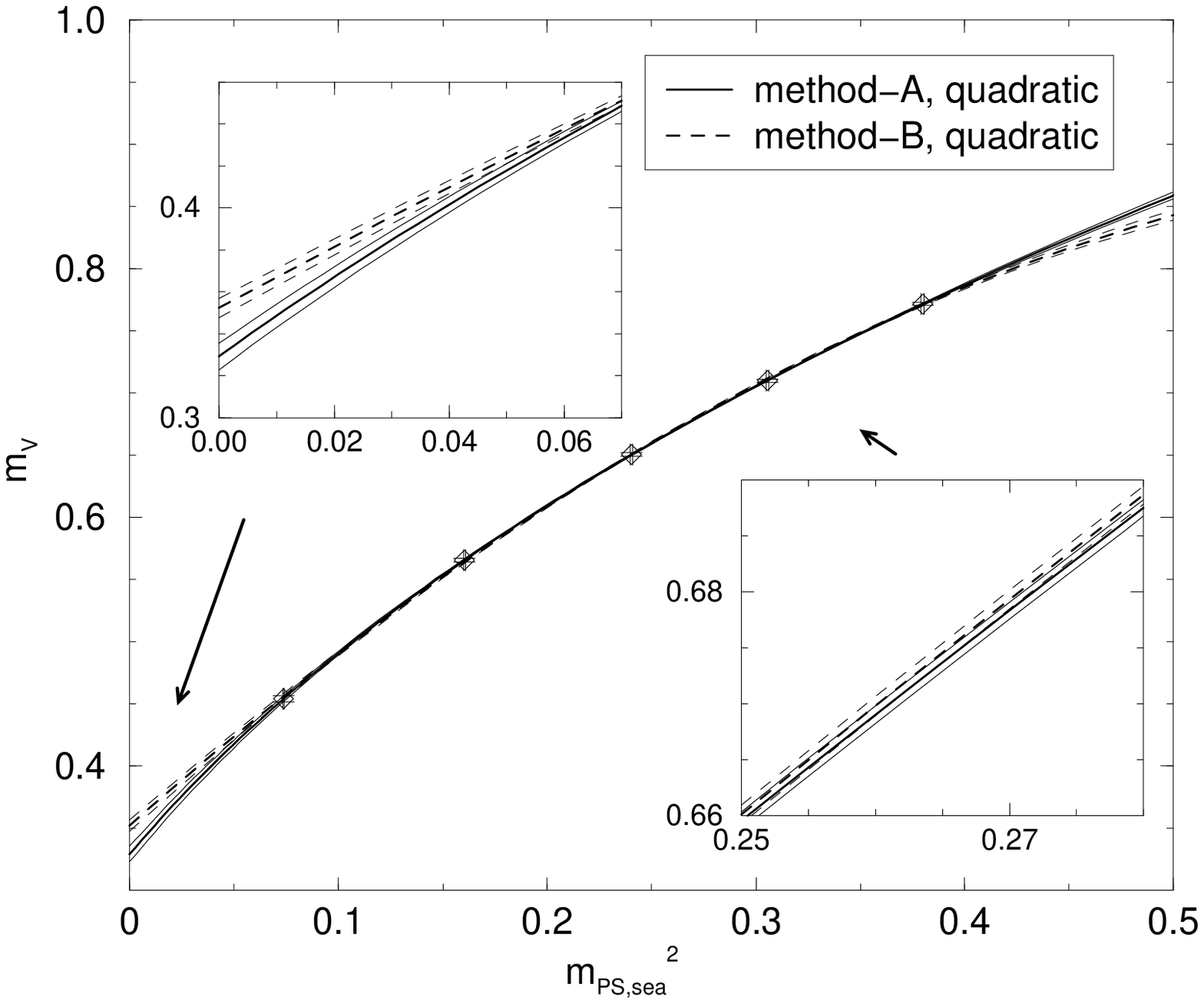}
   \includegraphics[width=70mm]{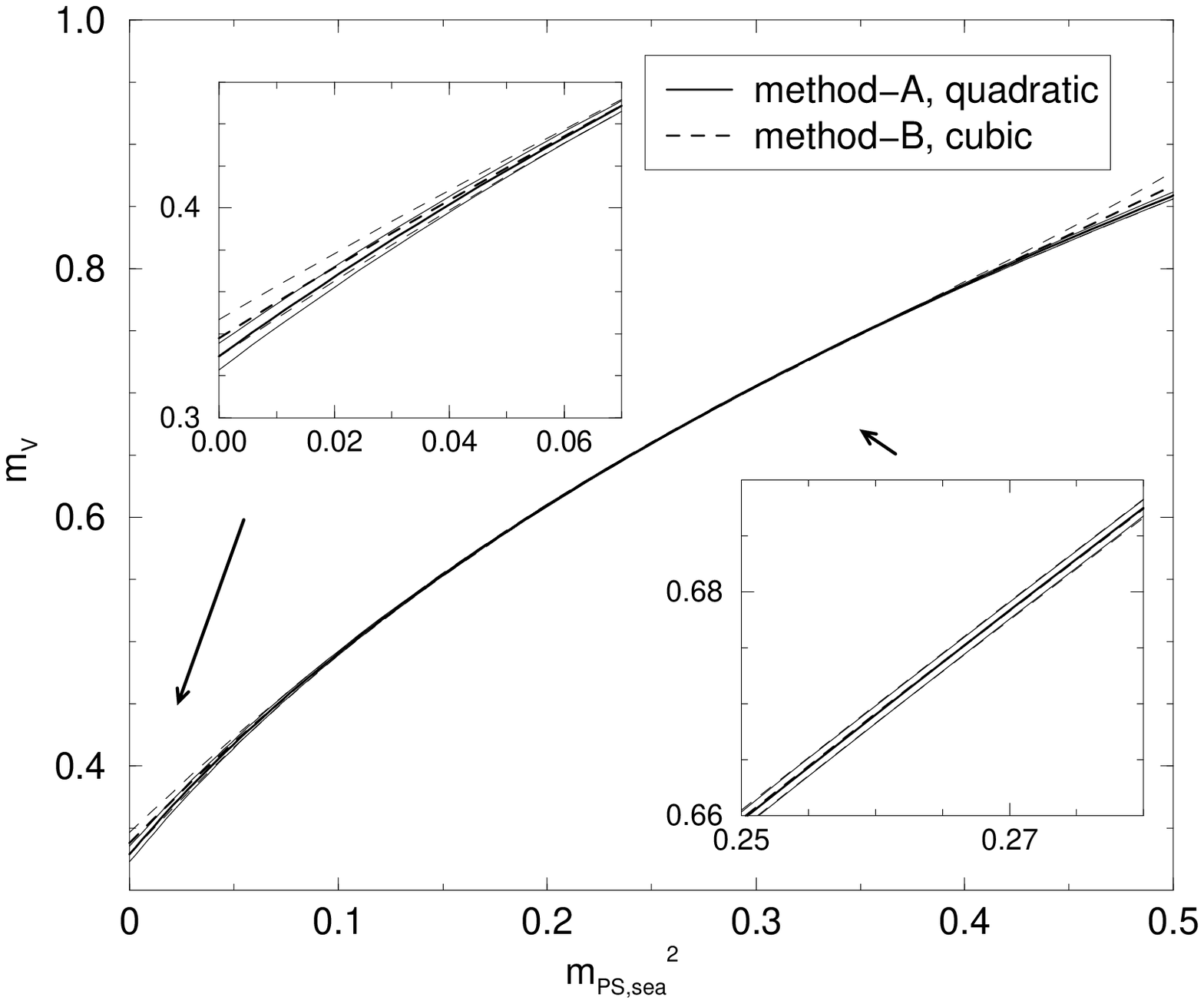}
   \caption
   {
      Comparison of diagonal fits in method-A and B.
   }
   \label{fig:chiralfit:wo_r0:compare-VPS}
\end{figure}

We come to conclude that 
the chiral expansions of meson masses in lattice units, 
Eqs.~(\ref{eqn:chiralfit:wo_r0:diag:PSK})
and (\ref{eqn:chiralfit:wo_r0:diag:VPS}),
have poor convergence properties compared to those in unit of $r_0$,
Eqs.~(\ref{eqn:chiralfit:w_r0:diag:PSK}) 
and (\ref{eqn:chiralfit:w_r0:diag:VPS})
in the method-A,
and the cubic term should not be ignored in the method-B.
We directly confirm this point 
in Fig.~\ref{fig:chiralfit:wo_r0:compare-VPS},
where the fit results for vector mesons from the method-A
are converted to lattice units using Eq.~(\ref{eqn:chiralfit:w_r0:r0inv_vs_Kinv})
and compared with the fits of method-B.
The cubic fit in the method-B shows a good consistency 
with the quadratic fit in the method-A,
while the quadratic fit in method-B does not.


The combined chiral fit including cubic terms is not very stable 
because it contains a number of free parameters.
In this study, therefore, we do not use the method-B
to extract the physical hadron spectrum.

\begin{figure}[htbp]
   \includegraphics[width=70mm]{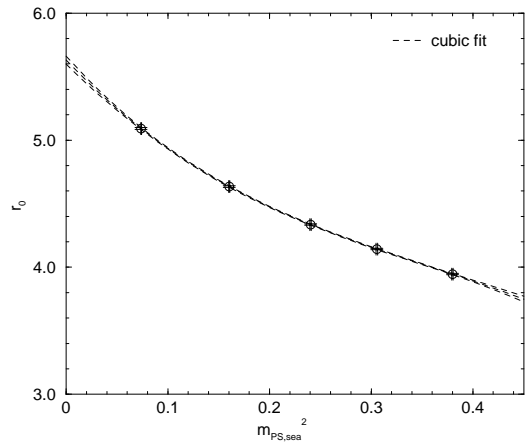}
   \caption
   {
      Chiral extrapolation of $r_0$ in terms of $m_{\rm PS,sea}^2$.
   }
   \label{fig:chiralfit:wo_r0:r0_vs_mPS2}
\end{figure}

Before we turn to details of the determination of the hadron spectrum, 
let us make additional comments on the failure of the method-B 
with our data.
%
%
Figure~\ref{fig:chiralfit:wo_r0:r0_vs_mPS2} 
shows the chiral fit of $r_0$ as a function of $m_{\rm PS,sea}^2$
\bea
   r_0(K_{\rm sea}) = A^{\prime}_{r_0}
                     +B^{\prime}_{r_0} \, m_{\rm PS,sea}^2
                     +C^{\prime}_{r_0} \, m_{\rm PS,sea}^4
                     +D^{\prime}_{r_0} \, m_{\rm PS,sea}^6.
\eea
Fit parameters in Table~\ref{tab:chiralfit:wo_r0:r0}
show that a large contribution of higher order terms is present 
also in this fit.
By substituting this parametrization of $r_0$ 
to Eqs.~(\ref{eqn:chiralfit:w_r0:diag:PSK}) and 
(\ref{eqn:chiralfit:w_r0:diag:VPS})
(diagonal fits in the method-A),
the large higher order corrections appear 
in Eqs.~(\ref{eqn:chiralfit:wo_r0:diag:PSK}) and 
(\ref{eqn:chiralfit:wo_r0:diag:VPS})
(diagonal fits in the method-B).
Conversely, why the method-A works well is that
large contributions of higher order terms 
in hadron masses and $r_0$
cancel with each other at least partially.

\begin{figure}[htbp]
   \includegraphics[width=70mm]{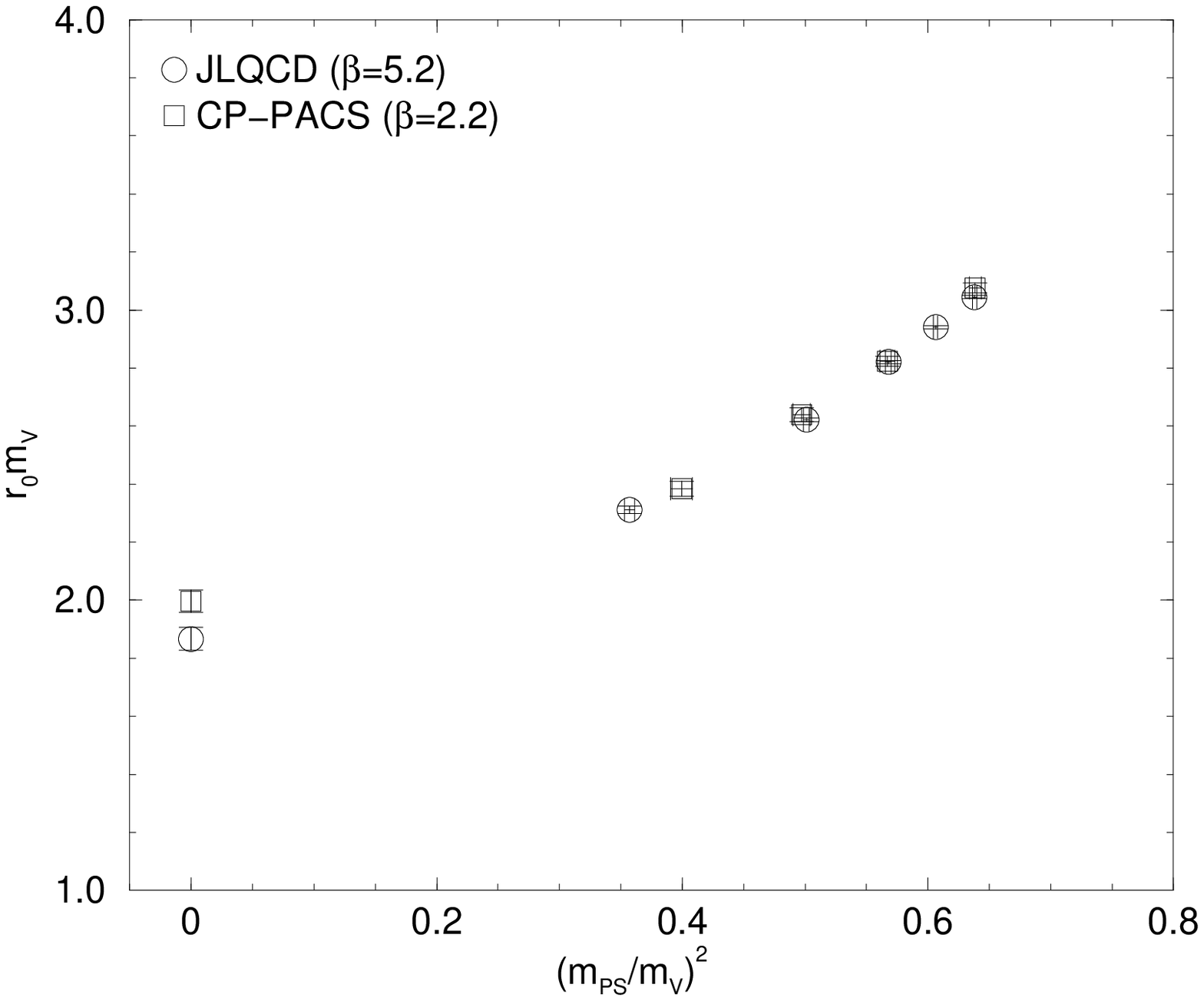}
   \includegraphics[width=70mm]{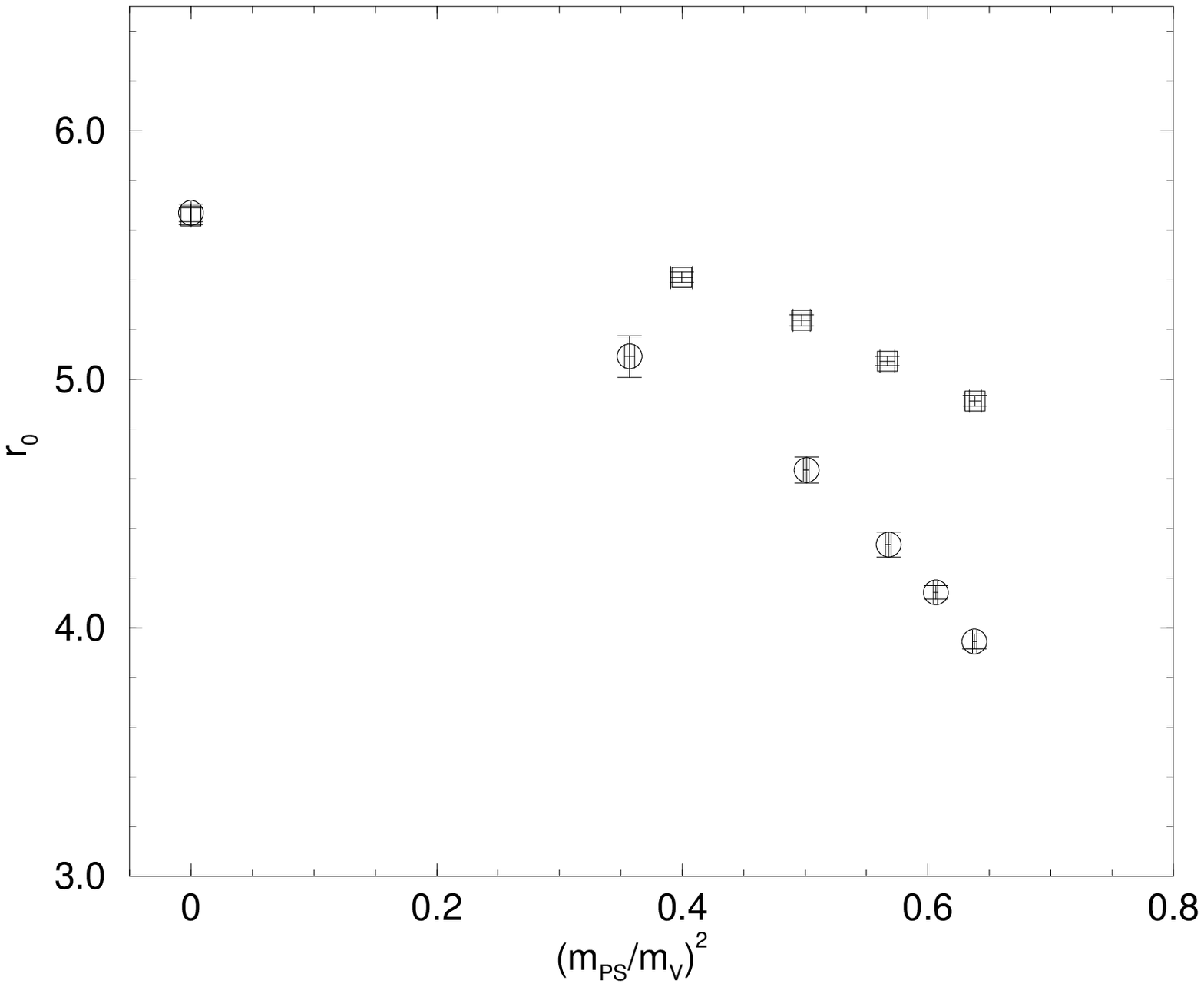}
   \caption
   {
      Comparison of sea quark mass dependence of $r_0 m_{\rm V}$ 
      (left figure) and $r_0$ in lattice units (right figure)
      with those in CP-PACS data.
      We estimate $r_0 m_{\rm V}$ in the chiral limit 
      in the CP-PACS data from linear fit 
      in terms of $(r_0 m_{\rm PS})^2$.
   }
   \label{fig:chiralfit:wo_r0:r0_scaling}
\end{figure}

We note that the method-B works well
in the CP-PACS's study~\cite{Spectrum.Nf2.CP-PACS}, 
where they took similar simulation parameters
but with different lattice actions, namely
the RG-improved gauge action
and the tadpole improved clover quark action.
We compare the CP-PACS data of $r_0 m_{\rm V}$ at $\beta\!=\!2.2$ 
and ours in Fig.~\ref{fig:chiralfit:wo_r0:r0_scaling}.
A good consistency in the whole range of the quark mass
suggests that two groups' data are in the scaling region.
However, 
we find that the CP-PACS data of $r_0$ in lattice units
show much milder dependence on the sea quark mass than ours.
This is the reason why the method-B works well in the CP-PACS study,
but does not with our data.
%
%
%
%
%
It is, of course, not surprising that 
different lattice actions lead to different sea quark mass dependences
of hadron masses and $r_0$ in the lattice units.
However, 
as discussed above,
the much stronger dependence with our choice of the
lattice action is practically problematic,
if one carries out the chiral extrapolation in lattice units.
%

\subsection{calculation of hadron spectrum}
\label{sec:chiralfit:spectrum}


The hadron spectrum at the physical quark mass is
determined as follows.
%
%
The pion and $\rho$ meson masses normalized by $r_0$ are
determined by tuning their ratio ${(r_0 m_{\pi})}/{(r_0 m_{\rho})}$
to its experimental value, i.e., by solving the equation,
\bea
   \frac{(r_0 m_{\pi})}
        { A^{\rm V}
         +\left(B^{\rm V}_{\rm s}+B^{\rm V}_{\rm v}\right)
          (r_0 m_{\pi})^2
         +C^{\rm V}_{\rm sv} (r_0 m_{\pi})^4
        }
   = 
   \frac{m_{\pi, \rm exp}}{m_{\rho, \rm exp}},
   \label{eqn:chiralfit:spectrum:fix_mud}
\eea 
where we denote the experimental value of hadron mass by
$m_{\rm had, exp}$, and 
$r_0$ represents the Sommer scale at the physical sea quark mass.
%
%
The hopping parameter corresponding to
the physical light quark mass, $K_{ud}$,
is fixed by solving 
\bea
   \left\{
      \mbox{ r.h.s of Eq.~(\ref{eqn:chiralfit:w_r0:PSK}) with  } 
      K_{\rm sea}\!=\!K_{\rm val,1}\!=\!K_{\rm val,2}\!=\!K_{ud}
   \right\}
   & = &
   (r_0 m_{\pi})^2.
\eea
%
%
Then we determine $r_0$,
which is required to convert $r_0 m_{\pi}$ and $r_0 m_{\rho}$ 
to $m_{\pi}$ and $m_{\rho}$,
from $K_{ud}$ and Eq.~(\ref{eqn:chiralfit:w_r0:r0inv_vs_Kinv}).

We test two meson mass inputs to fix the strange quark mass:
\bi
   \item In the first method, we use the kaon mass as input.
         The hopping parameter corresponding to the strange quark
         mass, $K_s$, is determined by solving            
         \bea
            \frac{\sqrt{
                   \mbox{ r.h.s of Eq.~(\ref{eqn:chiralfit:w_r0:PSK})}}
                 }
                 {r_0 m_{\rho}}
            = 
            \frac{m_{K, \rm exp}}{m_{\rho, \rm exp}},
         \label{eqn:chiralfit:spectrum:fix_ms:K}
         \eea
         where we set $K_{\rm sea}\!=\!K_{\rm val,1}\!=\!K_{ud}$
         and $K_{\rm val,2}\!=\!K_{s}$ in the r.h.s of 
         Eq.~(\ref{eqn:chiralfit:w_r0:PSK}).
         Then the mass of the ``$\eta_s$'' meson, 
         that is an unphysical $\bar{s}s$ PS meson,
         is determined from Eq.~(\ref{eqn:chiralfit:w_r0:PSK}),
         and used to calculate strange vector meson and strange baryon
         masses.
         We refer to this meson mass input as $K$-input.
   \item In the second method, we use the $\phi$ meson mass as input
         assuming that it is a pure $\bar{s}s$ vector meson.
         The $\eta_s$ meson mass is fixed from 
         \bea
            \frac{\mbox{ r.h.s of Eq.~(\ref{eqn:chiralfit:w_r0:VPS})}}
                 {r_0 m_{\rho}}
            = 
            \frac{m_{\phi, \rm exp}}{m_{\rho, \rm exp}},
         \label{eqn:chiralfit:spectrum:fix_ms:phi}
         \eea
         where we set 
         $m_{\rm PS,sea}\!=\!m_{\pi}$
         and 
         $m_{\rm PS,val,1}\!=\!m_{\rm PS,val,2}\!=\!m_{\eta_s}$.
         We determine $K_s$ from $m_{\eta_s}$ and 
         Eq.~(\ref{eqn:chiralfit:w_r0:PSK}).
         This input is called $\phi$-input.
\ei


The full spectrum of non-strange and strange hadrons is
determined by substituting $K_{ud}$, $m_{\pi}$, $K_s$, $m_{\eta_s}$
and $r_0$ to Eqs.~(\ref{eqn:chiralfit:w_r0:PSK}), 
(\ref{eqn:chiralfit:w_r0:VPS}), (\ref{eqn:chiralfit:w_r0:OPS:Sigma}),
(\ref{eqn:chiralfit:w_r0:OPS:Lambda}) and 
(\ref{eqn:chiralfit:w_r0:DPS}).
We use the lattice spacing $a$ determined from $m_{\rho}$
to convert the hadron masses in lattice units to 
those in physical units.
We note that this estimate of the scale $a$ is subject to 
a systematic uncertainty due to the use of 
the polynomial fitting forms for the chiral extrapolation.
However, if we use $r_0$ as the input to set the scale,
we obtain a consistent result for $a$ within errors.
The results of $K_{ud}$, $K_s$ and $a^{-1}$ are 
collected in Table~\ref{tab:chiralfit:spectrum:K}.


We repeat the above analysis using each of the alternative fit forms
Eqs.~(\ref{eqn:chiralfit:w_r0:alter:PSK}),
(\ref{eqn:chiralfit:w_r0:alter:VPS})
(\ref{eqn:chiralfit:w_r0:alter:OPS}) and 
(\ref{eqn:chiralfit:w_r0:alter:DPS}).
The largest deviation in the hadron spectrum among these 
analyses is taken as the systematic error due to the choice of the 
chiral fit forms.


For the chiral extrapolation of $r_0$, 
we find that an alternative form 
\bea
   r_0(K_{\rm sea}) = A_{r_0}^{\prime} 
                     +B_{r_0}^{\prime} \mu_{\rm PS,sea}^2
\eea
also describes our data well.
However, the hadron spectrum calculated using this fit
is completely consistent with those using
Eq.~(\ref{eqn:chiralfit:w_r0:r0inv_vs_Kinv}).
We therefore ignore the systematic error due to 
the choice of the fit form Eq.~(\ref{eqn:chiralfit:w_r0:r0inv_vs_Kinv}).


The systematic error of the measured value of $r_0(K_{\rm sea})$ leads
to an additional uncertainty in the result of the hadron spectrum.
We perform the calculation of the spectrum
with $r_0(K_{\rm sea})$ shifted
by its systematic error at one value of $K_{\rm sea}$.
This calculation is repeated for all $K_{\rm sea}$ and 
the largest deviation in the spectrum is included into 
the systematic error.

\subsection{chiral extrapolation in quenched QCD}

\begin{figure}[htbp]
   \includegraphics[width=70mm]{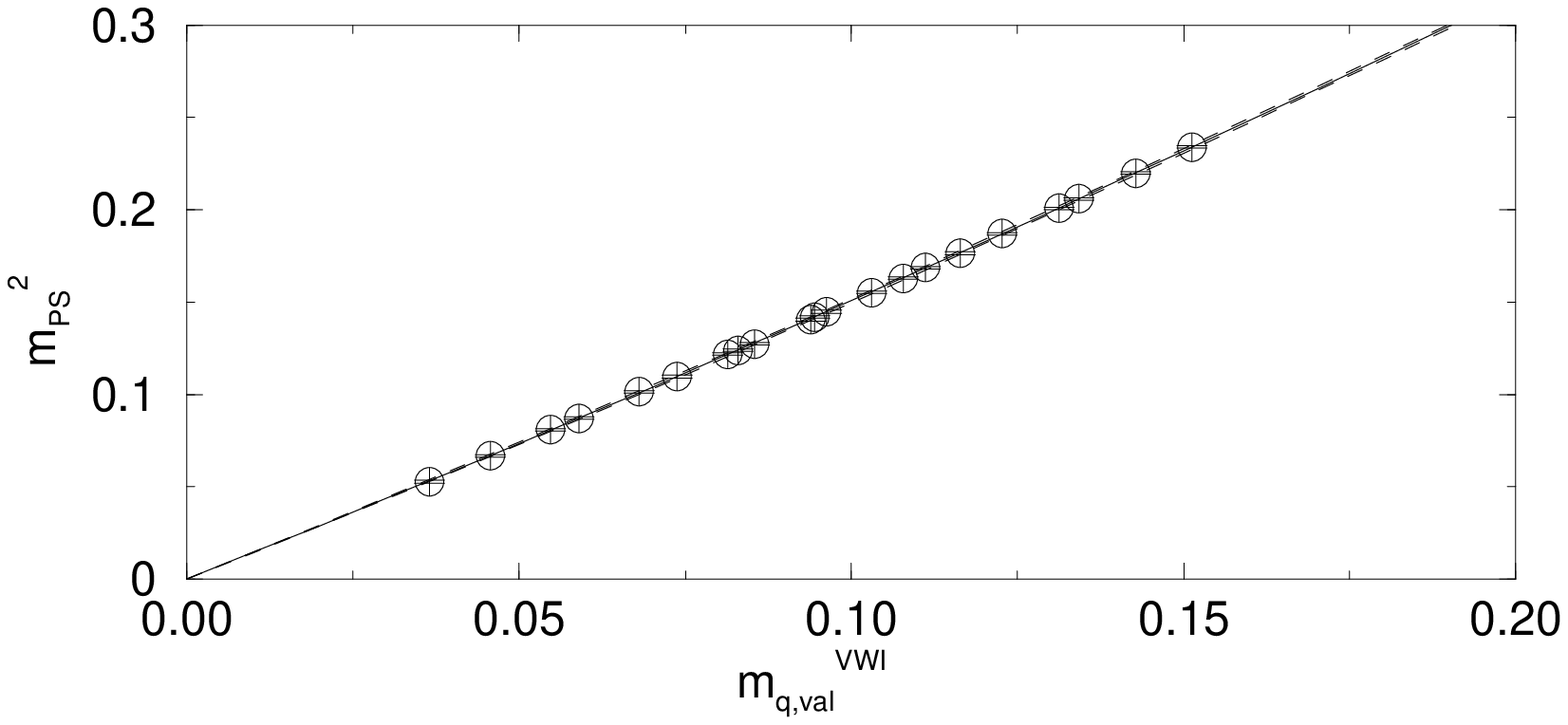}
   \includegraphics[width=70mm]{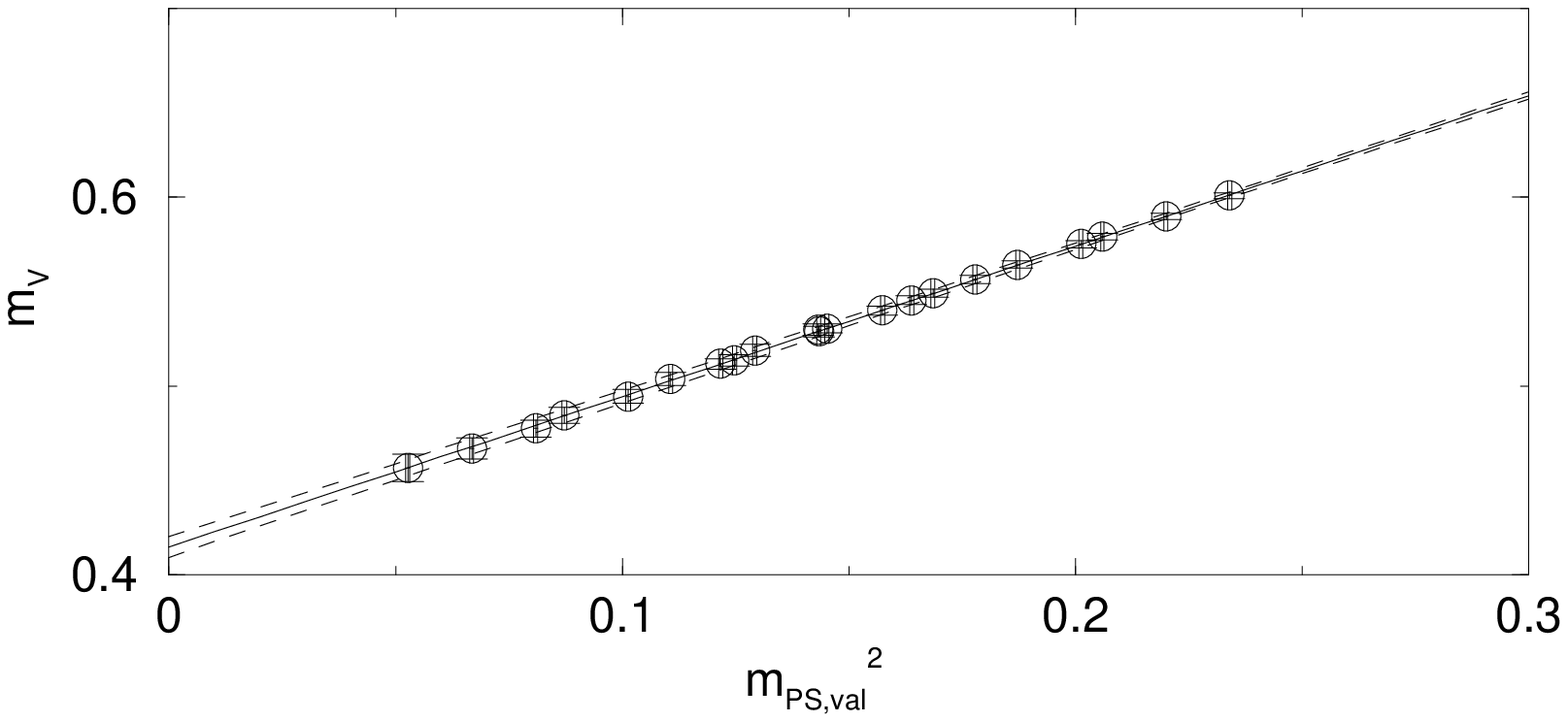}
   \includegraphics[width=70mm]{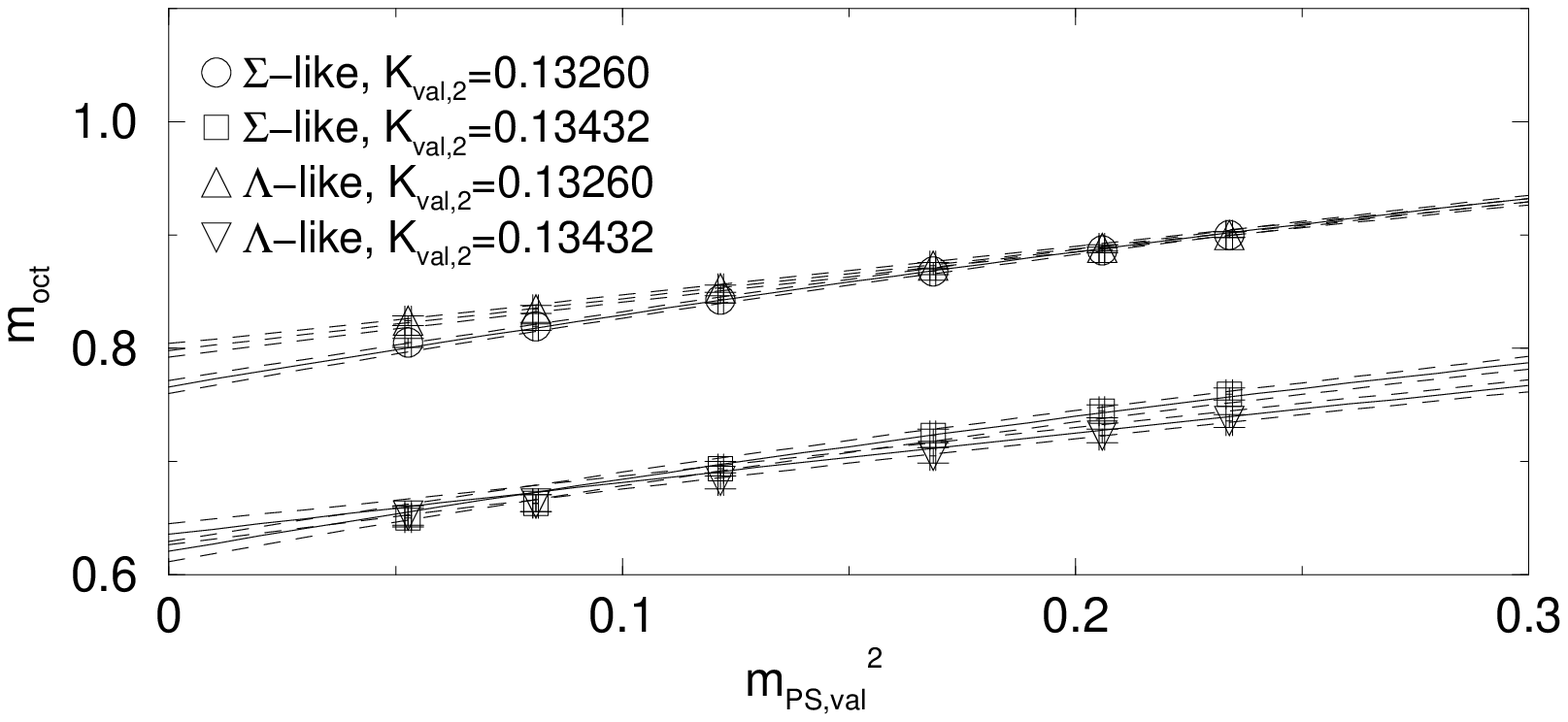}
   \includegraphics[width=70mm]{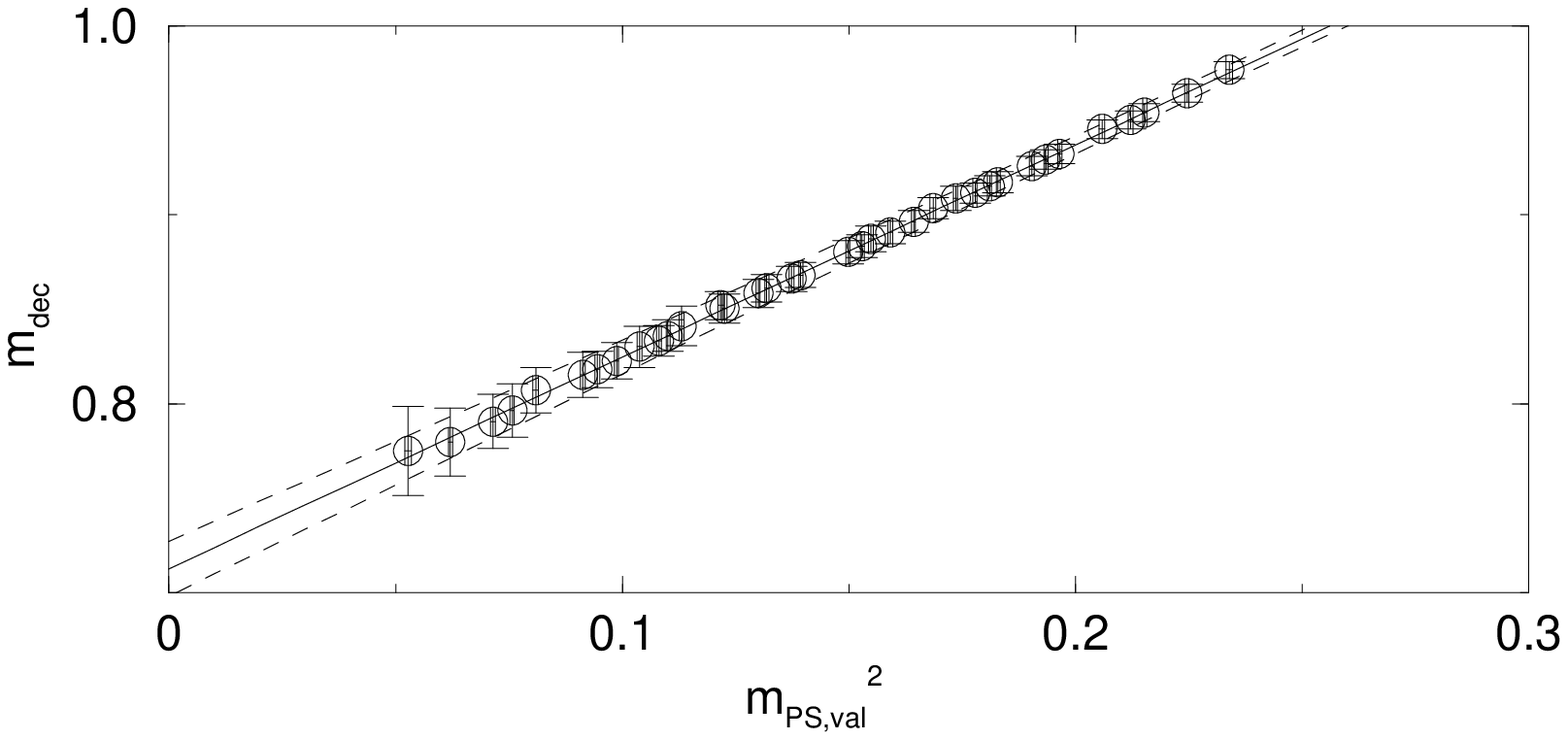}
   \caption
   {
      Chiral extrapolation of meson (top figures) and baryon masses
      (bottom figures) in quenched QCD. 
      For octet baryon masses, 
      we plot only data at $K_{\rm val,2}\!=\!0.13260$ 
      and 0.13432 for simplicity.
   }
   \label{fig:chiralfit:qQCD:fit}
\end{figure}

The chiral extrapolations in quenched QCD are performed 
using fit forms which are obtained 
from those used in the full QCD analysis
by dropping all terms describing the sea quark mass dependence.
%
%
Namely, fitting forms for meson masses are 
\bea
   m_{\rm PS}(K_{\rm val,1},K_{\rm val,2})^2
   & = &
   B^{\rm PS}_{\rm q} \, m_{\rm q,val}
  +C^{\rm PS}_{\rm q} \, m_{\rm q,val}^2,
   \label{eqn:chiralfit:qQCD:PSK}
   \\
   m_{\rm V}(K_{\rm val,1},K_{\rm val,2})
   & = & 
   A^{\rm V}_{\rm q}
  +B^{\rm V}_{\rm q} \, m_{\rm PS,val}^2.
   \label{eqn:chiralfit:qQCD:VPS}
\eea
%
%
The following fitting forms are used for baryon masses
\bea
   m_{{\rm oct}, \Sigma}(K_{\rm val,1},K_{\rm val,2},K_{\rm val,2})
   & = &
   A^{\rm O}_{\rm q} 
  +(F^{\rm O}_{\rm q}-D^{\rm O}_{\rm q}) \, m_{\rm PS,val,1}^2
  +2 F^{\rm O}_{\rm q} \, m_{\rm PS,val,2}^2
  +( C^{{\rm O}}_{\rm q} + C^{{\rm O},\Sigma}_{\rm q} ) \, 
   m_{\rm PS,val,1}^4
   \nn \\
   &   &
  +( C^{{\rm O}}_{\rm q} - C^{{\rm O},\Sigma}_{\rm q})  \,    
   m_{\rm PS,val,2}^4,
   \label{eqn:chiralfit:qQCD:OPS_Sigma}
   \\[3mm]
   m_{{\rm oct},\Lambda}(K_{\rm val,1},K_{\rm val,2},K_{\rm val,2})
   & = &
   A^{\rm O}_{\rm q} 
  +\left(F^{\rm O}_{\rm q}+\frac{D^{\rm O}_{\rm q}}{3}\right) \, 
   m_{\rm PS,val,1}^2
  +2 \left(F^{\rm O}_{\rm q}-\frac{2}{3}D^{\rm O}_{\rm q}\right)
                                          \, m_{\rm PS,val,2}^2
   \nn \\
   &   &
  +( C^{\rm O}_{\rm q} + C^{{\rm O},\Lambda}_{\rm q} ) \, 
   m_{\rm PS, val,1}^4
  +( C^{\rm O}_{\rm q} - C^{{\rm O},\Lambda}_{\rm q} ) \, 
   m_{\rm PS, val,2}^4,
   \label{eqn:chiralfit:qQCD:OPS_Lambda}
   \\[5mm]
   m_{\rm dec}(K_{\rm val,1}, K_{\rm val,2}, K_{\rm val,2})
   & = &
   A^{\rm D}_{\rm q} 
  +B^{\rm D}_{\rm q} \, m_{\rm PS, val}^2.
   \label{eqn:chiralfit:qQCD:DPS}
\eea
%
These forms fit to our data very well 
as shown in Fig.~\ref{fig:chiralfit:qQCD:fit}.
Fit parameters are summarized in Table~\ref{tab:chiralfit:qQCD:fit}.
%
%
%
The hadron spectrum is calculated in an analogous way
to that for full QCD.
Resulting values of $K_{ud}$, $K_s$ and $a^{-1}$ are 
summarized in Table~\ref{tab:chiralfit:qQCD:K}.


\section{Sea Quark Effects on Hadron Spectrum} 
\label{sec:SQE}

\subsection{Sea quark effects at simulated quark mass} 
\label{sec:SQE:J}

\begin{figure}[htbp]
   \includegraphics[width=70mm]{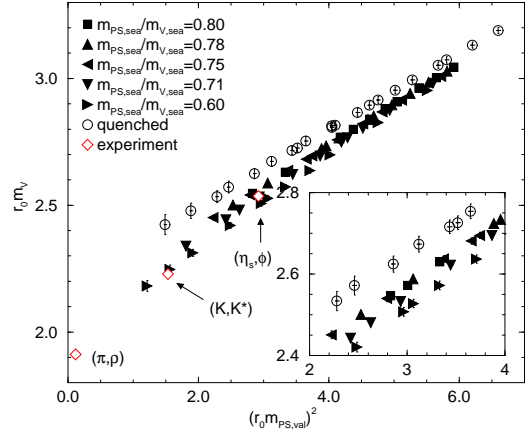}
   \caption
   {
      Vector meson mass as a function of PS meson mass squared 
      at each sea quark mass in full QCD and in quenched QCD.
      The experimental values of meson masses are also plotted  
      using our result $r_0\!=\!0.497$~fm,
      which is determined from 
      Eq.~(\ref{eqn:chiralfit:w_r0:r0inv_vs_Kinv})
      and $K_{ud}$ and $a$ in Table~\ref{tab:chiralfit:spectrum:K}.
   }
   \label{fig:SQE:J:mV_vs_mPS2-r0}
\end{figure}

Figure~\ref{fig:SQE:J:mV_vs_mPS2-r0} compares
the valence quark mass dependence of the vector meson mass 
at each sea quark mass in full QCD and in quenched QCD.
We observe that quenched data have a significantly 
smaller slope than experimental points. 
This leads to the underestimation of 
the $K^*$--$K$ hyperfine splitting in quenched QCD.

The slopes in full QCD data 
are clearly larger than in quenched QCD, 
and increase for decreasing sea quark mass.
This is reflected in a negative value of $C^{\rm V}_{\rm sv}$ 
for Eq.~(\ref{eqn:chiralfit:w_r0:VPS}) 
in Table~\ref{tab:chiralfit:w_r0:VPS}.
This sea quark effect leads to a better agreement of 
the meson spectrum in full QCD with experiment than in quenched QCD.

\begin{figure}[htbp]
   \includegraphics[width=70mm]{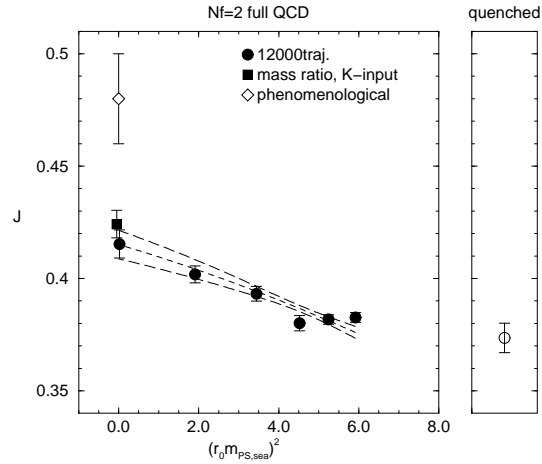}
   \caption
   {
      $J$ parameter defined by Eq.~(\ref{eqn:chiralfit:w_r0:J})
      in full (left panel) and quenched QCD (right panel).
      Dashed lines are reproduced from combined chiral fit,
      Eq.~(\ref{eqn:chiralfit:w_r0:VPS}).
      We also plot values calculated from an phenomenological 
      definition, Eq.~(\ref{eqn:SQE:J:J_massratio}), using 
      experimental spectrum (open diamond) and
      our results in Table~\ref{tab:SQE:had:meson}
      (filled square).
   }
   \label{fig:SQE:J:J}
\end{figure}

The $J$ parameter \cite{J} defined by 
\bea
   J = \left. m_{\rm V} \frac{d m_{\rm V}}{d m_{\rm PS}^2}
       \right|_{m_{\rm PS}/m_{\rm V}=m_{K}/m_{K^*}}
   \label{eqn:chiralfit:w_r0:J}
\eea
is useful to quantify the sea quark effect.
Numerical results of $J$ calculated from the partially 
quenched chiral fit Eq.~(\ref{eqn:chiralfit:w_r0:VPS_PQ})
are given in Table~\ref{tab:SQE:J}.
In Fig.~\ref{fig:SQE:J:J},
we plot $J$ in full QCD (filled circles) as a function of the sea quark mass
together with the quenched result (open circle in the right panel). 
We observe that $J$ in full QCD is close to the quenched value
at heavy sea quark masses 
corresponding to $m_{\rm PS}/m_{\rm V} \geq 0.75$
and increases as the sea quark masses decreases.

In the same figure, 
we also plot $J$ reproduced from the combined chiral fit 
Eq.~(\ref{eqn:chiralfit:w_r0:VPS}) (dashed lines).
The result is consistent with $J$ from the partially quenched fit,
as it should be,
and shows  a similar sea quark mass dependence.
We observe that $J$ extrapolated to the physical sea quark mass 
is closer to the phenomenological value \cite{J}
\bea
   J = m_{\rm K^*} \frac{ m_{K^*} - m_{\rho}  }
                        { m_K^2   - m_{\pi}^2 }
     = 0.48(2)
   \label{eqn:SQE:J:J_massratio}
\eea
than in quenched QCD.

Figure~\ref{fig:SQE:J:J}
also shows $J$ calculated 
from our results of the meson spectrum
(see Table~\ref{tab:SQE:had:meson}) using 
the above alternative definition Eq.~(\ref{eqn:SQE:J:J_massratio})
(filled square).
The result is in good agreement with other determinations, 
showing the magnitude of the sea quark effect in $J$
to be stable against the definition of $J$.

In Sec.~\ref{sec:FSE}, we pointed out that 
FSE decreases the slope $d m_{\rm V}/d m_{\rm PS}^2$.
This is confirmed numerically in $dm_{\rm V}/dm_{\rm PS}^2$ 
determined from the partially quenched chiral fit
Eq.~(\ref{eqn:chiralfit:w_r0:VPS_PQ}) 
at the lightest sea quark mass:
$dm_{\rm V}/dm_{\rm PS}^2$ is 
0.906(14) on $20^3$, 0.814(88) on $16^3$ and 0.68(30) on $12^3$.
The slope would be larger 
if we increase the spatial size beyond $20^3$.
Therefore, 
the observed sea quark mass dependence of the slope and $J$ is
a genuine effect of dynamical quarks and not an artifact of FSE.

\begin{figure}[htbp]
   \includegraphics[width=70mm]{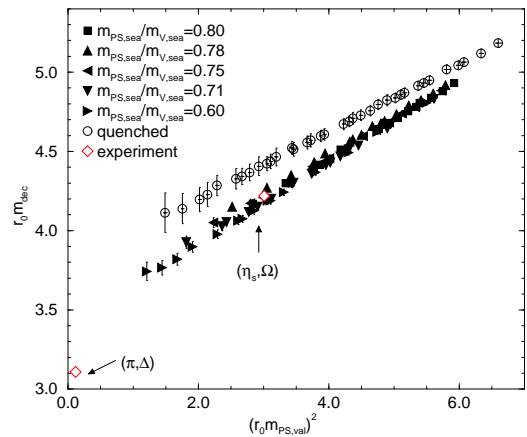}
   \caption
   {
      Decuplet baryon masses as a function of 
      PS meson mass squared at each sea quark mass 
      in full and quenched QCD.
   }
   \label{fig:SQE:J:mD_vs_mPS2-r0}
\end{figure}

A similar effect of sea quarks can be found in decuplet 
baryon masses as shown in Fig.~\ref{fig:SQE:J:mD_vs_mPS2-r0}.
However, a significant deviation still exists 
in the slope between full QCD data and 
the experimental spectrum.
%
%
%
%
We consider that 
a larger slope in full QCD is still partly masked by FSE 
on the $20^3$ volume;
volume as large as $30^3$ would be needed to reduce FSE
to a few \% level 
as discussed in Sec.~\ref{sec:FSE}.


We emphasize that 
the evidence of sea quark effects observed in this subsection
does not suffer from possibly large systematic errors 
due to the chiral extrapolation and 
the choice of inputs to fix the scale and quark masses:
the increase of the slopes,
$dm_{\rm V}/dm_{\rm PS}^2$ and $dm_{\rm dec}/m_{\rm PS}^2$,
is observed without any chiral extrapolation and inputs.
The sea quark mass dependence of $J$ is obtained 
by a short extrapolation or an interpolation 
to a relatively heavy valence quark mass
corresponding to $m_{\rm PS}/m_{\rm V}\!=\!m_{K}/m_{K^*}$.

\subsection{hadron spectrum}
\label{sec:SQE:meson}


The meson spectrum in full and quenched QCD 
is summarized in Table~\ref{tab:SQE:had:meson}.
%
%
Since our fitting functions to vector meson masses,
Eqs.~(\ref{eqn:chiralfit:w_r0:VPS}) and (\ref{eqn:chiralfit:qQCD:VPS}),
are linear in terms of the valence quark mass,
$m_{K^*}$ with $\phi$-input equals 
$\left(m_{\rho,\rm exp}+m_{\phi, \rm exp}\right)/2$
in both full and quenched QCD.
The deviation of this value from the experimental mass
$m_{K^*,\rm exp}$ is only 0.2\,\%.

\begin{figure}[htbp]
   \includegraphics[width=70mm]{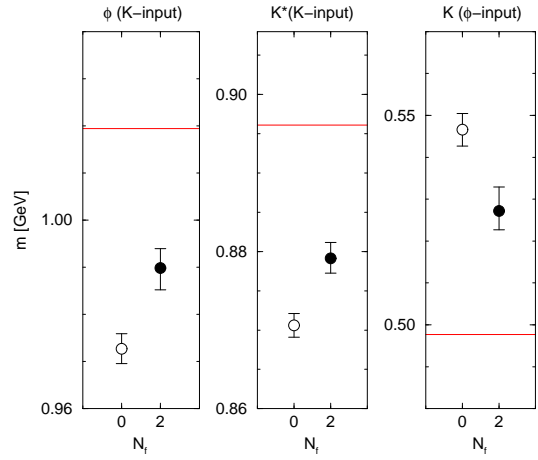}
   \caption
   {
      Comparison of strange meson masses between 
      full and quenched QCD.
      Experimental values are shown by horizontal lines.
   }
   \label{fig:SQE:had:meson}
\end{figure}

A clear difference between full and quenched QCD 
is observed in other meson masses 
as shown in Fig.~\ref{fig:SQE:had:meson}.
While the quenched meson spectrum shows a significant deviation
from experiment,
sea quark effects reduce the deviation by about 40\%.
%
%
This closer agreement of the meson spectrum with experiment 
is a consequence of the sea quark effects observed in the previous
subsection.

\begin{figure}
   \includegraphics[width=70mm]{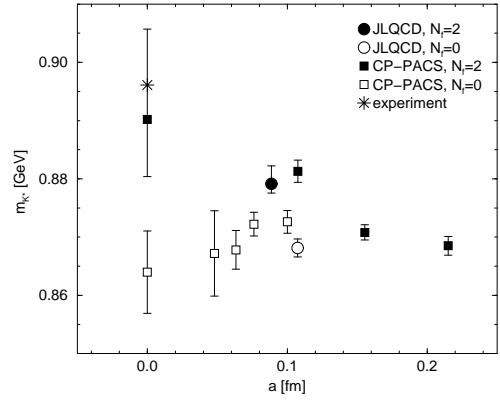}
   \caption
   {
      Mass of $K^*$ meson with $K$-input 
      as a function of lattice spacing.
   }
   \label{fig:SQE:had:scaling:meson}
\end{figure}

 
In Fig.~\ref{fig:SQE:had:scaling:meson}, 
$m_{K^*}$ with the $K$-input in full QCD 
is compared with
the CP-PACS results obtained with the RG-improved gauge 
and clover quark actions \cite{Spectrum.Nf2.CP-PACS}.
We observe that 
our $m_{K^*}$ is consistent with the CP-PACS' result 
at a similar lattice spacing,
and is at the lower edge of their estimate in the continuum limit.
In the same figure,
we also make a comparison in quenched QCD
with the CP-PACS results
obtained with the plaquette gauge and the Wilson quark actions
in Ref.~\cite{Spectrum.Nf0.CP-PACS}.
Two groups' results show a good agreement with each other.
%
%
%
%
%
%
These observations suggest that 
the scaling violation is small
in our data both in full and quenched QCD
and, hence,
the closer agreement of the meson spectrum in full QCD 
with experiment is a genuine effect of sea quarks.

\begin{figure}
   \includegraphics[width=70mm]{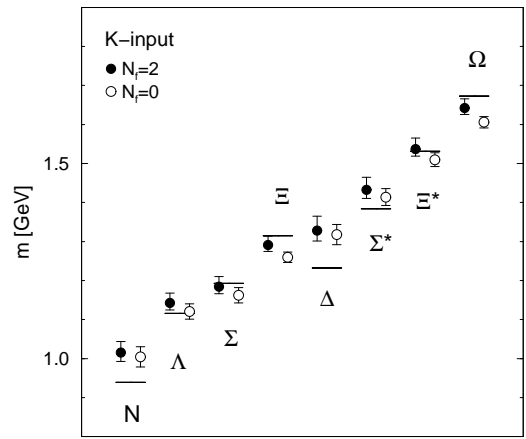}
   \caption
   {
      Baryon spectrum with $K$-input.
      Experimental values are shown by horizontal lines.
   }
   \label{fig:SQE:had:baryon}
\end{figure}

The baryon masses in full and quenched QCD 
are listed in Table~\ref{tab:SQE:had:baryon}.
These masses are compared with experiment 
in Fig.~\ref{fig:SQE:had:baryon}.
%
%
For heavier baryons, such as $\Sigma$, $\Xi$ and $\Omega$,
full QCD results show a closer agreement with experiment
than in quenched QCD.
%
%
The sea quark effect is, however, less clear for lighter baryons.
This is partly due to FSE in full QCD data
which is more pronounced for lighter valence quarks.



\begin{figure}
   \includegraphics[width=90mm]{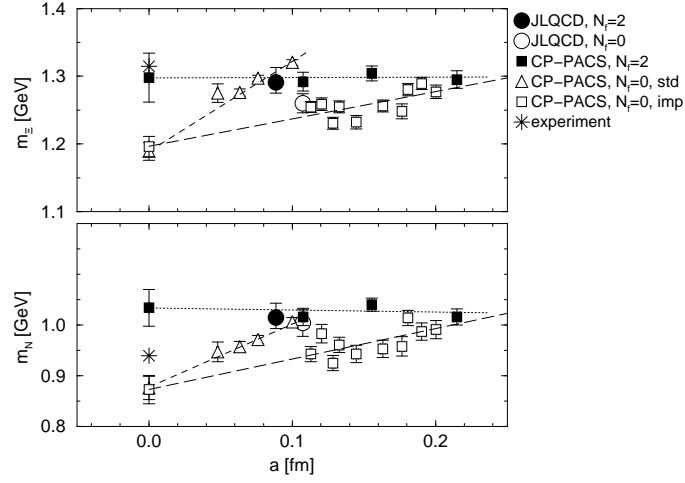}
   \caption
   {
      Nucleon mass (bottom panel) 
      and $\Xi$ baryon mass with $K$-input (top panel) 
      as a function of lattice spacing.
      Open triangles represent the CP-PACS results
      in quenched QCD using the standard plaquette gauge 
      and the Wilson quark actions,
      while open squares are obtained 
      with the renormalization group improved gauge 
      and the tadpole improved clover actions.
   }
   \label{fig:SQE:had:scaling:baryon}
\end{figure}

In Fig.~\ref{fig:SQE:had:scaling:baryon},
$m_N$ and $m_{\Xi}$ with the $K$-input
are compared with the CP-PACS results
\cite{Spectrum.Nf0.CP-PACS,Spectrum.Nf2.CP-PACS}.
While the full QCD results of two groups show a reasonable agreement
with each other, 
the CP-PACS results in the continuum limit in quenched QCD
are systematically smaller than ours.
This suggests that our quenched data has 
non-negligible scaling violation,
which is another source making sea quark effects less clear.
%
%
Therefore, further investigations of FSE in full QCD
and scaling violation in quenched QCD
are required to obtain a clear conclusion
on sea quark effects in the baryon spectrum.


We now turn to theoretical predictions 
which can be derived from our data. 
The first is the mass of the $\eta_{s}$ meson,
for which our full QCD data predict 
$m_{\eta_{s}}=0.6948(3)(+8/\!\!-\!\!1)(+2)$~GeV with the $K$-input, 
and $0.7381(46)(+57)(+40/\!\!-\!\!46)$~GeV with the $\phi$-input,
where the first error is statistical, 
and the second and third ones are
due to the choice of the fitting form and a systematic uncertainty
of $r_0$.
These results are to be compared with those in quenched QCD,
$0.6988(9)$~GeV ($K$-input) and $0.7719(58)$~GeV ($\phi$-input).
While the values themselves do not differ 
by going from quenched to full QCD,
the difference between the two inputs 
is reduced by about 40\% in full QCD.
This reflects the closer agreement of the meson spectrum in full QCD.

Another interesting prediction is the physical value of $r_0$.
Our full QCD simulation gives 
$r_0\!=\!0.497(6)(-9)(+11/\!\!-\!\!12)$~fm,
where 
the meaning of three errors are the same to those of $m_{\eta_s}$.
We note that this is close to the phenomenological estimate
in the original paper \cite{r0}, $r_0\!=\!0.49$~fm.
The quenched simulation gives 0.5702(75)(50)~fm, 
where the first error is statistical and the second 
comes from the systematic uncertainty of the measurement.
About a $14$~\% difference between full and quenched QCD
arises from the following two sea quark effects.
One is the difference of the physical value of $r_0$ itself
due to the change of the shape of the static quark potential.
The other source is the reduction of 
the quenching error in $m_{\rho}$ in full QCD,
which is used to fix the lattice scale.


\section{Decay constants}
\label{sec:f}


The PS meson decay constants are calculated 
using the fourth component of the improved axial vector current,
which is defined by 
\bea
   A_4^{\rm imp} 
   & = &
   A_4 + c_A \Delta_4 P
   \label{eqn:f:impA4}
\eea
with the symmetric lattice derivative $\Delta_4$.
We extract
the amplitude of the $\langle A_4^{\rm imp}(t) P(0)^{\dagger} \rangle$ 
correlator, $C_A^{\rm LS}$,  
by the correlated fit of the form 
\bea
   \langle A_4^{\rm imp}(t) P(0)^{\dagger} \rangle
   & = &
   C_A^{\rm LS} 
   \left\{ \exp \left[ -m_{\rm PS} \, t \right]
          -\exp \left[ -m_{\rm PS} \, (N_t-t) \right]
   \right\}
   \label{eqn:meas:mq:A4Pfit}
\eea
with $m_{\rm PS}$ fixed to the results given in Appendix~A.
We use the local operator for $A_4^{\rm imp}$, while
the double smearing is applied to $P$.
The amplitude of the $\langle P(t) P(0)^{\dagger} \rangle$ operator 
with the doubly smeared source and the sink operators, 
$C_P^{\rm SS}$,
is extracted assuming a single hyperbolic cosine form.

The renormalized decay constant is calculated by
\bea
   f_{\rm PS} 
   & = &
   2 K Z_A \left( 1 + b_A m_{\rm q,val} \right)
   C_A^{\rm LS} \sqrt{ \frac{2}{m_{\rm PS} C_P^{SS}}}.
   \label{eqn:f:renorm_f}
\eea
Since non-perturbatively determined values
for $Z_A$, $b_A$ and $c_A$ 
are not available for two-flavor QCD,
we adopt one-loop perturbative values
in Refs.~\cite{1loop_cSWcA.ALPHA,1loop_Z.Roma,1loop_b.ALPHA,1loop_Zb.Tsukuba,1loop_Zbc.Tsukuba,1loop_bc.Aoki}
with the tadpole improvement.
We calculate $\alpha_P(3.40/a)$ 
from the plaquette average $\langle U_P \rangle$
according to \cite{alpha_P.JLQCD,alpha_P}
\bea
   -\ln\left[ \langle U_P \rangle \right] 
   & = &
   \frac{4\pi}{3} \alpha_P(3.40/a)
   \left\{
      1 - \left( 1.1905 - 0.2266 N_f \right) \alpha_P
   \right\}.
   \label{eqn:f:alpha_P}
\eea
Then, $\alpha_P(3.40/a)$ is evolved to the optimum scale 
($q^*_{Z_A}\!=\!1.803/a$ for $Z_A$, 
 $q^*_{b_A}\!=\!2.289/a$ for $b_A$ and 
 $q^*_{c_A}\!=\!2.653/a$ for $c_A$ \cite{q_star}) 
using the universal two-loop beta function
and is used as the expansion parameter 
of tadpole improved perturbation theory.

\begin{figure}[htbp]
   \begin{center}
   \includegraphics[width=70mm]{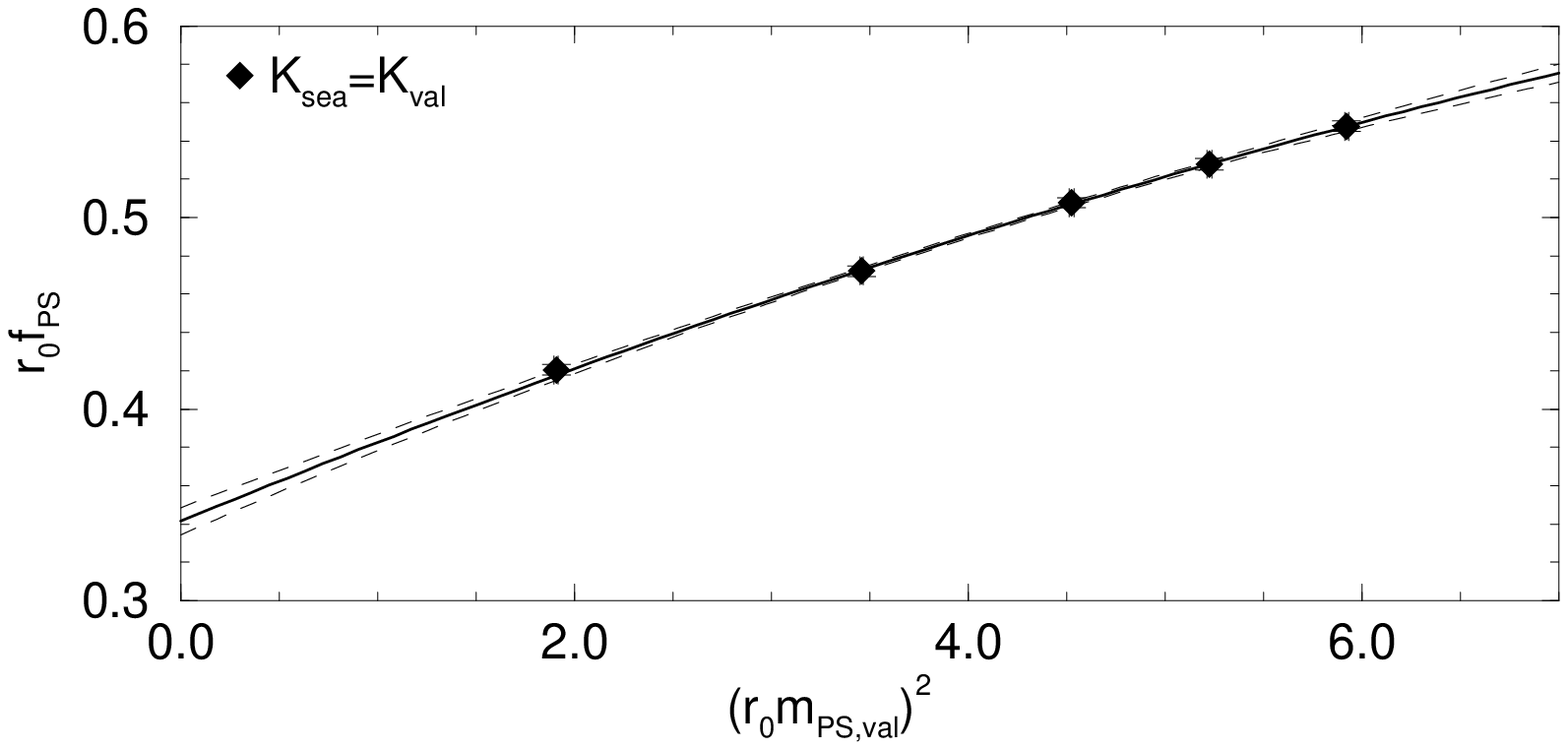}
   \includegraphics[width=70mm]{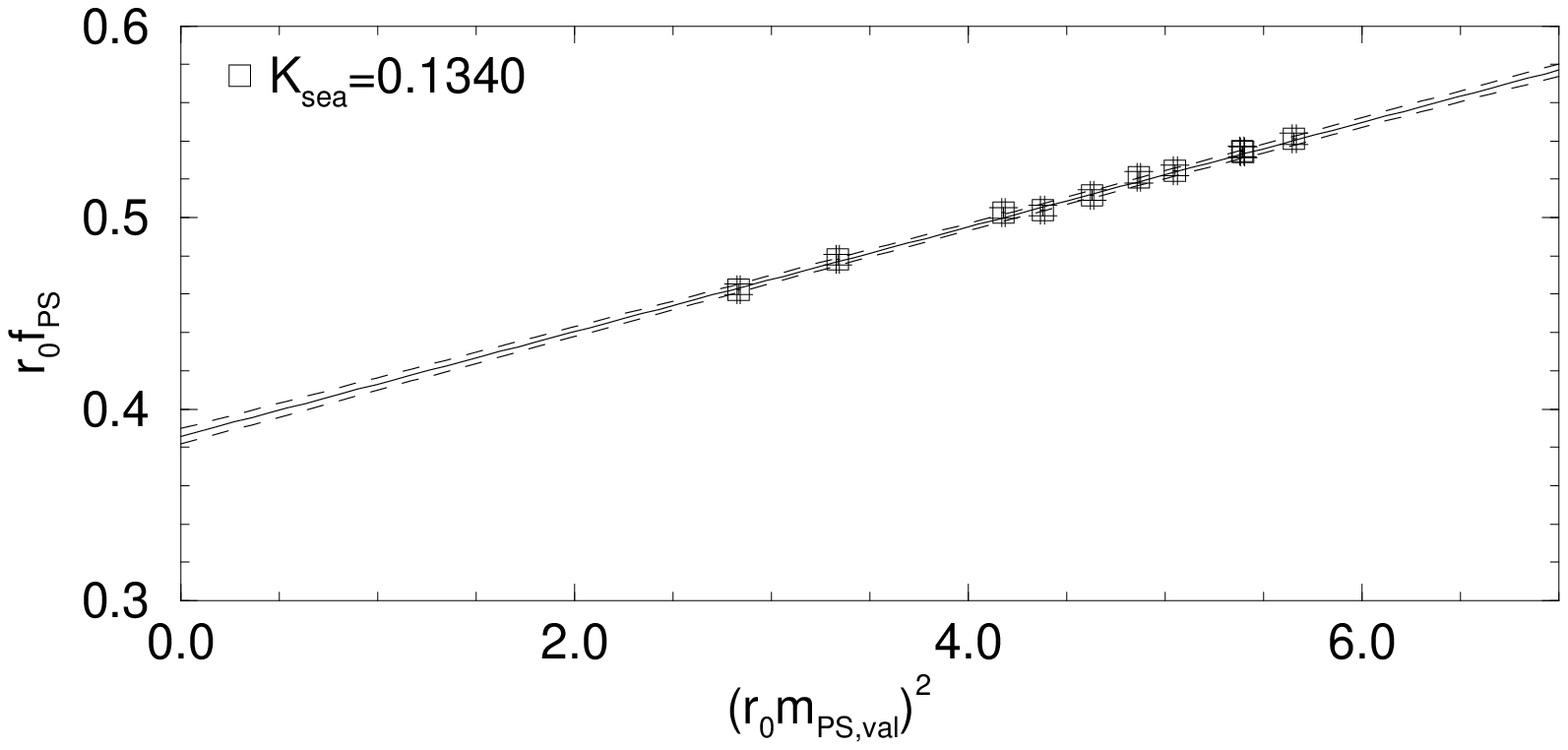}
   \end{center}
   \begin{center}
   \includegraphics[width=70mm]{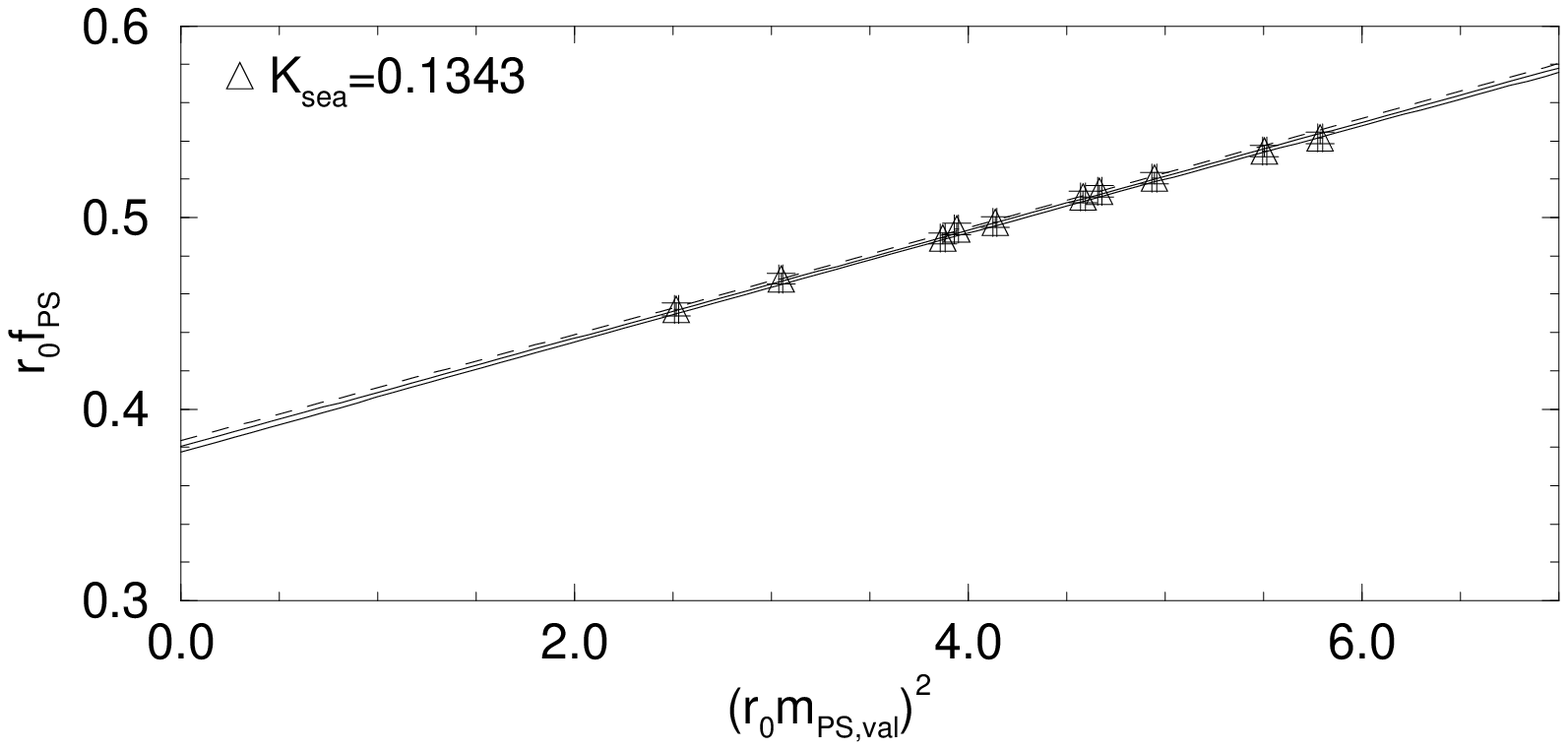}
   \includegraphics[width=70mm]{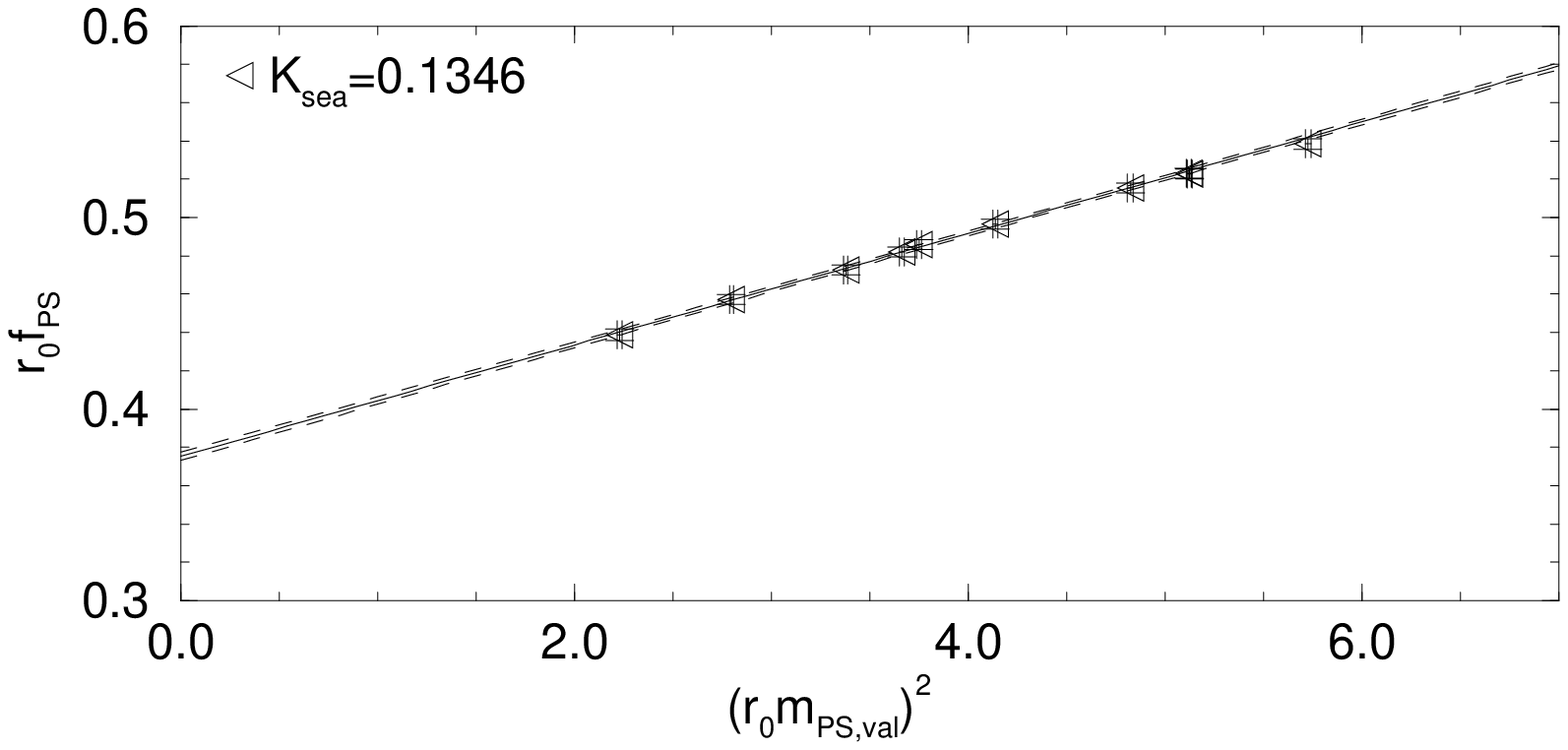}
   \vspace*{0mm}
   \end{center}
   \begin{center}
   \includegraphics[width=70mm]{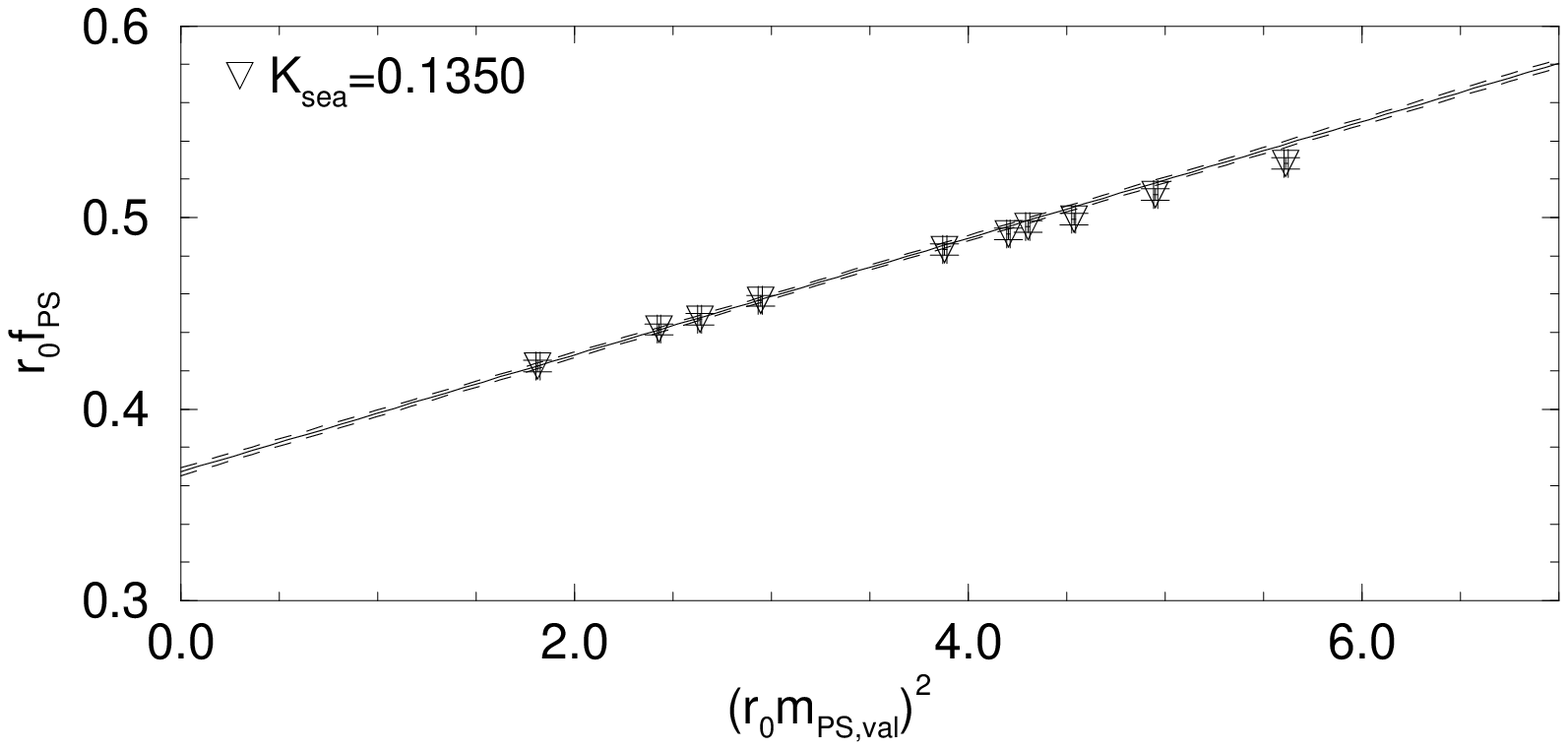}
   \includegraphics[width=70mm]{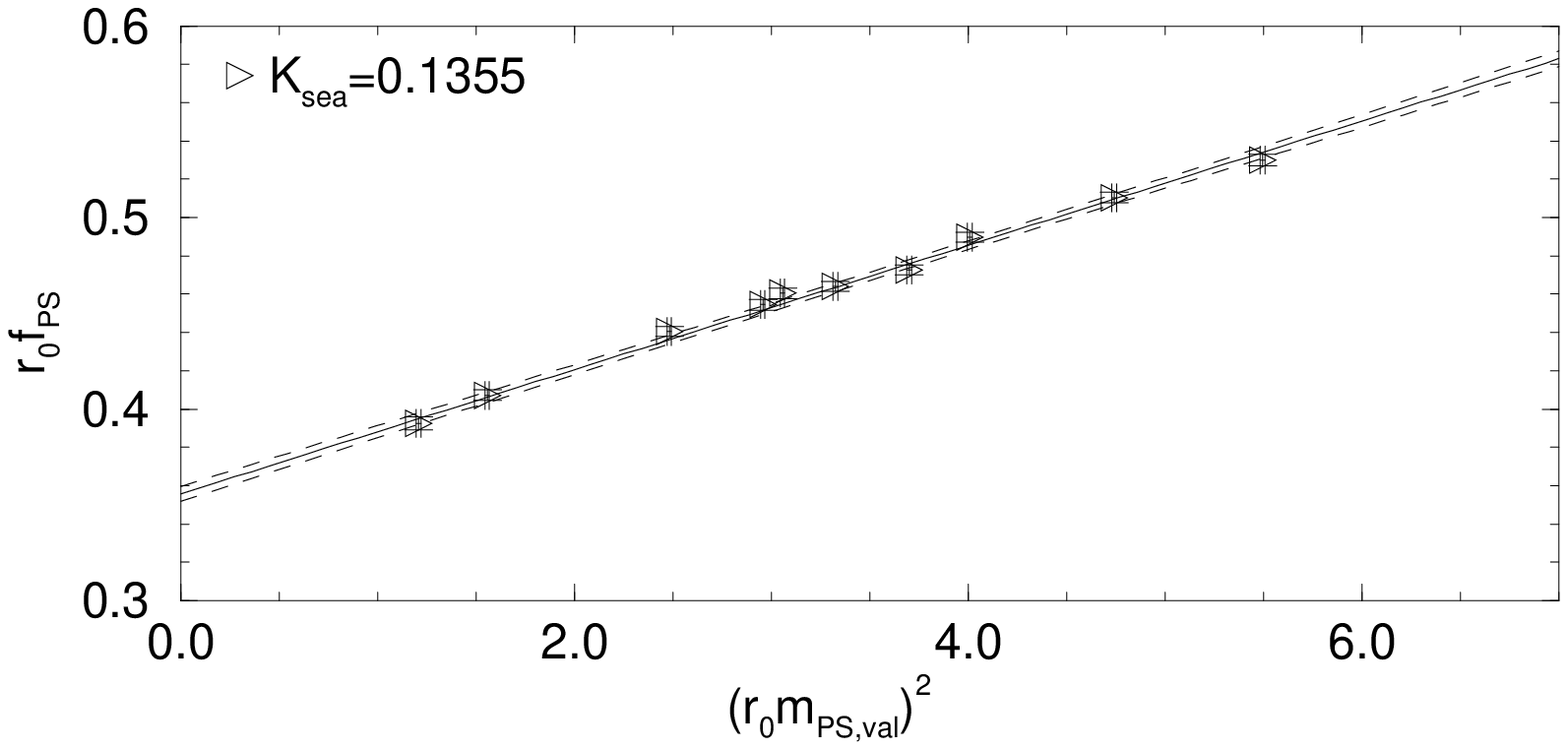}
   \end{center}
   \caption
   {
      Combined chiral extrapolation of PS meson decay constants.
   }
   \label{fig:chiralfit:w_r0:FPS}
\end{figure}

The consistent chiral extrapolation of the decay constant
should include the chiral logarithmic term 
as predicted by ChPT \cite{ChPT}.
However, our data do not show the characteristic curvature
of the chiral logarithm as discussed in Sec.\ref{sec:chiralfit:w_r0}
(and also in
Refs.~\cite{Spectrum.Nf2.JLQCD.lat01,Spectrum.Nf2.JLQCD.lat02.Hashimoto}).
We therefore use the following polynomial form 
for the chiral extrapolation 
leaving the problem of the chiral logarithm and 
associated uncertainty for future publication \cite{ChPT_test.Nf2.JLQCD}
\bea
   r_0(K_{\rm sea}) \, 
   f_{\rm PS}(K_{\rm sea};K_{\rm val,1},K_{\rm val,2})
   =
   A^{\rm f}
  +B^{\rm f}_{\rm s}  \, \mu_{\rm PS,sea}
  +B^{\rm f}_{\rm v}  \, \mu_{\rm PS,val}
  +C^{\rm f}_{\rm sv} \, \mu_{\rm PS,sea}
                      \, \mu_{\rm PS,val}.
   \label{eqn:f:chiralfit:fPS:Nf2}
\eea
The fit is plotted in Fig.~\ref{fig:chiralfit:w_r0:FPS}
with parameters summarized in Table~\ref{tab:chiralfit:w_r0:fPS:Nf2}.
Pion and kaon decay constants, $f_{\pi}$ and $f_K$, are calculated 
by tuning $\mu_{\rm PS,sea}$ to $(r_0 m_{\pi})^2$ and 
substituting $(r_0 m_{\pi})^2$ or $(r_0 m_{\eta_s})^2$
for $\mu_{\rm PS,val,{\it i}}$ $(i\!=\!1,2)$ in $\mu_{\rm PS,val}$.
The systematic errors due to the choice of the fitting function
and the uncertainty of $r_0$ are estimated in a way similar
to those described in Sec.~\ref{sec:chiralfit:spectrum}.
In the estimation of the former error,
we use
\bea
   r_0(K_{\rm sea}) \, 
   f_{\rm PS}(K_{\rm sea};K_{\rm val,1},K_{\rm val,2})
   & = & 
   \mbox{``r.h.s of Eq.~(\ref{eqn:f:chiralfit:fPS:Nf2})''}
   + C^{\rm f}_{\rm s} \mu_{\rm PS,sea}^2
   + C^{\rm f}_{\rm v} \mu_{\rm PS,val}^2
   \label{eqn:f:alter:PSK}
\eea
and Eqs.~(\ref{eqn:chiralfit:w_r0:alter:PSK})
and (\ref{eqn:chiralfit:w_r0:alter:VPS}) 
as the alternative fitting functions for the chiral extrapolation.

In quenched QCD, we use the chiral extrapolation form 
\bea
   f_{\rm PS}(K_{\rm val,1},K_{\rm val,2})
   & = & 
   A^{\rm f}_{\rm q}
  +B^{\rm f}_{\rm q} \, m_{\rm PS,val}^2.
   \label{eqn:f:chiralfit:fPS:qQCD}
\eea
and obtain parameters summarized 
in Table~\ref{tab:chiralfit:w_r0:fPS:qQCD}.
For $Z_A$, $b_A$ and $c_A$,
we test the one-loop perturbative value and 
the non-perturbative one in Ref.~\cite{NP_Zbc.LANL}.
%
%

\begin{figure}[htbp]
   \begin{center}
   \includegraphics[width=70mm]{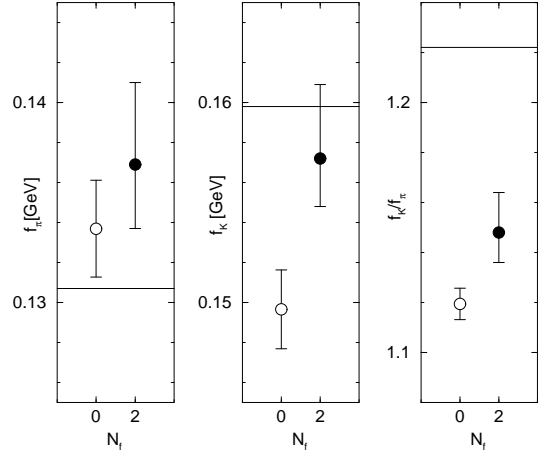}
   \end{center}
   \caption
   {
      Comparison of $f_{\pi}$ (left panel), $f_K$ (center panel) and 
      $f_{K}/f_{\pi}$ (right panel) between full and quenched QCD.
      We use $K$-input for $f_K$ and $f_K/f_{\pi}$.
      Experimental values are shown by horizontal lines.
   }
   \label{fig:f:SQE}
\end{figure}

Our results of the decay constants 
are summarized in Table~\ref{tab:f:result}.
A comparison between full and quenched QCD results
obtained by the one-loop matching 
is made in Fig.~\ref{fig:f:SQE}.
We observe that $f_{\pi}$ is consistent with 
the experimental value within two standard deviations 
in both full and quenched QCD.
While $f_K$ in quenched QCD is significantly smaller 
than the experimental value,
the deviation is reduced by sea quark effects
and the full QCD result becomes consistent with experiment.

The results obtained with one-loop renormalization factors
are subject to higher order corrections.
However, as shown in Table~\ref{tab:f:result}
for the quenched results,
the difference between the perturbative and non-perturbative matchings
is not large.
This is because the $O(a)$ correction to the improved current
in Eq.~(\ref{eqn:f:impA4}) is not large,
and the non-perturbative values for $Z_A$ and $b_A$ are close to 
those in tadpole improved perturbation theory.
We may therefore expect that the uncertainty due to the perturbative 
matching is small also in the full QCD results.

It is expected that various systematic uncertainties,
including the scaling violation, would partially cancel 
in the ratio $f_K/f_{\pi}$.
This expectation is supported by 
the good agreement of the quenched results
between the perturbative and non-perturbative matching,
which suggests that higher order corrections to the renormalization
factors almost cancel in the ratio.
Therefore the ratio is useful to discuss sea quark effects.
Comparison of this quantity shows
that the full QCD values are significantly closer to 
the experimental value $\simeq 1.22$ 
by about two standard deviations than in quenched QCD.


\section{Quark masses}
\label{sec:mq}



We calculate the up-down and strange quark masses
through the axial vector Ward identity (AWI).
The bare quark mass at simulation points is obtained by 
\bea
   m_{q}^{\rm AWI}
   & = &
   \frac{m_{\rm PS} C_A^{\rm LS}}{2 C_P^{\rm LS}},
   \label{eqn:mq:AWImq}
\eea
where $C_A^{\rm LS}$ and $C_P^{\rm LS}$ 
are the amplitudes of 
$\langle A_4^{\rm imp}(t) P(0)^{\dagger} \rangle$ 
and $\langle P(t) P(0)^{\dagger} \rangle$ with
the doubly smeared source and the local sink operators.

\begin{figure}[htbp]
   \begin{center}
   \includegraphics[width=70mm]{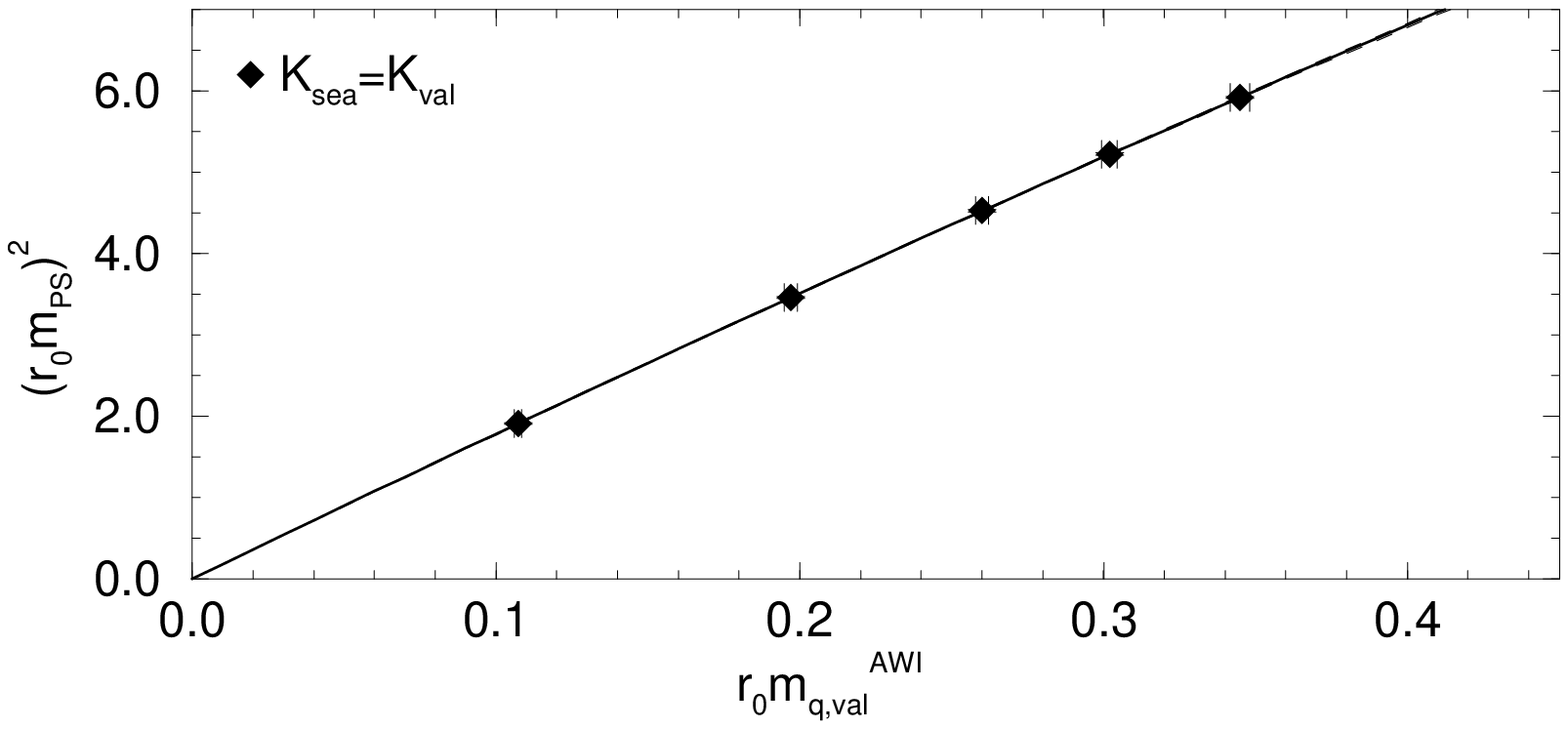}
   \includegraphics[width=70mm]{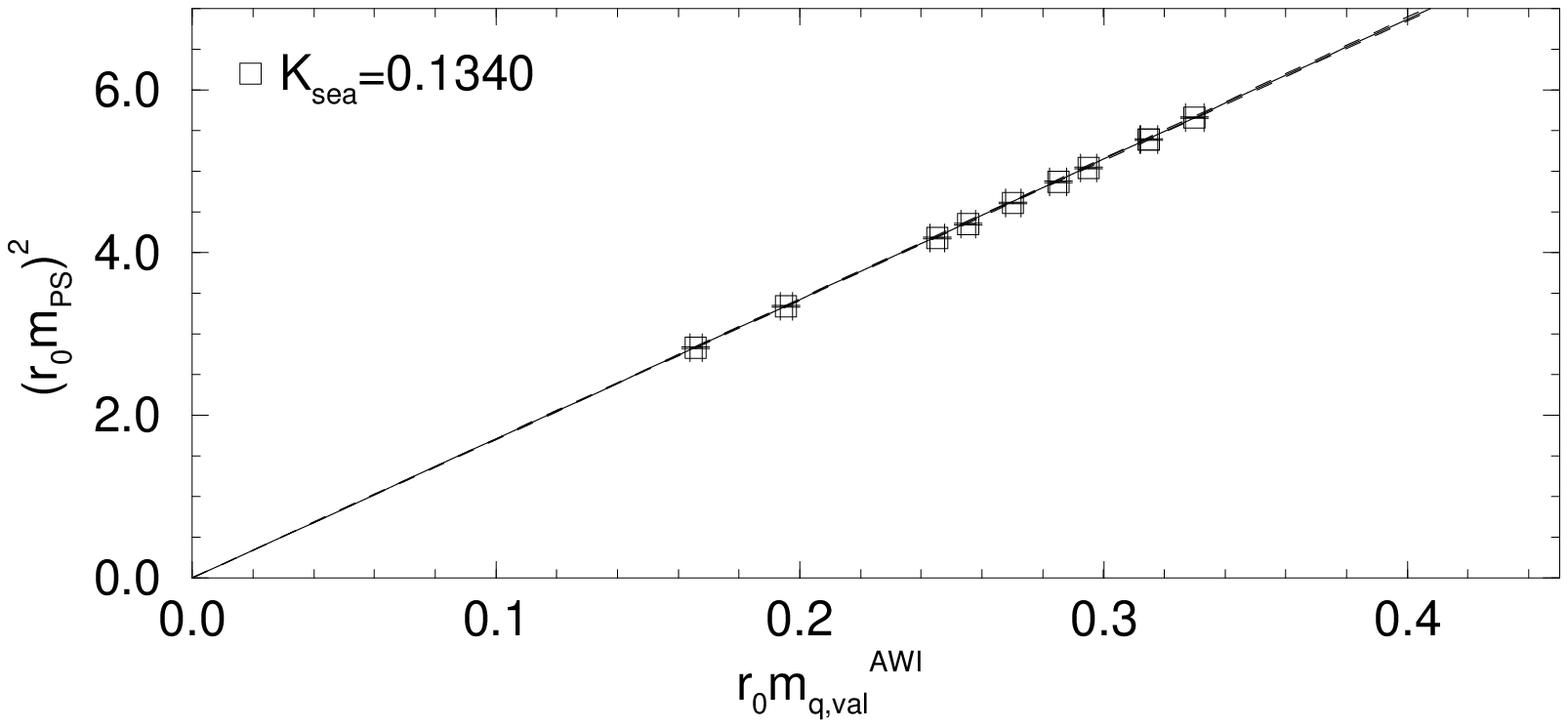}
   \end{center}
   \begin{center}
   \includegraphics[width=70mm]{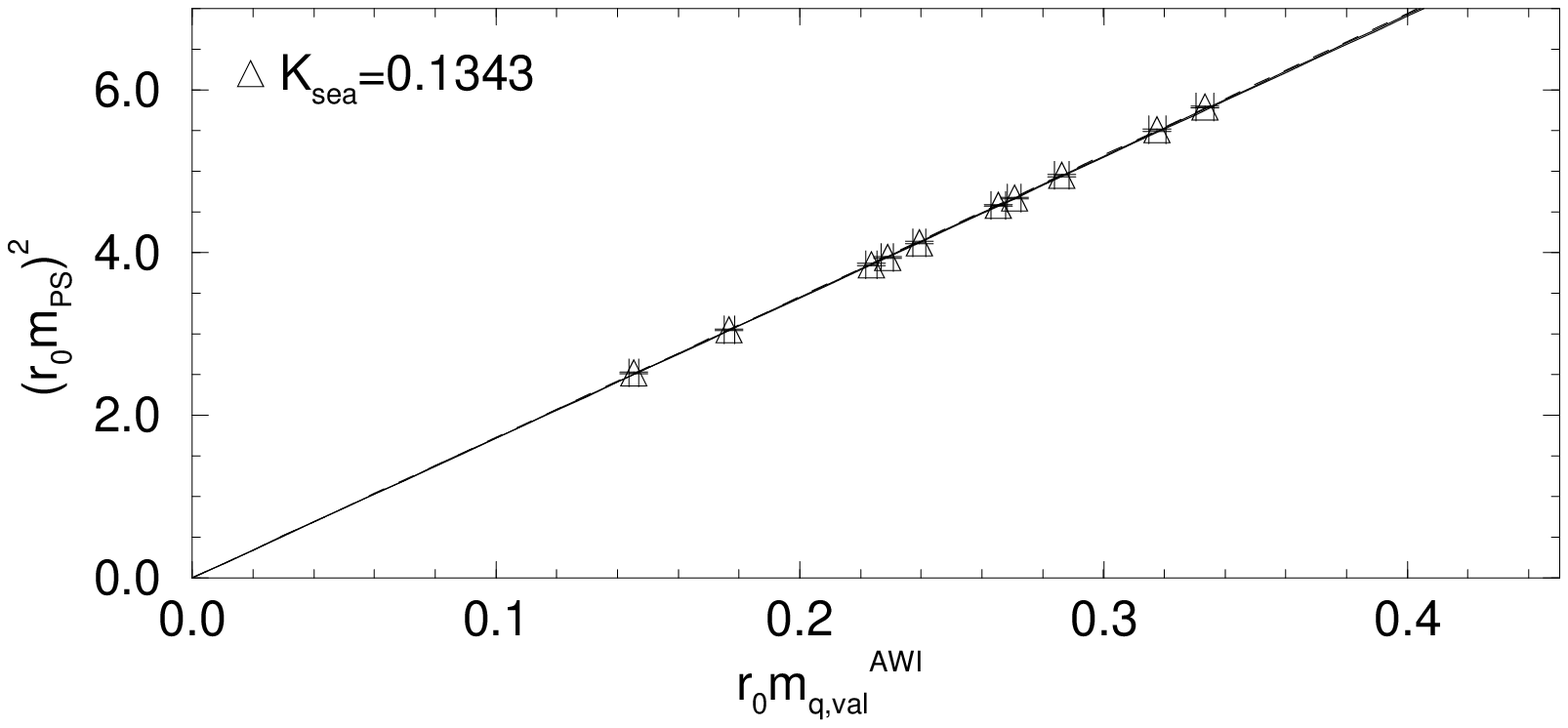}
   \includegraphics[width=70mm]{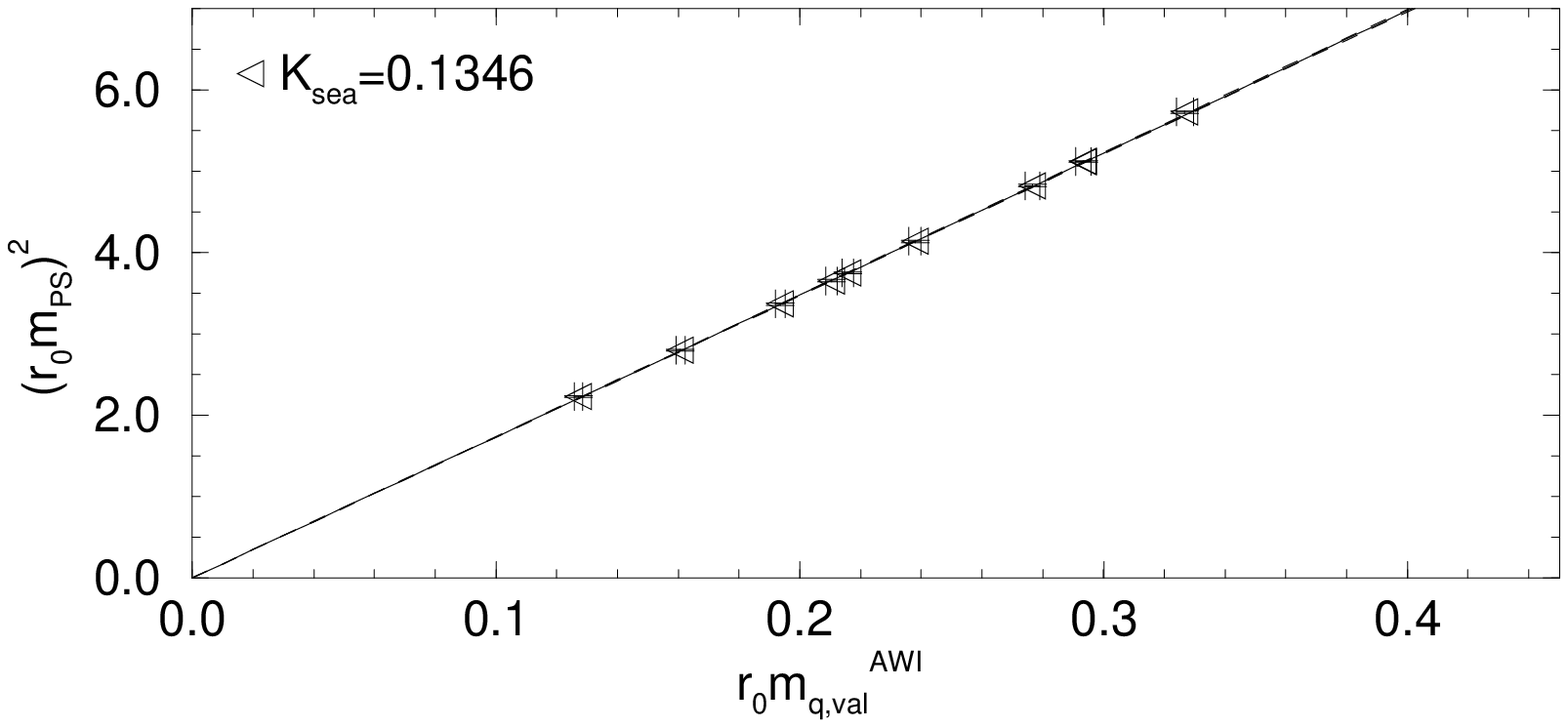}
   \vspace*{0mm}
   \end{center}
   \begin{center}
   \includegraphics[width=70mm]{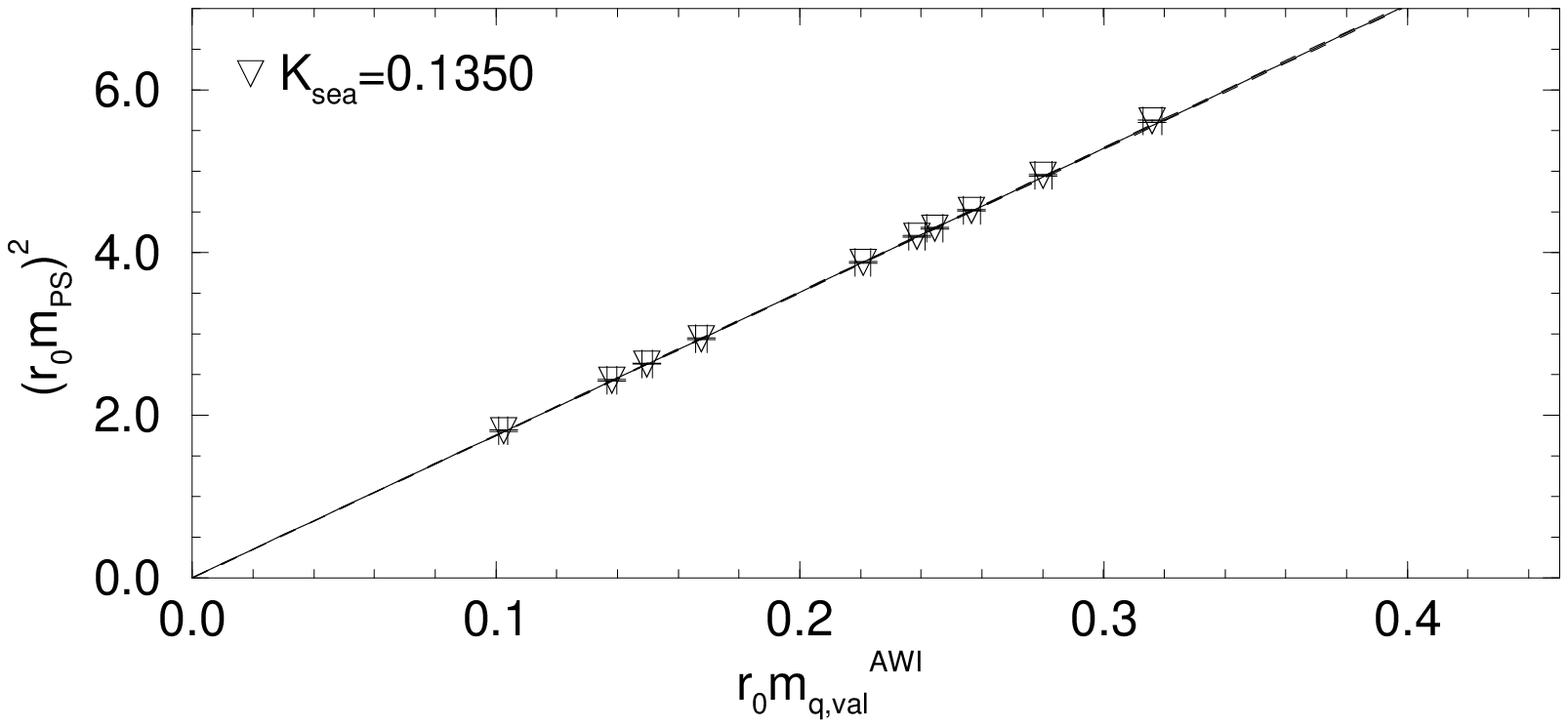}
   \includegraphics[width=70mm]{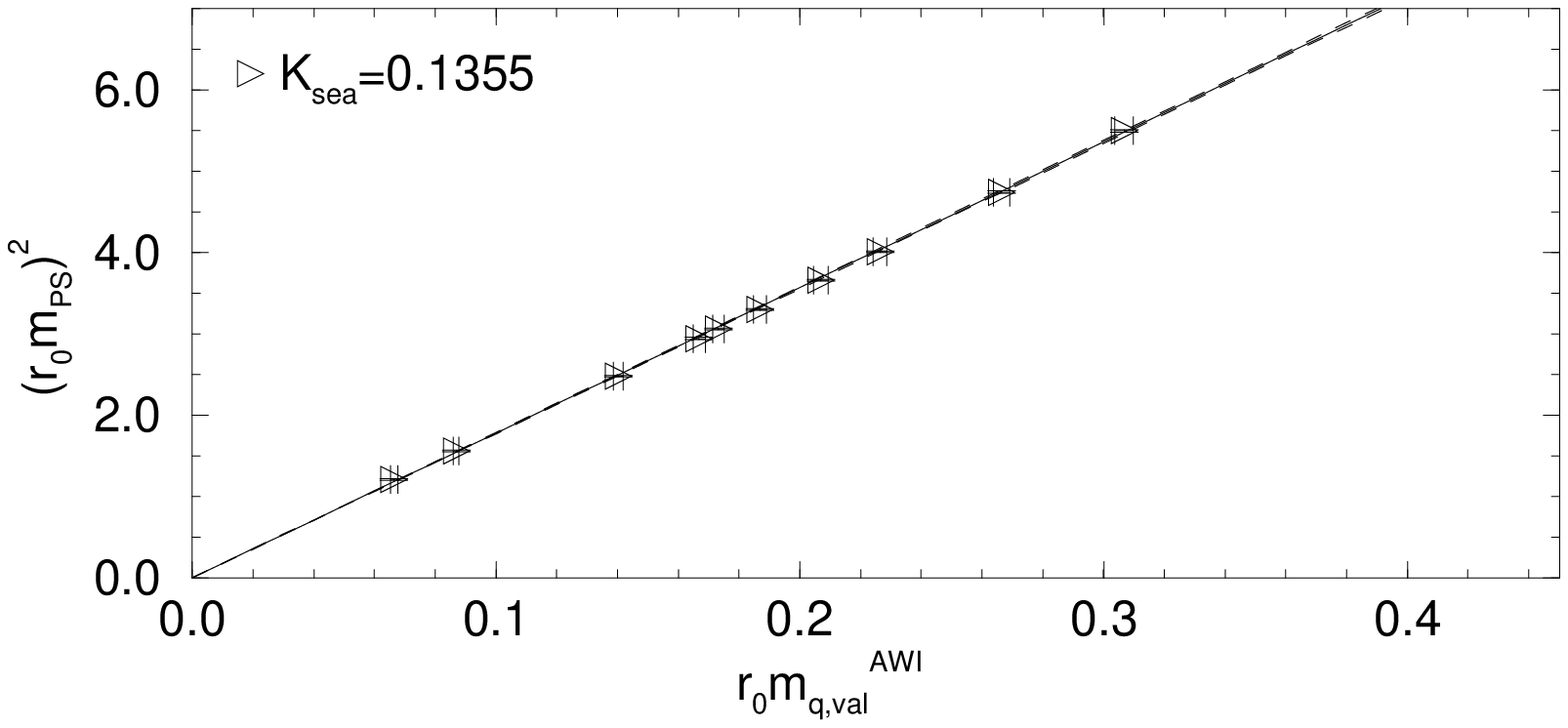}
   \end{center}
   \caption
   {
      Combined chiral extrapolation of PS meson masses
      in terms of AWI quark mass.
   }
   \label{fig:chiralfit:w_r0:PSq}
\end{figure}

We then carry out the chiral fit of the PS meson mass
as a function of the AWI bare quark mass. 
The fitting function is obtained 
from Eq.~(\ref{eqn:chiralfit:w_r0:PSK})
with the replacement of the VWI masses with the AWI ones.
We also drop all monomial terms in the sea quark mass,
since the PS meson mass vanishes in the chiral limit
$m_{\rm q,val}^{\rm AWI}\!=\!0$ 
even for non-zero sea quark masses \cite{PQChPT}.
The adopted form is 
\bea
   &   &
   \left( {r_0}(K_{\rm sea}) \,
          m_{\rm PS}(K_{\rm sea};K_{\rm val,1},K_{\rm val,2}) \right)^2
   \nn \\
   & = &
   B^{\rm PS,AWI}_{\rm v} \, \mu_{\rm q,val}^{\rm AWI}
  +C^{\rm PS,AWI}_{\rm v} \, (\mu_{\rm q,val}^{\rm AWI})^2
  +C^{\rm PS,AWI}_{\rm sv} \, \mu_{\rm q,sea}^{\rm AWI} 
                       \, \mu_{\rm q,val}^{\rm AWI},
   \label{eqn:mq:chiralfit:PSq:Nf2}
\eea
where 
\bea
   \mu_{\rm q,sea}^{\rm AWI} 
   & = & 
   r_0(K_{\rm sea}) \,  
   m_{\rm q}^{\rm AWI}(K_{\rm sea}; K_{\rm sea}, K_{\rm sea}),
   \\
   \mu_{\rm q,val}^{\rm AWI}
   & = & 
   r_0(K_{\rm sea}) \, m_{\rm q,val}^{\rm AWI},
   \\
   m_{\rm q,val}^{\rm AWI}
   & = & 
   \frac{1}{2}
   \left( m_{{\rm q,val},1}^{\rm AWI} +m_{{\rm q,val},2}^{\rm AWI}
   \right),
   \\
   m_{{\rm q,val},i}^{\rm AWI}
   & = & 
   m_{\rm q}^{\rm AWI}(K_{\rm sea}; K_{{\rm val},i}, K_{{\rm val},i}).
\eea
Our data and fit are shown in Fig.~\ref{fig:chiralfit:w_r0:PSq}.

We adopt the fit~(\ref{eqn:mq:chiralfit:PSq:Nf2}) 
because it is consistent with that in terms of the VWI quark mass
(Eq.~(\ref{eqn:chiralfit:w_r0:PSK})).
However, a function with fewer terms
\bea
   \left( {r_0}(K_{\rm sea}) \,
          m_{\rm PS}(K_{\rm sea};K_{\rm val,1},K_{\rm val,2}) \right)^2
   = 
   B^{\rm PS,AWI}_{\rm v} \, \mu_{\rm q,val}^{\rm AWI}
  +C^{\rm PS,AWI}_{\rm sv} \, \mu_{\rm q,sea}^{\rm AWI} 
                       \, \mu_{\rm q,val}^{\rm AWI}
   \label{eqn:mq:chiralfit:alter:PSq}
\eea
also gives an acceptable $\chi^2/{\rm dof}$.
We use this as an alternative fit
in our estimation of the systematic error
due to the choice of the fitting function (see below).
Parameters of these two fits are 
summarized in Table~\ref{tab:chiralfit:w_r0:PSq:Nf2}.

The bare AWI masses of the up-down and strange quarks
are fixed in a way analogous to that described 
in Sec.~\ref{sec:chiralfit:spectrum} 
by using Eq.~(\ref{eqn:mq:chiralfit:PSq:Nf2}) 
instead of Eq.~(\ref{eqn:chiralfit:w_r0:PSK}).
The matching to the $\overline{\rm MS}$ scheme
is made at the scale $\mu\!=\!2/a$ 
using the one-loop renormalization constant
\cite{1loop_Z.Roma,1loop_bc.Aoki,1loop_Zbc.Tsukuba,1loop_b.ALPHA,1loop_Zb.Tsukuba}
with the tadpole improvement.
We use $\alpha_P(q^*_{Z_A})$, $\alpha_P(q^*_{b_A})$ 
as the expansion parameter 
in the one-loop expression of $Z_A$ and $b_A$,
while we set $q^*\!=\!2/a$ for other coefficients
for which $q^*$ is not known.
The $\overline{\rm MS}$ quark mass is 
evolved to $\mu\!=\!2$~GeV
using the four-loop beta function
\cite{4loop.running.1,4loop.running.2}.

The VWI quark mass may differ from the AWI one
because of explicit violation of chiral symmetry
at finite lattice spacings. 
The difference between the AWI and VWI masses, therefore,
gives insight into the size of scaling violation in our results.
This leads us to repeat the calculation of quark masses 
using the VWI definition.
The bare quark mass is calculated from 
$K_c$, $K_{ud}$ and $K_s$ in Tables~\ref{tab:chiralfit:w_r0:PSK}
and \ref{tab:chiralfit:spectrum:K}.
The $\overline{\rm MS}$ mass is 
obtained by the one-loop matching at $\mu\!=\!2/a$ and 
the four-loop running to $\mu\!=\!2$~GeV.
The resulting AWI and VWI masses
are summarized in Table~\ref{tab:mq:Nf2}.

For a calculation of quark masses in quenched QCD,
the chiral extrapolation is carried out
using a quadratic fit form obtained from 
Eq.~(\ref{eqn:mq:chiralfit:PSq:Nf2})
by dropping the third term 
which represents the sea quark mass dependence
\bea
   m_{\rm PS}(K_{\rm val,1},K_{\rm val,2})^2
   = 
   B^{\rm PS,AWI}_{\rm q} \, \mu_{\rm q,val}^{\rm AWI}
  +C^{\rm PS,AWI}_{\rm q} \, (\mu_{\rm q,val}^{\rm AWI})^2.
    \label{eqn:mq:chiralfit:PSq:qQCD}
\eea
Obtained parameters are listed in 
Table~\ref{tab:chiralfit:w_r0:PSq:qQCD}.
We use either one-loop or non-perturbative
value in 
Refs.~\cite{NPcsw.Nf0.ALPHA,NP_ZA.ALPHA,NP_ZP.ALPHA,NP_bA-bP.ALPHA} 
for the renormalization factors.
Numerical results are summarized in Table~\ref{tab:mq:qQCD}.

\begin{figure}[htbp]
   \begin{center}
   \includegraphics[width=67mm]{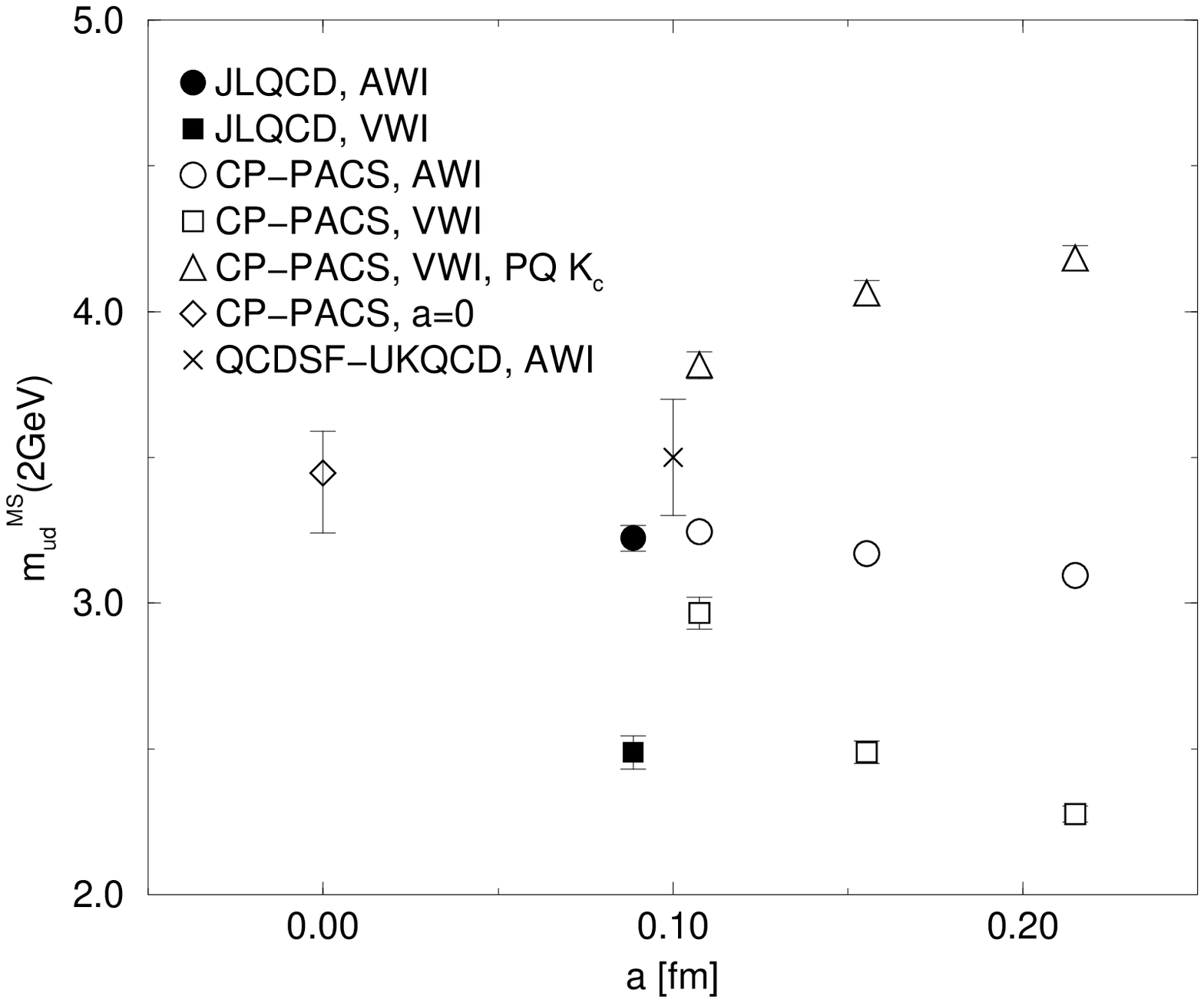}
   \includegraphics[width=70mm]{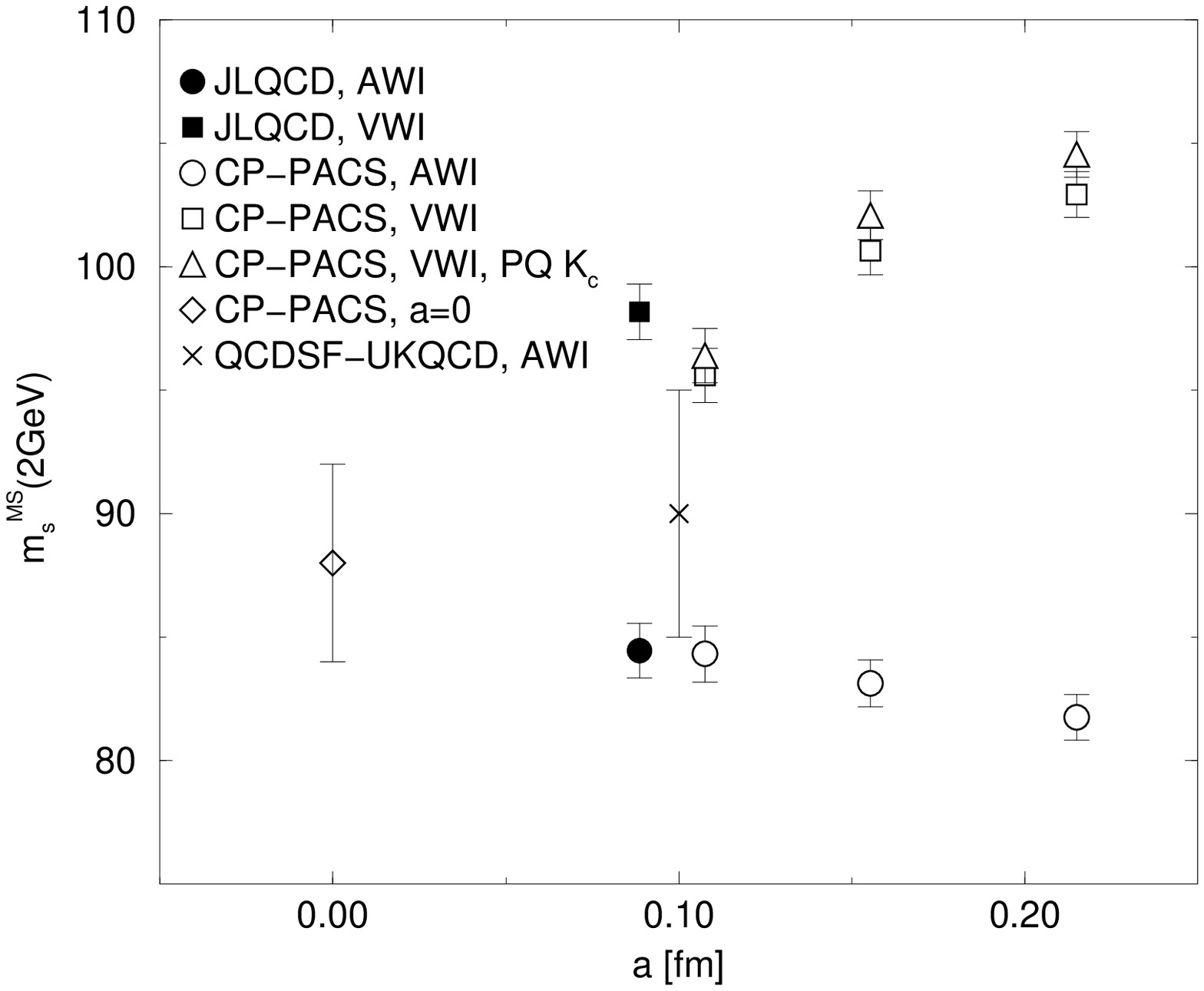}
   \end{center}
   \caption 
   {
      Comparison of 
      light (left figure) and strange quark mass with $K$-input 
      (right figure) in two-flavor QCD.
      Triangles represent the CP-PACS results of the VWI quark mass
      using $K_c$ determined by partially quenched chiral extrapolations. 
      The CP-PACS result in the continuum limit was obtained 
      by combined linear extrapolation of three data.
      We note that SESAM-T$\chi$L's results 
      in Ref.~\cite{Spectrum.Nf2.SESAM-TchiL}
      are consistent with these results 
      within large error arising from their continuum extrapolation.
   }
   \label{fig:mq:Nf2}
\end{figure}

%
%
In Figure~\ref{fig:mq:Nf2},
our full QCD results are compared with estimates 
by the CP-PACS \cite{Spectrum.Nf2.CP-PACS,mq.Nf2.CP-PACS} 
and the QCDSF-UKQCD Collaborations \cite{mq.Nf2.QCDSF}.
We observe good agreement for the AWI masses
among the three groups even at the finite lattice spacing of 
$a^{-1}\!\sim\!2$~GeV.
These results are consistent also with the CP-PACS's result 
extrapolated to the continuum limit.
This suggests that various uncertainties, 
   such as scaling violation and 
   higher order corrections to renormalization factors,
are likely to be canceled 
in the ratio defining the AWI mass (Eq.~(\ref{eqn:mq:AWImq})).

On the other hand,
such a cancellation is not expected in the VWI mass.
Indeed there is a sizable difference between our AWI and VWI 
results.
We also observe that our and the CP-PACS results of VWI $m_{ud}$ 
show a large discrepancy 
of about 18~\% (6 standard deviations).
These observations suggest that 
the scaling violation in our results of the VWI masses is not small.

%
%
In both full and quenched QCD, therefore,
we quote the AWI masses as the central value.
We adopt $K$-input for $m_s$.
The difference between $m_s$ with $K$- and $\phi$-inputs
is treated as the systematic error due to the choice of the
meson mass input to fix $m_s$.

Additional systematic errors due to the choice of the chiral fit form
and the uncertainty of the measured value of $r_0$
are included in our final results in full QCD.
These errors are estimated in a similar way 
to that described in Sec.~\ref{sec:chiralfit:spectrum}
by using Eqs.~(\ref{eqn:chiralfit:w_r0:alter:PSK}), 
(\ref{eqn:chiralfit:w_r0:alter:VPS}) 
and (\ref{eqn:mq:chiralfit:alter:PSq})
as alternative fit forms.

Adding all errors in quadrature, we obtain
\bea
   m_{ud}^{\overline{\rm MS}}(\mbox{2~GeV})
   & = &
   3.223\left(^{+0.046}_{-0.069}\right)
   \label{eqn:mq:mud:Nf2}
   \\
   m_{s}^{\overline{\rm MS}}(\mbox{2~GeV})
   & = &
   84.5\left(^{+12.0}_{-1.7}\right)
   \label{eqn:mq:ms:Nf2}
   \\
   m_{s}/m_{ud}
   & = &
   26.13\left(^{+3.65}_{-0.02}\right)
   \label{eqn:mq:ratio:Nf2}
\eea
in two-flavor QCD and 
\bea
   m_{ud}^{\overline{\rm MS}}(\mbox{2~GeV})
   & = &
   4.020\left(0.077\right)
   \label{eqn:mq:mud:Nf0}
   \\
   m_{s}^{\overline{\rm MS}}(\mbox{2~GeV})
   & = &
   104.1\left(^{+24.1}_{-1.6}\right)
   \label{eqn:mq:ms:Nf0}
   \\
   m_{s}/m_{ud}
   & = &
   25.90\left(^{+5.98}_{-0.13}\right).
   \label{eqn:mq:ratio:Nf0}
\eea
in quenched QCD using the one-loop matching.
The scaling violation is expected to be small 
and hence is ignored here.
This point, however, should be checked in future studies.

There is an additional uncertainty arising from 
the use of the perturbative value for $c_A$, $Z_{A,P}$, and $b_{A,P}$.
Comparing the quenched result of 
Eqs.~(\ref{eqn:mq:mud:Nf0})\,--\,(\ref{eqn:mq:ratio:Nf0})
with those obtained by the non-perturbative matching given by
\bea
   m_{ud}^{\overline{\rm MS}}(\mbox{2~GeV})
   & = &
   3.522\left(0.66\right)~\mbox{MeV}
   \label{eqn:mq:mud:Nf0:NP}
   \\
   m_{s}^{\overline{\rm MS}}(\mbox{2~GeV})
   & = &
   91.9\left(^{+21.3}_{-1.4}\right)~\mbox{MeV}
   \label{eqn:mq:ms:Nf0:NP}
   \\
   m_{s}/m_{ud}
   & = &
   26.08\left(^{+6.02}_{-0.13}\right),
   \label{eqn:mq:ratio:Nf0:NP}
\eea
we observe a systematic error of about 13~\%.
In quenched QCD 
the non-perturbative estimate of $Z_A/Z_P\!=\!1.19$ is very close 
to that in one-loop perturbation theory 1.22,
since higher order corrections in $Z_A$ and $Z_P$
partially cancel with each other.
The non-perturbative value $b_A\!-\!b_P\!=\!0.171$
deviates significantly from that at one-loop $-0.011$.
The $O(am_q)$ term, however, is a small correction in our data.
Most of the 13~\% difference 
originates from the large deviation between
the non-perturbative value $c_A\!=\!-0.083$ 
in Ref.~\cite{NPcsw.Nf0.ALPHA} 
and its one-loop value $-0.013$.
Therefore, a non-perturbative determination of $c_A$ in full QCD
is an important task toward a more precise calculation of 
the quark masses in future studies.

\begin{figure}[htbp]
   \begin{center}
   \includegraphics[width=67mm]{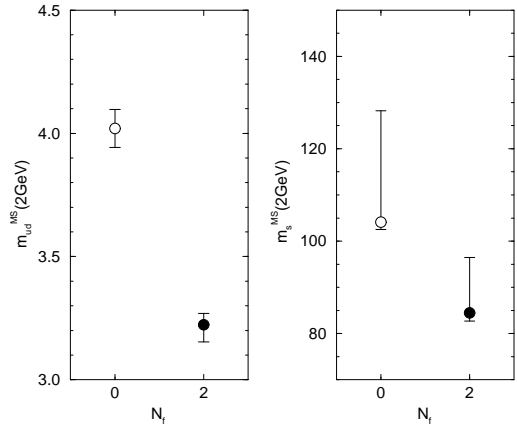}
   \end{center}
   \caption 
   {
      Comparison of light (left panel) and 
      strange quark mass (right panel) in full and quenched QCD.
   }
   \label{fig:mq:SQE}
\end{figure}

In Fig.~\ref{fig:mq:SQE}, 
we compare the quark masses in full and quenched QCD.
The chief observation is that 
sea quark effects reduce the light and strange quark masses 
by about 20~\%.
The magnitude of the sea quark effect is roughly consistent
with the CP-PACS observation in Refs.~\cite{Spectrum.Nf2.CP-PACS,mq.Nf2.CP-PACS}.

The quark mass ratio $m_s/m_{ud}$ in full QCD
is consistent with the quenched value, 
because the sea quark effects in $m_{ud}$ and $m_s$ 
almost cancel with each other in the ratio.
We note that $m_s/m_{ud} \! \simeq \! 26$
is in a good agreement with the estimate of 
one-loop chiral perturbation theory 24.4(1.5) \cite{mq.ChPT}.

Another important observation is that 
the deviation in $m_s$ between $K$- and $\phi$-inputs is 
reduced by effects of sea quarks:
the deviation is 21~\% ($\simeq 24$~MeV) in quenched QCD
and 13~\% ($\simeq 12$~MeV) in full QCD.
This reflects the closer agreement of the meson spectrum in full QCD
with experiment. 
The remaining deviation may be attributed to quenching of
strange quarks and scaling violation.


\section{Conclusions}
\label{sec:conclusion}


In this paper we have presented a high statistics study of 
the hadron spectrum and quark masses in two-flavor QCD using the 
plaquette gauge action and the fully $O(a)$-improved Wilson quark 
action. We find firm evidence of sea quark effects at the simulated 
quark masses: the slopes
$d m_{\rm V}/d m_{\rm PS}^2$ and $d m_{\rm dec}/d m_{\rm PS}^2$
are larger than in quenched QCD and the $J$ parameter increases 
for lighter sea quarks. 
These findings do not suffer 
from systematic errors arising from the chiral extrapolation
with respect to the sea quark mass, 
which is a major uncertainty
particularly in recent studies with the Wilson-type quark action.
Note that the use of a volume $La \! \geq \! 1.8$~fm
at smaller sea quark masses 
$m_{\rm PS,sea}/m_{\rm V,sea} \! \simeq \! 0.6$\,--\,0.7
is an important factor to control finite-size errors and 
in reaching our observations.

%
%

The sea quark effect observed at the simulated quark masses means
that the strange meson and baryon masses in full QCD 
show a better agreement with experiment than in quenched QCD.
A similar reduction of quenching errors is also observed 
in the ratio $f_{K}/f_{\pi}$.
We also find that the sea quark effects lead to 
about 20\,~\% reduction of quark masses.

For baryons finite-size effects are large for the volume we used,  
which render sea quark effects unclear for lighter baryons.
Further investigations on larger spatial volumes,
of the order of 3~fm at the lightest sea quark mass, 
are needed to observe sea quark effects in the light baryons. 

The present work is carried out at a single lattice spacing of
$a^{-1}\!=\!2.221(28)$~GeV.  
The $O(a)$-improved Wilson quark action we employed 
is designed to have reduced scaling violation, and experiences in 
quenched QCD \cite{scaling.Nf0.UKQCD}
support this expectation. 
Nonetheless scaling study of both the hadron spectrum 
and quark masses with this action
is needed 
to establish the sea quark effects
in the spectral quantities on a quantitative basis. 
%
%
%
%

Another important subject in future is 
simulations at much lighter sea quark mass, 
in particular below the $\rho \! \to \! \pi\pi$ threshold.
Such simulations will lead to 
a better control of the chiral extrapolation.
This would also give insight into the chiral logarithmic singularity
in the PS meson mass and decay constant,
which we have not observed in our data.


\begin{acknowledgments}
This work is supported by the Supercomputer Project No.79 (FY2002)
of High Energy Accelerator Research Organization (KEK),
and also in part by the Grant-in-Aid of the Ministry of Education
(Nos. 11640294, 12640253, 12740133, 13135204, 13640259, 13640260,
14046202, 14740173).
N.Y. is supported by the JSPS Research Fellowship.
\end{acknowledgments}

\appendix

\section{Hadron masses}

Measured hadron masses are summarized in
Tables~\ref{tab:meas:had:meson_12x48}\,--\,\ref{tab:meas:had:baryon_20x48}
for full QCD, 
and
Tables~\ref{tab:meas:had:meson_q12x48}\,--\,\ref{tab:meas:had:baryon_q20x48}
for quenched QCD.
Our choice of the fitting range and resulting value of 
$\chi^2/{\rm dof}$ are also shown in these tables.




\newpage


\begin{table}[htbp]
\caption
{
   Run parameters in simulations of two-flavor QCD.
   The step size $\Delta \tau$ is given by the inverse of the
   number of the molecular dynamics steps (\#MD).
   We denote the tolerance parameter in the stopping condition
   for the quark matrix inversion in calculations of the force
   and Hamiltonian by $\Delta_{\rm f}$ and $\Delta_H$, respectively.
   CPU time required per trajectory on the full machine 
   is written in units of minute.
   Number of measurement is denoted by $N_{\rm meas}$
   and the number of the exceptional configurations is written in 
   the bracket.
}
\begin{ruledtabular}
\begin{tabular}{llllll}
\multicolumn{6}{l}{$12^3{\times}48$,\ \  A-HMC}\\
\hline
$K_{\rm sea}$ & 
   0.1340             & 0.1343                & 0.1346          &
   0.1350             & 0.1355                \\
\hline
 \#MD                 &
   68                 & 72                    & 80              &
   110                & 175                   \\
accept.               & 
   0.667(8)           & 0.667(10)             & 0.668(10)       &
   0.772(6)           & 0.799(9)              \\
$\Delta_{\rm f}$      & 
   $10^{-8}$          & $10^{-8}$             & $10^{-8}$       &
   $10^{-8}$          & $10^{-8}$             \\
$\Delta_H$            & 
   $10^{-14}$         & $10^{-14}$            & $10^{-14}$      &
   $10^{-14}$         & $10^{-14}$            \\
$N_{\rm inv}$         & 
   64.1(2)            & 74.5(2)               & 88.6(3)         &
   117.6(5)           & 203.0(1.0)            \\
time/traj             & 
   0.291              & 0.351                 & 0.465           &
   0.829              & 2.20                  \\
\end{tabular}
\end{ruledtabular}
\label{tab:sim:param:param_all}
\end{table}


\begin{table}[htbp]
\caption{
  Parameters in Eq.~(\ref{eqn:meas:pot:fit2})
  and $r_0$ in full QCD.
  The first error is statistical.
  The second and third ones are the systematic error 
  due to the choice of $t_{\rm min}$ and $r_{\rm min}$.
}
\begin{ruledtabular}
\begin{tabular}{llllll}
\multicolumn{6}{l}{$12^3 \times 48$} \\
\hline
$K_{\rm sea}$ & $V_0$  & $\alpha$ & $g_c$ & $\sigma$ & $r_0$ \\
\hline
0.1340        & 0.549(19)(71)(37) & 0.301(13)(19)(21)  
              & 0.76(5)(24)(12)   & 0.0922(19)(30)(16) 
              & 3.826(24)(35)(27) \\
0.1343        & 0.552(14)(43)(47) & 0.288(9)(32)(44)  
              & 0.74(5)(8)(11)    & 0.0838(10)(49)(36)
              & 4.031(15)(76)(33) \\
0.1346        & 0.568(14)(23)(34) & 0.288(10)(16)(64) 
              & 0.67(4)(5)(8)     & 0.0772(15)(18)(32)
              & 4.200(29)(23)(42) \\
0.1350        & 0.603(10)(17)(13) & 0.296(8)(7)(19)   
              & 0.57(3)(7)(4)     & 0.0675(11)(14)(14)
              & 4.481(27)(59)(17) \\
0.1355        & 0.658(10)(14)(29) & 0.333(7)(24)(21)  
              & 0.47(3)(1)(7)     & 0.0514(11)(25)(25)
              & 5.059(45)(80)(77) \\
\hline
\hline
\multicolumn{6}{l}{$16^3 \times 48$} \\
\end{tabular}
\end{ruledtabular}
\label{tab:meas:pot:r_0:full}
\vspace{-2mm}
\end{table}

\begin{table}[htbp]
\caption{
  Parameters in Eq.~(\ref{eqn:meas:pot:fit2})
  and $r_0$ in quenched QCD.
}
\begin{ruledtabular}
\begin{tabular}{llllll}
lattice        & $V_0$   & $\alpha$ & $g_c$ & $\sigma$ & $r_0$ \\
\hline
$12^3\times48$ & 0.568(7)(17)(10)   & 0.276(5)(8)(13)   &
                 0.318(21)(36)(42)  & 0.0519(6)(17)(10) &
                 5.149(24)(72)(22)  \\
$16^3\times48$ & 0.592(4)(12)(4)    & 0.278(3)(9)(11)   &
                 0.324(15)(29)(27)  & 0.0497(4)(11)(2) 	&
                 5.255(15)(42)(15)  \\
$20^3\times48$ & 0.610(4)(5)(14)    & 0.287(3)(7)(18)   &
                 0.319(15)(9)(49)   & 0.0484(4)(2)(15) 	&
                 5.309(16)(1)(47)   \\
\end{tabular}
\end{ruledtabular}
\label{tab:meas:pot:r_0:qQCD}
\vspace{-2mm}
\end{table}

\vspace{10mm}
\begin{table}[htbp]
\caption{
  Autocorrelation time for plaquette ($\tau^{\rm cum}_{\rm plaq}$),
  PS meson propagator ($\tau^{\rm cum}_{\rm PS}$)
  and Wilson loop ($\tau^{\rm cum}_{\rm W}$) 
  for A-HMC and S-HMC simulations.
  All numbers are written in units of HMC trajectory.
}
\begin{ruledtabular}
\begin{tabular}{llllllll}
$K_{\rm sea}$             & 0.1340              & 0.1343
                          & 0.1346              & 0.1350
                          & 0.1355                        \\
\hline
A-HMC \\ 
\hline
$\tau^{\rm cum}_{\rm plaq}$
                          & 11.4(1.5)
                          & 35.4(9.8)
                          & 21.9(4.7)
                          & 10.3(2.4)
                          & 13.6(2.2)                     \\
$\tau^{\rm cum}_{\rm PS}$           
                          & 13.7(2.6)
                          & 13.5(3.1)
                          & 17.6(4.3)
                          & 7.7(0.9)
                          & 8.5(3.1)                      \\
$\tau^{\rm cum}_{\rm W}$ 
                          & 5.0(0.4)
                          & 5.9(0.8)
                          & 5.9(1.1)
                          & 4.6(0.7)
                          & 6.2(0.9)                      \\
\hline
\hline
S-HMC \\ 
\hline
$K_{\rm sea}$             & 0.1340              & 0.1343
                          & 0.1346              & 0.1350
                          & 0.1355                        \\
\hline
$\tau^{\rm cum}_{\rm plaq}$ 
                          & 32.6(6.1)
                          & 30.7(4.8)
                          & 20.0(3.1)
                          & 18.9(2.7)
                          & 12.0(1.3)                     \\
\end{tabular}
\end{ruledtabular}
\label{tab:meas:autocorr:tau_cum}
\vspace{-2mm}
\end{table}


\begin{table}[htbp]
\caption
{
   Parameters of diagonal fits to PS meson masses in method-A.
   We put ``--'' in some columns in this and the following tables 
   when the corresponding term is not included in the fit.
}
\begin{ruledtabular}
\begin{tabular}{llllll}
$\chi^2/{\rm dof}$      & 
$K_c$                   & $B^{\rm PS}_{\rm diag}$  &  
$C^{\rm PS}_{\rm diag}$ & $D^{\rm PS}_{\rm diag}$  \\
\hline
0.01                    & 
0.136026(11)            & 25.97(63)                &  
4.5(2.4)                & --                       \\
0.01                    & 
0.136025(43)            & 26.0(4.0)   &  4(31)     &  2(72)    \\
\end{tabular}
\end{ruledtabular}
\label{tab:chiralfit:w_r0:PSK_diag}
\end{table}

\begin{table}[htbp]
\caption
{
   Parameters of diagonal fits to vector meson masses in method-A.
}
\begin{ruledtabular}
\begin{tabular}{llllll}
$\chi^2/{\rm dof}$      & 
$A^{\rm V}_{\rm diag}$  & $B^{\rm V}_{\rm diag}$   &  
$C^{\rm V}_{\rm diag}$  & $D^{\rm V}_{\rm diag}$   \\
\hline
0.06                    & 
1.867(36)               & 0.248(17)                & 
-0.0083(19)             & --                       \\
0.04                    & 
1.90(10)                & 0.224(81)                & 
-0.002(21)              & -0.0005(17)              \\
\end{tabular}
\end{ruledtabular}
\label{tab:chiralfit:w_r0:VPS_diag}
\end{table}

\begin{table}[htbp]
\caption
{
   Parameters of combined chiral fits to PS meson masses
   in terms of VWI quark mass.
}
\begin{ruledtabular}
\begin{tabular}{llllllllll}
$\chi^2/{\rm dof}$ & 
$K_c$              & $B^{\rm PS}_{\rm s}$  & $B^{\rm PS}_{\rm v}$  &
                     $C^{\rm PS}_{\rm s}$  & $C^{\rm PS}_{\rm v}$  &  
                     $C^{\rm PS}_{\rm sv}$ & $C^{\rm PS}_{\rm vv}$ \\
\hline
0.33               & 
0.1360187(95)      & 9.64(51)              & 16.694(93)            &  
                     7.5(1.8)              & 2.39(13)              &  
                    -6.68(41)              & --                    \\
0.33               & 
0.1360185(95)      & 9.62(51)              & 16.751(95)            &  
                     7.6(1.8)              & 1.61(21)              &  
                    -6.78(41)              & 0.63(11)              \\
\end{tabular}
\end{ruledtabular}
\label{tab:chiralfit:w_r0:PSK}
\end{table}

\begin{table}[htbp]
\caption
{
   Parameters of combined chiral fits to vector meson masses.
}
\begin{ruledtabular}
\begin{tabular}{lllllllll}
$\chi^2/{\rm dof}$ & 
$A^{\rm V}$        & $B^{\rm V}_{\rm s}$  & $B^{\rm V}_{\rm v}$  &
                     $C^{\rm V}_{\rm s}$  & $C^{\rm V}_{\rm v}$  &  
                     $C^{\rm V}_{\rm sv}$ \\ 
\hline
0.55               & 
1.914(24)          & 0.0349(51)           & 0.1895(45)           &
                     --                   & --                   & 
                    -0.00554(96)          \\ 
0.23               & 
1.877(33)          & 0.051(14)            & 0.1953(55)           &
                    -0.0028(16)           &-0.00159(41)          & 
                    -0.00389(65)          \\ 
\end{tabular}
\end{ruledtabular}
\label{tab:chiralfit:w_r0:VPS}
\end{table}

\begin{table}[htbp]
\caption
{
   Parameters of partially quenched chiral fits
   to vector meson masses.
}
\begin{ruledtabular}
\begin{tabular}{lllll}
$K_{\rm sea}$ & $\chi^2/{\rm dof}$    & 
                $A^{\rm V}_{\rm PQ}$  & $B^{\rm V}_{\rm PQ}$ \\
\hline
0.1340        & 0.10(5)  & 0.5306(23) & 0.6361(53) \\
0.1343        & 0.08(5)  & 0.5067(23) & 0.6648(58) \\
0.1346        & 0.04(3)  & 0.4835(27) & 0.6940(84) \\
0.1350        & 0.14(10) & 0.4394(29) & 0.786(10)  \\
0.1355        & 0.27(20) & 0.3885(34) & 0.906(14)  \\
\end{tabular}
\end{ruledtabular}
\label{tab:chiralfit:w_r0:VPS_PQ}
\end{table}

\begin{table}[htbp]
\caption
{
   Parameters of combined chiral fits to octet baryon masses.
}
\begin{ruledtabular}
\begin{tabular}{llllllllll}
$\chi^2/{\rm dof}$ &
               $A^{\rm O}$           &  $B^{\rm O}_{\rm s}$  &  
               $F^{\rm O}_{\rm v}$   &  $D^{\rm O}_{\rm v}$  &
               $C^{\rm O}_{\rm s}$   &  $C^{\rm O}_{\rm v}$  &
               $C^{\rm O}_{\rm sv}$  &  $C^{\rm O}_{\rm vv}$ \\
& 
               $C^{{\rm O}, \Sigma }_{\rm v}$   &  
               $C^{{\rm O}, \Lambda}_{\rm v}$   &  
               $C^{{\rm O}, \Sigma }_{\rm sv}$  &  
               $C^{{\rm O}, \Lambda}_{\rm sv}$  \\
\hline
0.82      & 
               2.512(50)    &  0.018(23)    &
               0.1286(32)   & -0.0344(34)   &
               0.0001(28)   & -0.00621(39)  &
              -0.00148(54)  &  --           \\
          &          
               0.00047(23)  &  0.00374(26)  &
               0.00057(35)  &  0.00014(39)  \\
0.66      & 
               2.489(51)    &  0.020(23)    &
               0.1315(34)   & -0.0342(34)   &
              -0.0008(28)   & -0.00520(32)  &
              -0.00088(54)  & -0.00356(34)  \\
          &          
               0.00074(25)  &  0.00395(27)  &
               0.00044(35)  &  0.00006(39)  \\
\end{tabular}
\end{ruledtabular}
\label{tab:chiralfit:w_r0:OPS}
\end{table}

\begin{table}[htbp]
\caption
{
   Parameters of combined chiral fits to decuplet baryon masses.
}
\begin{ruledtabular}
\begin{tabular}{lllllllll}
$\chi^2/{\rm dof}$ & 
               $A^{\rm D}$          &  
               $B^{\rm D}_{\rm s}$  &  $B^{\rm D}_{\rm v}$  &
               $C^{\rm D}_{\rm s}$  &  $C^{\rm D}_{\rm v}$  &
               $C^{\rm D}_{\rm sv}$ \\
\hline
0.92        &  
               3.317(65)            &
               0.038(15)            &  0.268(13)            &
               --                   &  --                   &
              -0.0055(29)           \\
0.92        &  
               3.327(85)            &
               0.049(36)            &  0.253(16)            &
              -0.0012(43)           &  0.0020(13)           &
              -0.0059(19)           \\
\end{tabular}
\end{ruledtabular}
\label{tab:chiralfit:w_r0:DPS}
\end{table}

\begin{table}[htbp]
\caption
{
   Parameters of chiral extrapolation of $r_0^{-1}$.
}
\begin{ruledtabular}
\begin{tabular}{llllllllll}
$\chi^2/{\rm dof}$ &  $A_{r_0}$   &  $B_{r_0}$  \\
\hline
1.5                & -4.867(35)   &  0.6861(48) \\
\end{tabular}
\end{ruledtabular}
\label{tab:chiralfit:w_r0:r0inv_vs_Kinv}
\end{table}

\begin{table}[htbp]
\caption
{
   Parameters of diagonal fits to PS meson masses in method-B.
}
\begin{ruledtabular}
\begin{tabular}{llllll}
$\chi^2/{\rm dof}$       & 
$K^{\prime}_{c, \rm diag}$      & $B^{\prime \rm PS}_{\rm diag}$  &
$C^{\prime \rm PS}_{\rm diag}$  & $D^{\prime \rm PS}_{\rm diag}$  \\
\hline
2.5          &  
0.136020(10) &  4.67(11) &  39.1(1.3)  &  --        \\
1.9          &  
0.135976(23) &  5.47(42) &  15(13)     &  252(142)  \\
\end{tabular}
\end{ruledtabular}
\label{tab:chiralfit:wo_r0:PSK_diag}
\end{table}

\clearpage

\begin{table}[htbp]
\caption
{
   Parameters of diagonal fits to vector meson masses in method-B.
}
\begin{ruledtabular}
\begin{tabular}{lllllll}
$\chi^2/{\rm dof}$       & 
$A^{\prime \rm V}_{\rm diag}$   & $B^{\prime \rm V}_{\rm diag}$  &
$C^{\prime \rm V}_{\rm diag}$   & $D^{\prime \rm V}_{\rm diag}$  \\
\hline
1.4        &  
0.3523(45) &  1.481(38)  & -1.000(73) &  --        \\
0.1      &  
0.3379(89) &  1.73(13)   & -2.19(59)  &  1.71(80)  \\
\end{tabular}
\end{ruledtabular}
\label{tab:chiralfit:wo_r0:VPS_diag}
\end{table}

\begin{table}[htbp]
\caption
{
   Parameters of chiral extrapolation of $r_0$ 
   in terms of PS meson mass squared.
}
\begin{ruledtabular}
\begin{tabular}{lllllll}
$\chi^2/{\rm dof}$       & 
$A^{\prime}_{r_0}$   & $B^{\prime}_{r_0}$  &
$C^{\prime}_{r_0}$   & $D^{\prime}_{r_0}$  \\
\hline
0.01       &  
5.629(34)  & -8.41(56)   & 16.3(2.7)  & -15.2(3.8) \\
\end{tabular}
\end{ruledtabular}
\label{tab:chiralfit:wo_r0:r0}
\end{table}

\begin{table}[htbp]
\caption
{
   Hopping parameters corresponding to the light ($K_{ud}$)
   and the strange quark mass with $K$- ($K_s(K)$) and 
   $\phi$-input ($K_s(\phi)$)
   in full QCD.
   The lattice cutoff determined from $\rho$ meson mass
   is also written.
   Error is statistical only.
}
\begin{ruledtabular}
\begin{tabular}{llllll}
$K_{ud}$      & $K_s(K)$     & $K_s(\phi)$  & $a^{-1}$~[GeV]    \\
\hline
0.1359896(90) & 0.134857(29) & 0.134711(44) & 2.221(28)         \\
\end{tabular}
\end{ruledtabular}
\label{tab:chiralfit:spectrum:K}
\end{table}

\begin{table}[htbp]
\caption
{
   Fit parameters of chiral extrapolations in quenched QCD.
}
\begin{ruledtabular}
\begin{tabular}{l|lllllll}
                & $\chi^2/{\rm dof}$ & 
                  $K_{c, q}$         & $B^{\rm PS}_{\rm q}$  &
                                       $C^{\rm PS}_{\rm q}$ \\
$m_{\rm PS}^2$  & 0.18               & 
                  0.135315(12)       & 2.859(35)     & 3.09(31)   \\
\hline
                & $\chi^2/{\rm dof}$ & 
                  $A^{\rm V}_{\rm q}$& $B^{\rm V}_{\rm q}$   \\
$m_{\rm V}$     & 0.04               & 
                  0.4146(54)         & 0.797(22)     \\
\hline
                & $\chi^2/{\rm dof}$ & 
                  $A^{\rm O}_{\rm q}$      &
                  $F^{\rm O}_{\rm q}$      & $D^{\rm O}_{\rm q}$  &
                  $C^{\rm O}_{\rm q}$      & 
                  $C^{{\rm O},\Sigma}_{\rm q}$  &
                  $C^{{\rm O},\Lambda}_{\rm q}$ \\
$m_{\rm oct}$   & 0.66               & 
                  0.538(14)          & 0.521(47)     & -0.158(44)   &
                 -0.36(17)           &-0.059(92)     &  0.255(95)   \\
\hline
                & $\chi^2/{\rm dof}$ & 
                  $A^{\rm D}_{\rm q}$ & $B^{\rm D}_{\rm q}$  \\
$m_{\rm dec}$   & 0.03               & 
                  0.712(14)          & 1.123(59)     \\
\end{tabular}
\end{ruledtabular}
\label{tab:chiralfit:qQCD:fit}
\end{table}

\begin{table}[htbp]
\caption
{
   Hopping parameters corresponding to the light ($K_{ud}$)
   and the strange quark mass with $K$- ($K_s(K)$) and 
   $\phi$-input ($K_s(\phi)$)
   in quenched QCD.
}
\begin{ruledtabular}
\begin{tabular}{llllll}
$K_{ud}$      & $K_s(K)$     & $K_s(\phi)$  & $a^{-1}$~[GeV]    \\
\hline
0.135245(11)  & 0.133571(43) & 0.133214(76) & 1.834(24)         \\
\end{tabular}
\end{ruledtabular}
\label{tab:chiralfit:qQCD:K}
\end{table}


\begin{table}[htbp]
\caption
{
   $J$ parameter calculated by Eq.~(\ref{eqn:chiralfit:w_r0:J}).
   We obtain $J\!=\!0.4242(61)$ at $K_{\rm sea}\!=\!K_{ud}$
   with the alternative definition (\ref{eqn:SQE:J:J_massratio}).
}
\begin{ruledtabular}
\begin{tabular}{l|llllll|l}
$K_{\rm sea}$     & 0.1340 
                  & 0.1343
                  & 0.1346
                  & 0.1350
                  & 0.1355
                  & $K_{ud}$
                  & quenched \\ \hline
$J$               & 0.3827(22)
                  & 0.3818(21)
                  & 0.3801(34)
                  & 0.3932(32)
                  & 0.4018(37)
                  & 0.4153(62)
                  & 0.3735(66)
\end{tabular}
\end{ruledtabular}
\label{tab:SQE:J}
\end{table}

\clearpage

\begin{table}[htbp]
\caption
{
   Strange vector meson masses in GeV units in full and quenched QCD.
   The meson mass input to fix the strange quark mass
   is written in the brackets in the first column.
   The first error is statistical.
   The second and third ones are systematic error due to 
   the choice of chiral fit forms and systematic uncertainty of $r_0$.
   The deviation from the experimental spectrum is 
   denoted by $\Delta m$.
}
\begin{ruledtabular}
\begin{tabular}{l|ll|ll}
& $m(N_f\!=\!2)$ & $\Delta m(N_f\!=\!2)$ 
& $m(N_f\!=\!0)$ & $\Delta m(N_f\!=\!0)$ \\ \hline
${K^*}(K)$       &  $0.8791(14)(^{+3}_{-1})(^{+14}_{-12})$  
                 &  $-$1.9\%
                 &  0.8706(16)  & $-$2.9\% 
                 \\
${\phi}(K)$      &  $0.9899(28)(^{+5}_{-28})(^{+29}_{-24})$
                 &  $-$2.9\%
                 &  0.9727(32)  & $-$4.7\%
                 \\
${K}(\phi)$      &  $0.5272(31)(^{+40}_{-7})(^{+26}_{-31})$
                 &  +5.8\%
                 &  0.5466(39)  & +9.4\%
                 \\
${K^*}(\phi)$    &  $0.89390(0)(+173)(0)$
                 &  $-$0.2\%
                 &  0.89390(0) & $-$0.2\%
                 \\
\end{tabular}
\end{ruledtabular}
\label{tab:SQE:had:meson}
\end{table}
   
\begin{table}[htbp]
\caption
{
   Baryon spectrum in GeV units in full and quenched QCD.
}
\begin{ruledtabular}
\begin{tabular}{l|ll|ll}
& $m(N_f\!=\!2)$ & $\Delta m(N_f\!=\!2)$ 
& $m(N_f\!=\!0)$ & $\Delta m(N_f\!=\!0)$ \\ \hline
${N}$              &  $1.015(19)(^{+18}_{-9})(^{+10}_{-7})$
                   &  +7.7\%
                   &  1.004(26)  &  +6.7\%
                   \\
${\Sigma}(K)$      &  $1.185(17)(^{+16}_{-10})(^{+8}_{-5})$
                   &  $-$0.7\%
                   &  1.162(20)  &  $-$2.6\%
                   \\
${\Lambda}(K)$     &  $1.143(18)(^{+16}_{-9})(^{+9}_{-6})$
                   &  +2.4\%
                   &  1.120(20)  &  +0.4\%
                   \\
${\Xi}(K)$         &  $1.291(16)(+14)(^{+7}_{-4})$
                   &  $-$1.8\%
                   &  1.260(14)  &  $-$4.3\%
                   \\
${\Sigma}(\phi)$   &  $1.204(17)(+18)(^{+6}_{-3})$
                   &  +1.0\%
                   &  1.194(19)  &  +0.1\%
                   \\
${\Lambda}(\phi)$  &  $1.159(17)(+18)(^{+7}_{-4})$
                   &  +3.8\%
                   &  1.146(19)  &  +2.7\%
                   \\
${\Xi}(\phi)$      &  $1.325(16)(+18)(^{+3}_{-1})$
                   &  +0.7\%
                   &  1.315(11)  &  +0.02\%
                   \\
\end{tabular}
\end{ruledtabular}
\label{tab:SQE:had:baryon}
\end{table}

\begin{table}[htbp]
\caption
{
   Parameters of combined chiral fits to PS meson decay constant
   in full QCD.
}
\begin{ruledtabular}
\begin{tabular}{llllllllll}
$\chi^2/{\rm dof}$ & 
$A^{\rm f}$        & $B^{\rm f}_{\rm s}$   & $B^{\rm f}_{\rm v}$   &  
                     $C^{\rm f}_{\rm s}$   & $C^{\rm f}_{\rm v}$   & 
                     $C^{\rm f}_{\rm sv}$  \\
\hline
0.001               & 
0.3415(71)         & 0.0075(18)            & 0.0349(18)            &
                     --                    & --                    &
                    -0.00128(42)           \\
0.001               & 
0.344(11)          & 0.0019(55)            & 0.0398(18)            &
                     0.00046(70)           &-0.00105(19)           &
                    -0.00069(23)           \\
\end{tabular}
\end{ruledtabular}
\label{tab:chiralfit:w_r0:fPS:Nf2}
\end{table}

\begin{table}[htbp]
\caption
{
   Parameters of chiral extrapolation to PS meson decay constant
   in quenched QCD.
}
\begin{ruledtabular}
\begin{tabular}{llllllllll}
$\chi^2/{\rm dof}$ & 
$A^{\rm f}_{\rm q}$  & $B^{\rm f}_{\rm q}$ \\
\hline
0.06               & 
0.07221(94)        & 0.1246(42)            \\
\end{tabular}
\end{ruledtabular}
\label{tab:chiralfit:w_r0:fPS:qQCD}
\end{table}

\begin{table}[htbp]
\caption
{
   Pseudoscalar decay constants in full and quenched QCD
   calculated with one-loop and non-perturbative (NP) matchings.
   Decay constants are written in GeV units.
   The first error is statistical.
   The second and third ones for full QCD results
   are systematic error due to the choice of chiral fit forms
   and uncertainty of measured value of $r_0$.
}
\begin{ruledtabular}
\begin{tabular}{l|l|l|l}
& \multicolumn{2}{l|}{one-loop $c_A$, $Z_A$ and $b_A$} 
& {NP $c_A$, $Z_A$ and $b_A$} \\ 
\hline
& $N_f\!=\!2$
& $N_f\!=\!0$
& $N_f\!=\!0$ \\ \hline
$f_{\pi}$        &  $0.1372(31)(+25)\left(^{+11}_{-7}\right)$
                 &  0.1337(24)
                 &  0.1287(23)
                 \\
$f_K(K)$         &  $0.1576(23)(+28)\left(^{+8}_{-5}\right)$
                 &  0.1497(20)
                 &  0.1438(19)
                 \\
$f_K(\phi)$      &  $0.1603(21)(+29)\left(^{+5}_{-3}\right)$
                 &  0.1533(17)
                 &  0.1473(17)
                 \\
$f_K(K)/f_{\pi}$
                 &  $1.148(11)\left(^{+12}_{-5}\right)
                              \left(^{+2}_{-3}\right)$
                 &  1.1195(63)
                 &  1.1174(63)
                 \\
$f_K(\phi)/f_{\pi}$
                 &  $1.168(13)\left(^{+13}_{-3}\right)
                              \left(^{+4}_{-6}\right)$
                 &  1.1468(90)
                 &  1.1443(89)
                 \\
\end{tabular}
\end{ruledtabular}
\label{tab:f:result}
\end{table}


\begin{table}[htbp]
\caption
{
   Parameters of combined chiral fits to PS meson masses
   in terms of AWI quark mass in full QCD.
}
\begin{ruledtabular}
\begin{tabular}{llllllllll}
$\chi^2/{\rm dof}$ & 
$B^{\rm PS,AWI}_{\rm v}$  & 
$C^{\rm PS,AWI}_{\rm v}$  & $C^{\rm PS,AWI}_{\rm sv}$ \\
\hline
0.21                      &
18.08(12)                 &
0.33(19)                  & -3.01(38)                 \\
0.22                      &
18.13(11)                 &
--                        & -2.89(39)                 \\
\end{tabular}
\end{ruledtabular}
\label{tab:chiralfit:w_r0:PSq:Nf2}
\end{table}

\begin{table}[htbp]
\caption
{
   Quark masses in two-flavor QCD in $\overline{\rm MS}$ scheme
   at $\mu\!=\!2$~GeV.
   Values of $m_{ud}$ and $m_s$ are in MeV units.
   The meson mass input to fix $m_s$ is written in brackets.
   The quoted errors are statistical only.
}
\begin{ruledtabular}
\begin{tabular}{l|lllll}
     &  $m_{ud}$         &  $m_s(K)$            &  $m_s(\phi)$        
     &  $m_s(K)/m_{ud}$  &  $m_s(\phi)/m_{ud}$  \\ \hline
AWI  &  3.223(43)        &  84.5(1.1)           &  96.4(2.2)
     &  26.133(24)       &  29.78(38)           
     \\
VWI  &  2.488(57)        &  98.2(1.1)           & 110.4(2.3)
     &  39.36(73)        &  44.21(93)          
     \\
\end{tabular}
\end{ruledtabular}
\label{tab:mq:Nf2}
\end{table}

\begin{table}[htbp]
\caption
{
   Parameters of chiral fit to PS meson masses
   in terms of AWI quark mass in quenched QCD.
}
\begin{ruledtabular}
\begin{tabular}{llllllllll}
$\chi^2/{\rm dof}$ & 
$B^{\rm PS,AWI}_{\rm q}$  &  $C^{\rm PS,AWI}_{\rm q}$  \\ \hline
0.22               &
2.974(44)          & 1.31(61) \\
\end{tabular}
\end{ruledtabular}
\label{tab:chiralfit:w_r0:PSq:qQCD}
\end{table}

\begin{table}[htbp]
\caption
{
   Quark masses in quenched QCD
   in $\overline{\rm MS}$ scheme at $\mu\!=\!2$~GeV
   calculated in quenched QCD.
   Values of $m_{ud}$ and $m_s$ are in MeV units.
   Choice of input to fix $m_s$ is written in brackets.
}
\begin{ruledtabular}
\begin{tabular}{l|lllll}
     &  $m_{ud}$         &  $m_s(K)$            &  $m_s(\phi)$        
     &  $m_s(K)/m_{ud}$  &  $m_s(\phi)/m_{ud}$  \\ \hline
     &  \multicolumn{5}{l}{one-loop matching} \\ \hline
AWI  &  4.020(77)        &  104.1(1.6)          &  128.2(3.3)
     &  25.90(13)        &  31.88(56)           \\
VWI  &  4.628(77)        &  114.2(1.6)          &  136.9(3.2)     
     &  24.673(88)       &  29.57(42)           \\ \hline
     &  \multicolumn{5}{l}{non-perturbative matching} \\ \hline
AWI  &  3.522(66)        &  91.9(1.4)           &  113.1(2.9)
     &  26.08(13)        &  32.11(56)           \\ 
VWI  &  4.315(72)        &  106.5(1.5)          &  127.7(3.0)
     &  24.681(87)       &  29.58(42)           \\ 
\end{tabular}
\end{ruledtabular}
\label{tab:mq:qQCD}
\end{table}

\clearpage


\setlongtables
\setlength{\LTleft}{0pt}
\setlength{\LTright}{0pt}
\begin{longtable}{c@{\extracolsep{\fill}}cc|ccc|ccc}
   \caption
   {
      Meson masses on $12^3 \times 48$ lattice in $N_f\!=\!2$ full QCD.
   }
\endhead
\hline
\hline
$K_{\rm sea}$                 &
$K_{\rm val,1}$               &  
$K_{\rm val,2}$               &
[$t_{\rm min},t_{\rm max}$]   &
$\chi^2/{\rm dof}$            &
$m_{\rm PS}$                  &                  
[$t_{\rm min},t_{\rm max}$]   &
$\chi^2/{\rm dof}$            &
$m_{\rm V}$                  \\
\hline
0.1340 & 0.1340 & 0.1340 &
         [9,24] & 2.7 & 0.619(10) &
         [9,24] & 2.1 & 0.782(12) \\ 
0.1340 & 0.1343 & 0.1343 &
         [9,24] & 2.9 & 0.592(11) &
         [9,24] & 2.3 & 0.762(13) \\ 
0.1340 & 0.1346 & 0.1346 &
         [9,24] & 2.9 & 0.564(12) &
         [9,24] & 2.5 & 0.742(13) \\ 
0.1340 & 0.1350 & 0.1350 &
         [9,24] & 2.6 & 0.525(12) &
         [9,24] & 2.7 & 0.714(14) \\ 
0.1340 & 0.1355 & 0.1355 &
         [9,24] & 2.1 & 0.4709(95) &
         [9,24] & 2.6 & 0.679(14) \\ 
0.1340 & 0.1358 & 0.1358 &
         [9,24] & 1.9 & 0.4344(79) &
         [9,24] & 2.4 & 0.660(13) \\ 
0.1340 & 0.1340 & 0.1343 &
         [9,24] & 2.8 & 0.606(11) &
         [9,24] & 2.2 & 0.772(13) \\
\end{longtable}

\begin{longtable}{cccc|ccc|ccc|ccc}
   \caption
   {
      Baryon masses on $12^3 \times 48$ lattice in $N_f\!=\!2$ full QCD.
   } \\
   \hline
   \hline
   $K_{\rm sea}$                 &
   $K_{\rm val,1}$               &  
   $K_{\rm val,2}$               &
   $K_{\rm val,3}$               &
   [$t_{\rm min},t_{\rm max}$]   &
   $\chi^2/{\rm dof}$            &
   $m_{\Sigma}$                  &                  
   [$t_{\rm min},t_{\rm max}$]   &
   $\chi^2/{\rm dof}$            &
   $m_{\Lambda}$                 &
   [$t_{\rm min},t_{\rm max}$]   &
   $\chi^2/{\rm dof}$            &
   $m_{\rm dec}$                 \\
   \hline
\endfirsthead
   \caption{\it (Continued)} \\
   \hline
   \hline
   $K_{\rm sea}$                 &
   $K_{\rm val,1}$               &  
   $K_{\rm val,2}$               &
   $K_{\rm val,3}$               &
   [$t_{\rm min},t_{\rm max}$]   &
   $\chi^2/{\rm dof}$            &
   $m_{\Sigma}$                  &                  
   [$t_{\rm min},t_{\rm max}$]   &
   $\chi^2/{\rm dof}$            &
   $m_{\Lambda}$                 &
   [$t_{\rm min},t_{\rm max}$]   &
   $\chi^2/{\rm dof}$            &
   $m_{\rm dec}$                 \\
   \hline
\endhead
   \hline
\endfoot
0.1340 & 0.1340 & 0.1340 & 0.1340 &
         [10,24] & 1.7 & 1.153(12) &
         -- & -- & -- &
         [10,24] & 1.6 & 1.220(37) \\ 
0.1340 & 0.1343 & 0.1343 & 0.1343 &
         [10,24] & 1.7 & 1.122(16) &
         -- & -- & -- &
         [10,24] & 1.5 & 1.191(39) \\ 
0.1340 & 0.1346 & 0.1346 & 0.1346 &
         [10,24] & 1.7 & 1.089(24) &
         -- & -- & -- &
         [10,24] & 1.4 & 1.164(39) \\ 
0.1340 & 0.1350 & 0.1350 & 0.1350 &
         [10,24] & 1.6 & 1.040(33) &
         -- & -- & -- &
         [10,24] & 1.3 & 1.131(36) \\ 
0.1340 & 0.1355 & 0.1355 & 0.1355 &
         [10,24] & 1.8 & 0.971(41) &
         -- & -- & -- &
         [10,24] & 1.4 & 1.096(32) \\ 
0.1340 & 0.1358 & 0.1358 & 0.1358 &
         [10,24] & 1.8 & 0.922(43) &
         -- & -- & -- &
         [10,24] & 1.5 & 1.077(38) \\ 
0.1340 & 0.1340 & 0.1340 & 0.1343 &
         [10,24] & 1.8 & 1.142(13) &
         [10,24] & 1.6 & 1.143(13) &
         [10,24] & 1.6 & 1.209(37) \\ 
0.1340 & 0.1340 & 0.1340 & 0.1346 &
         [10,24] & 1.8 & 1.132(15) &
         [10,24] & 1.5 & 1.134(14) &
         [10,24] & 1.6 & 1.199(38) \\
\end{longtable}

\newpage
\begin{longtable}{c@{\extracolsep{\fill}}cc|ccc|ccc}
\caption
{
   Meson masses on $16^3 \times 48$ lattice in $N_f\!=\!2$ full QCD.
}
\endhead
\hline
\hline
$K_{\rm sea}$                 &
$K_{\rm val,1}$               &  
$K_{\rm val,2}$               &
[$t_{\rm min},t_{\rm max}$]   &
$\chi^2/{\rm dof}$            &
$m_{\rm PS}$                  &                  
[$t_{\rm min},t_{\rm max}$]   &
$\chi^2/{\rm dof}$            &
$m_{\rm V}$                  \\
\hline
0.1340 & 0.1340 & 0.1340 &
         [9,24] & 3.3 & 0.6200(21) &
         [9,24] & 2.0 & 0.7727(41) \\ 
0.1340 & 0.1343 & 0.1343 &
         [9,24] & 2.9 & 0.5917(20) &
         [9,24] & 1.6 & 0.7527(49) \\ 
0.1340 & 0.1346 & 0.1346 &
         [9,24] & 2.4 & 0.5623(19) &
         [9,24] & 1.4 & 0.7324(60) \\ 
0.1340 & 0.1350 & 0.1350 &
         [9,24] & 1.9 & 0.5212(22) &
         [9,24] & 1.6 & 0.7050(78) \\ 
0.1340 & 0.1355 & 0.1355 &
         [9,24] & 1.8 & 0.4656(27) &
         [9,24] & 2.5 & 0.672(12) \\ 
0.1340 & 0.1358 & 0.1358 &
         [9,24] & 2.1 & 0.4288(29) &
         [9,24] & 2.5 & 0.653(13) \\ 
0.1340 & 0.1340 & 0.1343 &
         [9,24] & 3.1 & 0.6060(20) &
         [9,24] & 1.8 & 0.7627(45) \\
\end{longtable}

\begin{longtable}{cccc|ccc|ccc|ccc}
   \caption
   {
      Baryon masses on $16^3 \times 48$ lattice 
      in $N_f\!=\!2$ full QCD.
   } \\
   \hline
   \hline
   $K_{\rm sea}$                 &
   $K_{\rm val,1}$               &  
   $K_{\rm val,2}$               &
   $K_{\rm val,3}$               &
   [$t_{\rm min},t_{\rm max}$]   &
   $\chi^2/{\rm dof}$            &
   $m_{\Sigma}$                  &                  
   [$t_{\rm min},t_{\rm max}$]   &
   $\chi^2/{\rm dof}$            &
   $m_{\Lambda}$                 &
   [$t_{\rm min},t_{\rm max}$]   &
   $\chi^2/{\rm dof}$            &
   $m_{\rm dec}$                 \\
   \hline
\endfirsthead
   \caption{\it (Continued)} \\
   \hline
   \hline
   $K_{\rm sea}$                 &
   $K_{\rm val,1}$               &  
   $K_{\rm val,2}$               &
   $K_{\rm val,3}$               &
   [$t_{\rm min},t_{\rm max}$]   &
   $\chi^2/{\rm dof}$            &
   $m_{\Sigma}$                  &                  
   [$t_{\rm min},t_{\rm max}$]   &
   $\chi^2/{\rm dof}$            &
   $m_{\Lambda}$                 &
   [$t_{\rm min},t_{\rm max}$]   &
   $\chi^2/{\rm dof}$            &
   $m_{\rm dec}$                 \\
   \hline
\endhead
   \hline
\endfoot
0.1340 & 0.1340 & 0.1340 & 0.1340 &
         [10,24] & 2.2 & 1.166(20) &
         -- & -- & -- &
         [10,24] & 5.1 & 1.256(20) \\ 
0.1340 & 0.1343 & 0.1343 & 0.1343 &
         [10,24] & 2.3 & 1.128(21) &
         -- & -- & -- &
         [10,24] & 3.8 & 1.224(17) \\ 
0.1340 & 0.1346 & 0.1346 & 0.1346 &
         [10,24] & 2.2 & 1.089(20) &
         -- & -- & -- &
         [10,24] & 2.8 & 1.192(15) \\ 
0.1340 & 0.1350 & 0.1350 & 0.1350 &
         [10,24] & 1.7 & 1.037(18) &
         -- & -- & -- &
         [10,24] & 2.0 & 1.149(13) \\ 
0.1340 & 0.1355 & 0.1355 & 0.1355 &
         [10,24] & 1.3 & 0.979(18) &
         -- & -- & -- &
         [10,24] & 1.4 & 1.096(13) \\ 
0.1340 & 0.1358 & 0.1358 & 0.1358 &
         [10,24] & 1.7 & 0.955(18) &
         -- & -- & -- &
         [10,24] & 1.2 & 1.067(16) \\ 
0.1340 & 0.1340 & 0.1340 & 0.1343 &
         [10,24] & 2.1 & 1.152(20) &
         [10,24] & 2.4 & 1.155(22) &
         [10,24] & 4.9 & 1.242(20) \\ 
0.1340 & 0.1340 & 0.1340 & 0.1346 &
         [10,24] & 2.0 & 1.138(20) &
         [10,24] & 2.7 & 1.144(23) &
         [10,24] & 4.3 & 1.231(19) \\
\end{longtable}

\newpage
\begin{longtable}{c@{\extracolsep{\fill}}cc|ccc|ccc}
\caption
{
   Meson masses on $20^3 \times 48$ lattice in $N_f\!=\!2$ full QCD.
}
\endhead
\hline
\hline
$K_{\rm sea}$                 &
$K_{\rm val,1}$               &  
$K_{\rm val,2}$               &
[$t_{\rm min},t_{\rm max}$]   &
$\chi^2/{\rm dof}$            &
$m_{\rm PS}$                  &                  
[$t_{\rm min},t_{\rm max}$]   &
$\chi^2/{\rm dof}$            &
$m_{\rm V}$                  \\
\hline
0.1340 & 0.1340 & 0.1340 &
         [9,24] & 0.8 & 0.61630(55) &
         [9,24] & 1.1 & 0.7715(12) \\ 
0.1340 & 0.1343 & 0.1343 &
         [9,24] & 0.7 & 0.58799(56) &
         [9,24] & 1.1 & 0.7507(12) \\ 
0.1340 & 0.1346 & 0.1346 &
         [9,24] & 0.7 & 0.55869(56) &
         [9,24] & 1.1 & 0.7297(13) \\ 
0.1340 & 0.1350 & 0.1350 &
         [9,24] & 0.7 & 0.51777(58) &
         [9,24] & 1.1 & 0.7016(13) \\ 
0.1340 & 0.1355 & 0.1355 &
         [9,24] & 0.7 & 0.46262(62) &
         [9,24] & 1.0 & 0.6665(15) \\ 
0.1340 & 0.1358 & 0.1358 &
         [9,24] & 0.8 & 0.42623(65) &
         [9,24] & 1.1 & 0.6454(17) \\ 
0.1340 & 0.1340 & 0.1343 &
         [9,24] & 0.7 & 0.60226(55) &
         [9,24] & 1.1 & 0.7611(12) \\
\end{longtable}

\begin{longtable}{cccc|ccc|ccc|ccc}
   \caption
   {
      Baryon masses on $20^3 \times 48$ lattice 
      in $N_f\!=\!2$ full QCD.
   } \\
   \hline
   \hline
   $K_{\rm sea}$                 &
   $K_{\rm val,1}$               &  
   $K_{\rm val,2}$               &
   $K_{\rm val,3}$               &
   [$t_{\rm min},t_{\rm max}$]   &
   $\chi^2/{\rm dof}$            &
   $m_{\Sigma}$                  &                  
   [$t_{\rm min},t_{\rm max}$]   &
   $\chi^2/{\rm dof}$            &
   $m_{\Lambda}$                 &
   [$t_{\rm min},t_{\rm max}$]   &
   $\chi^2/{\rm dof}$            &
   $m_{\rm dec}$                 \\
\endfirsthead
   \caption
   {
      {\it (Continued)}
   } \\
   \hline
   \hline
   $K_{\rm sea}$                 &
   $K_{\rm val,1}$               &  
   $K_{\rm val,2}$               &
   $K_{\rm val,3}$               &
   [$t_{\rm min},t_{\rm max}$]   &
   $\chi^2/{\rm dof}$            &
   $m_{\Sigma}$                  &                  
   [$t_{\rm min},t_{\rm max}$]   &
   $\chi^2/{\rm dof}$            &
   $m_{\Lambda}$                 &
   [$t_{\rm min},t_{\rm max}$]   &
   $\chi^2/{\rm dof}$            &
   $m_{\rm dec}$                 \\
   \hline
\endhead
   \hline
\endfoot
\hline
0.1340 & 0.1340 & 0.1340 & 0.1340 &
         [10,24] & 1.1 & 1.1566(26) &
         -- & -- & -- &
         [10,24] & 1.4 & 1.2492(36) \\ 
0.1340 & 0.1343 & 0.1343 & 0.1343 &
         [10,24] & 1.1 & 1.1202(26) &
         -- & -- & -- &
         [10,24] & 1.5 & 1.2175(37) \\ 
0.1340 & 0.1346 & 0.1346 & 0.1346 &
         [10,24] & 1.1 & 1.0831(26) &
         -- & -- & -- &
         [10,24] & 1.6 & 1.1856(40) \\ 
0.1340 & 0.1350 & 0.1350 & 0.1350 &
         [10,24] & 1.1 & 1.0325(27) &
         -- & -- & -- &
         [10,24] & 1.7 & 1.1428(45) \\ 
0.1340 & 0.1355 & 0.1355 & 0.1355 &
         [10,24] & 1.1 & 0.9666(28) &
         -- & -- & -- &
         [10,24] & 1.7 & 1.0890(54) \\ 
0.1340 & 0.1358 & 0.1358 & 0.1358 &
         [10,24] & 0.9 & 0.9255(31) &
         -- & -- & -- &
         [10,24] & 1.6 & 1.0561(63) \\ 
0.1340 & 0.1340 & 0.1340 & 0.1343 &
         [10,24] & 1.1 & 1.1436(26) &
         [10,24] & 1.1 & 1.1453(26) &
         [10,24] & 1.3 & 1.2360(36) \\ 
0.1340 & 0.1340 & 0.1340 & 0.1346 &
         [10,24] & 1.1 & 1.1305(26) &
         [10,24] & 1.1 & 1.1340(26) &
         [10,24] & 1.4 & 1.2253(37) \\
\end{longtable}

\newpage
\begin{longtable}{c@{\extracolsep{\fill}}c|ccc|ccc}
\caption
{
   Meson masses on $12^3 \times 48$ lattice in quenched QCD.
}
\endhead
\hline
\hline
$K_{\rm val,1}$               &  
$K_{\rm val,2}$               &
[$t_{\rm min},t_{\rm max}$]   &
$\chi^2/{\rm dof}$            &
$m_{\rm PS}$                  &                  
[$t_{\rm min},t_{\rm max}$]   &
$\chi^2/{\rm dof}$            &
$m_{\rm V}$                  \\
\hline
0.13260 & 0.13260 &
         [10,24] & 0.6 & 0.4906(17) &
         [10,24] & 1.2 & 0.6079(43) \\ 
0.13290 & 0.13290 &
         [10,24] & 0.6 & 0.4616(18) &
         [10,24] & 1.3 & 0.5875(48) \\ 
0.13331 & 0.13331 &
         [10,24] & 0.7 & 0.4201(20) &
         [10,24] & 1.3 & 0.5606(58) \\ 
0.13384 & 0.13384 &
         [10,24] & 0.9 & 0.3620(24) &
         [10,24] & 1.2 & 0.5285(79) \\ 
0.13432 & 0.13432 &
         [10,24] & 1.0 & 0.3018(30) &
         [10,24] & 0.9 & 0.502(12) \\ 
0.13465 & 0.13465 &
         [10,24] & 0.9 & 0.2519(38) &
         [10,24] & 0.4 & 0.484(20) \\ 
0.13260 & 0.13290 &
         [10,24] & 0.6 & 0.4763(18) &
         [10,24] & 1.3 & 0.5977(45) \\
\end{longtable}

\begin{longtable}{c@{\extracolsep{\fill}}cc|ccc|ccc|ccc}
\caption
{
   Baryon masses on $12^3 \times 48$ lattice in quenched QCD.
}
\endhead
\hline
\hline
$K_{\rm val,1}$               &  
$K_{\rm val,2}$               &
$K_{\rm val,3}$               &
[$t_{\rm min},t_{\rm max}$]   &
$\chi^2/{\rm dof}$            &
$m_{\Sigma}$                  &                  
[$t_{\rm min},t_{\rm max}$]   &
$\chi^2/{\rm dof}$            &
$m_{\Lambda}$                 &
[$t_{\rm min},t_{\rm max}$]   &
$\chi^2/{\rm dof}$            &
$m_{\rm dec}$                 \\
\hline
0.13260 & 0.13260 & 0.13260 &
         [11,24] & 0.7 & 0.9394(76) &
         -- & -- & -- &
         [11,24] & 0.7 & 1.014(11) \\ 
0.13260 & 0.13260 & 0.13290 &
         [11,24] & 0.7 & 0.9264(79) &
         [11,24] & 0.7 & 0.9294(78) &
         [11,24] & 0.7 & 1.003(11) \\ 
0.13260 & 0.13260 & 0.13331 &
         [11,24] & 0.8 & 0.9083(86) &
         [11,24] & 0.8 & 0.9159(83) &
         [11,24] & 0.7 & 0.989(12) \\ 
0.13260 & 0.13260 & 0.13384 &
         [11,24] & 0.9 & 0.8841(98) &
         [11,24] & 0.9 & 0.8983(94) &
         [11,24] & 0.7 & 0.970(13) \\ 
0.13260 & 0.13260 & 0.13432 &
         [11,24] & 1.1 & 0.861(12) &
         [11,24] & 0.9 & 0.882(11) &
         [11,24] & 0.8 & 0.952(14) \\ 
0.13260 & 0.13260 & 0.13465 &
         [11,24] & 1.4 & 0.841(14) &
         [11,24] & 0.8 & 0.869(13) &
         [11,24] & 1.0 & 0.937(16) \\ 
0.13290 & 0.13290 & 0.13260 &
         [11,24] & 0.8 & 0.9180(81) &
         [11,24] & 0.7 & 0.9148(82) &
         [11,24] & 0.7 & 0.993(12) \\
\end{longtable}

\newpage
\begin{longtable}{c@{\extracolsep{\fill}}c|ccc|ccc}
\caption
{
   Meson masses on $16^3 \times 48$ lattice in quenched QCD.
}
\endhead
\hline
\hline
$K_{\rm val,1}$               &  
$K_{\rm val,2}$               &
[$t_{\rm min},t_{\rm max}$]   &
$\chi^2/{\rm dof}$            &
$m_{\rm PS}$                  &                  
[$t_{\rm min},t_{\rm max}$]   &
$\chi^2/{\rm dof}$            &
$m_{\rm V}$                  \\
\hline
0.13260 & 0.13260 &
         [10,24] & 0.6 & 0.48454(94) &
         [10,24] & 1.3 & 0.6024(20) \\ 
0.13290 & 0.13290 &
         [10,24] & 0.6 & 0.45480(97) &
         [10,24] & 1.3 & 0.5805(23) \\ 
0.13331 & 0.13331 &
         [10,24] & 0.6 & 0.4119(10) &
         [10,24] & 1.2 & 0.5506(27) \\ 
0.13384 & 0.13384 &
         [10,24] & 0.6 & 0.3511(12) &
         [10,24] & 1.1 & 0.5121(36) \\ 
0.13432 & 0.13432 &
         [10,24] & 0.9 & 0.2871(15) &
         [10,24] & 0.9 & 0.4778(55) \\ 
0.13465 & 0.13465 &
         [10,24] & 1.3 & 0.2332(23) &
         [10,24] & 0.6 & 0.4531(88) \\ 
0.13260 & 0.13290 &
         [10,24] & 0.6 & 0.46982(95) &
         [10,24] & 1.3 & 0.5915(21) \\
\end{longtable}

\begin{longtable}{c@{\extracolsep{\fill}}cc|ccc|ccc|ccc}
\caption
{
   Baryon masses on $16^3 \times 48$ lattice in quenched QCD.
}
\endhead
\hline
\hline
$K_{\rm val,1}$               &  
$K_{\rm val,2}$               &
$K_{\rm val,3}$               &
[$t_{\rm min},t_{\rm max}$]   &
$\chi^2/{\rm dof}$            &
$m_{\Sigma}$                  &                  
[$t_{\rm min},t_{\rm max}$]   &
$\chi^2/{\rm dof}$            &
$m_{\Lambda}$                 &
[$t_{\rm min},t_{\rm max}$]   &
$\chi^2/{\rm dof}$            &
$m_{\rm dec}$                 \\
\hline
0.13260 & 0.13260 & 0.13260 &
         [11,24] & 0.9 & 0.9081(44) &
         -- & -- & -- &
         [11,24] & 1.7 & 0.9865(55) \\ 
0.13260 & 0.13260 & 0.13290 &
         [11,24] & 0.9 & 0.8948(45) &
         [11,24] & 0.9 & 0.8969(45) &
         [11,24] & 1.7 & 0.9746(56) \\ 
0.13260 & 0.13260 & 0.13331 &
         [11,24] & 0.9 & 0.8762(46) &
         [11,24] & 0.9 & 0.8817(47) &
         [11,24] & 1.7 & 0.9602(59) \\ 
0.13260 & 0.13260 & 0.13384 &
         [11,24] & 0.8 & 0.8512(49) &
         [11,24] & 0.9 & 0.8622(51) &
         [11,24] & 1.7 & 0.9415(64) \\ 
0.13260 & 0.13260 & 0.13432 &
         [11,24] & 0.7 & 0.8273(56) &
         [11,24] & 0.9 & 0.8454(56) &
         [11,24] & 1.8 & 0.9247(72) \\ 
0.13260 & 0.13260 & 0.13465 &
         [11,24] & 0.6 & 0.8107(67) &
         [11,24] & 0.8 & 0.8351(65) &
         [11,24] & 1.6 & 0.9146(86) \\ 
0.13290 & 0.13290 & 0.13260 &
         [11,24] & 0.9 & 0.8847(46) &
         [11,24] & 0.9 & 0.8824(46) &
         [11,24] & 1.7 & 0.9642(58) \\
\end{longtable}

\newpage
\begin{longtable}{c@{\extracolsep{\fill}}c|ccc|ccc}
\caption
{
   Meson masses on $20^3 \times 48$ lattice in quenched QCD.
}
\endhead
\hline
\hline
$K_{\rm val,1}$               &  
$K_{\rm val,2}$               &
[$t_{\rm min},t_{\rm max}$]   &
$\chi^2/{\rm dof}$            &
$m_{\rm PS}$                  &                  
[$t_{\rm min},t_{\rm max}$]   &
$\chi^2/{\rm dof}$            &
$m_{\rm V}$                  \\
\hline
0.13260 & 0.13260 &
         [10,24] & 0.7 & 0.48376(66) &
         [10,24] & 1.0 & 0.6008(16) \\ 
0.13290 & 0.13290 &
         [10,24] & 0.6 & 0.45385(69) &
         [10,24] & 1.0 & 0.5790(18) \\ 
0.13331 & 0.13331 &
         [10,24] & 0.4 & 0.41066(74) &
         [10,24] & 1.0 & 0.5492(21) \\ 
0.13384 & 0.13384 &
         [10,24] & 0.5 & 0.34897(96) &
         [10,24] & 1.1 & 0.5117(29) \\ 
0.13432 & 0.13432 &
         [10,24] & 0.6 & 0.2845(11) &
         [10,24] & 0.9 & 0.4773(43) \\ 
0.13465 & 0.13465 &
         [10,24] & 0.9 & 0.2296(15) &
         [10,24] & 1.0 & 0.4566(74) \\ 
0.13260 & 0.13290 &
         [10,24] & 0.6 & 0.46895(67) &
         [10,24] & 1.0 & 0.5899(17) \\
\end{longtable}

\begin{longtable}{c@{\extracolsep{\fill}}cc|ccc|ccc|ccc}
\caption
{
   Baryon masses on $20^3 \times 48$ lattice in quenched QCD.
}
\endhead
\hline
\hline
$K_{\rm val,1}$               &  
$K_{\rm val,2}$               &
$K_{\rm val,3}$               &
[$t_{\rm min},t_{\rm max}$]   &
$\chi^2/{\rm dof}$            &
$m_{\Sigma}$                  &                  
[$t_{\rm min},t_{\rm max}$]   &
$\chi^2/{\rm dof}$            &
$m_{\Lambda}$                 &
[$t_{\rm min},t_{\rm max}$]   &
$\chi^2/{\rm dof}$            &
$m_{\rm dec}$                 \\
\hline
0.13260 & 0.13260 & 0.13260 &
         [11,24] & 1.1 & 0.9001(27) &
         -- & -- & -- &
         [11,24] & 0.7 & 0.9765(44) \\ 
0.13260 & 0.13260 & 0.13290 &
         [11,24] & 1.1 & 0.8866(27) &
         [11,24] & 1.1 & 0.8885(27) &
         [11,24] & 0.7 & 0.9643(45) \\ 
0.13260 & 0.13260 & 0.13331 &
         [11,24] & 1.1 & 0.8678(28) &
         [11,24] & 1.1 & 0.8727(28) &
         [11,24] & 0.7 & 0.9501(47) \\ 
0.13260 & 0.13260 & 0.13384 &
         [11,24] & 0.9 & 0.8429(31) &
         [11,24] & 1.1 & 0.8525(31) &
         [11,24] & 0.8 & 0.9323(51) \\ 
0.13260 & 0.13260 & 0.13432 &
         [11,24] & 0.8 & 0.8192(34) &
         [11,24] & 0.9 & 0.8342(35) &
         [11,24] & 1.0 & 0.9173(55) \\ 
0.13260 & 0.13260 & 0.13465 &
         [11,24] & 0.6 & 0.8047(39) &
         [11,24] & 0.8 & 0.8240(42) &
         [11,24] & 1.3 & 0.9087(63) \\ 
0.13290 & 0.13290 & 0.13260 &
         [11,24] & 1.1 & 0.8759(28) &
         [11,24] & 1.1 & 0.8739(28) &
         [11,24] & 0.7 & 0.9539(46) \\
\end{longtable}


\clearpage

\begin{figure}[htbp]
   \includegraphics[width=70mm]{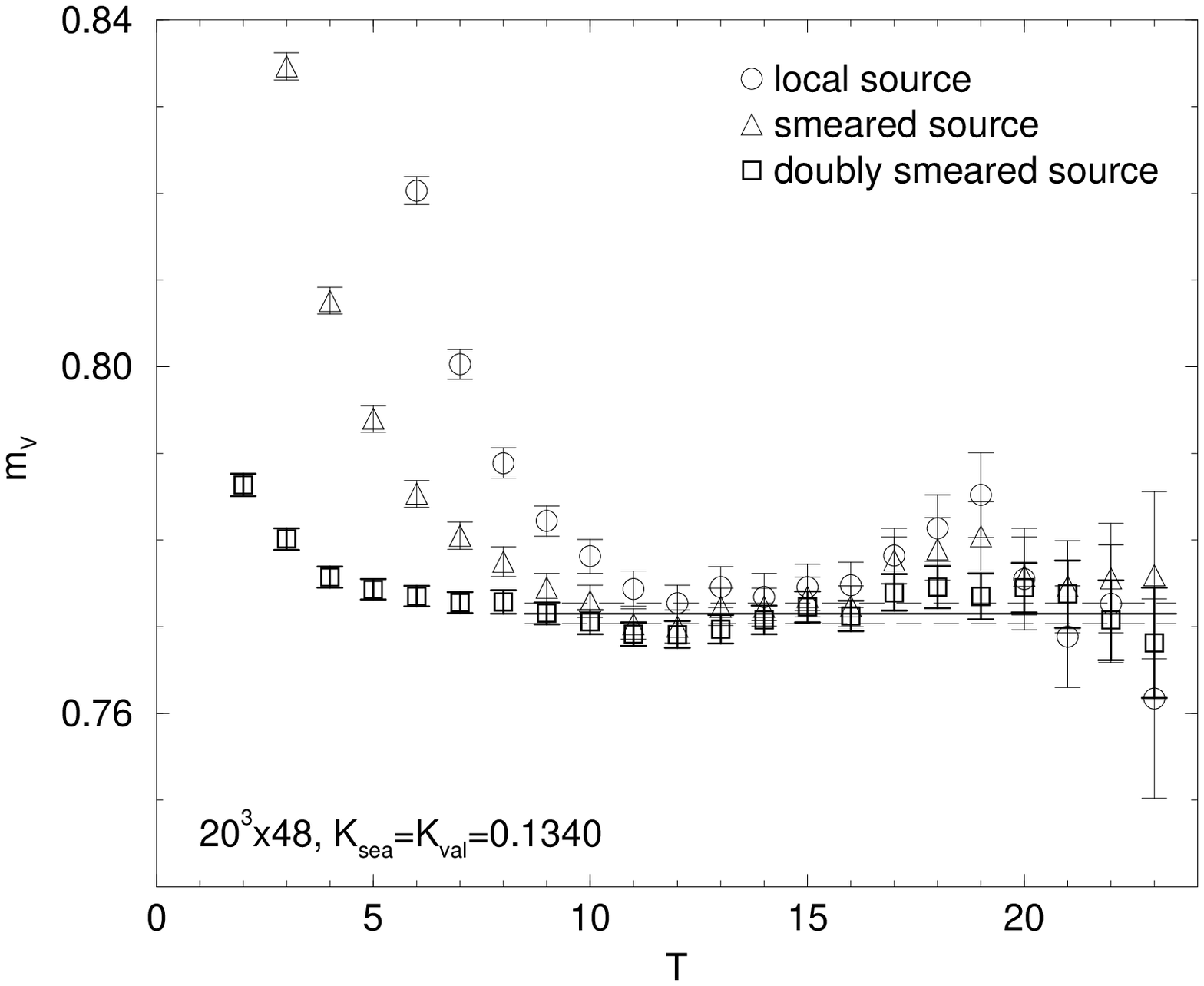}
   \includegraphics[width=70mm]{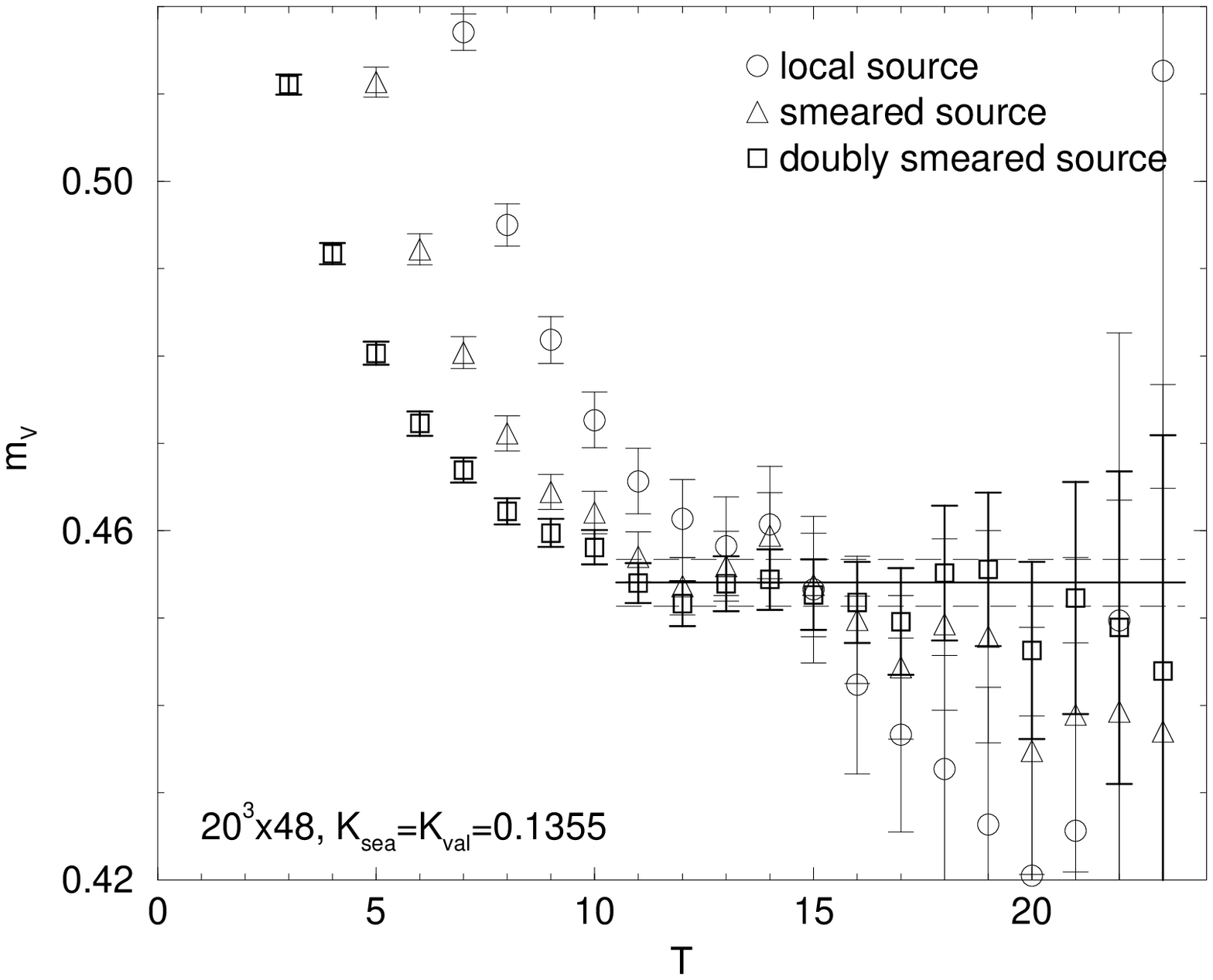}
   \caption
   {
      Effective mass of vector meson 
      at $K_{\rm sea}\!=\!K_{\rm val}\!=\!0.1340$ (left figure) 
      and 0.1355 (right figure) 
      on $20^3 \times 48$ lattice in full QCD.
      We use the local sink operator for all data.
   }
   \label{fig:meas:had:em_20x48_V}
\end{figure}

\begin{figure}[htbp]
   \includegraphics[width=70mm]{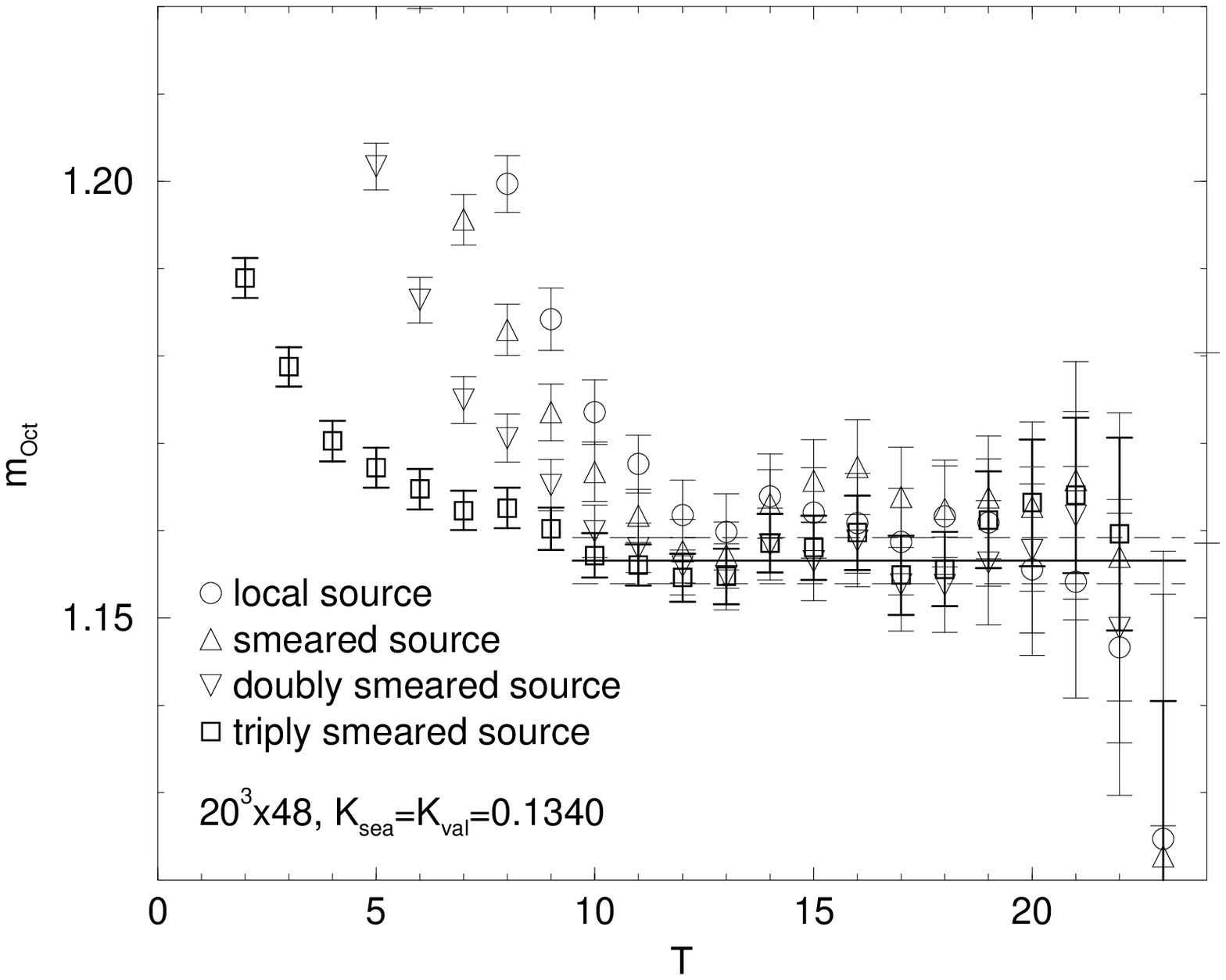}
   \includegraphics[width=70mm]{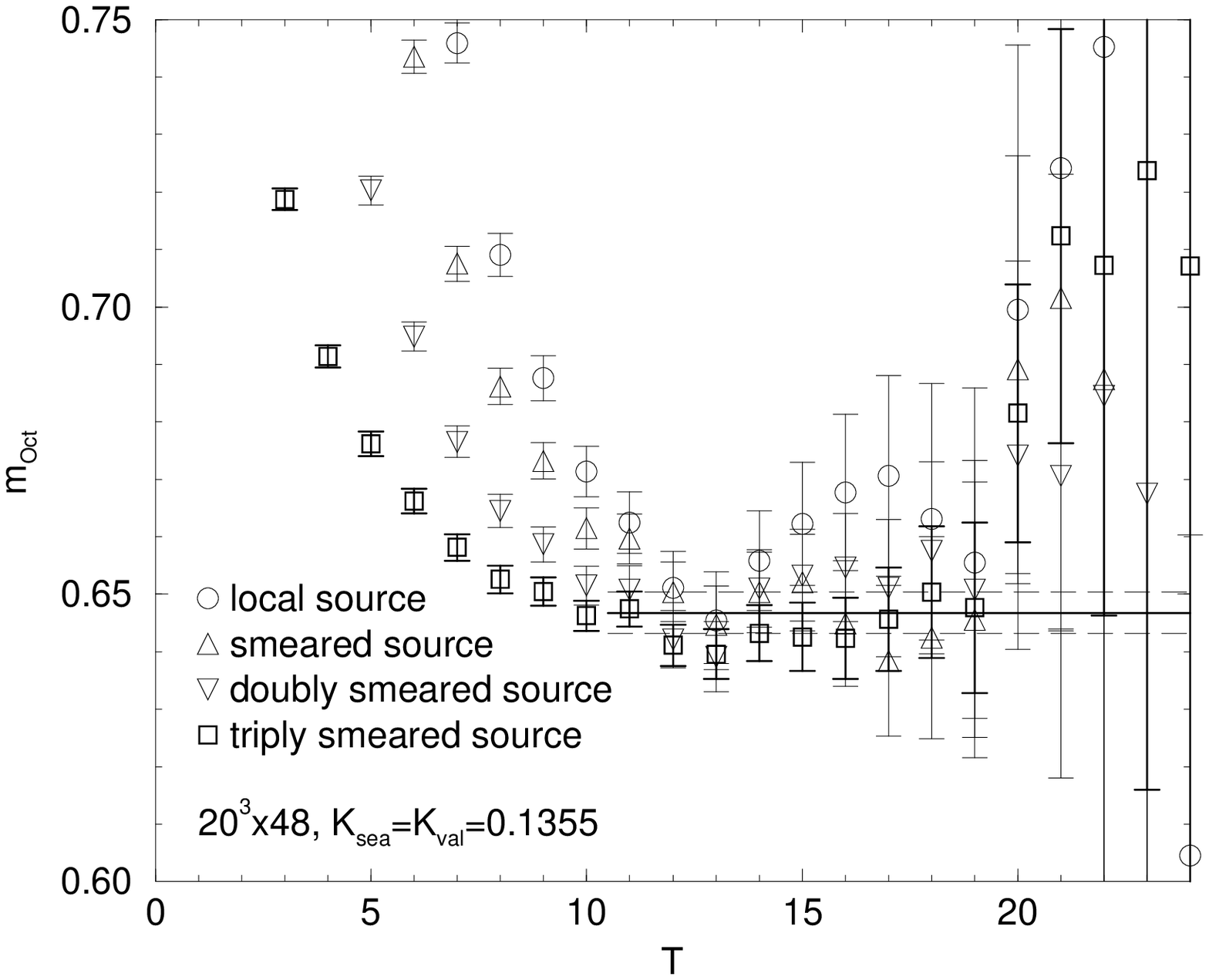}
   \caption
   {
      Effective mass of octet baryon 
      at $K_{\rm sea}\!=\!K_{\rm val}\!=\!0.1340$ (left figure) 
      and 0.1355 (right figure) 
      on $20^3 \times 48$ lattice in full QCD.
      We use the local sink operator for all data.
   }
   \label{fig:meas:had:em_20x48_Oct}
\end{figure}

%


\clearpage

\clearpage

\begin{thebibliography}{9}






\bibitem{Spectrum.review.lat00}
S.~Aoki, 
Nucl. Phys. B (Proc. Suppl.) {\bf 94}, 3 (2001).

\bibitem{KS-Spectrum.review.lat01}
D.~Toussaint, 
Nucl. Phys. B (Proc. Suppl.) {\bf 106}, 111 (2002).

\bibitem{Spectrum.review.lat01}
T.~Kaneko, 
Nucl. Phys. B (Proc. Suppl.) {\bf 106}, 133 (2002).

\bibitem{Spectrum.Nf0.CP-PACS}
S.Aoki {\it et al.} (CP-PACS Collaboration),
Phys. Rev. Lett. {\bf 84}, 238 (2000).

\bibitem{Spectrum.Nf2.SESAM}
N.~Eicker {\it et al.} (SESAM Collaboration),
Phys. Rev. D {\bf 59}, 014509 (1999).

\bibitem{Spectrum.Nf2.TchiL}
T.~Lippert, {\it et al.} (T$\chi$L Collaboration),
Nucl. Phys. B (Proc. Suppl.) {\bf 60A}, 311 (1998).

\bibitem{Spectrum.Nf2.SESAM-TchiL}
N.~Eicker, Th.~Lippert, B.~Orth and K.~Schilling
(SESAM-T$\chi$L Collaboration),
Nucl. Phys. B (Proc. Suppl.) {\bf 106}, 209 (2002).

\bibitem{Spectrum.Nf2.UKQCD.csw176}
C.R.~Allton {\it et al.} (UKQCD Collaboration),
Phys. Rev. D {\bf 60}, 034507 (1999).

\bibitem{Spectrum.Nf2.UKQCD}
C.R.~Allton {\it et al.} (UKQCD Collaboration),
Phys. Rev. D {\bf 65}, 054502 (2002).

\bibitem{mq.Nf2.CP-PACS}
A.~Ali Khan {\it et al.} (CP-PACS Collaboration),
Phys. Rev. Lett. {\bf 85}, 4674 (2000).

\bibitem{Spectrum.Nf2.CP-PACS}
A.~Ali Khan {\it et al.} (CP-PACS Collaboration),
Phys. Rev. D {\bf 65}, 054505 (2002).

\bibitem{Spectrum.Nf2.QCDSF}
H.~St\"uben, (QCDSF-UKQCD Collaboration),
Nucl. Phys. B (Proc. Suppl.) {\bf 94}, 273 (2001).

\bibitem{Spectrum.Nf2.MILC}
C.~Bernard {\it et al.} (MILC Collaboration),
Nucl. Phys. B (Proc. Suppl.) {\bf 73}, 198 (1999);
{\it ibid.} {\bf 60A}, 297 (1998) and 
references therein.


\bibitem{Spectrum.Nf3.MILC}
C.~Bernard {\it et al.} (MILC Collaboration),
Phys. Rev. D {\bf 64} 054506 (2001);
hep-lat/0208041.

\bibitem{Spectrum.Nf0.GF11}
F.~Butler, H.~Chen, J.~Sexton, A.~Vaccarino and D.~Weingarten,
Nucl. Phys. B {\bf 430}, 179 (1994).

\bibitem{RGaction}
Y.~Iwasaki, 
Nucl. Phys. B {\bf 258}, 141 (1985);
Univ. of Tsukuba report UTHEP-118 (1983),
unpublished.

\bibitem{SWaction}
B.~Sheikholeslami and R.~Wohlert,
Nucl. Phys. B {\bf 259}, 572 (1985).

\bibitem{tadpole_improvement}
G.P.~Lepage and P.B.~Mackenzie,
Phys. Rev. D {\bf 48}, 2250 (1993).


\bibitem{NPimprovement}
M.~L\"uscher, S.~Sint, R.~Sommer and P.~Weisz, 
Nucl. Phys. B {\bf 478}, 365 (1996).

\bibitem{Spectrum.Nf2.JLQCD.lat00}
S.~Aoki {\it et al.} (JLQCD Collaboration),
Nucl. Phys. B (Proc. Suppl.) {\bf 94}, 233 (2001).

\bibitem{Spectrum.Nf2.JLQCD.lat01}
S.~Aoki {\it et al.} (JLQCD Collaboration),
Nucl. Phys. B (Proc. Suppl.) {\bf 106}, 224 (2002).

\bibitem{Spectrum.Nf2.JLQCD.lat02.Kaneko}
T.~Kaneko {\it et al.} (JLQCD Collaboration),
hep-lat/0209057.

\bibitem{Spectrum.Nf2.JLQCD.lat02.Hashimoto}
S.~Hashimoto {\it et al.} (JLQCD Collaboration),
hep-lat/0209091.

\bibitem{FSE.Z3}
S.~Aoki {\it et al.},
Phys. Rev. D {\bf 50}, 486 (1994).

\bibitem{Even-Odd2}
K.~Jansen and C.~Liu,
Comput. Phys. Commun. {\bf 99}, 221 (1997).

\bibitem{PHMC.JLQCD}
S.~Aoki {\it et al.} (JLQCD Collaboration),
Phys. Rev. D {\bf 65}, 094507 (2002).






\bibitem{NPcsw.Nf2.ALPHA}
K.~Jansen and R.~Sommer (ALPHA Collaboration), 
Nucl. Phys. B {\bf 530}, 185 (1998).

\bibitem{NPcsw.Nf2.JLQCD}
JLQCD Collaboration, 
in preparation.


\bibitem{r0}
R.~Sommer, 
Nucl. Phys. B {\bf 411}, 839 (1994).

\bibitem{HMC.Duane}
S.~Duane, A.D.~Kennedy, B.J.~Pendleton, and D.~Roweth,
Phys. Lett. B {\bf 195}, 216 (1987).

\bibitem{HMC.Gottlieb}
S.~Gottlieb, W.~Liu, D.~Toussaint, R.L.~Renken and R.L.~Sugar,
Phys. Rev. D {\bf 35}, 2531 (1987).

\bibitem{Even-Odd.D}
T.A.~DeGrand, 
Comput. Phys. Commun. {\bf 52}, 161 (1988).

\bibitem{Even-Odd1}
X-Q.~Luo, 
Comput. Phys. Commun. {\bf 94}, 119 (1996).


\bibitem{BiCGStab}
H.~van der Vorst, 
SIAM J. Sc. Stat. Comp. {\bf 13}, 631 (1992).

\bibitem{NPcsw.Nf0.ALPHA}
M.~L\"uscher, S.~Sint, R.~Sommer, P.~Weisz and U.~Wolff,
Nucl. Phys. B {\bf 491}, 323 (1997).



\bibitem{Potential.Nf0.Bali}
G.S.~Bali and K.~Schilling, 
Phys. Rev. D {\bf 46}, 2636 1992.

\bibitem{deltaV}
C.B.~Lang and C.~Rebbi,
Phys. Lett. B {\bf 115}, 137 (1982).



\bibitem{FSE.expL}
M.~L\"uscher, 
Commun. Math. Phys. {\bf 104}, 177 (1986).

\bibitem{FSE.Vinv}
M.~Fukugita {\it et al.}, 
Phys. Lett. B {\bf 294}, 380 (1992);
Phys. Rev. D {\bf 47}, 4739 (1993).
 




\bibitem{ChPT}
J.~Gasser and H.~Leutwyler,
Annals Phys. {\bf 158}, 142 (1984);
Nucl. Phys. B {\bf 250}, 465 (1985).


\bibitem{PQChPT}
S.R.~Sharpe, 
Phys. Rev. D {\bf 56}, 7052 (1997).

\bibitem{ChPT_test.Nf2.JLQCD}
JLQCD Collaboration, 
in preparation.


\bibitem{ChPT.octet}
V.~Bernard, N.~Kaiser and U.G.~Meissner,
Z. Phys. C {\bf 60}, 111 (1993).



\bibitem{J}
P.~Lacock and C.~Michael (UKQCD Collaboration), 
Phys. Rev. D {\bf 52}, 5213 (1995).



\bibitem{1loop_cSWcA.ALPHA}
M.~L\"uscher and P.~Weisz,
Nucl. Phys. B {\bf 479}, 429 (1996).

\bibitem{1loop_Z.Roma}
E.~Gabrielli, G.~Martinelli, C.~Pittori, 
G.~Heatlie and C.T.~Sachrajda,
Nucl. Phys. B {\bf 362}, 475 (1991).

\bibitem{1loop_b.ALPHA}
S.~Sint, P.~Weisz,
Nucl. Phys. B {\bf 502}, 251 (1997).

\bibitem{1loop_Zb.Tsukuba}
S.~Aoki, K.~Nagai, Y.~Taniguchi and A.~Ukawa,
Phys. Rev. D {\bf 58}, 074505 (1998).

\bibitem{1loop_Zbc.Tsukuba}
Y.~Taniguchi and A.~Ukawa,
Phys. Rev. D {\bf 58}, 114503 (1998).

\bibitem{1loop_bc.Aoki}
S.~Aoki, R.~Frezzotti, P.~Weisz,
Nucl. Phys. B {\bf 540}, 501 (1999).


\bibitem{alpha_P.JLQCD}
K-I.~Ishikawa and S.~Hashimoto, private notes (2001).

\bibitem{alpha_P}
G.S.~Bali and P.~Boyle, hep-lat/0210033.

\bibitem{q_star}
J.~Harada, S.~Hashimoto, A.S.~Kronfeld and T.~Onogi,
hep-lat/0208004.


\bibitem{NP_Zbc.LANL}
T.~Bhattacharya, R.~Gupta, W.~Lee, S.~Sharpe,
Phys. Rev. D {\bf 63}, 074505 (2001);
Nucl. Phys. B (Proc.Suppl.) {\bf 106}, 789 (2002).






\bibitem{4loop.running.1}
K.G.~Chetyrkin, 
Phys. Lett. B404, 161 (1997).

\bibitem{4loop.running.2}
J.A.M.~Vermaseren, S.A.~Larin and T.~ van Ritbergen,
Phys. Lett. B405, 327 (1997).

\bibitem{NP_ZA.ALPHA}
M.~L\"uscher, S.~Sint, R.~Sommer and H.~Wittig,
Nucl. Phys. B {\bf 491}, 344 (1997).

\bibitem{NP_ZP.ALPHA}
S.~Capitani, M.~L\"uscher, R.~Sommer and H.~Wittig
(ALPHA Collaboration),
Nucl. Phys. B {\bf 544}, 669 (1999).

\bibitem{NP_bA-bP.ALPHA}
M.~Guagnelli {it et al} (ALPHA Collaboration), 
Nucl. Phys. B {\bf 595}, 44 (2001).

\bibitem{mq.Nf2.QCDSF}
D.Pleiter (QCDSF-UKQCD Collaboration),
Nucl. Phys. B (Proc. Suppl.) {\bf 94}, 265 (2001).

\bibitem{mq.ChPT}
H.~Leutwyler,
Phys. Lett. B {\bf 378}, 313 (1996).

\bibitem{scaling.Nf0.UKQCD}
K.C.~Bowler {\it et al.} (UKQCD Collaboration),
Phys. Rev. D {\bf 62}, 054506 (2000).




\end{thebibliography}
\end{document}